\def\hb{H$\beta$\/}
\def\oiii{[\ion{O}{III}]$\lambda\lambda$4959,5007\/}
\def\oiiionly{[\ion{O}{III}]\/}
\def\oiiiseven{[\ion{O}{III}]$\lambda$5007\/}
\def\civ{\ion{C}{IV}$\lambda$1549\/}
\def\siv{\ion{Si}{IV}$\lambda$1397\/}
\def\oiv{\ion{O}{IV}]$\lambda$1402\/}
\def\aliii{\ion{Al}{III}$\lambda$1860\/}
\def\heiiuv{\ion{He}{II}$\lambda$1640\/}
\def\feii{\ion{Fe}{II}\/}
\def\kms{km s$^{-1}$\/}
\def\feiii{\ion{Fe}{III}$\lambda$1914\/}
\def\ciii{\ion{C}{III}]$\lambda$1909\/}
\def\rfe{$R_{\ion{Fe}{II}}$\/}
\def\mbh{$M_{\rm BH}$\/}
\def\lledd{$L/L_{\rm Edd}$\/}
\def\civonly{\ion{C}{IV}\/}
\def\hbbc{H$\beta_{\rm{BC}}$\/}

\documentclass{aa}
\usepackage{ulem}
\usepackage{graphicx}
\usepackage{txfonts}
\usepackage{xcolor} 
\usepackage{rotating}
\usepackage{pdflscape}
\usepackage{multirow}

\begin{document} 
\defcitealias{Sulentic_2007}{S07}
\defcitealias{Marziani_2009}{M09}
\defcitealias{sulentic_2017}{S17}
\defcitealias{Vestergaard_2006}{VP06}
\defcitealias{Marziani_2019}{M19}
\defcitealias{Marziani_2003}{M03}

   \title{High-redshift quasars along the Main Sequence\thanks{Based on observations collected at ESO under programmes 083.B-0273(A) and 085.B-0162(A).}}

 %  \subtitle{I. Overviewing the $\kappa$-mechanism}

   \author{A. Deconto-Machado,
          \inst{1}
          %\fnmsep\thanks{adeconto@iaa.es}
          A. del Olmo Orozco \inst{1}
          \and
          P. Marziani \inst{2} 
          \and J. Perea \inst{1}
          \and G. M. Stirpe \inst{3}}

   \institute{Instituto de Astrofísica de Andalucía, IAA-CSIC, E-18008, Granada, Spain\\ \email{adeconto@iaa.es, chony@iaa.es, jaime@iaa.es}
   \and
  INAF, Osservatorio Astronomico di Padova, IT 35122, Padova, Italy\\ \email{paola.marziani@inaf.it}
  \and INAF, Osservatorio di Astrofisica e Scienza dello Spazio,
via Gobetti 93/3, 40129, Bologna, Italy\\ \email{giovanna.stirpe@inaf.it}
            }

   \date{Received \ldots; accepted \ldots}

% \abstract{}{}{}{}{} 
% 5 {} token are mandatory
 
  \abstract
   {The 4-Dimensional Eigenvector 1 empirical formalism (4DE1) and its Main Sequence (MS) for quasars emerged as a powerful tool to organise the quasar diversity since several key observational measures and physical parameters are systematically changing along it.}
   {The trends of the 4DE1 are very well established to explain all the diversity seen in low-redshift quasar samples. Nevertheless, the situation is by far less clear when dealing with high-luminosity and high-redshift sources. We aim to evaluate the behaviour of our 22 high-redshift ($2.2 \le z \le 3.7$) and high-luminosity ($47.39 \le L_{\textrm{bol}} \le 48.36$) quasars in the context of the 4DE1.}
   {Our approach involves studying quasar physics through spectroscopic exploration of UV and optical emission line diagnostics. We are using new  observations from the ISAAC instrument at ESO-VLT and mainly from the SDSS  to cover the optical and the UV rest-frames, respectively. 
   %Fluxes at radio frequencies are collected from FIRST and NVSS. 
   Emission lines are characterised both through a quantitative parametrisation of the line profiles, %by measuring centroid and widths at different fractional intensities
    and by decomposing the emission line profiles using multicomponent fitting routines.}
   {We provide spectrophotometric properties and line profile measurements for \hb{}+\oiii{}, as well as for \siv{}+\ion{O}{IV}]$\lambda$1402, \civ{}+\ion{He}{II}$\lambda$1640, 1900 \r{A} blend (including \aliii{}, \ion{Si}{III}]$\lambda$1892, and \ion{C}{III}]$\lambda$1909). Six out of the 22 objects present a significant blueshifted component on the H$\beta$ profile, and in 14/22  cases an H$\beta$ outflowing component associated to \oiiionly{} is detected. 
   The majority of \oiii{} emission line profiles show blueshifted  velocities larger than 250 km s$^{-1}$. 
   %\textbf{\textcolor{red}{Almost all of the \oiii{} profiles present strong and blueshifted semi-broad components, with the majority of the [OIII] emission lines showing peak velocity blue shifts larger than 250 km/s.} 
   We reveal extremely broad \oiii{} emission that is comparable to the width of \hb\ broad profile in some highly accreting quasars. \oiii{} and \civ{} blueshifts show very high amplitudes and a high degree of correlation.
   Line width and shift are correlated for both \oiii\ and \civ, suggesting that emission from outflowing gas is providing a substantial broadening to both lines.  Otherwise,   the links between \civ{} centroid velocity at half intensity ($c$(1/2)), Eddington ratio (L/L$_{\textrm{Edd}}$), and bolometric luminosity are found to be in agreement with previous studies of high-luminosity quasars.  
 
   %It seems that the Pop. B H$\beta$ profiles are wider than for Pop. A, while for \oiii{} emission lines it is the opposite. 
   
   %Some similarities are found for our sources between the optical and UV range: \oiii{} and \civ{} seem to follow a very similar relation between their respective FWHM and $c$(1/2) values.   
   %We have performed black hole mass estimations using the \civ{}, \aliii{}, and \hb{} emission lines, which indicate that the highest masses are found in Pop. B sources while the fastest accretors are Pop A. %A. Four out of the 22 sources are found to be radio-loud quasars and some UV and optical spectral characteristics are shared between them.
   }
   {Our analysis suggests that the behaviour of quasars of very high luminosity all along the main sequence is strongly affected by powerful outflows involving a broad range of spatial scales. 
      %FWHM and equivalent width are found to correlate with blueshift for both \oiii{} and \civ{} emission lines, in agreement with previous results. 
The main sequence correlations remain valid 
at high redshift and high luminosity even if a systematic increase in line width is observed.    Scaling laws based on UV \ion{Al}{III}$\lambda1860$ and \hb\ emission lines are equally reliable estimators of $M_{\rm{BH}}$. 
   %with respect to the ones derived from H$\beta$. 
   }  
   %  and the location of our sample in it 
   %The relation between profile features and the black hole mass ($M_{\rm{BH}}$) and L/L$_{\rm{Edd}}$ is confirmed.  
   %Our analysis  support that scaling laws based implies a displacement to higher values of both FWHM and $R_{\rm{\ion{Fe}{II}}}$.High-redshift quasars do present some differences that are uncommon at low $z$ in both optical and UV spectral ranges.  on the Main Sequence parameter space as well as on the 4DE1 in general is very peculiar and demands more samples dedicated on this kind of sources.}
   {}


   \maketitle
%
%-------------------------------------------------------------------
%\tableofcontents

\section{Introduction}
\par The nature of the many differences seen in quasar spectra has been a topic of interest for astronomers for many years and it is still a topic under discussion. One of the first successful attempts to define the systematic trends of quasar spectra was carried out by \cite{boroson_1992}. These authors have organised the relations between the optical and radio spectral ranges into an Eingenvector 1 scheme from studying the 80 quasars from the Palomar-Green sample \citep{Schmidt_1983} using principal component analysis (PCA). This scheme considers mainly the anticorrelation seen between the optical \ion{Fe}{II} strength and peak intensity of the \oiiiseven{} emission line, and the full width at half-maximum (FWHM) of the broad component of H$\beta$ (H$\beta_{\rm BC}$, typically $\gtrsim 4000$ km s$^{-1}$). A more overarching possibility for the arrangement of the individual characteristics found in the quasar spectra has been suggested by \cite{Sulentic_2000} and takes into account several observational measures in the optical, UV, and X-ray spectral ranges as well as physical parameters such as outflow relevance and accretion mode. According to those authors, one can organise the quasar diversity into a fourth dimensional correlation space known as Eigenvector 1 (4DE1).

\par One of the four key parameters considered by the 4DE1 is the FWHM of the Hydrogen \hb{} broad component. Since the emission of Balmer lines like \hb{} are thought to be arising from the quasar Broad Line Region (BLR), this parameter can be used to measure black hole mass assuming gas to be virialised \citep[and references therein]{collin_1988,small_1992,marziani_1996,McLure_2002,McLure_2004, Sulentic_2006, Assef_2011, shen_2013,Gaskell_2018}. 

The 4DE1 also considers as a parameter the ratio between the intensities of the blend of \ion{Fe}{II} emission lines at 4570\r{A} and \hb{}\  ($R_{\ion{Fe}{II}}=I(\ion{Fe}{II}\lambda4570)/I(\rm H\beta)$), which can be used for the estimation of physical properties of the BLR such as the ionisation state, the electron density, and the column density of the ionised gas \citep{ferland_2009,panda_2020}. %The $R_{\ion{Fe}{II}}$ parameter is also related to many of the optical and UV spectral line measures \citep{kovacevic_2010,Marziani_2010,shapovalova_2012}. 

\par The other parameters of the 4DE1 are the blueshifts of high-ionisation lines (HIL) with respect to the quasar systemic redshift and the soft X-ray photon index ($\Gamma_{\rm soft}$). HIL like the \civ{} emission line for instance are considered a strong diagnostic of outflows and there is some evidence that the optical and UV properties are related \citep{Bachev_2004, Du_2016, sniegowska_2020}. Similarly, the equivalent width of the optical \ion{Fe}{II} contribution is seen as a measure of the thermal emission from the accretion disc \citep{Singh_1985, Walter_1993}. 
The $R_{\ion{Fe}{II}}$, the soft X-ray photon index and the line width of \hb{} are also significantly correlated among themselves \citep{Wang_1996, Boller_1996,sulenticetal00c}. 

\par Exploration of the distribution of low-redshift ($z\!\lesssim\!0.8$) quasars in the optical plane of the 4DE1, defined by FWHM(H$\beta_{\rm BC}$) \textit{vs.} $R_{\ion{Fe}{II}}$ and called Main Sequence (MS) of quasars, gave rise to the identification of two main populations with very significant differences between their spectra \citep{zamfir_2010}. Population A quasars show low FWHM (usually $\le 4000$ km s$^{-1}$) and a wide range of $R_{\ion{Fe}{II}}$, while Population B sources present a very wide range of FWHM(H$\beta_{\rm BC}$) but usually small $R_{\ion{Fe}{II}}$ ($\le 1.$). \cite{Marziani2018} summarise more than 15 years of discussion on the empirical parametrisation of the quasar properties and their trends as well as on how physical parameters related to the accretion rate and feedback seem to be changing along the MS, going from quasars with low \ion{Fe}{II} emission and high black hole mass ($\gtrsim 10^9$ M$_{\odot}$, Pop. B) to the extreme Pop. A, xA \citep{martinezaldama_2018}, with strong \ion{Fe}{II} emission ($R_{\rm{\ion{Fe}{II}}}>1$) and strong blueshifts in High-Ionisation Lines (HIL) as \civ{} indicating strong wind effects on the quasars \citep[``wind-dominated'' sources, ][]{richards_2011}.

\par Eddington ratio together with the orientation effect are seen as key properties in the MS context \citep{Sulentic_2000, marziani_2001, boroson_2002, shen_2014, Sun_2015}.   The $R_{\ion{Fe}{II}}$ parameter together with the H$\beta_{\rm BC}$ FWHM can be associated with $L/L_{\rm Edd}$ (\citealt{marziani_2001, Panda_2019}). Low values of the \lledd\ (typically $\lesssim$ 0.2) are usually found in Pop. B sources, while Pop. A sources are found to be high accretors (in extreme cases reaching \lledd\ $\gtrsim 1$).  Consequently, sources with the highest $R_{\ion{Fe}{II}}$ are thought to be the ones with the highest Eddington ratio. %, as is the case of PHL 1092 which is one of the narrow-line Seyfert 1s (NLSy1s) galaxy that presents the strongest \ion{Fe}{II} emission detected until now, with a $R_{\ion{Fe}{II}}=2.58$ and $L/L_{\rm Edd}=1.24$ \citep{marinello_2020}. 
Consistent results are also reported by \cite{Du_2016}, which denote the strong correlation found between Eddington ratio, shape of the broad profile of the H$\beta$ emission line, and the flux ratio $I$(H$\beta$)/$I$(\ion{Fe}{II}) as the fundamental plane of accreting black holes. According to those authors, the shape of both Lorentzian and Gaussian profiles may reveal details on the BLR dynamic and may be different depending on the Eddington ratio \citep[c.f. ][]{Collin_2006,KOllatschny_2011}. Regarding the width of the emission line profiles, the broadest sources that are more Gaussian-like due to a redward asymmetry can be seen as the sum of two Gaussian (usually a BC and a very broad component, VBC) and the sources which present a VBC, usually with FWHM $> 7000$ km s$^{-1}$, are the ones which show lower Eddington ratio \citep{Marziani_2003, Marziani_2019}. However, the line width  is believed to be highly influenced by source orientation and the Eddington ratio \citep{marziani_2001,McLure_2002,Zamfir_2008,Panda_2019}. Since it is likely that \hb{} lines are emitted by flattened systems, orientation may affect the FWHM(H$\beta$), going from broader (bigger $\theta$, with $\theta$ indicating the inclination of the source with respect to the line-of-sight) to narrower (smaller $\theta$) profiles.

%\par  
%\sout{A more detailed review about accretion rates in supermassive black hole is given by \cite{marziani_2006}. While the relations for the black hole mass estimation are well established for low-redshift sources, the situation is not so clear for the high-redshift range. A recent review about the past, present, and future of scaling relations of galaxies and AGN is given by \cite{Donofrio_2021}. }

Of special relevance is also the issue of sources that are radio-loud. Only about 10\% of quasars are strong emitters in radio. \cite{zamfir_2010} have analysed $\sim 470$ low-redshift (z\,<\,0.8) quasars from the SDSS DR5 \citep{mccarthy_2007} and found that in general the radio-loud sources are located in the Pop. B domain of the MS, while the radio-quiet are found in both populations. The location of the radio-loud (RL) quasars seems to indicate different properties with respect to a large fraction of the radio-quiet (RQ) sources. However, the fact that the RQs are distributed in both populations A and B complicates the interpretation. For instance, radio-loud and a large fraction of radio-quiet quasars both present strong asymmetries towards red wavelengths in the emission line profiles \citep{marziani_1996,punsly_2010}. The paucity  of radio-loud Population A sources at low-$z$\ implies that  the Eddington ratio and the black hole mass distributions are different for radio-quiet and radio-loud sources matched in redshift and luminosity \citep{woourry02,Marziani_2003,fraix-burnetetal17}. In the two cases, the radio-quiet quasars are the ones that usually present smaller masses and larger Eddington ratio. %Details that justify these considerations are explained in recent reviews about the MS \citep[e.g.][]{marziani_2018,marziani_2021}. 
This is not necessarily true for sources at high-redshift \citep{sikoraetal07, Marinello_2020_2, Diana_2022}.

%Recently a relatively high fraction of intermediate radio-emitters has been found in the extreme Pop. A quasars \citep{delolmo_2021}. \cite{Ganci_2019} have shown that Pop. A spectral bins present radio powers that are associated to emission mechanisms due to star formation processes and extreme Pop. A present high values of both radio power and star formation rate.

%However,  the behaviour of sources at high-luminosity in the high-redshift range is still unclear.   

\par The 4DE1 formalism and especially the MS represent the most effective way to distinguish quasars according to their BLR structural and kinematic differences. It has been extensively analysed in samples at $z<0.8$ \citep[e.g.,][]{zamfir_2010, negrete_2018}. Trends between the optical plane of the 4DE1 and the $L/L_{\rm Edd}$ (for instance, sources with strong $L/L_{\rm Edd}$ are usually found to have strong \rfe{}) are also seen in high-luminosity high- and intermediate-redshift sources \citep[$z\gtrsim 2$ and $L\gtrsim 47$,][]{ Yuan_2003,Netzer_2004, Sulentic_2004}. However, high-$z$ quasar samples that have been studied  including NIR observations of the \hb\ spectral range  are relatively few \citep[e.g.,][]{mcintoshetal99,Capellupo_2015,coatman_2016,bischettietal17,vietrietal18,vietrietal20,Matthews_2021}.  One of the studies of high-$z$ quasars under this context has been performed by \citet[][hereafter \citetalias{Marziani_2009}]{Marziani_2009}. They analyse the optical region of 53 Hamburg-ESO sources in a redshift range $z\approx 0.9 - 3.0$ using VLT ISAAC spectra \citep[][\citetalias{Marziani_2009}]{Sulentic_2004,Sulentic_2006}. Additional UV spectra were obtained for some of these sources and the results are reported by \citet[][hereafter S17]{sulentic_2017}. The authors found that both Pop. A and Pop. B quasars present evidences of significant outflows at high redshift while at low $z$ only Pop. A sources tend to show strong contribution of outflowing gas. Extreme Pop. A quasars (xA) in a redshift range of $z=2.0-2.9$ and with an averaged bolometric luminosity of $\log L \sim 47$ [erg s$^{-1}$] have been analysed in details on the UV region by \cite{martinezaldama_2018} using GTC spectra. These authors found that the xA sources at high $z$\ share the same characteristics of the sources at low redshift, albeit with the higher outflow velocities (reaching values of $\sim 4000$ km s$^{-1}$).

\par In this paper new observations from VLT/ISAAC are reported for 22 high-redshift and high-luminosity quasars and the data analysis of the UV and optical regions along the Main Sequence is performed and discussed. Our goal is to improve the sampling of the MS and the understanding of high-$z$, high-luminosity quasars. In order to do so, we take advantage of previous high- and low-$z$ samples and perform a comparison between the different data under the 4DE1 context. Details on the sample and on the observations are presented in Sect. \S \ref{sample} and \S \ref{obs}. The procedures and the approach followed during the line decomposition are presented in detail on Sect. \S \ref{analysis}. Results on the complete analysis of both optical (\hb{}+\oiii{}) and UV (\ion{Si}{IV}$\lambda$1392, \civ{}, and the 1900 \r{A} blend) regions are reported in Sect. \S \ref{results} and additional discussions are provided in Sect. \S \ref{discussion}. In Sect. \S \ref{conclusions} we list the main conclusions of our work.

%--------------------------------------------------- One column table
   \begin{table*}[t!]
   \centering
      \caption[]{Source identification.}
      \
         \label{tab:source_id}
        \resizebox{\linewidth}{!}{
            \begin{tabular}{@{}lccccccccccc@{}}
            \hline
            \hline
            \noalign{\smallskip}
            Source &  RA (J2000) & DEC (J2000) & z & $\delta$z & $m^{(b)}$ & Band & $M_{\mathrm{i}}$ & $m_{\mathrm{VCV10}}$ & $f_{\textrm{Radio}}$ (mJy)$^{(h)}$ & Survey & Radio Class.\\
            (1) & (2) & (3) & (4) & (5) & (6) & (7) & (8) & (9) & (10) & (11) & (12)\\
            \noalign{\smallskip}
            \hline
            \noalign{\smallskip}
            HE 0001-2340 & 00 03 44.95 & -23 23 54.7 & 2.2651$^{(a)}$ & 0.0036 & 14.78 & H & -29.68 & 16.70$^{(c)}$  & $<2.70$ & NVSS & RQ\\
            $[\rm HB89]$ 0029+073 & 00 32 18.37 & +07 38 32.4 & 3.2798$^{(a)}$ & 0.0055 & 15.12 & K & -29.26 & 17.44$^{(d)}$ & $<0.86$ & FIRST & RQ\\
            CTQ 0408 & 00 41 31.49 &  -49 36 12.4 & 3.2540 & 0.0048 &14.02 & K & -29.86 & 16.10$^{(d)}$ & 7.26 & SUMSS & RI \\
            SDSSJ005700.18+143737.7 & 00 57 00.19 & +14 37 37.7 & 2.6638 & 0.0023 & 15.70 & H & -29.06  &  17.96$^{(e)}$ & $<2.70$ & NVSS & RQ\\
            H 0055-2659 & 00 57 57.92 & -26 43 14.1 & 3.6599 & 0.0062 & 15.50 & K & -30.74 & 17.47$^{(f)}$ & $<2.70$ & NVSS & RQ\\
            SDSSJ114358.52+052444.9 & 11 43 58.52 & +05 24 44.9 & 2.5703 & 0.0038 & 15.54 & H & -29.47 & 17.27$^{(e)}$ & $<0.96$ & FIRST & RQ \\
            SDSSJ115954.33+201921.1 & 11 59 54.33 & +20 19 21.1 & 3.4277 & 0.0068 & 15.13 & K & -29.92 & 17.92$^{(e)}$ & $<0.97$ & FIRST & RQ\\
            SDSSJ120147.90+120630.2 & 12 01 47.91 & +12 06 30.2 & 3.5136 & 0.0063  & 14.60 & K & -29.76 & 18.16$^{(e)}$ & $<0.98$ & FIRST & RQ\\
            SDSSJ132012.33+142037.1 & 13 20 12.34 & +14 20 37.1 & 2.5356 & 0.0023 & 15.52 & H & -28.87 & 17.82$^{(e)}$ & $<0.98$ & FIRST & RQ\\
            SDSSJ135831.78+050522.8 & 13 58 31.79 & +05 05 22.7 & 2.4627 & 0.0020 & 15.40 & H & -29.21 & 17.33$^{(e)}$ & $<0.94$ & FIRST & RQ\\
            Q 1410+096 & 14 13 21.05 & +09 22 04.8 & 3.3240 & 0.0029 & 14.83 & K & -29.44 & 17.80$^{(g)}$  & $<0.99$ & FIRST & RQ\\
            SDSSJ141546.24+112943.4 & 14 15 46.23 & +11 29 43.4 & 2.5531 & 0.0043 & 14.53 & H & -29.40 &  17.23$^{(e)}$& 7.80 & NVSS & RI\\
            B1422+231 & 14 24 38.10 & +22 56 01.0 & 3.6287 & 0.0031 & 12.66 & K & -29.85 & 15.84$^{(e)}$ & 273.42 & FIRST & RL\\
            SDSSJ153830.55+085517.0 & 15 38 30.55 & +08 55 17.1 & 3.5554 & 0.0060 & 14.72 & K & -30.07 &  17.98$^{(e)}$ & $<1.02$ & FIRST & RQ\\
            SDSSJ161458.33+144836.9 & 16 14 58.34 & +14 48 36.9 & 2.5698 & 0.0022 & 15.23 & H & -29.41 & 17.43$^{(e)}$ & $<0.94$ & FIRST & RQ\\
            PKS 1937-101 & 19 39 57.30 & -10 02 41.0 & 3.7908 & 0.0032 & 13.81 & K & -30.40 & 17.00$^{(g)}$ & 838.30 & NVSS & RL\\
            PKS 2000-330 & 20 02 24.00 & -32 51 47.0 & 3.7899 & 0.0033 & 15.15 & K & -30.99 & 17.30$^{(g)}$ & 446.00 & NVSS & RL\\
            SDSSJ210524.49+000407.3 & 21 05 24.47 & +00 04 07.3 & 2.3445 & 0.0020 & 14.59 & H & -29.96 & 16.98$^{(e)}$ & 3.20 & NVSS & RI\\
            SDSSJ210831.56-063022.5 & 21 08 31.56 & -06 30 22.6 & 2.3759$^{(a)}$ & 0.0016 & 15.77 & H & -29.11 & 17.41$^{(e)}$ & $<1.18$ & FIRST & RQ\\
            SDSSJ212329.46-005052.9 & 21 23 29.46 & -00 50 52.9 & 2.2800 & 0.0017 & 14.61 & H & -29.76 & 16.62$^{(e)}$ & $<0.96$ & FIRST & RQ\\
            PKS 2126-15 & 21 29 12.10 & -15 38 42.0 & 3.2987 & 0.0042 & 14.22 & K & -29.88 & 17.00$^{(g)}$ & 589.70 & NVSS & RL\\
            SDSSJ235808.54+012507.2 & 23 58 08.62 & +01 24 34.8 & 3.4009 & 0.0029 & 14.84 & K & -29.58 &  17.50$^{(d)}$ & $<0.96$ & FIRST & RQ\\
           \noalign{\smallskip}
            \hline
         \end{tabular}
             }
    \tablefoot{$^{(a)}$ Redshift estimation based on the $[\ion{O}{III}]\lambda 5007$. $^{(b)}$ 2MASS magnitude. $^{(c)}$ $m_{\rm{O}}$. $^{(d)}$ $m_{\rm{R}}$. $^{(e)}$ $m_{\rm{B}}$. $^{(f)}$ Photographic $m_{\rm{B}}$. $^{(g)}$ $m_{\rm{V}}$. $^{(h)}$ Values with $<$ are upper limits.}
   \end{table*}
%--------------------------------------------------------------------
\section{Sample}
\label{sample}
\par The sample consists of 22 quasars with high redshift, going from $z=2.2$ to $z=3.7$, and high luminosity (47.39 $\le  \log L_{\textrm{bol}}\le$ 48.36 [erg s$^{-1}$]), including both radio-loud and radio-quiet sources that were observed under the ESO programmes 083.B-0273(A) and 085.B-0162(A). These sources were selected from the Hamburg-ESO survey (HE,  \citealt{Wisotzki_2000}), which consists of a flux limited ($m_{\rm{B}}\approx 17.5$), color-selected survey with a redshift range $0 \lesssim z \lesssim 3.2$. Our sample presents a redshift that allows for the detection and observation of the \hb{}+\oiii{} region through the transparent window in the near-infrared with the ISAAC spectrograph at VLT \citep{Sulentic_2006,sulentic_2017}. There is also a cut in $\delta$ at +25 degrees due to the geographic location of the telescope. Table \ref{tab:source_id} presents the main properties of our sample, reporting the source identification according to the different catalogues (Col. 1); right ascension and declination at J2000 coordinates (Cols. 2 and 3, respectively); redshift estimated as explained in Section \ref{sub:redshift} (Col. 4); redshift uncertainties (Col. 5); the $H$- or $K$-band (depending on the range of the spectrum) apparent magnitude $m_\mathrm{H}$ or $m_\mathrm{K}$ from the 2-MASS catalogue (Col. 6); the respective band (H or K, Col. 7); the $i$-band absolute magnitude $M_\mathrm{i}$ (Col. 8) estimated for our data; the apparent magnitude from \citet{veron_2010} (Col. 9); the radio flux in mJy (Col. 10) in the frequency of the survey listed in Col. 11; radio classification according \citet{Ganci_2019} is shown in Col. 12  and explained in section \S\ref{sub:radiodata}.   
%In order to calculate $M_\mathrm{i}$ we use the following equation:
%\begin{equation}
%    M_{\mathrm{i}}=m_{\mathrm{i}}-5\log d_{\mathrm{L}}-k_{\rm i}-A_{\rm i}+5,
%\end{equation}
%in which $d_{\mathrm{L}}$ is the luminosity distance, $k$ is the 
The $k$-correction used is the one available on \cite{Richards_2006} for sources with similar redshift and the galactic extinction were collected from the DR16 catalogue \citep{Lyke_2020}.  The luminosity distance was computed from the redshift using the approximation of \citet{Sulentic_2006}, valid for $\Omega_\mathrm{M}=0.3$, $\Omega_{\Lambda}=0.7$, and H$_{\rm{0}}=70$ km s$^{-1}$ Mpc$^{-1}$. 

\par For all sources, we have computed synthetic magnitudes $H$ or $K$  from our flux-calibrated spectra, and we have compared them  with the 2MASS $H$- and $K$-band apparent magnitudes. In cases in which the difference between our $m$ synthetic magnitudes and the 2MASS catalogue magnitudes is higher than 0.5 mag we use the $m$ from 2MASS (these sources are identified in Table \ref{tab:source_id}). Also, in one very special source (B1422+231) a magnification correction is applied lowering the flux by a factor 6.6, which leads to a change in the magnitude, since this source is a gravitationally lensed quasar with four images (\cite{patnaik_1992,Assef_2011}; see Appendix \ref{app_uv} for notes on individual objects). Apart from the well-known lensed quasar  B1422+231, our sample includes other quasars known or suspected to undergo micro-lensing: SDSSJ141546.24+112943.4 \citep{sluse_2012, takahashi_2014} and some  candidates, as $[\rm HB89]$ 0029+073 \citep{Jaussen_1995}, but we did not apply any magnitude correction for lensing on these sources.

\begin{figure}[t!]
    \centering
    \includegraphics[width=0.9\columnwidth]{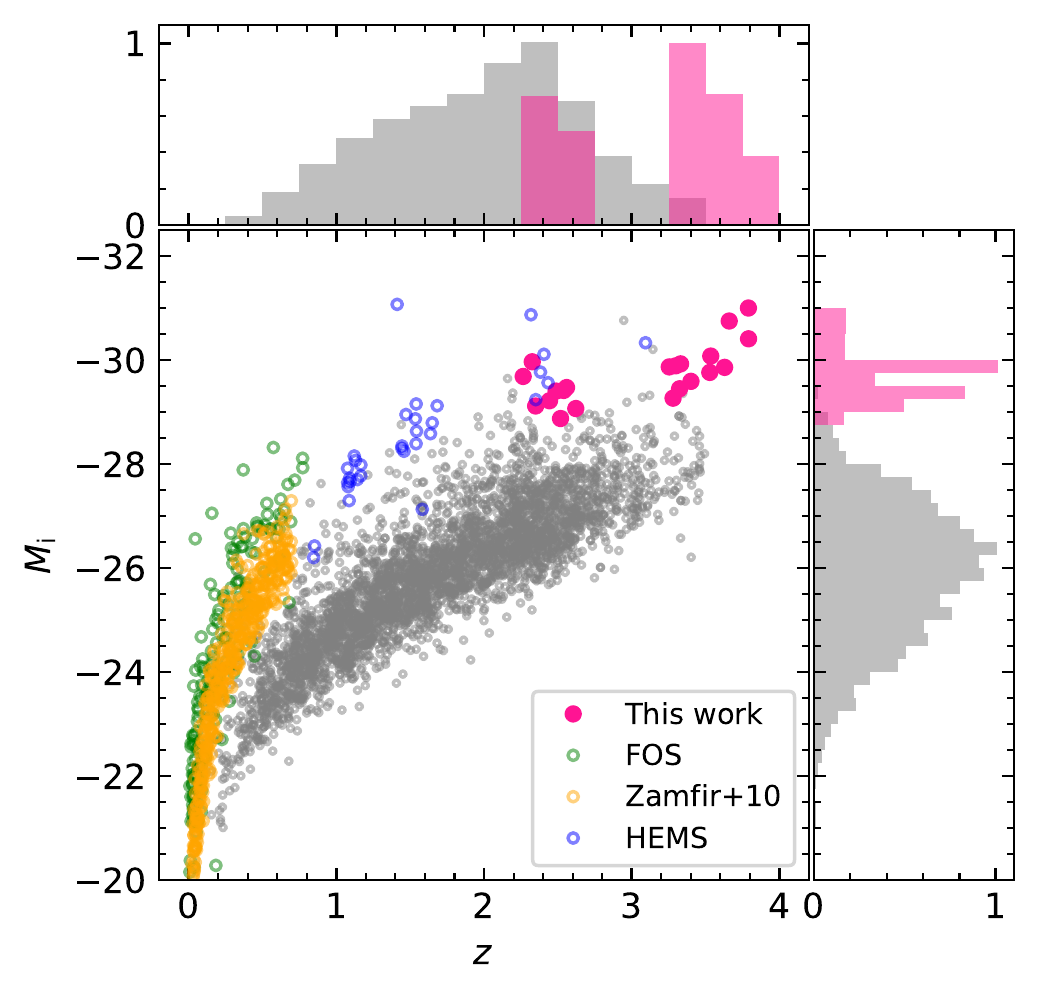}
    \caption{Location of the complete sample in the Hubble diagram (pink spots). Gray dots represent a random subsample of the DR16 catalogue from \cite{Lyke_2020} and green, yellow, and blue dots represent the comparison samples defined in Section \ref{sub:comp_samples}.}
    \label{fig:mag_z}
\end{figure}

\subsection{Redshift estimations and sample location in the Hubble diagram}
\label{sub:redshift}

\par Redshift measurements were  based on the H$\beta$\ emission line profile, and were obtained from the observed wavelength of the \hb\ narrow component (FWHM $\lesssim 1000$ km s$^{-1}$). The values are reported in Table \ref{tab:source_id}. In the case of HE 0001-2340 and SDSSJ210831.56-063022.5 the redshift was estimated through the [\ion{O}{iii}]$\lambda 5007$ emission line, due to the difficulty of isolating a narrow component of H$\beta$. Before performing the spectral analysis, described in section \ref{analysis}, both optical and UV spectra are set at rest-frame using the IRAF task \texttt{dopcor}. %However, in both cases, the location of the peak of the H$\beta$\ profile is in agreement with the vacuum wavelength rest-frame.  The obtained redshift values were applied to both the optical and UV ranges. 

\par Fig. \ref{fig:mag_z} shows the location of the objects of the sample (pink spots) in the $i$-band absolute magnitude $M_{\rm{i}}$ \textit{vs.} redshift plane. Our sample as well as the comparison samples (defined in Section \ref{sub:comp_samples}) are located at the high end of the luminosity distribution when compared with the SDSS DR16 data (grey spots, \citealt{Lyke_2020}). Moreover, some of our sources are well located within the region with the highest values of the SDSS redshift distribution for quasars.

\subsection{Comparison samples}
\label{sub:comp_samples}

\par %In order to perform a more general study of our data, we selected comparison samples from related papers. 
We have chosen comparison samples that include high- and low-redshift, as well as high- and low-luminosity sources that were studied in the recent literature, as follows:
\begin{itemize}
\item \textit{Low-$z$ SDSS sources:} \cite{Zamfir_2008} and  \cite{zamfir_2010} analyse $\sim$ 470 low-redshift ($z\le 0.7$)  quasars with high signal-to-noise (S/N $\sim 29$) ratio optical spectra observed  with the Sloan  Digital  Sky  Survey data release (DR) 5 \citep{mccarthy_2007}, including both radio-loud and radio-quiet sources. This sample covers a wide range of bolometric luminosity, going from $\log L_{\rm{bol}}= 10^{43}$ to $\log L_{\rm{bol}}= 10^{47}$ erg s$^{-1}$.

\item \textit{High-luminosity Hamburg-ESO sample (hereafter, HEMS)}:  \cite{Sulentic_2004, Sulentic_2006} and \citetalias{Marziani_2009} obtained, using the VLT-ISAAC camera, H$\beta$ region measurements for a sample of 53 high-redshift high-luminosity objects, selected from the Hamburg-ESO  survey. The sources are extremely luminous (47$\lesssim L_{\textrm{bol}}\lesssim$ 48) and are located in a redshift range of $0.9 \lesssim z \lesssim 3.1$. We use these data as comparison sample at high $L$, for both H$\beta$ and [\ion{O}{III}] lines.  For the UV region we use the results provided by \citetalias{sulentic_2017}, who studied 28 quasars of the previous sample, and from which they obtained \ion{C}{IV}$\lambda1549$ observations with the VLT and  TNG telescopes.  The redshift range of the HEMS is $1.4 \le z \le 3.1$ and a typical bolometric luminosity is $L_{\rm bol} \gtrsim 10^{47.5}$ erg s$^{-1}$. {The  location of the HEMS in the Hubble diagram is similar to the one of our data (Fig. \ref{fig:mag_z})} in terms of absolute magnitudes, although a large fraction of HE sources is with $1 \lesssim z \lesssim 2$. %Both samples are therefore of uncommonly high luminosity even for $z \gtrsim 1$. 

 \item \textit{Low-luminosity FOS data:} The low-luminosity Faint Object Spectrograph (FOS) comparison sample was selected from \citet[][S07 from now on]{Sulentic_2007}, where the authors analysed the \ion{C}{IV} line parameters. \citet[][hereafter M03]{Marziani_2003} provide measurements on the H$\beta$ emission line for most sources in this sample. The typical bolometric luminosity of this sample is $\log L_{\rm{bol}} \sim 45.6$ [erg s$^{-1}$] and a redshift range $z \le 0.5$.
 
\end{itemize}

 %-----------------------------------------------------------------
\section{Observations and Data}
\label{obs}
%\subsection{Near Infrared Observations}

%\par Spectroscopic observations in the near infrared were performed with the ISAAC camera and the VLT-U1 telescope at the European Southern Observatory (ESO). The ISAAC spectrograph was operated in service mode between 2009 April and 2010 August under ESO programmes 0.83.B-0273(A) and 0.85.B-0162(A). All spectra were obtained with a slit width of 0.6 arcsec. In the redshift range we consider, the high-ionization lines in the UV in the quasar rest-frame are redshifted into the optical domain, while the low-ionization  Balmer line H$\beta$ and \ion{Fe}{ii}$\lambda4570$ (or \ion{Fe}{ii}$\lambda5130$ blend) are shifted into the near infrared (sZ, Z, J, H). Typical S/N values of the NIR spectra are $\gtrsim 50$.

%--------------------------------------------------- One column table
   \begin{table}[t!]
      \caption[]{Log of optical observations with VLT/ISAAC.}
         \label{tab:obs_log}
     
        \resizebox{\linewidth}{!}{
         \begin{tabular}{lccccccc}
            \hline
            \hline
            \noalign{\smallskip}
     Source           &  Date obs. & Band & DIT & N$_{\rm{exp}}$ & Airmass & Averaged\\
        &           (start) & &        (s)      & & start-end & seeing\\
            (1) & (2) & (3) & (4) & (5) & (6) & (7)\\ 
            \noalign{\smallskip}
            \hline
            \noalign{\smallskip}

HE 0001-2340  &     2010-07-06 & sH  & 180 & 12  &  1.05-1.02 & 1.03\\
               &   2010-07-24 & sH &  180 & 12  &  1.03-1.01 & 0.55\\
                &   2010-08-04 & sH &  180 & 12  &  1.06-1.12 & 1.44\\
$[\rm{HB89}]$ 0029+073  &  2009-09-03 & sK &  180 & 20  &  1.43-1.25 & 1.08\\
                  & 2010-07-08 & sK &  180 & 20  &  1.25-1.19 & 0.49\\
                   & 2010-08-04 & sK  & 180 & 20  &  1.19-1.18 & 1.31\\
CTQ 0408           & 2009-07-09 & sK  & 180 &  8  &  1.11-1.12 & 0.51\\
                   & 2009-08-31 & sK  & 180 &  8  &  1.11-1.12 & 3.05\\
SDSSJ005700.18+143737.7 &   2009-09-04 & sH &  180 & 28 &   1.44-1.98 & 1.99\\
H 0055-2659      &  2009-09-02 & sK  & 180 & 40  &  1.01-1.23 & 1.25\\
SDSSJ114358.52+052444.9  &  2009-07-08 & sH &  180 & 20 &   1.34-1.67 & 0.83\\
SDSSJ115954.33+201921.1   & 2010-04-05 & sK &  175 & 24 &   2.35-1.69 & 1.08\\
SDSSJ120147.90+120630.2  &  2010-04-15 & sK &  180 & 20 &   1.92-1.49 & 1.23\\
SDSSJ132012.33+142037.1  &  2009-05-04 & sH &  180 & 24 &   1.36-1.28 & 1.37\\
SDSSJ135831.78+050522.8  &  2010-04-18 & sH &  180 & 20 &   1.23-1.18 & 0.73 \\
Q 1410+096      &   2010-04-20 & sK  & 180 & 28  &  1.45-1.23 & 0.83\\
SDSSJ141546.24+112943.4  &  2009-05-04 & sH &  180 & 20  &  1.24-1.26 & 0.50\\
B1422+231      &   2009-04-13 & sK &  180 & 12  &  1.50-1.48 & 0.98\\
SDSSJ153830.55+085517.0  &  2010-04-05 & sK &  180 & 20  &  1.20-1.24 & 1.14\\
SDSSJ161458.33+144836.9  &  2009-07-08 & sH &  180 & 20  &  1.33-1.30 & 1.35\\
                  & 2009-09-22 & sH &  180 & 20  &  1.73-2.05 & 2.24\\
PKS 1937-101      & 2009-09-03 & sK &  180 & 12  &  1.13-1.23 & 1.07\\
PKS 2000-330      & 2009-08-31 & sK &  180 & 20  &  1.43-1.89 & 2.03\\
SDSSJ210524.49+000407.3  &  2010-07-23 & sH  & 180 & 12  &  1.29-1.45& 1.80\\
SDSSJ210831.56-063022.5  &  2010-06-19 & sH  & 175 & 24  &  1.07-1.18 & 0.93\\
SDSSJ212329.46-005052.9  &  2010-07-08 & sH  & 180 & 12  &  1.12-1.18 & 0.48\\
              &     2010-07-24 & sH  & 180 & 12  &  1.41-1.27 & 0.90\\
PKS 2126-15    &    2009-09-01 & sK  & 180 &  12  &  1.84-2.36 & 1.41\\
                &   2010-06-09 & sK &  180 & 16  &  1.26-1.12 & 1.07\\
SDSSJ235808.54+012507.2   &     2010-06-21 & sK &  180 & 20  &  1.25-1.13 & 1.01\\          \noalign{\smallskip}
            \hline
         \end{tabular}
        }
     
     %\tablefoot{Columns: (1) Source identification. (2) Observation date. (3) Grating. (4) Central wavelength of the filter, in units of $\mu$m.}
   \end{table}
%-----------------------------------------------------------------
%Present including the sources not yet studies, the full sample, presenting the redshift and luminosity and indicating which are radio-loud or radio-quiet.

\subsection{Near Infrared Observations and data reduction}
\par Spectra were taken in service mode in 2009 and 2010, with the infrared spectrometer ISAAC, mounted at the Nasmyth B focus of VLT-U1 (Antu) until August 2009, and later at the Nasmyth A focus of VLT-U3 (Melipal) at the ESO Paranal Observatory. Table \ref{tab:obs_log} summarises the NIR observations, listing the date of observation (Col. 2), used grating (Col. 3), individual Detector Integration Time (DIT) in seconds (Col. 4), number of exposures with single exposure time equal to DIT (Col. 5), the range of air mass of the observations (Col. 6), and the averaged seeing (Col. 7). When the source was observed more than once (HE 0001-2340, [HB89] 0029+073, CTQ 0408, SDSSJ161458.33+144836.9, SDSSJ212329.46-005052.9, and PKS 2126-15), we combine the individual spectra to obtain a median weighted final spectrum for the analysis. Spectral resolutions are FWHM $= 16$ \r{A} in the $H$ band and FWHM $= 22$ \r{A} in the $K$ band.

Reductions were performed using standard IRAF routines. Several spectra of each source were taken, nodding the telescope between two positions (nodding amplitude = 20''). Each obtained frame was divided by the flat field provided by the ESO automated reduction pipeline. 1D spectra were extracted using the IRAF program ``apall''. Cosmic ray hits were eliminated by interpolation using a median filter, comparing the affected spectrum with the other spectra of the same source. 

\par For each position along the slit, a 1D xenon/argon arc spectrum was extracted from the calibration lamp frame, using the same extraction parameters as the corresponding target spectrum. The wavelength calibration was well modeled by 3rd order Chebyshev polynomial fits to the positions of 15-30 lines, with rms residuals of 0.3\,\r{A} in sH and 0.6\,\r{A} in sK. The wavelength calibration is usually affected by a small 0-order offset caused by grism and telescope movement, because the arc lamp frames were obtained in daytime. A correction for these shifts was obtained by measuring the centroids of 2–3 OH sky against the arc calibration and calculating the average difference, which reached at most 2-3 pixels in either direction. Once matched with the corresponding arc calibrations, the individual spectra of each source were rebinned to a common linear wavelength scale and stacked.

\par The spectra of the atmospheric standard stars were extracted and wavelength-calibrated in the same way. All clearly identifiable stellar features (H and HeI absorption lines) were eliminated from the stellar spectra by spline interpolation of the surrounding continuum intervals. Each target spectrum was then divided by its corresponding standard star spectrum in order to correct for the atmospheric absorption features. This was achieved with the IRAF routine ``telluric'', which allows to optimise the correction with slight adjustments in shift and scaling of the standard spectrum. %The shape of the continuum of the standard star was eliminated from the spectrum of each target by multiplying the latter with an artificial black-body continuum corresponding to the temperature of the star, determined on the basis of its tabulated spectral type. 
The correct flux calibration of each spectrum was achieved by scaling it according to the magnitude of the standard star and to the ratio of the respective DITs. Because the seeing often exceeded the width of the slit, significant light loss occurred, and therefore the absolute flux scale of the spectra is not to be considered as accurate. However, in this long-wavelength range we consider the light losses to be independent of wavelength (i.e., we assume that differential atmospheric refraction is negligible), and they should therefore not affect the relative calibration of the spectra. We carried out an a posteriori evaluation of the absolute flux calibration uncertainty  performing a comparison between the $H/K$-band magnitudes estimated by convolving the $H/K$\ filter with the observed spectrum and the $H/K$\ magnitudes in NED. The  differences are smaller than 0.5 mag, except for SDSSJ132012.33+142037.1, SDSSJ141546.24+112943.4, and SDSSJ153830.55+085517.0 where a correction  factor $\sim 0.880\pm0.176$ mag was applied.

 \begin{table}[t!]
      \caption[]{UV spectra information.}
         \label{tab:uv_obs_log}
     
        \resizebox{\linewidth}{!}{
         \begin{tabular}{lcc}
            \hline
            \hline
            \noalign{\smallskip}
            Source &  UV & Comments\\
              & Spect. &\\
            (1) & (2) & (3)\\ 
            \noalign{\smallskip}
            \hline
            \noalign{\smallskip}
            SDSSJ005700.18+143737.7 & BOSS &\\
            SDSSJ114358.52+052444.9 & BOSS & \\
            SDSSJ115954.33+201921.1 & SDSS &\\
            SDSSJ120147.90+120630.2 & BOSS &\\
            SDSSJ132012.33+142037.1 & BOSS &\\
            SDSSJ135831.78+050522.8 & BOSS &\\
            Q 1410+096 & SDSS & BAL\\
            SDSSJ141546.24+112943.4 & SDSS & BAL\\
            SDSSJ153830.55+085517.0 & BOSS & BAL$^{(a)}$\\
            SDSSJ161458.33+144836.9 & BOSS &\\
            PKS 2000-330 & \cite{barthel_1990} &\\
            SDSSJ210524.49+000407.3 & SDSS & BAL\\
            SDSSJ210831.56-063022.5 & SDSS &\\
            SDSSJ212329.46-005052.9 & BOSS &\\
            SDSSJ235808.54+012507.2 & BOSS &\\          
            \noalign{\smallskip}
            \hline
         \end{tabular}
        }
     
     \tablefoot{$^{(a)}$ In this case, there is only a small but significant absorption close to \siv{} and do not affect the other UV emission lines that we are considering in the fittings.}
   \end{table}

\subsection{UV}
\par We have found useful UV spectra (i.e., which include at least one of the three UV regions of our interest) for the 15/22 sources that are listed in Table \ref{tab:uv_obs_log}, where we also list in Col. 2 the database/reference from which each spectrum was obtained and in Col. 3 the Broad Absorption Lines (BAL) quasars. We report the BAL sources as the analysis of the region in which the broad absorptions are located should be taken with care, once it demands the addition of absorption components in the fitting routine. The other seven quasars do not have useful UV spectra, either because they are old spectra not digitally available and have low S/N not suitable for accurate profile fitting, or because they do not include any of the three UV regions we want to analyse (\siv{}+\ion{O}{IV}$\lambda$1402, \civ{}+\heiiuv{}, and the 1900 \r{A} blend). For the UV spectral range (observed in the optical domain at the redshift of this sample), the spectra were collected mainly from the SDSS DR16 database \citep[and references therein]{Ahumada_2020}. For one source (PKS 2000-330), the UV spectrum was digitalized from \cite{barthel_1990}. Four out of the 22 quasars (Q 1410+096, SDSSJ141546.24+112943.4, SDSSJ153830.55+085517.0 and SDSSJ210524.49+000407.3) are classified as BAL quasars, due to strong absorption lines \citep{Gibson_2009, Scaringi_2009, Allen_2011, welling_2014, Bruni_2019,Yi_2020}. 

\subsection{Radio data}
\label{sub:radiodata}
\par The radio fluxes presented in Table \ref{tab:source_id} were collected from the 1.4-GHz NRAO VLA Sky Survey (NVSS, \cite{Condon_1998}) and from the VLA FIRST Survey \citep{Gregg_1996,Becker_1995}. The flux in radio is reported for seven of our sources and in the case the object is not detected we provide an upper limit to the flux, which corresponds to the detection limit ($\approx$ 5 times the rms in both FIRST and NVSS catalogue) at the position of the source. In the case of CTQ 0408, there is no coverage either in the FIRST survey or in the NVSS survey. The radio flux for this source is obtained at 408MHz from the Sidney University Molonglo Sky Survey catalogue \citep{Mauch_2003}.  The radio classification is shown in Col. 12 of Table \ref{tab:source_id} and it was determined following \citet{Ganci_2019} through the estimation of a modified rest-frame radio loudness parameter $R_{\rm{K}}={f_{\rm{\nu,radio}}/f_{\rm{\nu,optical}}}$ \citep{Kellermann_1989}, defined as the ratio between the specific flux at 1.4 GHz and in the $g$-band. Accordingly, our sample is separated into three different ranges: radio-quiet (RQ; $R_{\rm{K}}<10$), radio-intermediate (RI; $10 \leq R_{\rm{K}}<70$), and radio-loud (RL; $R_{\rm{K}}\ge70$).
 Of the seven sources with radio detection, three of them (CTQ 0408, SDSSJ141546.24+112943.4, and SDSSJ210524.49+000407.3) are classified as radio-intermediate. The radio-loud sources from our sample are B1422+231, PKS 1937-101, PKS 2000-330, and PKS 2126-15, about 20\% of the sources of this sample. We plan to carry out in a forth-coming paper a similar spectroscopic study focused on RL quasars, to supplement those  listed in Table \ref{tab:source_id}. 
%comparison with the RQ sample, of the RL sources observed in a third run with similar redshift and luminosity ranges.

\section{Spectral Analysis}
\label{analysis}
%-----------------------------------------------------------------
\subsection{Optical range: multicomponent fitting}

\subsubsection{\hb{}}

The multicomponent fits were performed, after the spectra were set at rest-frame, using the \textsc{specfit} routine from \textsc{IRAF} \citep{kriss_1994}. This routine allows for simultaneous minimum-$\chi^2$ fit of the continuum approximated by a power-law and the spectral line components yielding FWHM, peak wavelength, and intensity of all line components. In the optical range we fit the H$\beta$ profile as well as the [\ion{O}{III}]$\lambda\lambda 4959,5007$ emission lines and the \ion{Fe}{II} multiplets for the 22 objects. 

\par Our approach includes the continuum (a power law), a semi-empirical scalable \ion{Fe}{II} emission template and the emission line components to fit the H$\beta$+[\ion{O}{III}]$\lambda\lambda 4959,5007$ region. The \ion{Fe}{II} template used is almost identical to the one of \cite{boroson_1992}. The specfit routine allows for the scaling, broadening, and shifting of the \ion{Fe}{II} template to model the observed features \citep{Marziani_2003}. %\ion{Fe}{II} contribution is also fitted in order to estimate the R$_{\rm{\ion{Fe}{II}}}$ parameter, which consists of the ratio between the intensities of the \ion{Fe}{II} blend at 4570\r{A} and H$\beta$. 
\par
\begin{figure}
    \centering
    \includegraphics[width=\columnwidth]{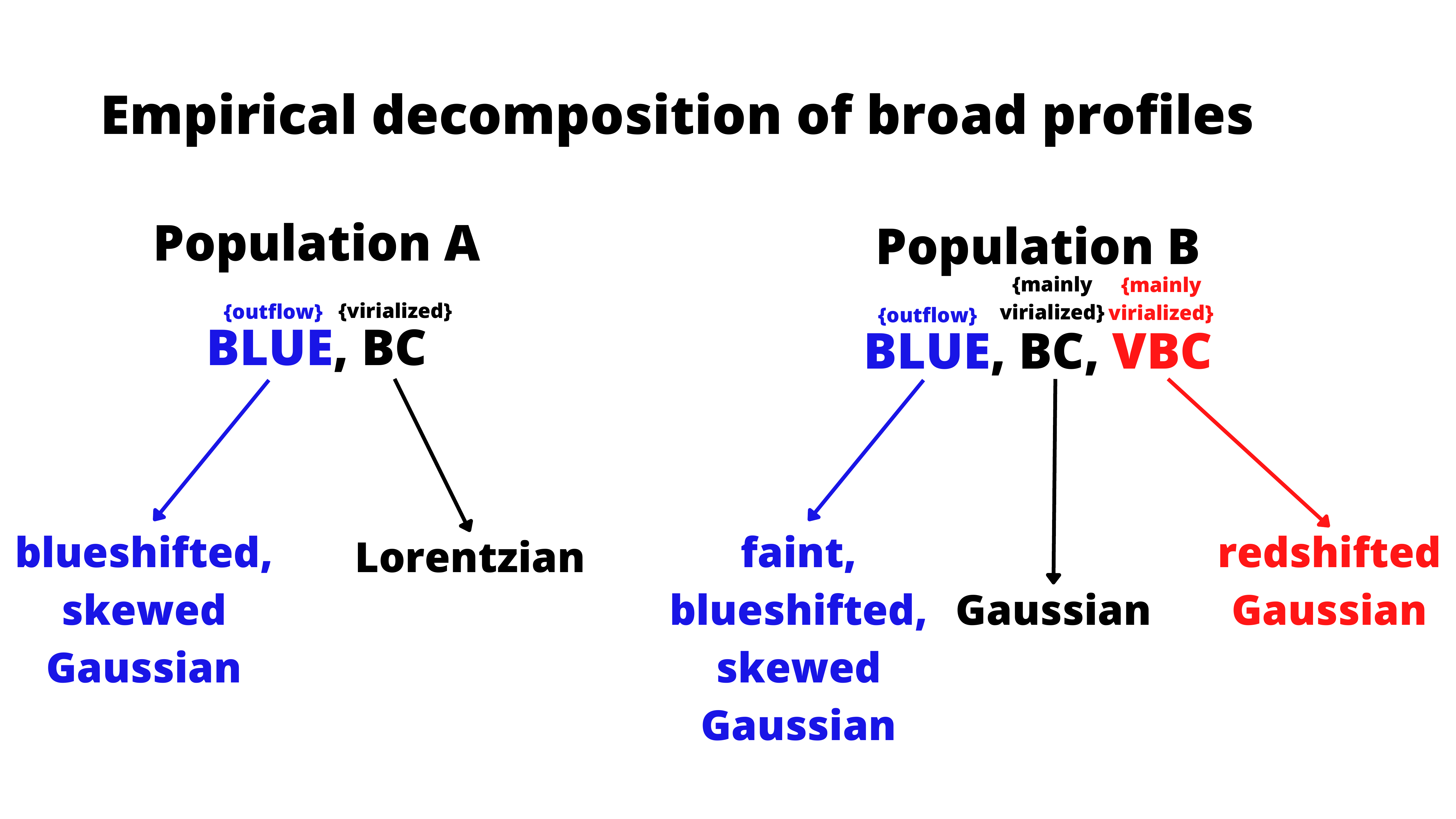}
    \caption{Sketch of the approach followed {for} broad profile fittings.}
    \label{fig:broad_profiles}
\end{figure}

 A sketch to illustrate the decomposition approach performed on the fittings of emission lines broad profiles is presented in Fig. \ref{fig:broad_profiles}. The fit of the \hb{} full profile takes into account three main components:
\begin{itemize}
    \item Broad component (BC): this component is kept symmetric, set almost always at rest-frame, and presents a FWHM range that goes from $\gtrsim 3000$ km s$^{-1}$ for some Pop. A and $\lesssim 6000$ km s$^{-1}$ for Pop. B quasars. The profile changes depending on the quasar population, presenting a Lorentzian-like shape for Pop. A and a Gaussian-like one for Pop. B sources \citep{Sulentic_2002,Marziani_2003, zamfir_2010, Cracco_2016}; %It is believed to be emitted within the virialized BLR;
    
    \item Blueshifted component (BLUE): %it is detected mainly in Pop. A quasars, but also in Pop. B sources at high and intermediate redshift \citepalias{Marziani_2009}. 
    our first assumption is to model this component by only a blue-shifted Gaussian (symmetric or skewed) profile  with FWHM and shift similar to the \oiii\ semi-broad component (SBC, explained in \S \ref{sec:oiii_specanalysis}). When the H$\beta$ profile does not correspond to the blueshifted SBC of \oiii{} profile, it is included in the fitting an additional Gaussian blue-shifted component that may not belong to NLR emission (BLUE). BLUE is believed, unlike the SBC,  to be associated with emission of outflowing gas from within the BLR  \citep{negrete_2018};
        \item Very broad component (VBC): it is clearly observed in Population B sources \citep{sulentic_2017,wolf_2020}. It is always represented by a redshifted Gaussian profile and it is thought to be related with the high-ionisation virialised region closest to the continuum source \citep{ Peterson_1986,snedden_2007S, wang_2011}. This component can easily achieve FWHM $\gtrsim$ 10000 km s$^{-1}$.
\end{itemize}

\par In addition, we include a narrow component (NC) superimposed to the broad emission line profile and it is fitted as an unshifted Gaussian. %In Pop. B sources, we maximize the FWHM of \hb{} including broader components and, in cases like these, it is also taken into account when deciding the best fitting model.

The Population classification depends on the FWHM of the full profile. Depending on the population assignment, different components are included and they assume different line shapes. An exhaustive analysis of the fittings were performed and in borderline cases (i.e., with line width close to the boundary between Pop. A and B), the final conclusion on the population of the source is based on the $\chi^2$ of the fitting (i.e., the fittings with the minimum $\chi^2$ are selected).
%The selected fittings are always the ones with lower residuals in the region of interest and the ones that are in agreement with our approach (described below). 
%\subsection{\oiii{}}
\subsubsection{\oiii{}}
\label{sec:oiii_specanalysis}
\par An approach similar to the one used for \hb\ is also followed for the [\ion{O}{III}]$\lambda\lambda 4959,5007$ emission line profiles. The [\ion{O}{III}]$\lambda\lambda 4959,5007$ full profile is assumed to be well represented by a Gaussian narrow (FWHM $\lesssim 1200$ km s$^{-1}$) component set at rest-frame or shifted to the blue and a semi-broad component (SBC, FWHM typically $\lesssim 3000$ km s$^{-1}$) that usually appears more shifted to the blue \citep[$\sim$ 1000 km/s,][]{zhang_2011,Marziani_2016, Marziani_2022}. The NC is modelled as a Gaussian profile for the two populations and in a first approach the NC of both H$\beta$ and \oiii{} share the same line width. The blueshifted contributions can be modelled by one or more Gaussian profiles and in some cases the Gaussian needs to be asymmetric towards the blue to account for the line shape, i.e., to be a \textit{skewed} Gaussian. The use of a skewed Gaussian has a physical motivation, as it might be associated with bipolar outflow emission in which the receding side of the outflow is obscured. Apart from that, both [\ion{O}{iii}]$\lambda 4959$ and [\ion{O}{iii}]$\lambda 5007$ emission line profiles are assumed to have the same FWHM and shifts, and the intensity ratio between the two lines is kept fixed at 1:3 \citep{dimitrijevic_2007}.

\subsection{UV range: multicomponent fitting}

\par Fits are performed for three different regions centred on \ion{Si}{IV}$\lambda1397$+\ion{O}{IV}]$\lambda1402$, \ion{C}{IV}$\lambda1549$+\ion{He}{II}$\lambda1640$, 
and the 1900 \r{A} blend that includes the \ion{Al}{III}$\lambda 1860$ doublet, \ion{Si}{III}]$\lambda 1892$, \ion{Si}{II} $\lambda 1816$, and \ion{C}{III}]$\lambda 1909$. UV fittings are presented for 15 sources. The fits were not carried out in cases where the emission lines are strongly affected by BALs. The fittings of the absorption lines are performed in sources in which the presence of these lines allows to clearly see the emission line profile  (as in the case of mini-BALs; \citealt{Sulentic_2006a}). %\par We set the UV spectra at rest-frame using the IRAF task \texttt{dopcor}, applying the redshift from the H$\beta$ narrow component. 
The UV blends are fit following the population assignments from the H$\beta$ spectral range. The only exception is SDSSJ153830.55+085517.0, where its UV spectrum cannot be fitted in agreement with its classification as Pop. B in the H$\beta$ region, as highlighted in the Appendix \ref{app_uv}. %The final choices of the best UV fits are done based on the lowest residuals and $\chi^2$\ values.  

\subsubsection{1900 \r{A} blend}
%-----------------------------------------------------------------
In the UV region corresponding to the 1900 $\AA$ blend, the fittings are performed considering the \ion{Al}{III}$\lambda 1860$ doublet, \ion{Si}{III}]$\lambda 1892$, \ion{Si}{II} $\lambda 1816$, and \ion{C}{III}]$\lambda 1909$ emission line profiles.  \ion{Fe}{III} is especially relevant on the red side of the 1900 \r{A} blend and its modelling has been performed using the \cite{Vestergaard_2001} empirical template, and following the same approach as for the optical \ion{Fe}{II}. 
%, the 1900\r{A} emission lines \textbf{(\aliii{}, \ion{Si}{III}]$\lambda 1892$, and \ion{C}{III}]$\lambda 1909$)
Broad line components of the 1900 \AA\ blend  in Pop. A sources are fitted by Lorentzian profiles while in Pop. B by Gaussian functions (as is the case of \hb{} in the optical spectral range), keeping the same (or at least a comparable) FWHM to the broad component of H$\beta$.%In four cases, it was necessary to include an additional broad blueshifted component (BLUE) to account for contributions in \ion{Al}{III}$\lambda1860$ doublet that cannot be well fitted by only a BC at rest-frame.} In this case, the \ion{Si}{III}]$\lambda$1892 profiles are also allowed to be fitted with an additional BLUE. 
%\ion{Fe}{III} contributions are also expected, specially on the red side of \ion{C}{III}]$\lambda 1908$ \citep{martinezaldama_2018}.

\par In several Pop. A and more frequently in xA quasars, the \ion{Fe}{III} template and the \ion{C}{III}]$\lambda 1909$ broad component do not reproduce well the shape of the red side of the 1900 \r{A} blend. A better model of the \ion{C}{III}]$\lambda 1909$ region is achieved assuming an additional contribution mostly likely due to the \ion{Fe}{III} line at 1914\r{A}. The same approach was employed in \cite{martinezaldama_2018}.

\par {Regarding the Pop. B sources, the  \ion{C}{III}]$\lambda 1909$ emission line is well represented by the same combination applied for Pop. B H$\beta$: VBC+BC+NC. The VBC FWHM is expected to be $\ge 4000$ km s$^{-1}$, and the NC FWHM $\sim 1000$ km s$^{-1}$ (there is only one source that shows a significant NC in \ion{C}{III}])}. Differently from \ion{C}{III}]$\lambda 1909$, lines such as the \ion{Al}{III}$\lambda 1860$ doublet, \ion{Si}{III}]$\lambda 1892$, and \ion{Si}{II} $\lambda 1816$ are assumed not to present a very broad component \citep{Buendia-rios}. They are  fit by only one BC. 
%or, at most, by a combination BC+BLUE, independently of the quasar population. 

%----------------

\subsubsection{\ion{C}{IV}$\lambda1549$+\ion{He}{II}$\lambda1640$}

\par In order to fit the \ion{C}{IV}$\lambda1549$+\ion{He}{II}$\lambda1640$ blend and contaminant lines within it, we use the approach    followed by \citetalias{sulentic_2017}. As in the previous fits, we represent Pop. A profiles of \ion{C}{IV}$\lambda1549$ and \ion{He}{II}$\lambda1640$ by a broad and a blueshifted component. Pop. B are represented by the combination VBC+BC+BLUE. The broad components of \ion{C}{IV}$\lambda1549$ and \ion{He}{II}$\lambda1640$ are fixed at rest-frame and the other components are left free to vary in wavelength. Nevertheless, the FWHM and shapes of \ion{He}{II}$\lambda1640$ BC and BLUE are restricted to be equal (or comparable) to the corresponding ones of \ion{C}{IV}$\lambda1549$. These constraints are physically motivated since both \civ{} and \heiiuv{} are expected to be emitted from the same regions, because of the similar ionisation potential of the parent ionic species. An additional condition is that the broad component (BC) of \civ{} and \ion{He}{II}$\lambda1640$ FWHM should be equal or larger to the one of \hb{}, following previous works \citep[e.g.][and references therein]{sulentic_2017}. 

\subsubsection{\ion{Si}{IV}$\lambda1397$+\ion{O}{IV}]$\lambda1402$}

\par For the \ion{Si}{IV}$\lambda1397$+\ion{O}{IV}]$\lambda1402$ emission lines region, we follow the same steps as in the \ion{C}{IV}+\ion{He}{II} blend. The  \ion{Si}{IV}$\lambda1397$+\ion{O}{IV}]$\lambda1402$ feature profile is similar to the one of \ion{C}{IV}$\lambda1549$, and  is therefore expected to present BC and blueshifted component similar to those of \ion{C}{IV}$\lambda1549$. The broad component in this case is also kept at rest-frame and any other necessary component is free to change in all the parameters.

\subsection{Analysis of the full profile parameters for the optical and UV regions}
\par Apart from the multicomponent analysis, the H$\beta$ and [\ion{O}{III}] emission line profiles are also characterised through the parametrisation of the full profile by measuring centroids and widths at different fractional intensities (1/4, 1/2, 3/4, and 9/10), as well as asymmetry and kurtosis indexes in order to provide a quantitative description independent of the \textsc{specfit} modelling. In the case of H$\beta$, the full profile (H$\beta_{\textrm{full}}$) for Pop. B quasars  is represented by BC+VBC plus BLUE when detected. For Pop. A quasars, full H$\beta$ profile consists of BC plus BLUE if an additional blue component is present. For [\ion{O}{iii}] we consider NC and one (or more, if needed) blueshifted SBC. Full profile parameters for the UV region are provided for the \civ{} broad emission line, excluding NC. %As mentioned previously, the 1900\r{A} blend and the \heiiuv{} line show profiles similar to those of \civ{} and the \aliii{} blueshifted excess (BLUE) is not detected in all cases. For sources in which only the BC is present, the main parameters such as shifts and FWHM are provided by the {\tt specfit} analysis. 

\subsection{Error estimates}

\par Uncertainties in the multicomponent fits were estimated by running Markov Chain Monte Carlo (MCMC) simulations, following the approach described in \citet{Marziani_2022apj} for both optical and UV spectral ranges. {Observed spectra were modeled using the components employed in the best fit, with a Markov Chain to sample the domain around the minimum $\chi^2$. The dispersion in the posterior distribution of a parameter was assumed to be its 1$\sigma$\ confidence interval.  } For the optical range, the errors are around the order of 10\% for the FWHM of the H$\beta_{\rm{BC}}$ and [\ion{O}{III}]$_{\rm{SBC}}$. Larger uncertainties ($\sim 30\%$) are found for the narrow components of both lines and for the blueshifted component of H$\beta$. Flux uncertainties for strong or sharp  emission lines are $\sim 10\%$, while typical errors for the continuum and \ion{Fe}{II} are $\sim 5\%$ and $\sim 15\%$\  respectively, if \feii\ is reasonably strong.
%Especially for the very broad components, the uncertainties should be taken as formal errors and are computed assuming that the best fit is the correct one and that the functional form that describes the component is correct.}
%In order to evaluate the errors we variate the intensity of the continuum in a 5\% of its flux

\par For the UV, the FWHM uncertainties are between 10\% and 15\% and errors on intensity measurements  are usually $ \lesssim 10\%$ for the strongest emission line components.

\begin{figure}[t]
    \centering
    \includegraphics[width=0.9\columnwidth]{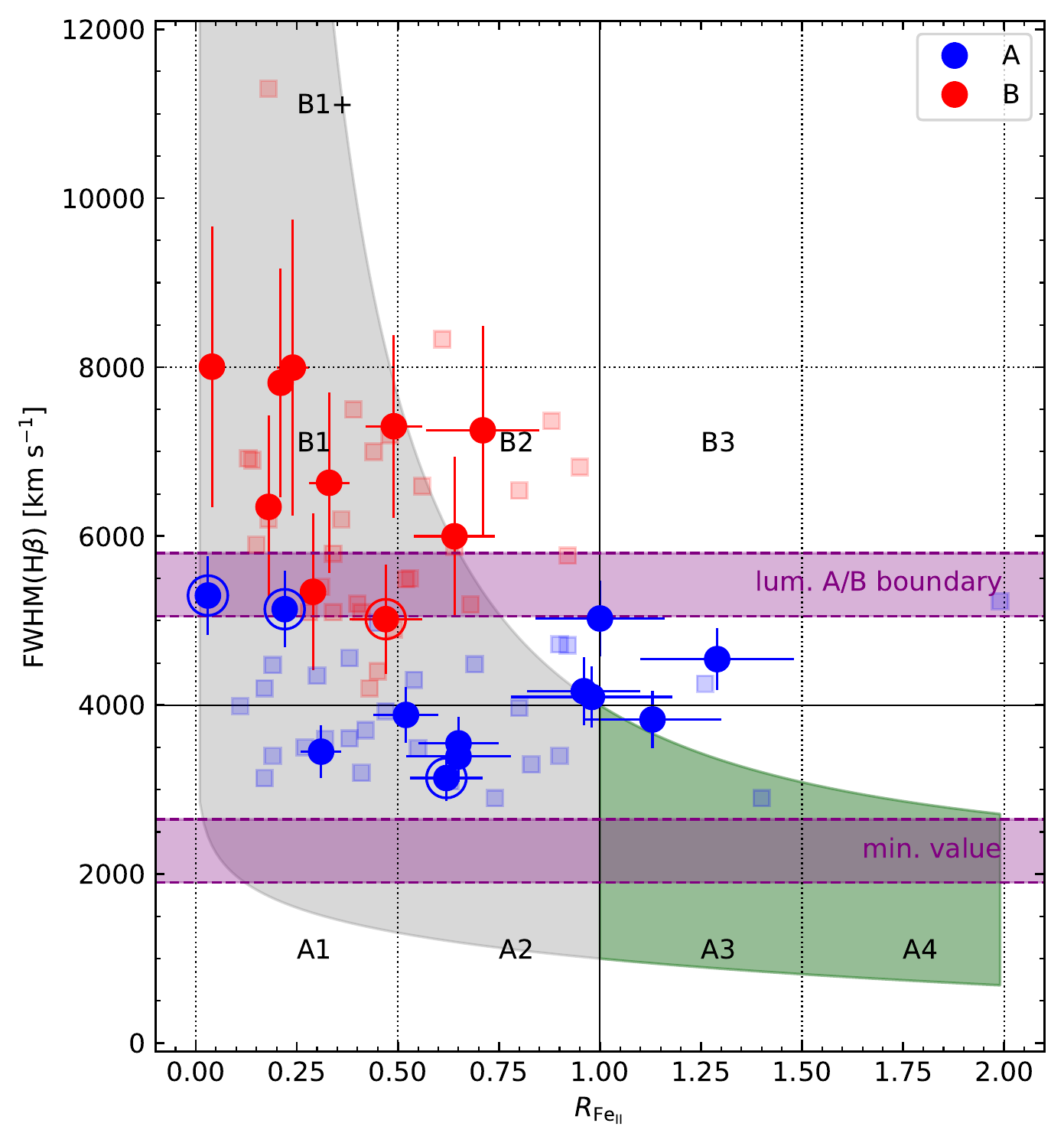}
    \caption{Location of our sources through the optical plane of the 4DE1. Pop. A quasars are represented by blue circles and Pop. B by red circles. The radio-loud sources from the sample are surrounded by a circle in the plot. Grey and green regions indicate the distribution of the Main Sequence of quasars across the 4DE1 optical plane in a low-$z$ context \citep{Marziani2018}. Spectral Types (ST) of the MS are labelled according to \citet{Sulentic_2002}. Blue and red squares indicate high-redshift Pop. A and Pop. B quasars from \citetalias{Marziani_2009}, respectively. Purple-shaded areas show the luminosity-dependent boundary between Pops. A and B and the minimum FWHM value for sources with $47 \le L_{\mathrm{bol}} \le 48$ and assuming $\alpha=0.67$ \citep{Marziani_2009}.}
    \label{fig:optical_plane}
\end{figure}

\section{Results}
\label{results}
%-----------------------------------------------------------------
\subsection{Location in the optical plane of Main Sequence}

\par After performing the multicomponent fitting, we can locate the sample in the MS optical plane, using the FWHM(H$\beta_{\rm full}$) as well as the flux ratio of \ion{Fe}{II}$\lambda$4570 and \hb{}, $R_\mathrm{\ion{Fe}{II}}$. Fig. \ref{fig:optical_plane} shows the location of the sources and a comparison between our sample and low- and high-$z$ samples. The grey- and green-shaded areas in the plot indicate the location of low-redshift quasars on the MS, with the xA sources situated on the green shadow.   

\par Our sample shows a slight displacement in the direction of increasing FWHM(H$\beta$), if compared to low-$z$ samples \citep[e.g.,][ and the shaded area in Fig. \ref{fig:optical_plane}]{zamfir_2010}. There are some Pop. A sources that present a FWHM(H$\beta_{\rm{full}}$) $\sim 5000$ km s$^{-1}$. The Pop. A/B boundary at FWHM(H$\beta$) $ =4000$ km s$^{-1}$ is reasonable when considering low redshift and, consequently, lower luminosity ranges than those of high-$z$ quasars \citep{Sulentic_2004}. However, at high luminosity there is a significant effect on the H$\beta$ profile width that may shift the boundary between Pop. A and B by more than 1000 km s$^{-1}$ \citepalias{Marziani_2009}. %A similar displacement is also seen in the high-$z$ sample from \citetalias{Marziani_2009}, shown in Fig. \ref{fig:optical_plane} as squared symbols, and has also been reported before by \citetalias{sulentic_2017} for example. 
Up and bottom purple shadows in Fig. \ref{fig:optical_plane} indicate, respectively, the Pop. A/B boundary and the minimum FHWM(H$\beta$) found in a $47 \leq \log L_{\rm{bol}} \leq 48$ range, representative of our sample. Both boundaries were determined following \citetalias{Marziani_2009}.

From the 22 sources of our sample, 12 are classified as Pop. A and 10 as Pop. B quasars. We found three sources to belong to spectral type (ST) A1; six Pop. A2; three Pop. A3 (extreme A); eight to be Pop. B1; and two Pop B2. The four radio-loud objects (B1422+231, PKS 1937-101, PKS 2000-330, and PKS 2126-15) are identified in Figure \ref{fig:optical_plane} by the blue and red symbols surrounded by a open circle. Three of them are classified as Pop. A and only PKS 2126-15 is a Pop. B quasar. Of the three radio-intermediate, two (SDSSJ141546.24+112943.4, ST B2; SDSSJ210524.49+000407.3,  A3) show significant \feii\ and an overall spectrum associated with high accretion rates \citep{Ganci_2019,delolmo_2021,marziani_2021}. Only CTQ 0408 is classified as B1, with an optical spectrum resembling the powerful jetted sources at low $z$. The orientation may strongly affect the classification of the sources, especially in the cases when the FWHM(\hb{}) is at the border line of the A/B boundary. Orientation effects could be particularly important in the case of core-dominated RL sources, as the pole-on orientation may imply a narrowing of low-ionization emission line widths  \citep{willsbrowne86,marziani_2001,sulenticetal03,rokakietal03,Zamfir_2008, Bisogni_2017}.

There is just one case of a blatant disagreement between the classification deduced from the optical and UV spectra of the same source: SDSSJ153830.55+085517.0, which is optically classified as Pop. B but presents clearly Pop. A-like profiles in the UV spectrum (see \S \ref{SDSSJ1538} and Fig. \ref{fig:1538_UV} in the Appendix \ref{app_uv}). In this case it was necessary to fit the optical as Pop. B and the UV as Pop. A. %\sout{A more debatable example is PKS 2000-330, a Population A quasar according to the optical spectrum, and that presents a Pop. B-like UV spectrum with flux intensity of \siv{} much less than \civ{}. In the case of PKS 2000-330, we performed the fits of both spectral regions following the optical population classification.} 

\begin{table}[t!]
      \caption[]{Spectrophotometric measurements on H$\beta$.}
         \label{tab:specphot_HB}
      \resizebox{\linewidth}{!}{
                 \begin{tabular}{lccccc}
            \hline
            \hline
            \noalign{\smallskip}
             Source & Spectral &$f_{\lambda,5100}$ $^{(a)}$ & $W$(H$\beta$) & $W$([\ion{O}{III}]) & $R_{\ion{Fe}{II}}$\\
             & Type & & & [\r{A}] &\\
             (1) & (2) & (3) & (4) & (5) & (6)\\
            \noalign{\smallskip}
            \hline
            \noalign{\smallskip}
            \multicolumn{6}{c}{Population A}\\
            \noalign{\smallskip}
            \hline
            \noalign{\smallskip}
SDSSJ005700.18+143737.7	&	A3	&	2.5	$\pm$	0.3	&	53	$\pm$	6	&	4.5	$\pm$	0.5	&	1.13	$\pm$	0.17	\\
SDSSJ132012.33+142037.1	&	A1	&	1.2	$\pm$	0.1	&	74	$\pm$	9	&	19.8	$\pm$	2.4	&	0.31	$\pm$	0.05	\\
SDSSJ135831.78+050522.8	&	A2	&	2.2	$\pm$	0.3	&	68	$\pm$	8	&	16.3	$\pm$	2.0	&	0.65	$\pm$	0.10	\\
Q 1410+096	&	A2	&	3.0	$\pm$	0.4	&	57	$\pm$	7	&	13.7	$\pm$	1.6	&	0.65	$\pm$	0.13	\\
B1422+231	&	A1	&	4.0	$\pm$	0.5	&	78	$\pm$	9	&	9.3	$\pm$	1.1	&	0.22	$\pm$	0.03\\	
SDSSJ161458.33+144836.9	&	A2	&	2.3	$\pm$	0.3	&	83	$\pm$	9	&	16.8	$\pm$	2.0	&	0.52	$\pm$	0.08	\\
PKS 1937-101	&	A1	&	7.6	$\pm$	0.9	&	62	$\pm$	7	&	13.5	$\pm$	1.6	&	0.03	$\pm$	0.01	\\
PKS 2000-330	&	A2	&	2.8	$\pm$	0.3	&	59	$\pm$	7	&	15.1	$\pm$	1.8	&	0.62	$\pm$	0.09	\\
SDSSJ210524.49+000407.3	&	A3	&	3.9	$\pm$	0.5		&	31	$\pm$	4	&	0.4	$\pm$	3.0	&	1.00	$\pm$	0.16	\\
SDSSJ210831.56-063022.5	&	A3	&	1.4	$\pm$	0.2		&	44	$\pm$	6	&	4.7	$\pm$	0.6	&	1.29	$\pm$	0.19	\\
SDSSJ212329.46-005052.9	&	A2	&	4.4	$\pm$	0.5		&	34	$\pm$	5	&	5.5	$\pm$	0.7	&	0.96	$\pm$	0.14	\\
SDSSJ235808.54+012507.2	&	A2	&	4.7	$\pm$	0.6		&	63	$\pm$	8	&	10.7	$\pm$	1.3	&	0.98	$\pm$	0.20	\\

           \noalign{\smallskip}
            \hline
            \noalign{\smallskip}
            \multicolumn{6}{c}{Population B}\\
            \noalign{\smallskip}
            \hline
\noalign{\smallskip}
HE 0001-2340	&	B1	&	1.6	$\pm$	0.2		&	77	$\pm$	9	&	10.6	$\pm$	1.3	&	0.33	$\pm$	0.05	\\
$[\rm HB89]$ 0029+073	&	B1	&	2.4	$\pm$	0.3	&	62	$\pm$	7	&	7.8	$\pm$	0.9	&	0.18	$\pm$	0.02	\\
CTQ 408	&	B1	&	6.9	$\pm$	0.8	&	53	$\pm$	6	&	2.5	$\pm$	4.0	&	0.49	$\pm$	0.07	\\
H 0055-2659	&	B1	&	1.9	$\pm$	0.2	&	92	$\pm$	11	&	40.6	$\pm$	4.9	&	0.29	$\pm$	0.03	\\
SDSSJ114358.52+052444.9	&	B2	&	1.1	$\pm$	0.1	&	82	$\pm$	10	&	5.4	$\pm$	0.6	&	0.64	$\pm$	0.10	\\
SDSSJ115954.33+201921.1	&	B1	&	2.5	$\pm$	0.3	&	78	$\pm$	9	&	15.2	$\pm$	1.8	&	0.04	$\pm$	0.01	\\
SDSSJ120147.90+120630.2	&	B1	&	4.3	$\pm$	0.5	&	117	$\pm$	14	&	21.0	$\pm$	2.5	&	0.24	$\pm$	0.04	\\
SDSSJ141546.24+112943.4	&	B2	&	2.1	$\pm$	0.3	&	108	$\pm$	13	&	33.9	$\pm$	4.1	&	0.71	$\pm$	0.14	\\
SDSSJ153830.55+085517.0	&	B1	&	1.1	$\pm$	0.1	&	94	$\pm$	11	&	5.5	$\pm$	0.7	&	0.21	$\pm$	0.03	\\
PKS 2126-15	&	B1	&	4.8	$\pm$	0.6	&	93	$\pm$	11	&	10.0	$\pm$	1.2	&	0.47	$\pm$	0.09	\\
            \noalign{\smallskip}
            \hline
         \end{tabular}}
    
     \tablefoot{$^{(a)}$ In units of $10^{-15}$ erg s$^{-1}$ cm$^{-2}$ \r{A}$^{-1}$.}
   \end{table}

\begin{sidewaystable*}
      \caption[]{Results from \textsc{specfit} analysis on H$\beta$.}
         \label{tab:specfit_HB}
         \resizebox{\linewidth}{!}{
                       \begin{tabular}{lccccccccccccccccccccccccccccccc}
            \hline
            \hline
            \noalign{\smallskip}
            \noalign{\smallskip}
              & \multicolumn{7}{c}{H$\beta$ full broad profile}  & \multicolumn{15}{c}{Full broad profile (BLUE+BC+VBC)} &  \multicolumn{9}{c}{Narrow profile (SBC+NC)}\\
              \noalign{\smallskip}
              \cline{2-8} \cline{10-22} \cline{24-32}
              \noalign{\smallskip}
               & & & & & & & & & &  & \multicolumn{3}{c}{BLUE$^{(a)}$} & & \multicolumn{3}{c}{BC} & & \multicolumn{3}{c}{VBC}  & & & & \multicolumn{3}{c}{SBC}  & & \multicolumn{3}{c}{NC}\\
              \noalign{\smallskip}
              \cline{12-14} \cline{16-18} \cline{20-22} \cline{26-28} \cline{30-32}
              \noalign{\smallskip}
            Source & FWHM & A. I. & Kurt. & c($\frac{1}{4}$) & c($\frac{1}{2}$)  & c($\frac{3}{4}$) & c($\frac{9}{10}$) & & $I_{\mathrm{tot}}^{(b)}$ & & $I/I_{\mathrm{tot}}$ & FWHM & Shift & & $I/I_{\mathrm{tot}}$ & FWHM & Shift & & $I/I_{\mathrm{tot}}$ & FWHM & Shift & & $I_{\mathrm{tot}}^{(c)}$ & & $I/I_{\mathrm{tot}}$ & FWHM & Shift & & $I/I_{\mathrm{tot}}$ & FWHM & Shift\\
             & [km s$^{-1}$] &  &  & [km s$^{-1}$] & [km s$^{-1}$] & [km s$^{-1}$] & [km s$^{-1}$]  & & & & & [km s$^{-1}$] & [km s$^{-1}$] & & & [km s$^{-1}$] & [km s$^{-1}$] &  &  & [km s$^{-1}$] & [km s$^{-1}$] & & & & & [km s$^{-1}$] & [km s$^{-1}$] & & & [km s$^{-1}$] & [km s$^{-1}$]\\
             (1) & (2) & (3) & (4) & (5) & (6) & (7) & (8) & &(9) & & (10) & (11) &  (12) & & (13) & (14) & (15) & & (16) & (17) &  (18) & & (19) & &(20) &  (21) & (22) & & (23) & (24) & (25) \\
            \noalign{\smallskip}
            \hline
            \noalign{\smallskip}
            \multicolumn{24}{c}{Population A}\\
            \noalign{\smallskip}
            \hline
            \noalign{\smallskip}
     SDSSJ005700.18+143737.7	&	3830	$\pm$	342	&	-0.09	$\pm$	0.01	&	0.33	$\pm$	0.01	&	-359	$\pm$	237	&	-326	$\pm$	83	&	-137	$\pm$	107	&	-82	$\pm$	118	&	&	1.66	$\pm$	0.20	&	&	0.08	$\pm$	0.03	&	2562	$\pm$	359	&	-2154	$\pm$	548	&	&	0.92	$\pm$	0.06	&	3255	$\pm$	346	&	0	$\pm$	10	&	&	…			&	…			&	…			&	&	1.89	$\pm$	0.23	&	&	0.00			&	…			&	…			&	&	1.00	$\pm$	0.32	&	873	$\pm$	93	&	0	$\pm$	10	\\
SDSSJ132012.33+142037.1	&	3450	$\pm$	314	&	-0.11	$\pm$	0.03	&	0.31	$\pm$	0.01	&	-476	$\pm$	323	&	-215	$\pm$	72	&	-138	$\pm$	93	&	-125	$\pm$	110	&	&	0.91	$\pm$	0.11	&	&	0.12	$\pm$	0.05	&	7519	$\pm$	1053	&	-1745	$\pm$	584	&	&	0.88	$\pm$	0.05	&	3224	$\pm$	343	&	-41	$\pm$	14	&	&	…			&	…			&	…			&	&	1.13	$\pm$	0.14	&	&	0.00			&	…			&	…			&	&	1.00	$\pm$	0.32	&	623	$\pm$	66	&	-22	$\pm$	17	\\
SDSSJ135831.78+050522.8	&	3548	$\pm$	314	&	0.00	$\pm$	0.12	&	0.33	$\pm$	0.01	&	-1	$\pm$	297	&	0	$\pm$	60	&	0	$\pm$	97	&	0	$\pm$	120	&	&	1.58	$\pm$	0.19	&	&	0.00			&	…			&	…			&	&	0.92	$\pm$	0.06	&	3548	$\pm$	377	&	0	$\pm$	10	&	&	…			&	…			&	…			&	&	1.49	$\pm$	0.18	&	&	1.00	$\pm$	0.32	&	5497	$\pm$	770	&	-1712	$\pm$	572	&	&	0.00			&	…			&	…			\\
Q 1410+096	&	3394	$\pm$	299	&	0.00	$\pm$	0.08	&	0.33	$\pm$	0.01	&	54	$\pm$	279	&	54	$\pm$	58	&	54	$\pm$	92	&	54	$\pm$	112	&	&	1.79	$\pm$	0.21	&	&	0.00			&	…			&	…			&	&	1.00	$\pm$	0.06	&	3394	$\pm$	361	&	54	$\pm$	58	&	&	…			&	…			&	…			&	&	0.37	$\pm$	0.04	&	&	1.00	$\pm$	0.32	&	4581	$\pm$	320	&	-3144	$\pm$	1051	&	&	0.00			&	…			&	…			\\
B1422+231	&	5136	$\pm$	452	&	0.00	$\pm$	0.06	&	0.33	$\pm$	0.01	&	-1	$\pm$	422	&	-1	$\pm$	89	&	0	$\pm$	139	&	0	$\pm$	169	&	&	3.20	$\pm$	0.38	&	&	0.00			&	…			&	…			&	&	1.00	$\pm$	0.06	&	5135	$\pm$	546	&	0	$\pm$	10	&	&	…			&	…			&	…			&	&	0.57	$\pm$	0.07	&	&	0.00			&	…			&	…			&	&	1.00	$\pm$	0.32	&	1113	$\pm$	118	&	0	$\pm$	10	\\
SDSSJ161458.33+144836.9	&	3885	$\pm$	349	&	-0.06	$\pm$	0.11	&	0.33	$\pm$	0.01	&	-241	$\pm$	327	&	-98	$\pm$	73	&	-51	$\pm$	106	&	-38	$\pm$	130	&	&	1.89	$\pm$	0.21	&	&	0.09	$\pm$	0.03	&	7224	$\pm$	506	&	-1713	$\pm$	571	&	&	0.91	$\pm$	0.06	&	3721	$\pm$	396	&	0	$\pm$	10	&	&	…			&	…			&	…			&	&	0.54	$\pm$	0.06	&	&	0.70	$\pm$	0.25	&	2445	$\pm$	170	&	-1709	$\pm$	569	&	&	0.30	$\pm$	0.09	&	439	$\pm$	47	&	0	$\pm$	10	\\
PKS 1937-101	&	5298	$\pm$	466	&	0.00	$\pm$	0.07	&	0.33	$\pm$	0.01	&	56	$\pm$	434	&	56	$\pm$	92	&	56	$\pm$	144	&	56	$\pm$	174	&	&	5.15	$\pm$	0.62	&	&	0.00			&	…			&	…			&	&	1.00	$\pm$	0.06	&	5296	$\pm$	563	&	56	$\pm$	92	&	&	…			&	…			&	…			&	&	1.57	$\pm$	0.19	&	&	0.29	$\pm$	0.09	&	2618	$\pm$	183	&	-555	$\pm$	183	&	&	0.71	$\pm$	0.22	&	852	$\pm$	91	&	-47	$\pm$	-77	\\
PKS 2000-330	&	3230	$\pm$	294	&	-0.06	$\pm$	0.07	&	0.32	$\pm$	0.01	&	-175	$\pm$	306	&	-30	$\pm$	61	&	1	$\pm$	89	&	9	$\pm$	108	&	&	1.92	$\pm$	0.22	&	&	0.05	$\pm$	0.01	&	5499	$\pm$	385	&	-3127	$\pm$	1031	&	&	0.95	$\pm$	0.06	&	3138	$\pm$	334	&	28	$\pm$	54	&	&	…			&	…			&	…			&	&	2.31	$\pm$	0.28	&	&	0.36	$\pm$	0.11	&	1499	$\pm$	105	&	-599	$\pm$	197	&	&	0.63	$\pm$	0.20	&	1082	$\pm$	115	&	-46	$\pm$	-89	\\
SDSSJ210524.49+000407.3	&	5026	$\pm$	446	&	0.00	$\pm$	0.06	&	0.33	$\pm$	0.01	&	-1	$\pm$	420	&	0	$\pm$	85	&	0	$\pm$	137	&	0	$\pm$	169	&	&	1.28	$\pm$	0.15	&	&	0.00			&	…			&	…			&	&	1.00	$\pm$	0.06	&	5026	$\pm$	534	&	0	$\pm$	10	&	&	…			&	…			&	…			&	&	0.40	$\pm$	0.05	&	&	0.00			&	…			&	…			&	&	1.00	$\pm$	0.32	&	1407	$\pm$	150	&	0	$\pm$	10	\\
SDSSJ210831.56-063022.5	&	4545	$\pm$	200	&	0.00	$\pm$	0.04	&	0.33	$\pm$	0.01	&	93	$\pm$	372	&	93	$\pm$	79	&	93	$\pm$	124	& 93	$\pm$	149	&	&	0.60	$\pm$	0.08	&	&	0.00			&	…			&	…			&	&	0.91	$\pm$	0.06	&	4543	$\pm$	483	&	93	$\pm$	16	&	&	…			&	…			&	…			&	&	0.68	$\pm$	0.08	&	&	0.89	$\pm$	0.28	&	2597	$\pm$	364	&	-2018	$\pm$	347	&	&	0.11	$\pm$	0.03	&	1403	$\pm$	149	&	0	$\pm$	10	\\
SDSSJ212329.46-005052.9	&	4165	$\pm$	366	&	0.00	$\pm$	0.03	&	0.33	$\pm$	0.01	&	-1	$\pm$	342	&	-1	$\pm$	72	&	0	$\pm$	200	&	-1	$\pm$	231	&	&	1.60	$\pm$	0.22	&	&	0.00			&	…			&	…			&	&	0.89	$\pm$	0.05	&	4164	$\pm$	443	&	0	$\pm$	10	&	&	…			&	…			&	…			&	&	2.40	$\pm$	0.29	&	&	0.86	$\pm$	0.27	&	4969	$\pm$	696	&	-1844	$\pm$	323	&	&	0.14	$\pm$	0.04	&	1188	$\pm$	126	&	0	$\pm$	10	\\
SDSSJ235808.54+012507.2	&	4098	$\pm$	360	&	0.03	$\pm$	0.06	&	0.35	$\pm$	0.01	&	0	$\pm$	336	&	0	$\pm$	70	&	0	$\pm$	111	&	-1	$\pm$	136	&	&	2.88	$\pm$	0.35	&	&	0.00			&	…			&	…			&	&	1.00	$\pm$	0.06	&	4098	$\pm$	436	&	0	$\pm$	10	&	&	…			&	…			&	…			&	&	3.76	$\pm$	0.45	&	&	0.38	$\pm$	0.12	&	2463	$\pm$	172	&	-1030	$\pm$	340	&	&	0.61	$\pm$	0.19	&	954	$\pm$	101	&	0	$\pm$	10	\\

\noalign{\smallskip}
            \hline
            \noalign{\smallskip}
Median & 3991 $\pm$ 1141 & 0.00 $\pm$ 0.06 & 0.33 $\pm$ 0.01 & -1 $\pm$ 205 & -0.5 $\pm$ 61 & 0 $\pm$ 27 & 0 $\pm$ 51 & &	1.72 $\pm$ 0.65 &	& 0.08 $\pm$ 0.02 &	6362 $\pm$ 2533  & -1949 $\pm$ 660 & &	0.93 $\pm$ 0.09 & 3909 $\pm$ 1305 & 0$\pm$ 34 & & … & … & … & & 1.31 $\pm$ 1.43 & & 0.78 $\pm$ 0.54 & 2607 $\pm$ 2219 & -1710 $\pm$ 965 & & 0.67 $\pm$ 0.62 & 1018 $\pm$ 312 & 0 $\pm$ 16\\
\noalign{\smallskip}
            %\hline
          
            \hline
            \noalign{\smallskip}
            \multicolumn{24}{c}{Population B}\\
            \noalign{\smallskip}
            \hline
            \noalign{\smallskip}
HE 0001-2340	&	6632	$\pm$	1065	&	0.16	$\pm$	0.22	&	0.37	$\pm$	0.01	&	1757	$\pm$	873	&	1134	$\pm$	153	&	960	$\pm$	354	&	909	$\pm$	471	&	&	1.30	$\pm$	0.16	&	&	0.00			&	…			&	…			&	&	0.42	$\pm$	0.06	&	5063	$\pm$	366	&	681	$\pm$	92	&	&	0.58	$\pm$	0.06	&	10982	$\pm$	1955	&	2538	$\pm$	342	&	&	0.33	$\pm$	0.04	&	&	0.34	$\pm$	0.11	&	3772	$\pm$	264	&	-414	$\pm$	136	&	&	0.65	$\pm$	0.21	&	1067	$\pm$	77	&	0	$\pm$	10$^{(d)}$	\\
$[\rm HB89]$ 0029+073	&	6347	$\pm$	1085	&	0.18	$\pm$	0.05	&	0.37	$\pm$	0.01	&	932	$\pm$	757	&	292	$\pm$	138	&	51	$\pm$	339	&	-19	$\pm$	442	&	&	1.54	$\pm$	0.18	&	&	0.07	$\pm$	0.01	&	4995	$\pm$	837	&	-1216	$\pm$	575	&	&	0.33	$\pm$	0.05	&	4508	$\pm$	326	&	-310	$\pm$	147	&	&	0.60	$\pm$	0.06	&	9609	$\pm$	1710	&	1430	$\pm$	193	&	&	1.54	$\pm$	0.18	&	&	0.00			&	…			&	…			&	&	1.00	$\pm$	0.32	&	1024	$\pm$	74	&	0	$\pm$	10	\\
CTQ 408	&	7301	$\pm$	1081	&	0.10	$\pm$	0.09	&	0.40	$\pm$	0.02	&	724	$\pm$	910	&	306	$\pm$	179	&	197	$\pm$	390	&	164	$\pm$	528	&	&	3.90	$\pm$	0.47	&	&	0.00			&	…			&	…			&	&	0.52	$\pm$	0.08	&	6050	$\pm$	437	&	0	$\pm$	10	&	&	0.48	$\pm$	0.05	&	13243	$\pm$	2357	&	1963	$\pm$	264	&	&	2.54	$\pm$	0.30	&	&	0.00			&	…			&	…			&	&	1.00	$\pm$	0.32	&	1500	$\pm$	108	&	0	$\pm$	10	\\
H 0055-2659	&	5342	$\pm$	925	&	0.36	$\pm$	0.13	&	0.30	$\pm$	0.03	&	2249	$\pm$	1085	&	580	$\pm$	101	&	380	$\pm$	283	&	329	$\pm$	377	&	&	1.85	$\pm$	0.22	&	&	0.00			&	…			&	…			&	&	0.27	$\pm$	0.04	&	4609	$\pm$	333	&	-224	$\pm$	39	&	&	0.73	$\pm$	0.07	&	13355	$\pm$	2377	&	2527	$\pm$	340	&	&	0.82	$\pm$	0.10	&	&	0.00			&	…			&	…			&	&	1.00	$\pm$	0.32	&	1500	$\pm$	108	&	0	$\pm$	10	\\
SDSSJ114358.52+052444.9	&	5999	$\pm$	942	&	0.24	$\pm$	0.15	&	0.34	$\pm$	0.05	&	1485	$\pm$	1408	&	354	$\pm$	133	&	210	$\pm$	319	&	170	$\pm$	431	&	&	1.00	$\pm$	0.12	&	&	0.00			&	…			&	…			&	&	0.52	$\pm$	0.08	&	5007	$\pm$	362	&	0	$\pm$	10	&	&	0.48	$\pm$	0.05	&	13341	$\pm$	2375	&	4095	$\pm$	552	&	&	0.34	$\pm$	0.04	&	&	0.47	$\pm$	0.15	&	1830	$\pm$	128	&	-1277	$\pm$	421	&	&	0.53	$\pm$	0.17	&	1304	$\pm$	94	&	0	$\pm$	10$^{(d)}$	\\
SDSSJ115954.33+201921.1	&	8006	$\pm$	1659	&	0.29	$\pm$	0.10	&	0.33	$\pm$	0.01	&	2389	$\pm$	1025	&	950	$\pm$	133	&	469	$\pm$	422	&	359	$\pm$	536	&	&	2.05	$\pm$	0.25	&	&	0.00			&	…			&	…			&	&	0.36	$\pm$	0.05	&	5458	$\pm$	395	&	-35	$\pm$	15	&	&	0.64	$\pm$	0.06	&	12924	$\pm$	2300	&	3287	$\pm$	443	&	&	0.42	$\pm$	0.05	&	&	0.72	$\pm$	0.23	&	2660	$\pm$	186	&	-1215	$\pm$	401	&	&	0.27	$\pm$	0.08	&	995	$\pm$	72	&	-101	$\pm$	14	\\
SDSSJ120147.90+120630.2	&	7995	$\pm$	1748	&	0.28	$\pm$	0.11	&	0.33	$\pm$	0.01	&	2331	$\pm$	947	&	1124	$\pm$	140	&	540	$\pm$	421	&	414	$\pm$	519	&	&	4.31	$\pm$	0.52	&	&	0.00			&	…			&	…			&	&	0.30	$\pm$	0.04	&	4986	$\pm$	360	&	9	$\pm$	10	&	&	0.70	$\pm$	0.07	&	12190	$\pm$	2170	&	2850	$\pm$	384	&	&	6.33	$\pm$	0.76	&	&	0.05	$\pm$	0.01	&	1879	$\pm$	131	&	-246	$\pm$	81	&	&	0.95	$\pm$	0.30	&	1879	$\pm$	136	&	-1	$\pm$	10	\\
SDSSJ141546.24+112943.4	&	7253	$\pm$	1237	&	0.23	$\pm$	0.11	&	0.36	$\pm$	0.02	&	1769	$\pm$	952	&	736	$\pm$	145	&	447	$\pm$	387	&	367	$\pm$	510	&	&	1.59	$\pm$	0.19	&	&	0.00			&	…			&	…			&	&	0.45	$\pm$	0.07	&	5544	$\pm$	401	&	14	$\pm$	13	&	&	0.55	$\pm$	0.05	&	11523	$\pm$	2051	&	3144	$\pm$	423	&	&	1.31	$\pm$	0.16	&	&	0.29	$\pm$	0.09	&	1000	$\pm$	70	&	-784	$\pm$	258	&	&	0.71	$\pm$	0.22	&	1162	$\pm$	84	&	14	$\pm$	13	\\
SDSSJ153830.55+085517.0	&	7816	$\pm$	1354	&	0.25	$\pm$	0.11	&	0.35	$\pm$	0.02	&	2090	$\pm$	1081	&	800	$\pm$	152	&	478	$\pm$	415	&	390	$\pm$	548	&	&	1.16	$\pm$	0.14	&	&	0.00			&	…			&	…			&	&	0.47	$\pm$	0.07	&	6038	$\pm$	437	&	9	$\pm$	12	&	&	0.53	$\pm$	0.05	&	12642	$\pm$	2250	&	3869	$\pm$	521	&	&	0.50	$\pm$	0.06	&	&	0.00			&	…			&	…			&	&	1.00	$\pm$	0.32	&	1000	$\pm$	72	&	-1	$\pm$	10	\\
PKS 2126-15$^{(e)}$	&	5018	$\pm$	648	&	0.50	$\pm$	0.09	&	0.22	$\pm$	0.01	&	2104	$\pm$	414	&	81	$\pm$	51	&	-303	$\pm$	168	&	-713	$\pm$	90	&	&	4.28	$\pm$	0.51	&	&	0.09	$\pm$	0.02	&	4662 $\pm$ 781	&	-4717 $\pm$ 2970		&	&	0.35	$\pm$	0.05	&	4074	$\pm$	295	&	-589	$\pm$	371	&	&	0.56	$\pm$	0.05	&	9992	$\pm$	1779	&	3585	$\pm$	483	&	&	3.34	$\pm$	0.40	&	&	0.17	$\pm$	0.05	&	1345	$\pm$	94	&	-426	$\pm$	140	&	&	0.83	$\pm$	0.26	&	922	$\pm$	67	&	0	$\pm$	10	\\

\noalign{\smallskip}
    \hline
    \noalign{\smallskip}
Median & 6942 $\pm$ 1601 & 0.24 $\pm$ 0.09 & 0.34 $\pm$ 0.03  & 1929 $\pm$ 659 & 658 $\pm$ 594 & 413 $\pm$ 275 & 344 $\pm$ 218 & & 1.72 $\pm$ 2.07 & & 0.08 $\pm$ 0.01 & 4828 $\pm$ 166 & -2966 $\pm$ 1750 & & 0.39 $\pm$ 0.13 & 5035 $\pm$ 819 & 0 $\pm$ 186 & & 0.57 $\pm$ 0.09 & 12416 $\pm$ 2046 & 2997 $\pm$ 980 & & 1.06 $\pm$ 1.85 & & 0.31 $\pm$ 0.23 & 1854 $\pm$ 998  & -605 $\pm$ 690 & & 0.89 $\pm$ 0.33 & 1114 $\pm$ 445 & 0 $\pm$ 1\\
\noalign{\smallskip}
            \hline
              \noalign{\smallskip}
         \end{tabular}
}
    
     \tablefoot{$^{(a)}$ In these columns, we report additional blueshifted components that were included in the H$\beta$ profile.
     $^{(b)}$ In units of $10^{-13}$ erg s$^{-1}$ cm$^{-2}$.
     $^{(c)}$ In units of $10^{-14}$ erg s$^{-1}$ cm$^{-2}$.
     $^{(d)}$ There is an additional contribution due to small blueshifted components associated with narrow line emission.
     $^{(e)}$ Full profile centroid velocities and the presence of a BLUE H$\beta$ component in this case are affected by the location of the continuum.}
   \end{sidewaystable*}

\subsection{Optical}
\subsubsection{H$\beta$}

\par Appendix \ref{app_uv} shows the full VLT-ISAAC optical spectra and their respective H$\beta$+[\ion{O}{III}]$\lambda\lambda 4959,5007$ fittings. Spectrophotometric measurements on H$\beta$  are presented in Table \ref{tab:specphot_HB}. The ST of each source is listed in Col. 2; the specific continuum flux at 5100\r{A} (in rest-frame) is in Col. 3; equivalent width (EW) of the \hb{} and \oiii{} full profiles in Cols. 4 and 5, respectively; and the \ion{Fe}{II} prominence parameter ($R_{\ion{Fe}{ii}}$) is listed in Col. 7.  

\par Table \ref{tab:specfit_HB} reports  measurements on the H$\beta$ profile. First we  present the parameters obtained through the analysis of the H$\beta$ full broad profile, including FWHM(H$\beta_{\rm{full}}$) (Col. 2), asymmetry index (Col. 3), kurtosis (Col. 4), and the centroid velocity shifts at $\frac{1}{4}$, $\frac{1}{2}$, $\frac{3}{4}$, and $\frac{9}{10}$ fractional intensity (Cols. 5, 6, 7, and 8, respectively). In Col. 9 we list the full H$\beta$ line flux (i.e., the flux for all broad line components, BC and VBC and BLUE whenever appropriate). For each broad component isolated with the \textsc{specfit} analysis we report, from Col. 10 to 18, flux normalised by the total flux ($I/I_{\mathrm{tot}}$), FWHM, and velocity shift. Col. 19 shows the total flux of the narrow profile (SBC+NC) and from Cols. 20 to 25 we report $I/I_{\mathrm{tot}}$, FWHM, and shift for these components. Additionally, we provide, in the last row of Pop. A and B respectively, the  median values of the measurements, together with the interquartile range. 

\par The centroid velocities are close to the rest-frame wavelength for the majority of Pop. A and only two of the Pop. A quasars present $c$(1/2) strongly shifted towards the blue with an averaged value of $\approx -271$ km s$^{-1}$. These sources are SDSSJ005700-18+14737.7 (ST A3) and  SDSSJ132012.33+142037.1 (A1), both of them presenting a clear blueshifted contribution in H$\beta$ profile. In the case of Pop. B quasars, the centroid velocities are significantly shifted towards the red wing of the profile in every fractional intensity for the majority of the sources, with an averaged $c$(1/2) value of $\sim 640$ \kms{}. Even higher values are found for $c$(1/4): Pop. B present an averaged value of $\approx$1780 km s$^{-1}$ while Pop. A have $c$(1/4) $\approx$ -149 \kms{}. %The more remarkable object in this context is the Pop. B PKS 2126-15, which present a very peculiar \hb{} profile, with a peak shifted by more than 700 \kms{}. 

\par Pop. B sources show FWHM(H$\beta_{\rm{full}}$) values much larger than those of Pop. A, usually $\sim 6000$ km s$^{-1}$. This difference is a direct consequence of the definition of Pop. A and B.  The Pop. B asymmetry index is positive and very significantly different from 0, since the contribution of the VBC  -- that  reaches FWHM values $\gtrsim$ 10000 km s$^{-1}$ -- in all cases  represents  $\gtrsim 50$\% of the full emission line profile. In other words, differently from Pop. A sources (in which only a symmetric BC is enough to represent the full profile in the majority of the cases), a second redshifted component is always needed to reproduce the strong red wing of the observed \hb\ profile. 

\begin{figure*}[ht!]
\includegraphics[width=0.325\linewidth]{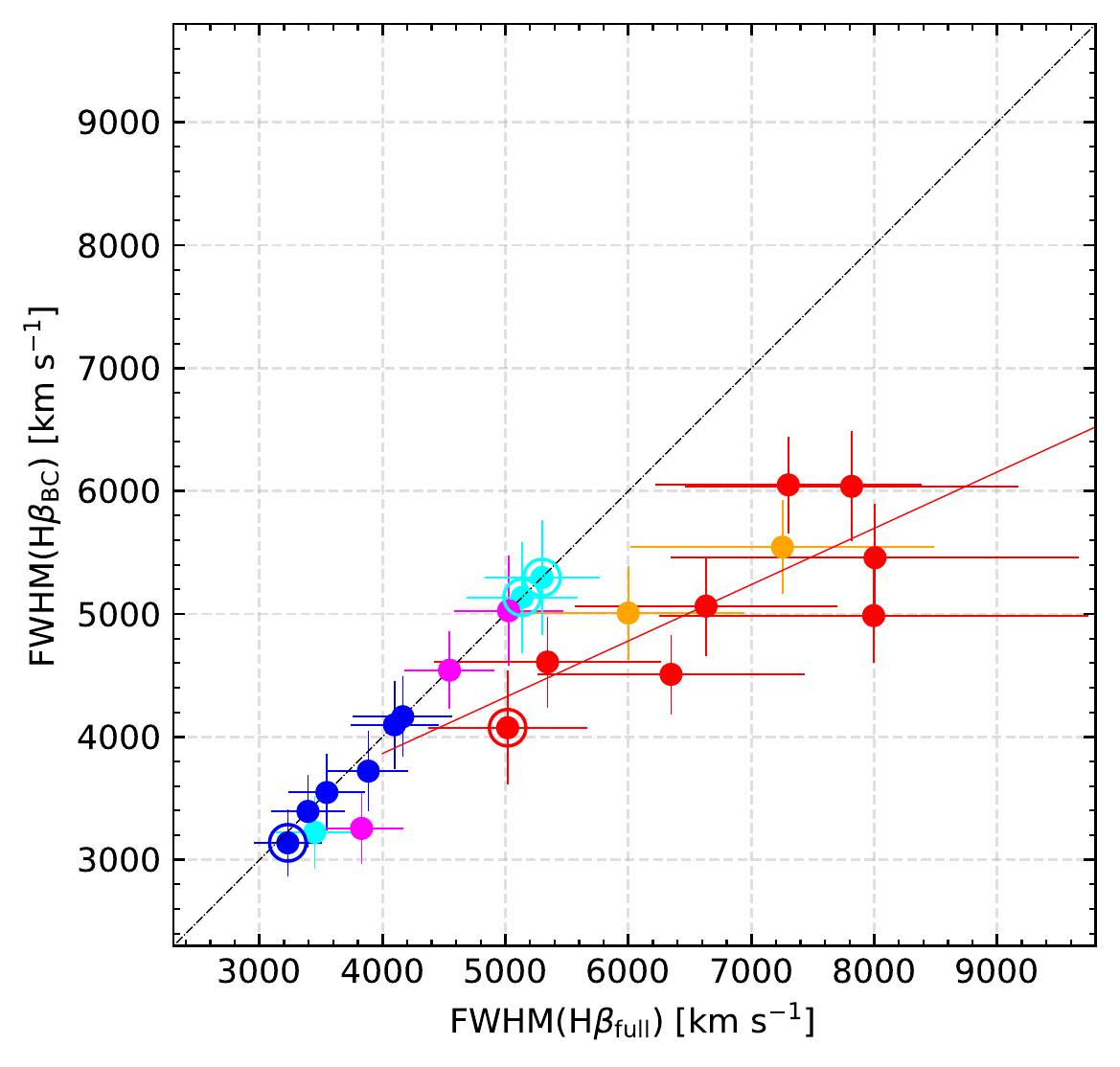}
\includegraphics[width=0.33\linewidth]{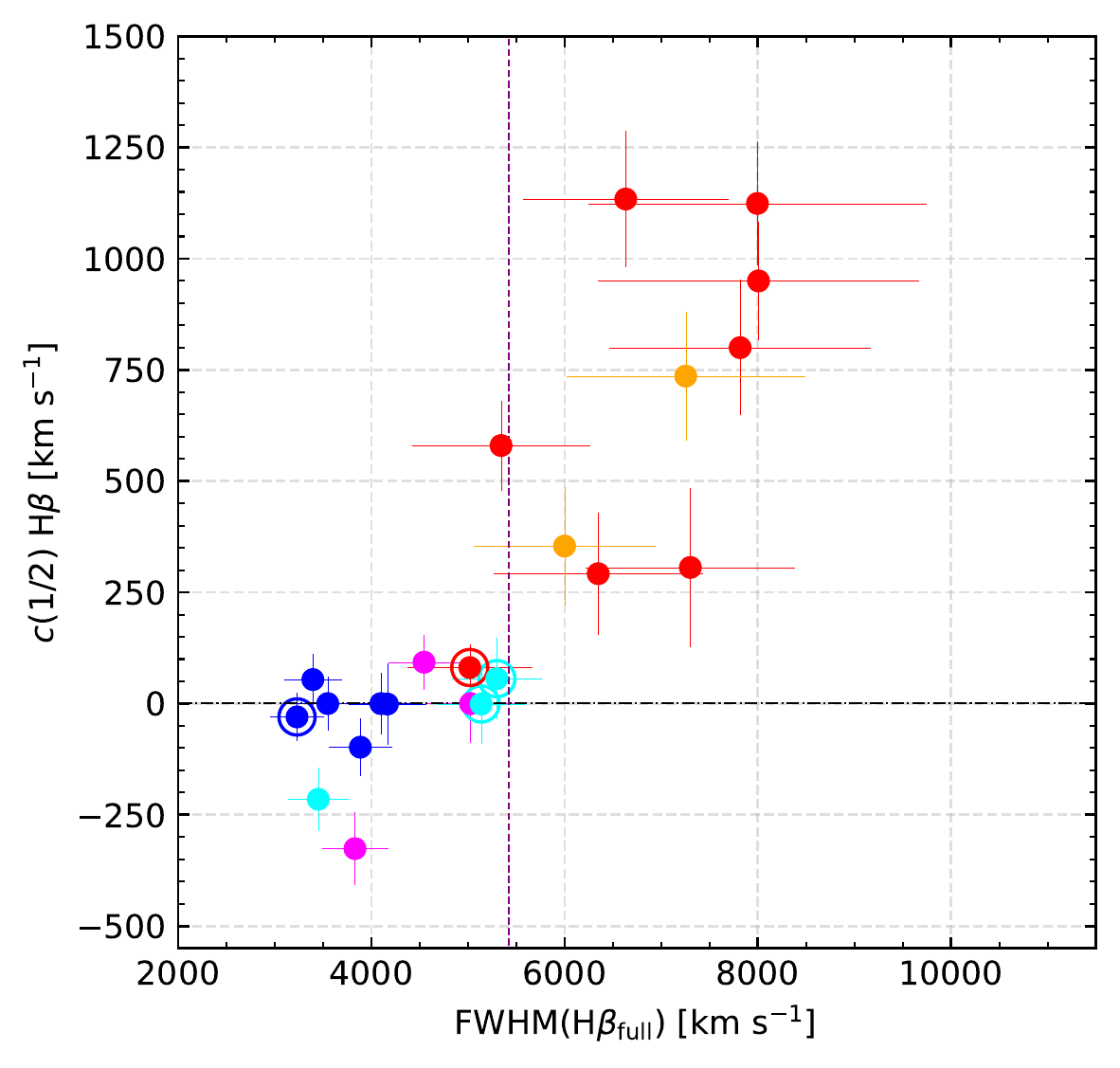}
\includegraphics[width=0.337\linewidth]{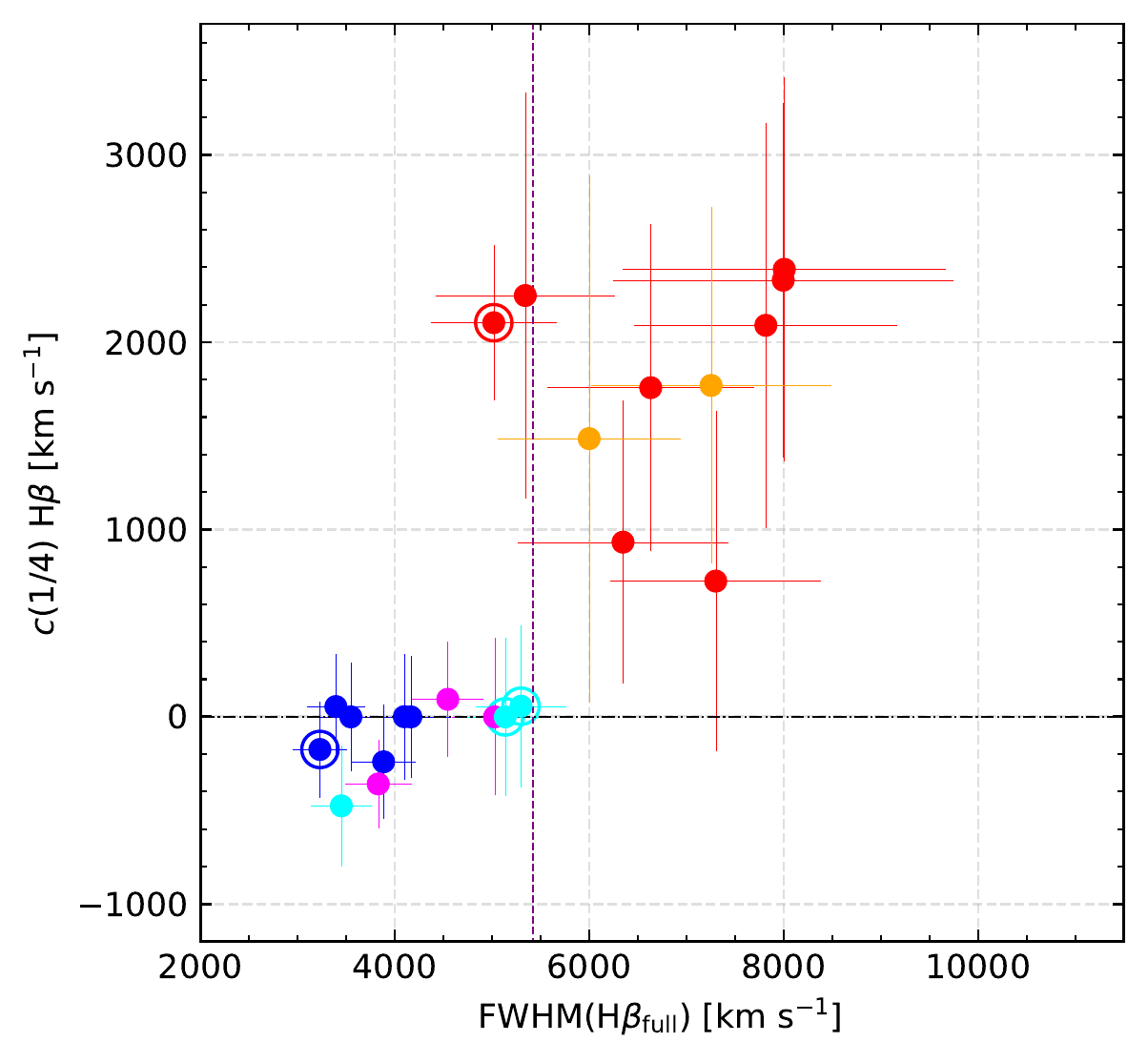}
\caption{\textit{Left panel: }FWHM(H$\beta_{\rm BC}$) \textit{vs.} FWHM of the full H$\beta$ profile. \textit{Central panel:} velocity centroid at 1/2 intensity $c$(1/2) \textit{vs.} FWHM(H$\beta_{\rm full}$). \textit{Right panel:} velocity centroid at 1/4 flux intensity $c$(1/4) \textit{vs.} FWHM(H$\beta_{\rm full}$). Symbols surrounded by open circles show the radio-loud sources and filled circles represent the radio-quiet from the sample. Red line represent the linear regression performed for Pop. B. Horizontal dashed lines in the central and right panels indicate $c$(1/4) $=0$ and $c$(1/2) $=0$ for \hb{}, respectively. Purple vertical dashed lines shows the boundary between Pop. A and Pop. B. The many MS spectral type (ST) are represented in different colours: A1 (\textcolor{cyan}{$\bullet$}); A2 (\textcolor{blue}{$\bullet$}); A3 (\textcolor{magenta}{$\bullet$}); B1 (\textcolor{red}{$\bullet$}); and B2 (\textcolor{orange}{$\bullet$}).}
\label{fig:HB_BC_Full}
\end{figure*}

\par  Semi-broad \hb\ blueshifts could be mainly associated with the \oiii{} SBC. However in some cases an additional BLUE H$\beta$ component is needed (see Table \ref{tab:specfit_HB}). PKS 2126-15  presents a huge BLUE blueshift  of $\approx -4700$ km s$^{-1}$, associated with a boxy termination of the H$\beta$ blue wing. A second peculiarity of this source is that the \hb\ BC is significantly shifted to the blue, down to half maximum, at variance with all other Pop. B sources.  
%In Table \ref{tab:specfit_HB}, we include only the blue \hb{} components that are additional to the one we imposed in correspondence to the one \oiii{}. 

\par The \hb\ NC is weak especially in Pop. A ($\lesssim 0.02$\ of the \hb\ line flux) although it is  observed in all cases save two Pop. A2 sources (SDSSJ132012.33+142037.1 and Q 1410+096). In Pop. B, the narrow component is somewhat stronger but always $\lesssim 0.05$\ of the line flux. Given its weakness,  the NC is therefore not significantly affecting the flux of any broad component measured in this paper. 

% In general, the narrow component reaches values $\lesssim$ 1000 km s$^{-1}$.

\paragraph{Relation between FWHM(H$\beta_{\rm{full}}$) and FWHM(H$\beta_{\rm{BC}}$).}   This work confirms previous studies \citep[e.g.][and references therein]{Sulentic_2006,Marziani_2009} that showed that the full \hb{} profile of Pop. A can be accounted for mostly by the BC, and that a redshifted VBC seems to be absent in Population A but present in Pop. B sources: Table \ref{tab:specfit_HB}  lists  centroid and asymmetry index values for Pop. B sources of $1000 \lesssim c(\frac{1}{4}) \lesssim 2000$ \kms\ and $\approx 0.2$, respectively. For Pop. A these values are $\lesssim 0$ \kms\ and 0.

The left plot of Fig. \ref{fig:HB_BC_Full} shows the relation between the FWHM of the BC and the FWHM of the full profile of H$\beta$ for the two populations. Pop. A and Pop. B sources  follow different trends for the ratio $\xi(\rm H\beta)=$ FWHM(H$\beta_{\rm BC}$)/FWHM(H$\beta_{\rm full}$). For Pop. A quasars, we obtain $\xi(\rm H\beta)= 1.00$ for all sources but four (SDSSJ005700.18+143737.7, SDSSJ132012.33+142037.1, SDSSJ161458.33+144836.9, and PKS2000-330, with $\xi(\rm H\beta)\approx 0.9$ due to the presence of a blueshifted component). The $\xi(\rm H\beta)$ ratio for Pop. B is only $\approx$ 0.76, indicating that the Pop. B \hb{} full profile is less representative of the BC than the Pop. A sources. Similar results were also shown by \cite{Marziani_2013}: they obtain $\xi(\rm H\beta)=1.00$ for Pop. A and $\sim 0.78$ for Pop. B quasars.
In our sample Pop. B sources  usually present a very strong and wide  VBC component that accounts for $\approx 0.57$ of the profile, with a mean FWHM of 11240 km s$^{-1}$.  Meanwhile, as mentioned, the Pop. A sources in general are well represented by only a BC.
This also seems to be true for the radio-loud Pop. A sources from the sample. %In these three cases, the radio-loud objects can be well represented by only a broad component or, if not, with a small contribution of blueshifted components. 

%The mean value of FWHM(H$\beta_{\rm{full}}$) for the Pop. B sources is 5959 km s$^{-1}$, larger than for the Pop. A (4097 km s$^{-1}$), as expected from the selection according to Eq. \ref{eq:ab}.  A similar behaviour is occurring in the BC, which presents a mean value of FWHM equal to 4154 km s$^{-1}$ for Pop. A and 4696 km s$^{-1}$ for Pop. B. As mentioned, the contribution of the H$\beta$ BC to the full profile is more significant in Pop. A than in Pop. B sources. 

\par Clear distinctions between Pop. A and Pop. B are also seen in the center and left panels of Fig. \ref{fig:HB_BC_Full}, which present the $c$(1/2) and $c$(1/4) \textit{vs.} FWHM(H$\beta_{\rm{full}}$) relations, respectively. Pop. A sources show in general no shift, within the uncertainties, or a negative value of velocity centroids in the case of the 4 sources with a BLUE component (with average centroid values at 1/2 and 1/4 of -167 and -312 km\ s$^{-1}$, respectively). On the other hand, Pop. B sources always present positive velocities for $c$(1/2) and $c$(1/4), with a mean value of $658$ and $1929$ km\ s$^{-1}$, respectively, as a consequence of a very strong VBC. The VBC individually has a median velocity shift of $\sim3000$ km s$^{-1}$, in a range from $-1500$ to $4000$ km s$^{-1}$. These results are in complete agreement with previous results \citep[e.g.][and references therein]{wolf_2020}.\\

%\textbf{NOT CLEAR: In these cases, five out of the twelve Pop. A sources usually show negative values of velocity centroids $c$(1/2) and c(1/4), with an average value of -142 and -107 km s$^{-1}$. This is a consequence of a significant contribution of blueshifted components to the full profile. However, the majority of the Pop. A quasars of our sample do not present blueshifted components and consequently do not present shifts ($c$(1/2) $\sim 0$). The cases in which Pop. A sources present $c$(1/2) and/or $c$(1/4) $>0$ are within the uncertainties in the rest-frame of the sources. A case in point is Q 1410+096 (which has a $c$(1/2) $=54$ km s$^{-1}$). On the other hand, Pop. B sources always present $c$(1/2) $> 0$, with a mean value $c$(1/4) $=1419$ and $c$(1/2) $= 454$ km s$^{-1}$ as a consequence of very strong VBC. In Fig. \ref{fig:HB_BC_Full}, the radio-loud quasars seem to share the characteristics of the population they belong (i.e., they are located in the same area of the parameter planes as radio-quiet). }\chonyblue{The three Pop A RL have no shift  or positive but well inside the uncertainties, and the Pop B RL is the one with less redshifted $c$(1/2) and $c$(1/4) but it is the one with the problem of the location of the continuum that produce a probably "false" BLUE component in H$\beta$. Better wait to the RL sample - paper II to see the behaviour}

 \begin{table*}[ht!]
      \caption[]{Measurements on the [\ion{O}{III}]$\lambda5007$ full line profile.}
         \label{tab:spec_OIII}
      \resizebox{\linewidth}{!}{
                 \begin{tabular}{lccccccccccccc}
            \hline
            \hline
            \noalign{\smallskip}
             & \multicolumn{5}{c}{[\ion{O}{III}]$\lambda5007$ full profile} & & \multicolumn{3}{c}{SBC} & & \multicolumn{3}{c}{NC}\\
              \cline{2-6} \cline{8-10} \cline{12-14}
              \noalign{\smallskip}
            Source &  FWHM &  A. I. & Kurtosis & $c$(1/2) & $c$(9/10) & & $I/I_{\mathrm{tot}}$ & FWHM & Shift & & $I/I_{\mathrm{tot}}$ & FWHM & Shift\\
             & [km s$^{-1}$] &  &   &  [km s$^{-1}$] & [km s$^{-1}$] &  & & [km s$^{-1}$] & [km s$^{-1}$] &  &  & [km s$^{-1}$] & [km s$^{-1}$]\\
             (1) & (2) & (3) & (4) &  (5) & (6) & & (7) & (8) & (9) & & (10) & (11) & (12)\\
            \noalign{\smallskip}
            \hline
            \noalign{\smallskip}
            \multicolumn{14}{c}{Population A}\\
            \noalign{\smallskip}
            \hline
            \noalign{\smallskip}
        SDSSJ005700.18+143737.7$^{(a)}$	&	764::		&	-0.01::		&	0.45::		&	-89::		&	-88::		&	&	0.27::		&	1214::		&	-1856::		&	&	0.732::		&	761::		&	-87::	\\
SDSSJ132012.33+142037.1	&	2124	$\pm$	147	&	-0.54	$\pm$	0.01	&	0.23	$\pm$	0.01	&	-770	$\pm$	62	&	-221	$\pm$	36	&	&	0.64	$\pm$	0.06	&	2275	$\pm$	235	&	-1429	$\pm$	115	&	&	0.357	$\pm$	0.114	&	895	$\pm$	95	&	-149	$\pm$	22	\\
SDSSJ135831.78+050522.8	&	4320	$\pm$	297	&	-0.43	$\pm$	0.05	&	0.19	$\pm$	0.10	&	-1895	$\pm$	91	&	-630	$\pm$	59	&	&	0.90	$\pm$	0.08	&	5422	$\pm$	561	&	-2080	$\pm$	100	&	&	0.096	$\pm$	0.031	&	1000	$\pm$	106	&	-534	$\pm$	80	\\
Q 1410+096	&	3363	$\pm$	283	&	-0.52	$\pm$	0.01	&	0.28	$\pm$	0.01	&	-1404	$\pm$	107	&	-656	$\pm$	72	&	&	0.73	$\pm$	0.07	&	5573	$\pm$	577	&	-1255	$\pm$	96	&	&	0.267	$\pm$	0.085	&	1379	$\pm$	147	&	-260	$\pm$	39	\\
B1422+231	&	1535	$\pm$	99	&	-0.18	$\pm$	0.01	&	0.43	$\pm$	0.01	&	-217	$\pm$	26	&	-88	$\pm$	47	&	&	0.32	$\pm$	0.03	&	1128	$\pm$	117	&	-850	$\pm$	102	&	&	0.680	$\pm$	0.218	&	1117	$\pm$	119	&	27	$\pm$	4	\\
SDSSJ161458.33+144836.9	&	2616	$\pm$	160	&	-0.49	$\pm$	0.01	&	0.28	$\pm$	0.02	&	-1452	$\pm$	54	&	-800	$\pm$	50	&	&	0.75	$\pm$	0.07	&	2759	$\pm$	286	&	-1875	$\pm$	70	&	&	0.245	$\pm$	0.078	&	1006	$\pm$	107	&	-667	$\pm$	100	\\
PKS 1937-101	&	1046	$\pm$	79	&	-0.15	$\pm$	0.09	&	0.35	$\pm$	0.02	&	-155	$\pm$	17	&	-135	$\pm$	31	&	&	0.57	$\pm$	0.05	&	2657	$\pm$	275	&	-484	$\pm$	53	&	&	0.432	$\pm$	0.138	&	832	$\pm$	88	&	-120	$\pm$	18	\\
PKS 2000-330	&	1314	$\pm$	89	&	-0.13	$\pm$	0.01	&	0.42	$\pm$	0.01	&	-425	$\pm$	22	&	-371	$\pm$	41	&	&	0.42	$\pm$	0.04	&	1500	$\pm$	155	&	-827	$\pm$	43	&	&	0.578	$\pm$	0.185	&	1082	$\pm$	115	&	-263	$\pm$	39	\\
SDSSJ210524.49+000407.3	&	1365::		&	0.00::		&	0.45::		&	201::		&	202::		&	&	1.00::		&	1361::		&	203::		&	&	0.000		&	…			&	…			\\
SDSSJ210831.56-063022.5	&	3665::	&	0.00::		&	0.46::		&	-1243::		&	-1242::		&	&	1.00::		&	3679::		&	-1244::		&	&	0.000	&	…	&	…	\\
SDSSJ212329.46-005052.9	&	3656	$\pm$	247	&	-0.04	$\pm$	0.04	&	0.39	$\pm$	0.02	&	-2885	$\pm$	58	&	-2584	$\pm$	119	&	&	0.88	$\pm$	0.08	&	4719	$\pm$	488	&	-2376	$\pm$	48	&	&	0.115::		&	2043::		&	174::		\\
SDSSJ235808.54+012507.2	&	1870	$\pm$	224	&	0.00	$\pm$	0.03	&	0.46	$\pm$	0.01	&	-943	$\pm$	78	&	-942	$\pm$	49	&	&	1.00	$\pm$	0.09	&	2464	$\pm$	255	&	-1053	$\pm$	87	&	&	0.000	&	…			&	…\\			

\noalign{\smallskip}
    \hline
    \noalign{\smallskip}
Median & 1997 $\pm$ 2045 & -0.14 $\pm$ 0.42 & 0.40 $\pm$ 0.17 & -856 $\pm$ 1214 & -500 $\pm$ 707 & & 0.68 $\pm$ 0.49  & 2560 $\pm$ 2032 & -1249 $\pm$ 1016 & & 0.312 $\pm$ 0.493 & 1082 $\pm$ 427 & -120 $\pm$ 288\\
\noalign{\smallskip}
            \hline
            \noalign{\smallskip}
            \multicolumn{14}{c}{Population B}\\
            \noalign{\smallskip}
            \hline
            \noalign{\smallskip}
    HE 0001-2340	&	1029	$\pm$	145	&	-0.07	$\pm$	0.10	&	0.40	$\pm$	0.02	&	-24	$\pm$	30	&	-12	$\pm$	64	&	&	0.43	$\pm$	0.03	&	2768	$\pm$	90	&	-443	$\pm$	554	&	&	0.565	$\pm$	0.202	&	894	$\pm$	65	&	-1	$\pm$	10	\\
$[\rm HB89]$ 0029+073	&	1866	$\pm$	271	&	-0.47	$\pm$	0.01	&	0.38	$\pm$	0.01	&	-922	$\pm$	97	&	-562	$\pm$	114	&	&	1.00	$\pm$	0.06	&	3431	$\pm$	168	&	-397	$\pm$	42	&	&	0.000	&	…			&	…			\\
CTQ 408	&	1497	$\pm$	190	&	0.00	$\pm$	0.01	&	0.46	$\pm$	0.01	&	-524	$\pm$	37	&	-524	$\pm$	97	&	&	1.00	$\pm$	0.09	&	1500	$\pm$	401	&	-524	$\pm$	37	&	&	0.000	&	…			&	…			\\
H 0055-2659	&	1405	$\pm$	325	&	-0.47	$\pm$	0.06	&	0.25	$\pm$	0.01	&	-198	$\pm$	103	&	-91	$\pm$	81	&	&	0.50	$\pm$	0.07	&	3204	$\pm$	412	&	-599	$\pm$	312	&	&	0.496	$\pm$	0.177	&	912	$\pm$	66	&	69	$\pm$	10	\\
SDSSJ114358.52+052444.9	&	2577	$\pm$	251	&	0.05	$\pm$	0.03	&	0.54	$\pm$	0.02	&	-968	$\pm$	52	&	-1079	$\pm$	186	&	&	0.78	$\pm$	0.03	&	1982	$\pm$	83	&	-1295	$\pm$	70	&	&	0.220	$\pm$	0.078	&	1304	$\pm$	94	&	-1	$\pm$	10	\\
SDSSJ115954.33+201921.1	&	1233	$\pm$	186	&	-0.20	$\pm$	0.10	&	0.37	$\pm$	0.03	&	-347	$\pm$	46	&	-302	$\pm$	76	&	&	0.55	$\pm$	0.07	&	2684	$\pm$	204	&	-499	$\pm$	66	&	&	0.448	$\pm$	0.160	&	995	$\pm$	72	&	-261	$\pm$	39	\\
SDSSJ120147.90+120630.2	&	2584	$\pm$	200	&	-0.10	$\pm$	0.01	&	0.61	$\pm$	0.01	&	-594	$\pm$	49	&	-504	$\pm$	126	&	&	0.72	$\pm$	0.05	&	1879	$\pm$	196	&	-978	$\pm$	81	&	&	0.280	$\pm$	0.100	&	1129	$\pm$	82	&	244	$\pm$	37	\\
SDSSJ141546.24+112943.4	&	1193	$\pm$	204	&	-0.37	$\pm$	0.01	&	0.38	$\pm$	0.01	&	-116	$\pm$	70	&	55	$\pm$	75	&	&	0.47	$\pm$	0.04	&	1499	$\pm$	111	&	-415	$\pm$	250	&	&	0.526	$\pm$	0.188	&	700	$\pm$	51	&	151	$\pm$	23	\\
SDSSJ153830.55+085517.0	&	1520	$\pm$	192	&	0.00	$\pm$	0.01	&	0.46	$\pm$	0.01	&	-545	$\pm$	38	&	-545	$\pm$	97	&	&	1.00	$\pm$	0.09	&	1523	$\pm$	101	&	-545	$\pm$	38	&	&	0.000	&	…			&	…			\\
PKS 2126-15	&	1212	$\pm$	167	&	-0.22	$\pm$	0.03	&	0.41	$\pm$	0.01	&	-441	$\pm$	44	&	-355	$\pm$	75	&	&	0.30	$\pm$	0.09	&	1332	$\pm$	272	&	-988	$\pm$	99	&	&	0.702	$\pm$	0.250	&	937	$\pm$	68	&	-285	$\pm$	43	\\

\noalign{\smallskip}
  \hline
  \noalign{\smallskip}
Median & 1451 $\pm$ 562 & -0.15 $\pm$ 0.31 & 0.40 $\pm$ 0.08 & -482 $\pm$ 346 & -429 $\pm$ 396 & & 0.52 $\pm$ 0.32 & 1930 $\pm$ 1241 & -534 $\pm$ 426 & & 0.364 $\pm$ 0.463 & 937 $\pm$ 159 & -1 $\pm$ 241\\
\noalign{\smallskip}
 \hline
            \noalign{\smallskip}
           
         \end{tabular}
}
    
    \tablefoot{$^{a}$ [\ion{O}{iii}]$\lambda$5007 location is not clear for this source since it is placed at the red end of the observed spectrum. In this case, the measurements were performed with [\ion{O}{iii}]$\lambda$4959.}
   \end{table*}

\subsubsection{[\ion{O}{III}]$\lambda\lambda 4959,5007$}

\par High-ionisation lines like \oiii{} are seen as one of the main detectors of outflowing gas in radio-quiet sources \citep{zamanovetal02,Komossa_2008,zhang_2011,Marziani_2016}.  In the case of radio-loud sources, narrow-line outflowing gas has been associated with jets through the blueshifted line components \citep{Capetti_1996, Axon_2000, Bicknell_2002, Kauffmann_2008, Best_2012, Reynaldi_2013, Jarvis_2019, Berton_2021}.  Outflows are also  detected in \hb{}. Usually, the blueshifted components on the H$\beta$ profile  are related to those found for the \oiii{} lines  \citep{Carniani_2015,Cresci_2015,Brusa_2015,Marziani_2022}. %Nevertheless, we have six cases in which we do not detect the BLUE component of \hb, most likely because its \textbf{\oiiiseven{}$_{\rm{BLUE}}$/\hb{}$_{\rm{BLUE}}$} ratio might be  $\gg$ 1. Other cases demanded an additional blueshifted component in the \hb\ profile (BLUE) related to the BLR emission. 
Results obtained for [\ion{O}{iii}]$\lambda 5007$ full profile and individual components are reported in Table \ref{tab:spec_OIII}. We present FWHM (Col. 2), asymmetry (Col. 3), kurtosis (Col. 4), and the centroid velocities at 1/2 and 9/10 intensities (Cols. 5 and 6, respectively) for the full profile.  Relative intensities, FWHM, and shifts are reported for the  [\ion{O}{iii}]$\lambda 5007$ blueshifted components (Cols. 7 to 9) and for the narrow components (Cols. 10 to 12) of each quasar. It is difficult to distinguish the relative contribution of the two components in the majority of the cases. One of the main reasons for this is that in many sources the full [\ion{O}{iii}]$\lambda 5007$ emission is shifted to the blue, implying a shift of [\ion{O}{iii}]$\lambda 5007$\ NC with respect to the rest-frame.  Some sources like SDSSJ135831.78+050522.8 and SDSSJ161458.33+144836.9 present a NC strongly shifted to the blue, reaching shifts of $\approx -530$ and $-670$ km s$^{-1}$ respectively, comparable to the shifts found for the semi-broad component (see e.g.  Figs. \ref{fig:1358_UV} and \ref{fig:1614_UV} in the Appendix \ref{app_uv}).This is consistent with outflowing gas dominating the \oiiiseven{} luminosity in luminous quasars \citep{shen_2014,bischettietal17,Zakamska_2016}. %\chonyblue{It would be good to mention here the corresponding figures of the appendix. In particular when we want to emphasise a special feature such as the fact that the [OIII] is dominated by the SBC.} 

The remarkable feature of the \oiii\ profiles is  a very intense and blueshifted SBC, such as the one of SDSSJ135831.78+050522.8 (Fig. \ref{fig:1358_UV}), which presents a blueshifted SBC that accounts for the full \oiiiseven{} profile with a FWHM $\approx 5422$ km s$^{-1}$ and a shift of $\approx -2080$ km s$^{-1}$. In this case, the blueshifted SBC corresponds to 90\% of the full profile and can be interpreted as a strong indicative of outflowing gas. SDSSJ135831.78+050522.8, along with Q 1410+096 (Fig. \ref{fig:1410_UV} in the Appendix \ref{app_uv}) which shows a very similar profile, requires a strong and broad \oiiiseven\ to account for the flux on the red side of \hb. In our high luminosity sample, for 15/22 objects ($\sim$ 70\%) the blueshifted SBC accounts for more than 50\% of the total intensity of the \oiiionly{} profile. In fact,  in 6 of these sources, the \oiiionly{} consists of exclusively a blueshifted SBC.

%On average, the sources classified as Pop. A quasars present a wider FWHM([\ion{O}{III}]$_{\rm{full}}$) than the ones classified as Pop. B.  
Very small blueshifted components are found in low-redshift \oiii\ profiles of Pop. B AGN \citep{Zamfir_2008, Sulentic_2004} and even in Population A, shifts at peak above 250 \kms\ are very rare (the so-called ``blue-outliers''; \citealt{zamanovetal02}). At high-redshift, Pop. B \oiii\ profiles do present significant blueshifted SBC components but they still present different properties when compared to Pop. A sources: in our sample, the full profiles of Pop. A present a [\ion{O}{III}] FWHM $ \approx 2000$ km s$^{-1}$ while Pop. B profiles have FWHM $ \approx 1450$ km s$^{-1}$. The more remarkable difference is seen in the shifts at  9/10 and 1/2 intensities of the full profile ($c$(1/2) $\approx -850$ km\ s$^{-1}$ for Pop. A and $\approx -480$ km s$^{-1}$ for Pop. B). However, the A.I. of the two populations are almost the same, indicating that the lines present similar profile shapes.  The \oiii\ emission of both Pop. A and Pop. B appears to be strongly affected (if not dominated) by outflowing gas.

\par  An important consequence for redshift estimates is that the \oiii{} lines should be avoided when considering high-$z$ quasars. In addition, three Pop. A (SDSSJ210524.49+000407.3, SDSSJ210831.56-063022.5, and SDSSJ235808.54+012507.2) and three Pop. B sources ([HB89] 0029+073, CTQ 0408, and SDSSJ153830.55+085517.0) present [\ion{O}{iii}]$\lambda 5007$ full profiles that can be well represented only by the blueshifted component (see the respective spectra in Figs. \ref{fig:2355_UV}, \ref{fig:0029_HB}, \ref{fig:1801_HB}, and \ref{fig:1538_UV} in the Appendix \ref{app_uv}).

%A systematic comparison between \hb{} and \oiii{} profiles is done in Section \ref{section_hb_oiii}.

\begin{figure}
    \centering
    \includegraphics[width=0.8\columnwidth]{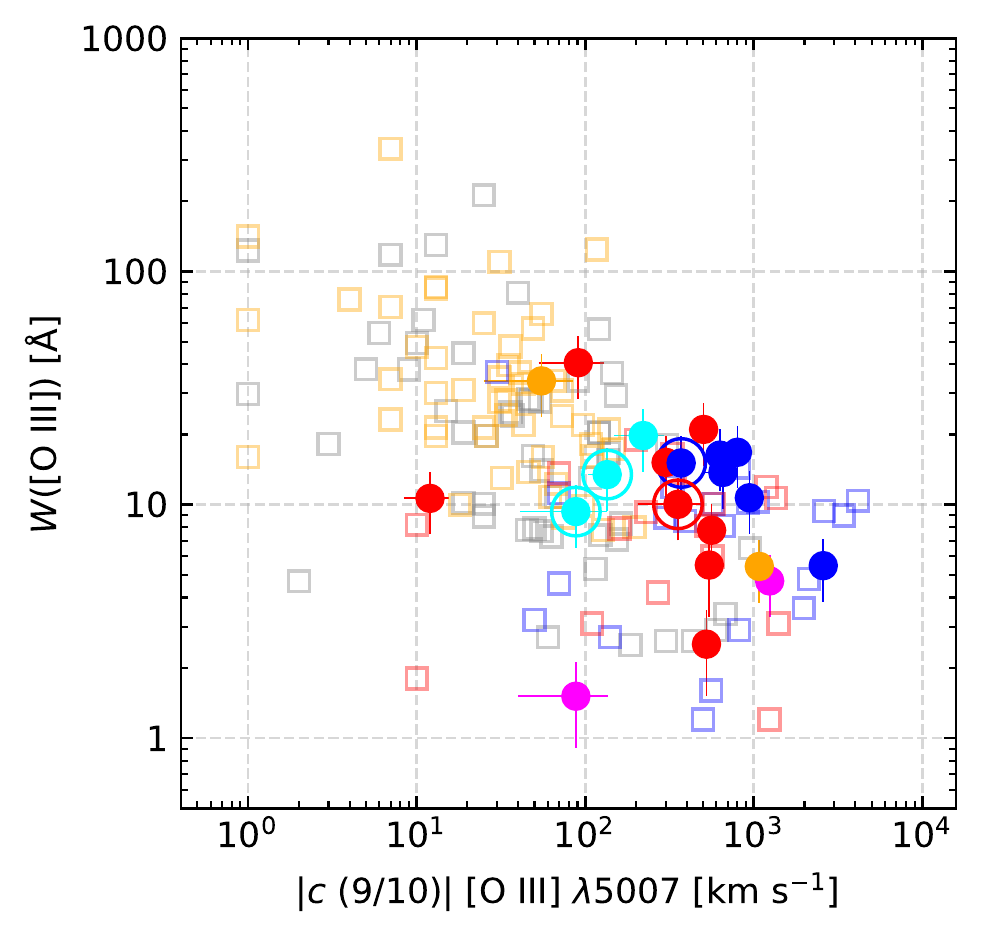}
    \caption{$W$(\oiiiseven) \textit{versus} $c$(9/10). Blue and red squares indicate high-redshift Pop. A and Pop. B quasars from \citetalias{Marziani_2009} while grey and orange open squares represent low-$z$ sources from \cite{Marziani_2003}, respectively. Color scheme for our sample as in Fig. \ref{fig:HB_BC_Full}. Error bars refer to 1$\sigma$ level of confidence and were estimated only for our sample.}
    \label{fig:oiii_wcfwhm}
\end{figure}

\begin{figure}
    \centering
    \includegraphics[width=0.8\columnwidth]{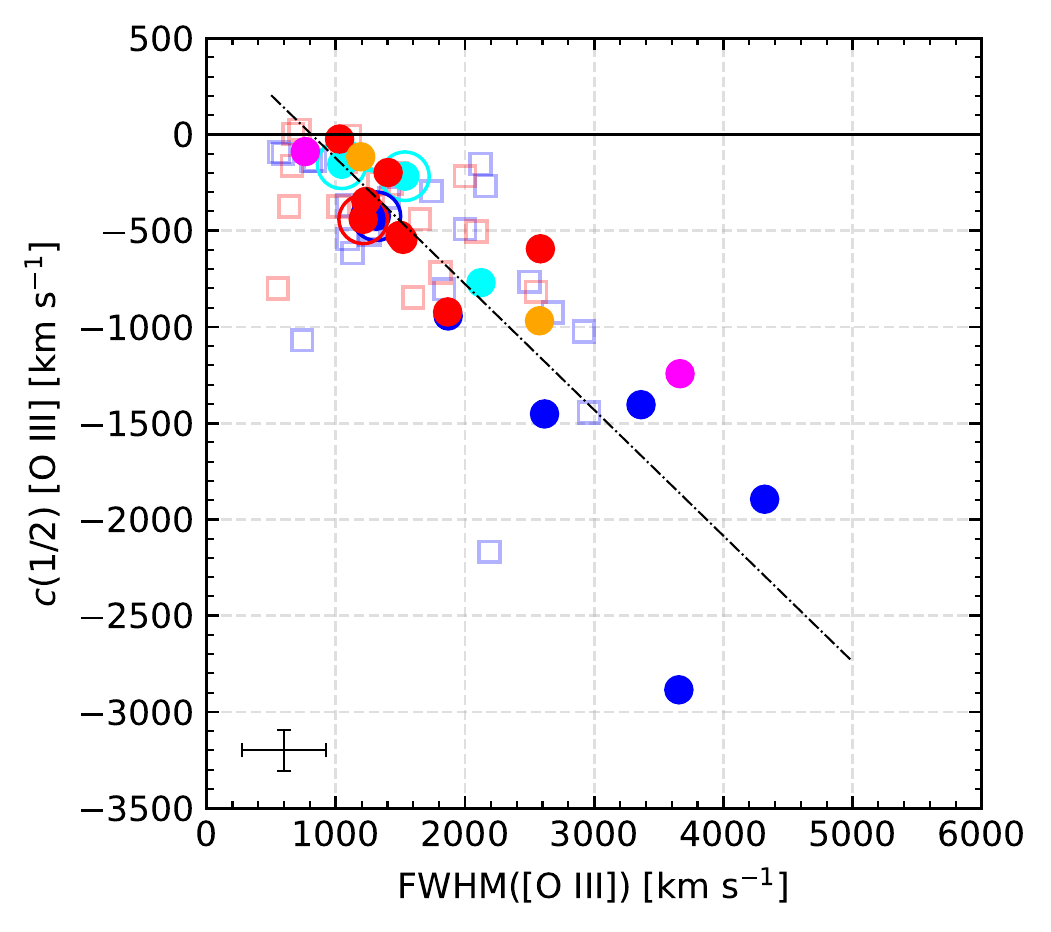}
    \caption{FWHM of the full profile of \oiii{} versus centroid at 1/2 flux intensity (c(1/2)) for the \oiii{} emission line. Blue and red squares indicate high-redshift sources from \citetalias{Marziani_2009}. Dashed line represents a linear regression. Color scheme as in Fig. \ref{fig:HB_BC_Full}. Error bars refer to 1$\sigma$ level of confidence and were estimated only for our sample.}
    \label{fig:oiii_c12_fwhm}
\end{figure}

Fig. \ref{fig:oiii_wcfwhm} shows the relation between $W$([\ion{O}{III}]) and |$c$(9/10)| of \oiiiseven{}. For our sample, the $W$([\ion{O}{III}]) ranges from $\approx$ 50\AA\  for some Pop. B sources to $< 1\r{A}$ for xAs, reaching the detection limit. It shares the same location as the high-redshift HE data. No clear difference between the location of Pop. A and Pop. B is confirmed by the median values of equivalent widths, 12\AA\ \textit{vs.} 10\AA\ for Pop. A and B, respectively (the average $W$\ is affected by the two A3 extremely faint sources with $W$([\ion{O}{III}]) $\lesssim 1$\AA). At low $z$\ (and low luminosity),  sources have considerably larger $W$ values but lower velocity centroids than the high-$z$ sample. Our sample  is located at the low end of the $W$([\ion{O}{III}]) distribution of low-$z$ sources as expected for luminous quasars \citep{shen_2014}, and in a shift range that at low $z$ is the exclusive domain of the rare, namesake ``blue outliers'' \citep{marziani_2016a}.  Fig. \ref{fig:oiii_wcfwhm} also shows that the radio-loud quasars from our sample are very tightly grouped together with moderate blueshifts for high-$L$\ quasars ($\lesssim 500$ \kms\ at 0.9 fractional intensity).

\par A comparison between $c$(1/2) and FWHM of the \oiiiseven{} profiles is shown in Fig. \ref{fig:oiii_c12_fwhm}. There is a strong correlation between the blueshift of [\ion{O}{III}], parameterised by the centroid at 1/2 intensity,  and the FWHM. The dashed line in the plot traces the linear regression between both parameters, derived from an unweighted least squares fit: 
\begin{equation}
    c(1/2)_{\rm [\ion{O}{III}]}\approx(-0.65\pm 0.08)\ \textrm{FWHM}({\rm [\ion{O}{III}]})+(530\pm 205), 
    \label{eq:oiii}
\end{equation}
Table \ref{tab:linreg} lists this linear relation together with the other that involve also \civ{}. We report the fitted parameters (Cols. 1 and 2), the method used (Col. 3), the linear correlation coefficients (Cols. 4 and 5), the rms (Col. 6), the Pearson $r$\ score (Col. 7),  and its associated null hypothesis probability value (Col. 8).
%with a rms error of $\approx 400$ \kms{} and a correlation coefficient of 0.86. 
Equation \ref{eq:oiii} confirms several previous works \citep{Komossa_2008,Marziani_2016} and  justifies the interpretation of the \oiiiseven\ profile in terms of a blueshifted semibroad component and of a narrow component \citep{zhang_2011}.
\par 

%\chony{We need to say something here. Important relation of the paper}\\\chony{I suggest to change the order of Fig. 4 and 5 and put this relation,}\\\chony{after the description of the table, when we talk about blue outliers}

%\chonyblue{Mention Marziani et al. 2016, that paper indicates the relation between the blueshift of [OIII], the presence of the SBC as characteristics of the outflows, and the Eddintong ratio. Also Komossa et al (2008) suggested a similar trend for Seyfert galaxies.} \chonyblue{We can mention that by joining \textbf{HEMS} and the  measures of the objects in this sample we obtained this clear correlation} 

\begin{table*}[t!]
    \centering
     \caption{Linear relations ($y=a+b*x$) between different emission line properties. %We also list RMS error and Pearson's correlation coefficient.
     }
    % \resizebox{\linewidth}{!}{
    \begin{tabular}{lccccccr}
    \hline
    \hline
    \noalign{\smallskip}
    y & x & Method & $a\pm \delta a$ & $b \pm \delta b$ & RMSE & CC & $\rho$-value \\
      (1) & (2) & (3) & (4) & (5) & (6) & (7) & (8)\\
      \noalign{\smallskip}
    \hline
    \noalign{\smallskip}
    $c(1/2)_{\rm{[\ion{O}{iii}]}}$ & $\rm{FWHM}(\rm{[\ion{O}{iii}]})$ & Least squares & $530\pm 205$ & $-0.65 \pm 0.08$ & 402 & -0.86 &  $4.1 \times 10^{-7}$\\
    $c(1/2)_{\rm{\ion{C}{IV}}}$ & $\rm{FWHM}(\rm{\ion{C}{IV}})$ & Least squares & $2000\pm 737^{(a)}$ & $-0.68 \pm 0.10^{(a)}$ & 429 & -0.90$^{(a)}$ & $1.1 \times 10^{-4}$ $^{(a)}$\\
    $\log W(\textrm{\ion{C}{IV}})$ & $\log |c(1/2)|_{\rm{\ion{C}{IV}}}$ & Bisector & $3.21 \pm 0.17^{(a)}$ & $-0.58 \pm 0.05^{(a)}$ & 0.28 & -0.64$^{(a)}$ & $2.1 \times 10^{-12}$ $^{(a)}$\\
    $\log W(\textrm{\ion{C}{IV}})$ & $\log \textrm{FWHM(\ion{C}{IV})}$ & Bisector & $8.31\pm 0.71^{(a)}$ & $-1.84 \pm 0.19^{(a)}$ & 0.31 & -0.49$^{(a)}$ & $2.5 \times 10^{-7}$ $^{(a)}$\\
    $c(1/2)_{\rm{\ion{C}{IV}}}$ & $c(1/2)_{\rm{[\ion{O}{III}]}}$ & Orthogonal & $-1239 \pm 276$ & $1.50\pm 0.31$ & 1086 & 0.45 & $4\times 10^{-3}$\\
    \noalign{\smallskip}
          \hline
    \end{tabular}%}
    \tablefoot{$^{(a)}$ Only for Pop. A.}
    \label{tab:linreg}
\end{table*}

\begin{table*}[t!]
      \caption[]{Spectrophotometric measurements on the UV region.}
         \label{tab:specphot_1900A}
      \resizebox{\linewidth}{!}{
                 \begin{tabular}{lccccccccccccc}
            \hline
            \hline
            \noalign{\smallskip}
             Source & $f_{\lambda,1750}$ $^{(a)}$ & $f$(\ion{Si}{IV}+\ion{O}{IV})$^{(b)}$ & W(\ion{Si}{IV}+\ion{O}{IV}) & $f$(\ion{C}{IV})$^{(b)}$ & $W$(\ion{C}{IV}) & $f$(\ion{He}{II})$^{(b)}$ & $W$(\ion{He}{II}) & $f$(\ion{Al}{III})$^{(b)}$  &  $W$(\ion{Al}{III}) & $f$(\ion{C}{III}])$^{(b)}$ & $W$(\ion{C}{III}]) & $f$(\ion{Si}{III}])$^{(b)}$ & $W$(\ion{Si}{III}])\\
             & & & [\r{A}] & & [\r{A}] & & [\r{A}] & & [\r{A}] & & [\r{A}] & & [\r{A}]\\
             (1) & (2) & (3) & (4) & (5) & (6) & (7) & (8) & (9) & (10) & (11) & (12) & (13) & (14)\\
            \noalign{\smallskip}
            \hline
            \noalign{\smallskip}
            \multicolumn{14}{c}{Population A}\\
            \noalign{\smallskip}
            \hline
            \noalign{\smallskip}
SDSSJ005700.18+143737.7$^{(e)}$	&	5.9	$\pm$	0.9	&	1.29	$\pm$	0.19	&	16	$\pm$	2	&	1.58	$\pm$	0.24	&	22	$\pm$	3	&	0.33	$\pm$	0.05	&	5	$\pm$	1	&	0.41	$\pm$	0.06	&	8	$\pm$	1	&	0.23	$\pm$	0.03	&	5	$\pm$	1	&	0.42	$\pm$	0.06	&	8	$\pm$	1	\\
SDSSJ132012.33+142037.1	&	7.9	$\pm$	1.2	&	1.77	$\pm$	0.27	&	18	$\pm$	2	&	3.66	$\pm$	0.55	&	40	$\pm$	5	&	0.81	$\pm$	0.12	&	10	$\pm$	1	&	0.31	$\pm$	0.05	&	4	$\pm$	1	&	0.57	$\pm$	0.09	&	8	$\pm$	1	&	0.75	$\pm$	0.11	&	11	$\pm$	1	\\
SDSSJ135831.78+050522.8	&	13.6	$\pm$	2.0	&	3.78	$\pm$	0.57	&	20	$\pm$	2	&	5.43	$\pm$	0.81	&	34	$\pm$	4	&	1.31	$\pm$	0.20	&	9	$\pm$	1	&	0.89	$\pm$	0.13	&	7	$\pm$	1	&	1.01	$\pm$	0.15	&	9	$\pm$	1	&	1.52	$\pm$	0.23	&	13	$\pm$	2	\\
Q 1410+096	&	14.1	$\pm$	2.1	&	2.45	$\pm$	0.37	&	15	$\pm$	2	&	3.50	$\pm$	0.53	&	23	$\pm$	3	&	1.50	$\pm$	0.23	&	10	$\pm$	1	&	0.96	$\pm$	0.14	&	7	$\pm$	1	&	1.17	$\pm$	0.18	&	9	$\pm$	1	&	1.18	$\pm$	0.18	&	9	$\pm$	1	\\
SDSSJ161458.33+144836.9	&	15.4	$\pm$	2.3	&	2.90	$\pm$	0.44	&	14	$\pm$	2	&	5.74	$\pm$	0.86	&	31	$\pm$	4	&	0.94	$\pm$	0.14	&	6	$\pm$	1	&	0.57	$\pm$	0.09	&	4	$\pm$	1	&	1.41	$\pm$	0.21	&	11	$\pm$	1	&	1.32	$\pm$	0.20	&	10	$\pm$	1	\\
PKS 2000-330$^{(c)}$	&	-			&	3.20	$\pm$	0.48	&	4	$\pm$	2	&	4.50	$\pm$	0.68	&	19	$\pm$	2	&	-			&	-			&	-			&	-			&	-			&	-			&	-			&	-			\\
SDSSJ210524.47+000407.3$^{(d)(e)}$	&	21.8	$\pm$	3.3	&	-			&	-			&	1.50	$\pm$	0.23	&	6	$\pm$	1	&	0.75	$\pm$	0.11	&	3	$\pm$	1	&	2.19	$\pm$	0.33	&	10	$\pm$	1	&	0.51	$\pm$	0.08	&	2	$\pm$	1	&	1.97	$\pm$	0.30	&	9	$\pm$	1	\\
SDSSJ210831.56-063022.5$^{(e)}$	&	14.3	$\pm$	2.1	&	2.47	$\pm$	0.37	&	14	$\pm$	2	&	2.25	$\pm$	0.34	&	13	$\pm$	2	&	0.14	$\pm$	0.02	&	1	$\pm$	1	&	1.01	$\pm$	0.15	&	7	$\pm$	1	&	0.31	$\pm$	0.05	&	2	$\pm$	1	&	0.94	$\pm$	0.14	&	7	$\pm$	1	\\
SDSSJ212329.46-005052.9	&	23.9	$\pm$	3.6	&	3.58	$\pm$	0.54	&	9	$\pm$	1	&	6.09	$\pm$	0.91	&	19	$\pm$	2	&	0.85	$\pm$	0.13	&	3	$\pm$	1	&	1.77	$\pm$	0.27	&	9	$\pm$	1	&	0.59	$\pm$	0.09	&	3	$\pm$	1	&	1.89	$\pm$	0.28	&	10	$\pm$	1	\\
SDSSJ235808.54+012507.2	&	7.5	$\pm$	1.1	&	1.52	$\pm$	0.23	&	12	$\pm$	1	&	2.55	$\pm$	0.38	&	25	$\pm$	3	&	0.71	$\pm$	0.11	&	8	$\pm$	1	&	1.91	$\pm$	0.29	&	5	$\pm$	1	&	0.47	$\pm$	0.07	&	8	$\pm$	1	&	0.61	$\pm$	0.09	&	10	$\pm$	1	\\

\noalign{\smallskip}
            \hline
            \noalign{\smallskip}
            \multicolumn{14}{c}{Population B}\\
            \noalign{\smallskip}
            \hline
\noalign{\smallskip}
SDSSJ114358.52+052444.9	&	20.2	$\pm$	3.0	&	2.28	$\pm$	0.34	&	8	$\pm$	1	&	5.86	$\pm$	0.88	&	24	$\pm$	3	&	1.42	$\pm$	0.21	&	6	$\pm$	1	&	0.71	$\pm$	0.11	&	4	$\pm$	1	&	1.30	$\pm$	0.20	&	7	$\pm$	1	&	0.98	$\pm$	0.15	&	5	$\pm$	1	\\
SDSSJ115954.33+201921.1	&	24.3	$\pm$	3.6	&	3.64	$\pm$	0.55	&	12	$\pm$	1	&	5.55	$\pm$	0.83	&	20	$\pm$	2	&	0.92	$\pm$	0.14	&	4	$\pm$	1	&	1.14	$\pm$	0.17	&	5	$\pm$	1	&	1.94	$\pm$	0.29	&	9	$\pm$	1	&	0.62	$\pm$	0.09	&	3	$\pm$	1	\\
SDSSJ120147.90+120630.2	&	16.3	$\pm$	2.4	&	3.44	$\pm$	0.52	&	15	$\pm$	2	&	7.57	$\pm$	1.14	&	37	$\pm$	4	&	1.63	$\pm$	0.24	&	9	$\pm$	1	&	0.67	$\pm$	0.10	&	5	$\pm$	1	&	3.00	$\pm$	0.45	&	22	$\pm$	3	&	1.13	$\pm$	0.17	&	8	$\pm$	1	\\
SDSSJ141546.24+112943.4$^{(d)}$	&	16.8	$\pm$	2.5	&	-			&	-			&	-			&	-			&	-			&	-			&	1.59	$\pm$	0.24	&	7	$\pm$	1	&	5.10	$\pm$	0.77	&	23	$\pm$	3	&	2.83	$\pm$	0.42	&	13	$\pm$	2	\\
SDSSJ153830.55+085517.0$^{(f)}$	&	17.4	$\pm$	2.6	&	4.92	$\pm$	0.74	&	19	$\pm$	2	&	5.34	$\pm$	0.80	&	24	$\pm$	3	&	0.64	$\pm$	0.10	&	3	$\pm$	1	&	1.91	$\pm$	0.29	&	12	$\pm$	1	&	0.94	$\pm$	0.14	&	6	$\pm$	1	&	1.27	$\pm$	0.19	&	8	$\pm$	1	\\

            \noalign{\smallskip}
            \hline
         \end{tabular}
}
    
     \tablefoot{$^{(a)}$ In units of $10^{-15}$ erg s$^{-1}$ cm$^{-2}$ \r{A}$^{-1}$. $^{(b)}$ In units of $10^{-13}$ erg s$^{-1}$ cm$^{-2}$. $^{(c)}$ The UV spectrum does not cover the 1900 \r{A} blend. $^{(d)}$ Presents broad absorption lines in the regions of \civ{}+\heiiuv{} and \siv{}+\oiv{}. $^{(e)}$ We consider BLUE+BC when estimating fluxes and equivalent widths of \aliii{} and \ion{Si}{III}]$\lambda1892$. $^{(f)}$ This source is optically classified as Pop. B, but the UV spectrum fits perfectly with the one of a Pop. A quasar.}
   \end{table*}

\begin{sidewaystable*}[htp!]

      \caption[]{Measurements on the 1900 \r{A} blend.}
         \label{tab:meas_1900A}
      \resizebox{\linewidth}{!}{
                 \begin{tabular}{lccccccccccccccccccccccccccccccc}
            \hline
            \hline
            \noalign{\smallskip}
             & \multicolumn{3}{c}{\ion{Al}{III}$\lambda$1860 BLUE} & & \multicolumn{3}{c}{\ion{Al}{III}$\lambda$1860 BC} & & \multicolumn{3}{c}{\ion{Si}{III}]$\lambda$1892 BLUE} & &\multicolumn{3}{c}{\ion{Si}{III}]$\lambda$1892 BC} & & \multicolumn{3}{c}{\ion{C}{III}]$\lambda$1909 BC} & & \multicolumn{3}{c}{\ion{C}{III}]$\lambda$1909 VBC} & & \multicolumn{3}{c}{\ion{C}{III}]$\lambda$1909 NC} & & 
            \multicolumn{3}{c}{\ion{Fe}{III}$\lambda$1914}\\
            \noalign{\smallskip}
              \cline{2-4} \cline{6-8} \cline{10-12} \cline{14-16} \cline{18-20} \cline{22-24} \cline{26-28} \cline{30-32}
              \noalign{\smallskip}
           Source & $I/I_{\rm{tot}}$ & FWHM & Shift & & $I/I_{\rm{tot}}$ & FWHM & Shift & & $I/I_{\rm{tot}}$ & FWHM & Shift & & $I/I_{\rm{tot}}$ & FWHM & Shift & & $I/I_{\rm{tot}}$ & FWHM & Shift & & $I/I_{\rm{tot}}$ & FWHM & Shift & & $I/I_{\rm{tot}}$ & FWHM & Shift & & $I/I_{\rm{tot}}$ & FWHM & Shift \\
            & & [km s$^{-1}$] & [km s$^{-1}$] & & & [km s$^{-1}$] & [km s$^{-1}$] & & & [km s$^{-1}$] & [km s$^{-1}$] & & & [km s$^{-1}$] & [km s$^{-1}$] & & & [km s$^{-1}$] & [km s$^{-1}$] & & & [km s$^{-1}$] & [km s$^{-1}$]& & & [km s$^{-1}$] & [km s$^{-1}$] & & & [km s$^{-1}$] & [km s$^{-1}$]\\
            (1) & (2) & (3) & (4) & & (5) & (6) & (7) & & (8) & (9) & (10) & & (11) & (12) & (13) & & (14) & (15) & (16) & & (17) & (18) & (19) & & (20) & (21) & (22) & & (23) & (24) & (25)\\
            \noalign{\smallskip}
            \hline
            \noalign{\smallskip}
            \multicolumn{32}{c}{Population A}\\
            \noalign{\smallskip}
            \hline
            \noalign{\smallskip}
SDSSJ005700.18+143737.7	&	0.47	$\pm$	0.04	&	3268	$\pm$	396	&	-1686	$\pm$	202	&	&	0.53	$\pm$	0.04	&	2792	$\pm$	283	&	0	$\pm$	10	&	&	0.36	$\pm$	0.07	&	3268	$\pm$	392	&	-1686	$\pm$	-202	&	&	0.64	$\pm$	0.13	&	2792	$\pm$	335	&	-3	$\pm$	11	&	&	1.00	$\pm$	0.18	&	2400	$\pm$	288	&	0	$\pm$	10	&	&	…			&	…			&	…			&	&	…			&	…			&	…			&	&	1.00	$\pm$	0.12	&	2792	$\pm$	335	&	1494	$\pm$	179	\\
SDSSJ132012.33+142037.1	&	0.27	$\pm$	0.02	&	1731	$\pm$	210	&	-967	$\pm$	116	&	&	0.72	$\pm$	0.06	&	2029	$\pm$	206	&	0	$\pm$	10	&	&	0.00			&	…			&	…			&	&	0.08	$\pm$	0.02	&	3030	$\pm$	364	&	0	$\pm$	10	&	&	0.92	$\pm$	0.17	&	3030	$\pm$	364	&	0	$\pm$	10	&	&	…			&	…			&	…			&	&	…			&	…			&	…			&	&	1.00	$\pm$	0.12	&	3030	$\pm$	364	&	1876	$\pm$	225	\\
SDSSJ135831.78+050522.8	&	0.00			&	…			&	…			&	&	1.00	$\pm$	0.08	&	4225	$\pm$	428	&	0	$\pm$	10	&	&	0.00			&	…			&	…			&	&	1.00	$\pm$	0.21	&	4225	$\pm$	507	&	0	$\pm$	10	&	&	1.00	$\pm$	0.18	&	2720	$\pm$	326	&	0	$\pm$	10	&	&	…			&	…			&	…			&	&	…			&	…			&	…			&	&	1.00	$\pm$	0.12	&	4231	$\pm$	508	&	1033	$\pm$	124	\\
Q 1410+096	&	0.00			&	…			&	…			&	&	1.00	$\pm$	0.08	&	2604	$\pm$	264	&	279	$\pm$	33	&	&	0.00			&	…			&	…			&	&	1.00	$\pm$	0.21	&	2604	$\pm$	312	&	279	$\pm$	33	&	&	1.00	$\pm$	0.18	&	2604	$\pm$	312	&	278	$\pm$	58	&	&	…			&	…			&	…			&	&	…			&	…			&	…			&	&	1.00	$\pm$	0.12	&	1607	$\pm$	193	&	1253	$\pm$	150	\\
SDSSJ161458.33+144836.9	&	0.00			&	…			&	…			&	&	1.00	$\pm$	0.08	&	3374	$\pm$	342	&	0	$\pm$	10	&	&	0.00			&	…			&	…			&	&	1.00	$\pm$	0.21	&	3374	$\pm$	405	&	0	$\pm$	10	&	&	1.00	$\pm$	0.18	&	3319	$\pm$	398	&	0	$\pm$	10	&	&	…			&	…			&	…			&	&	…			&	…			&	…			&	&	0.00			&	…			&	…			\\
SDSSJ210524.49+000407.3	&	0.61	$\pm$	0.05	&	4365	$\pm$	529	&	-2003	$\pm$	240	&	&	0.39	$\pm$	0.03	&	3672	$\pm$	372	&	0	$\pm$	10	&	&	0.23	$\pm$	0.04	&	3415	$\pm$	410	&	-2002	$\pm$	-240	&	&	0.77	$\pm$	0.16	&	4169	$\pm$	500	&	0	$\pm$	10	&	&	1.00	$\pm$	0.18	&	2577	$\pm$	309	&	0	$\pm$	10	&	&	…			&	…			&	…			&	&	…			&	…			&	…			&	&	1.00	$\pm$	0.12	&	3672	$\pm$	441	&	771	$\pm$	93	\\
SDSSJ210831.56-063022.5	&	0.36	$\pm$	0.03	&	3181	$\pm$	386	&	-1951	$\pm$	234	&	&	0.64	$\pm$	0.05	&	4173	$\pm$	423	&	0	$\pm$	10	&	&	0.21	$\pm$	0.04	&	3187	$\pm$	382	&	-1951	$\pm$	-234	&	&	0.79	$\pm$	0.17	&	4173	$\pm$	501	&	0	$\pm$	10	&	&	1.00	$\pm$	0.18	&	3156	$\pm$	379	&	0	$\pm$	10	&	&	…			&	…			&	…			&	&	…			&	…			&	…			&	&	1.00	$\pm$	0.12	&	2177	$\pm$	261	&	2102	$\pm$	252	\\
SDSSJ212329.46-005052.9	&	0.00			&	…			&	…			&	&	1.00	$\pm$	0.08	&	5122	$\pm$	519	&	0	$\pm$	10	&	&	0.00			&	…			&	…			&	&	1.00	$\pm$	0.21	&	5122	$\pm$	615	&	0	$\pm$	10	&	&	1.00	$\pm$	0.18	&	3125	$\pm$	375	&	0	$\pm$	10	&	&	…			&	…			&	…			&	&	…			&	…			&	…			&	&	0.00			&	…			&	…			\\
SDSSJ235808.54+012507.2	&	0.00			&	…			&	…			&	&	1.00	$\pm$	0.08	&	3243	$\pm$	329	&	0	$\pm$	10	&	&	0.00			&	…			&	…			&	&	1.00	$\pm$	0.21	&	3243	$\pm$	389	&	0	$\pm$	10	&	&	1.00	$\pm$	0.18	&	3243	$\pm$	389	&	0	$\pm$	10	&	&	…			&	…			&	…			&	&	…			&	…			&	…			&	&	1.00	$\pm$	0.12	&	3204	$\pm$	384	&	111	$\pm$	13	\\

\noalign{\smallskip}
\hline
            \noalign{\smallskip}
Median & 0.00 $\pm$ 0.36 & 3224 $\pm$ 723 & -1818 $\pm$ 457 & & 1.00 $\pm$ 0.36 & 3374 $\pm$ 1381 & … & & 0.23 $\pm$ 0.07 & 3268 $\pm$ 114 & -1951 $\pm$ 158 & & … & 3374 $\pm$ 1381 & …  & & … & 2760 $\pm$ 552 & … & & … & … & … & & … & … & … & & … & 2298 $\pm$ 1224 & 1143 $\pm$ 597\\
\noalign{\smallskip}

            \hline
            \noalign{\smallskip}
            \multicolumn{32}{c}{Population B}\\
            \noalign{\smallskip}
            \hline
            \noalign{\smallskip}
SDSSJ114358.52+052444.9	&	0.00			&	…			&	…			&	&	1.00	$\pm$	0.08	&	3967	$\pm$	402	&	0	$\pm$	10	&	&	0.00			&	…			&	…			&	&	1.00	$\pm$	0.18	&	3967	$\pm$	476	&	-3	$\pm$	11	&	&	0.50	$\pm$	0.10	&	3967	$\pm$	476	&	0	$\pm$	10	&	&	0.50	$\pm$	0.05	&	7038	$\pm$	1253	&	2842	$\pm$	369	&	&	0.02	$\pm$	0.01	&	1000	$\pm$	178	&	0	$\pm$	10	&	&	0.00			&	…			&	…			\\
SDSSJ115954.33+201921.1	&	0.00			&	…			&	…			&	&	1.00	$\pm$	0.08	&	4999	$\pm$	506	&	0	$\pm$	10	&	&	0.00			&	…			&	…			&	&	1.00	$\pm$	0.18	&	4999	$\pm$	600	&	0	$\pm$	10	&	&	0.64	$\pm$	0.13	&	4999	$\pm$	600	&	0	$\pm$	10	&	&	0.36	$\pm$	0.03	&	6563	$\pm$	1168	&	1998	$\pm$	260	&	&	0.01	$\pm$	0.01	&	733	$\pm$	130	&	0	$\pm$	10	&	&	0.00			&	…			&	…			\\
SDSSJ120147.90+120630.2	&	0.00			&	…			&	…			&	&	1.00	$\pm$	0.08	&	3999	$\pm$	405	&	0	$\pm$	10	&	&	0.00			&	…			&	…			&	&	1.00	$\pm$	0.18	&	3999	$\pm$	480	&	0	$\pm$	10	&	&	0.70	$\pm$	0.14	&	3999	$\pm$	480	&	0	$\pm$	10	&	&	0.30	$\pm$	0.03	&	7046	$\pm$	1254	&	2207	$\pm$	287	&	&	0.01	$\pm$	0.01	&	999	$\pm$	178	&	0	$\pm$	10	&	&	0.00			&	…			&	…			\\
SDSSJ141546.24+112943.4	&	0.00			&	…			&	…			&	&	1.00	$\pm$	0.08	&	5105	$\pm$	517	&	0	$\pm$	10	&	&	0.00			&	…			&	…			&	&	1.00	$\pm$	0.18	&	5105	$\pm$	613	&	0	$\pm$	10	&	&	0.64	$\pm$	0.13	&	4101	$\pm$	492	&	0	$\pm$	10	&	&	0.36	$\pm$	0.03	&	7365	$\pm$	1311	&	3134	$\pm$	407	&	&	0.10	$\pm$	0.01	&	1062	$\pm$	189	&	0	$\pm$	10	&	&	0.00			&	…			&	…			\\
SDSSJ153830.55+085517.0$^{(a)}$	&	0.00			&	…			&	…			&	&	1.00	$\pm$	0.08	&	3085	$\pm$	313	&	0	$\pm$	10	&	&	0.00			&	…			&	…			&	&	1.00	$\pm$	0.18	&	3085	$\pm$	370	&	0	$\pm$	10	&	&	1.00	$\pm$	0.20	&	3104	$\pm$	372	&	28	$\pm$	16	&	&	…			&	…			&	…			&	&	0.00			&	…			&	…			&	&	1.00	$\pm$	0.12	&	3141	$\pm$	377	&	783	$\pm$	94	\\

            \noalign{\smallskip}
            \hline
            \noalign{\smallskip}
Median & … & … & … & & … & 3999 $\pm$ 1032 & … & & … & … & … & & … & 3999 $\pm$ 1032 & …  & & 0.64 $\pm$ 0.06 & 3999 $\pm$ 134 & … & & 0.36 $\pm$ 0.05 & 7042 $\pm$ 206 & 2524 $\pm$ 760 & & … & 999 $\pm$ 83 & … & & … & … & …\\
\noalign{\smallskip}

            \hline
         \end{tabular}
}
  \tablefoot{$^{(a)}$ UV line profiles of this source were fitted with Lorentzian shapes. See note in the Appendix \ref{app_uv} (Fig. \ref{fig:1538_UV})}  
    % \tablefoot{Columns: (1) Source identification. (2) Observation date. (3) Grating. (4) Central wavelength of the filter, in units of $\mu$m.}
   \end{sidewaystable*}

%\subsubsection{Error estimates in the optical range}

%\par \textbf{Uncertainties were estimated running  Markov Chain Monte Carlo (MCMC) simulations (see \citealt{Marziani_2022apj} for a full description of the method)  for  one source (SDSSJ235808.54+012507.2) in the spectral range covering H$\beta$+\oiii{}.  The errors are around the order of 10\% for the FWHM of the H$\beta_{\rm{BC}}$ and [\ion{O}{III}]$_{\rm{SBC}}$. Larger uncertainties ($\sim 30\%$) are found for the narrow components of both lines and for the blueshifted component of H$\beta$. Flux uncertainties for strong or sharp  emission lines are $\sim 10\%$, while typical errors for the continuum and \ion{Fe}{II} are $\sim 5$ and $\sim 15\%$\  respectively, if \feii\ is reasonably strong.}

%-----------------------------------------------------------------
\subsection{UV}

\par The fits for the UV emission lines are shown in Appendix \ref{app_uv} together with the full UV spectra of the 15 available sources  (10 Pop. A and 5 Pop. B). Spectrophotometric measurements in the UV region are reported in Table \ref{tab:specphot_1900A} and concern the 1900 \r{A} blend, \ion{C}{iv}$\lambda1549$+\ion{He}{II}$\lambda1640$, as well as flux intensity and equivalent width of \ion{Si}{IV}+\ion{O}{IV} (Cols. 3-4), \ion{C}{IV} (Cols. 5-6), \ion{He}{II} (Cols. 7-8), \ion{Al}{III} (Cols. 9-10), \ion{C}{III}] (Cols. 11-12), and \ion{Si}{III} (Cols 13-14).

%For sources  PKS 2000-330, SDSSJ210524.47+000407.3, and SDSSJ141546.24+112943.4 we do not report neither flux intensity nor equivalent width of the emission lines due to the difficulty  in measuring their full profile.

\subsubsection{1900 \r{A} blend}

\par Table \ref{tab:meas_1900A} presents the measurements resulting from the \texttt{specfit} analysis of the 1900 \r{A} blend. Intensities, FWHM, and shifts are shown for the emission line profiles and their components, whenever the profile presents more than just one component. Cols. 2 to 4 list relative intensity, FWHM and shift of the blueshifted component of \ion{Al}{III}$\lambda1860$; Cols. 5 to 7 show the broad component of \ion{Al}{III}$\lambda1860$; Cols. 8 to 10 for \ion{Si}{III}]$\lambda$1892 BLUE; Cols. 11 to 13 for \ion{Si}{III}]$\lambda$1892 BC; Cols. 14 to 16, the BC of \ion{C}{III}]$\lambda$1909; Cols. 17 to 19, the VBC of \ion{C}{III}]$\lambda$1909; Cols. 20 to 22, the NC of \ion{C}{III}]$\lambda$1909;  and, finally, Cols. 23 to 25 for a \ion{Fe}{III}]$\lambda$1914 emission line component. 

%In this blend, all emission lines that were analysed were set at rest-frame wavelength in the wide majority of cases; however  BLUE and VBC were left free to shift in wavelength in some cases.  In the

The \ion{Al}{III}$\lambda1860$ contribution can be well fitted assuming only a BC for all Pop. A sources but four: SDSSJ005700.18+143737.7, SDSSJ210524.49+000407.3, SDSSJ210831.56-063022.5, and SDSSJ132012.33+142037.1. These same three sources are the ones that are located in the A3 region of the quasar MS. In these cases, we need to add a BLUE component with $3000 \le $ FWHM $ \le 4500$ km s$^{-1}$ in the \ion{Al}{III}$\lambda$1860 doublet and in the \ion{Si}{III}]$\lambda$1892  emission line profile. We have imposed the same FWHM for both components. %For SDSSJ005700.18+143737.7 and SDSSJ210524.49+000407.3, we present in the Appendix two different fittings for the 1900\AA\ blend, one with and the other without a blueshifted component in order to show that an additional blueshifted component is needed (Figs. \ref{fig:0057_UV} and \ref{fig:2105_UV}). In  the latter case the \aliii\ profiles are significantly blueshifted from  the  rest-frame. For SDSSJ210831.56-063022.5 we show only the fitting with the blueshifted component, due to the presence of a prominent excess at the blue side of the 1900\AA\ blend (see Fig. \ref{fig:2108_UV}). 

\par %The \ion{Si}{III}]$\lambda$1892 line profile was fit by only one component in both Pop. A and Pop. B (except in the three cases discussed above when a BLUE was added). 
The Pop. A \ion{C}{III}]$\lambda$1909  profile can be well reproduced by a strong BC at rest-frame in combination with a component that accounts for the \ion{Fe}{III}$\lambda$1914 contributions in its red wing. The motivation to include the \ion{Fe}{III}$\lambda$1914 line resides in the selective enhancement due to Lyman $\alpha$ fluorescence that is well-known to affect the UV \ion{Fe}{II} emission and \ion{Fe}{III} emission as well \citep{sigutpradhan98,sigutetal04}. Of the \ion{Fe}{II} features the UV multiplet 191 at $\lambda$1785 is known to be enhanced by Ly$\alpha$ fluorescence: a strong \feii\ UV multiplet 191 may suggest a strong \ion{Fe}{III}$\lambda$1914 line. 
There is an overall consistency between the presence of the \ion{Fe}{III}$\lambda$1914 line and the detection of the \feii\ feature (8 out of 10 Pop. A sources with UV suitable data have both).
However, the relative contribution of the \ion{C}{III}]$\lambda$1909 and \ion{Fe}{III}$\lambda$1914 remains difficult to ascertain because the two lines are severely blended together and some \ion{Fe}{III}$\lambda$1914 emission is already included in the \ion{Fe}{III} template. %SDSSJ132012.33+142037.1, SDSSJ161458.33+144836.9, and SDSSJ212329.46-005052.9 do not show evidence of the \ion{Fe}{III} component. 

For Pop. B sources, \aliii{} and \ion{Si}{III}]$\lambda$1892 are well represented by only one component and in the five cases these emission lines  share the same FWHM and are centred at the respective rest-frame wavelength. However, the \ion{C}{III}]$\lambda$1909  VBC is systematically less broad than the \hb\ VBC: average FWHM is $\approx 7000$ \kms\ for \ion{C}{III}] \textit{versus} $\approx 12000$ \kms  for \hb.

%SDSSJ153830.55+085517.0 
%With the exception of this very special case, there are no \ion{Fe}{III}$\lambda$1914 components for Pop. B sources. It is an expected behaviour in this quasar population \citep{sulentic_2014}. 
   \begin{sidewaystable*}[htp!]
      \caption[]{Measurements on the \ion{C}{IV}$\lambda1549$+\ion{He}{II}$\lambda1640$ full line profiles and \textsc{specfit} analysis.}
         \label{tab:meas_CIV}
      \resizebox{\linewidth}{!}{
                 \begin{tabular}{lccccccccccccccccccccccccccccccc}
            \hline
            \hline
            \noalign{\smallskip}
            & & & & & & & & \multicolumn{12}{c}{\ion{C}{IV}$\lambda1549$} & \multicolumn{12}{c}{\ion{He}{II}$\lambda1640$}\\
            \noalign{\smallskip}
            \cline{10-20} \cline{22-32}
            \noalign{\smallskip}
             & \multicolumn{8}{c}{\ion{C}{IV}$\lambda1549$ full profile} & \multicolumn{3}{c}{BLUE} & & \multicolumn{3}{c}{BC} & & \multicolumn{3}{c}{VBC} & & \multicolumn{3}{c}{BLUE} & & \multicolumn{3}{c}{BC} & & \multicolumn{3}{c}{VBC}\\
             \noalign{\smallskip}
              \cline{2-8} \cline{10-12} \cline{14-16} \cline{18-20} \cline{22-24} \cline{26-28} \cline{30-32}
              \noalign{\smallskip}
            Source &  FWHM & A. I. & Kurt. & c($\frac{1}{4}$) & c($\frac{1}{2}$) & c($\frac{3}{4}$) & c($\frac{9}{10}$) & & $I/I_{\mathrm{tot}}$ & FWHM & Shift & & $I/I_{\mathrm{tot}}$ & FWHM & Shift & & $I/I_{\mathrm{tot}}$ & FWHM & Shift & & $I/I_{\mathrm{tot}}$ & FWHM & Shift & & $I/I_{\mathrm{tot}}$ & FWHM & Shift & & $I/I_{\mathrm{tot}}$ & FWHM & Shift\\
             & [km s$^{-1}$] &  &  & [km s$^{-1}$] & [km s$^{-1}$] & [km s$^{-1}$] & [km s$^{-1}$] & & & [km s$^{-1}$] & [km s$^{-1}$] &  &  & [km s$^{-1}$] & [km s$^{-1}$] &  &  & [km s$^{-1}$] & [km s$^{-1}$] &  &  & [km s$^{-1}$] & [km s$^{-1}$] &  &  & [km s$^{-1}$] & [km s$^{-1}$] &  &  & [km s$^{-1}$] & [km s$^{-1}$]\\
             (1) & (2) & (3) & (4) & (5) & (6) & (7) & (8) & & (9) & (10) & (11) & & (12) & (13) & (14) & & (15) & (16) & (17) & & (18) & (19) & (20) & & (21) & (22) & (23) & & (24) & (25) & (26)\\
            \noalign{\smallskip}
            \hline
            \noalign{\smallskip}
            \multicolumn{32}{c}{Population A}\\
            \noalign{\smallskip}
            \hline
            \noalign{\smallskip}
            
          SDSSJ005700.18+143737.7	&	8648	$\pm$	562	&	-0.38	$\pm$	0.01	&	0.40	$\pm$	0.01	&	-4951	$\pm$	334	&	-3994	$\pm$	181	&	-3272	$\pm$	272	&	-2604	$\pm$	326	&	&	0.81	$\pm$ 	0.05	&	13187	$\pm$ 	1846	&	-2318	$\pm$ 	278	&	&	0.19	$\pm$ 	0.05	&	2808	$\pm$ 	298	&	0	$\pm$ 	10	&	&	…			&	…			&	…			&	&	0.67	$\pm$ 	0.04	&	13187	$\pm$ 	1846	&	-4249	$\pm$ 	510	&	&	0.33	$\pm$ 	0.02	&	2808	$\pm$ 	298	&	0	$\pm$	10	&	&	…			&	…			&	…			\\
SDSSJ132012.33+142037.1	&	5549	$\pm$	399	&	-0.36	$\pm$	0.01	&	0.37	$\pm$	0.01	&	-2956	$\pm$	263	&	-2407	$\pm$	124	&	-1830	$\pm$	177	&	-1419	$\pm$	190	&	&	0.72	$\pm$ 	0.04	&	8739	$\pm$ 	1223	&	-1446	$\pm$ 	174	&	&	0.28	$\pm$ 	0.03	&	3278	$\pm$ 	348	&	0	$\pm$	10	&	&	…			&	…			&	…			&	&	0.60	$\pm$ 	0.04	&	9837	$\pm$ 	1377	&	-3388	$\pm$ 	407	&	&	0.40	$\pm$ 	0.02	&	3278	$\pm$ 	348	&	0	$\pm$	10	&	&	…			&	…			&	…			\\
SDSSJ135831.78+050522.8	&	7365	$\pm$	655	&	-0.27	$\pm$	0.03	&	0.30	$\pm$	0.01	&	-3188	$\pm$	416	&	-2460	$\pm$	190	&	-1808	$\pm$	256	&	-1556	$\pm$	187	&	&	0.52	$\pm$ 	0.03	&	11983	$\pm$ 	1678	&	-1909	$\pm$ 	229	&	&	0.48	$\pm$ 	0.02	&	4404	$\pm$ 	468	&	0	$\pm$	10	&	&	…			&	…			&	…			&	&	0.55	$\pm$ 	0.03	&	11983	$\pm$ 	1678	&	-3386	$\pm$ 	406	&	&	0.45	$\pm$ 	0.03	&	4404	$\pm$ 	468	&	0	$\pm$	10	&	&	…			&	…			&	…			\\
Q 1410+096	&	6311	$\pm$	552	&	-0.18	$\pm$	0.05	&	0.33	$\pm$	0.01	&	-2604	$\pm$	374	&	-1746	$\pm$	179	&	-1052	$\pm$	172	&	-1691	$\pm$	90	&	&	0.41	$\pm$ 	0.03	&	10881	$\pm$ 	1523	&	-1526	$\pm$ 	183	&	&	0.59	$\pm$ 	0.01	&	3293	$\pm$ 	350	&	0	$\pm$	10	&	&	…			&	…			&	…			&	&	0.53	$\pm$ 	0.03	&	9369	$\pm$ 	1312	&	-7880	$\pm$ 	946	&	&	0.47	$\pm$ 	0.03	&	3292	$\pm$ 	350	&	0	$\pm$	10	&	&	…			&	…			&	…			\\
SDSSJ161458.33+144836.9	&	5790	$\pm$	554	&	-0.27	$\pm$	0.03	&	0.30	$\pm$	0.02	&	-2779	$\pm$	327	&	-2394	$\pm$	139	&	-1894	$\pm$	190	&	-1519	$\pm$	168	&	&	0.59	$\pm$ 	0.04	&	9333	$\pm$ 	1307	&	-1609	$\pm$ 	193	&	&	0.41	$\pm$ 	0.01	&	3816	$\pm$ 	406	&	0	$\pm$	10	&	&	…			&	…			&	…			&	&	0.66	$\pm$ 	0.04	&	9556	$\pm$ 	1338	&	-5483	$\pm$ 	658	&	&	0.34	$\pm$ 	0.02	&	3816	$\pm$ 	406	&	0	$\pm$	10	&	&	…			&	…			&	…			\\
PKS 2000-330	&	4950	$\pm$	346	&	-0.33	$\pm$	0.01	&	0.39	$\pm$	0.01	&	-2373	$\pm$	226	&	-1923	$\pm$	104	&	-1465	$\pm$	154	&	-1155	$\pm$	169	&	&	0.70	$\pm$ 	0.04	&	7340	$\pm$ 	1028	&	-1300	$\pm$ 	156	&	&	0.30	$\pm$ 	0.02	&	3142	$\pm$ 	334	&	0	$\pm$	10	&	&	…			&	…			&	…			&	&	…			&	…			&	…			&	&	…			&	…			&	…			&	&	…			&	…			&	…			\\
SDSSJ210524.49+000407.3	&	8443	$\pm$	592	&	-0.15	$\pm$	0.01	&	0.43	$\pm$	0.01	&	-5060	$\pm$	297	&	-4695	$\pm$	134	&	-4452	$\pm$	228	&	-4175	$\pm$	279	&	&	0.89	$\pm$ 	0.05	&	10418	$\pm$ 	1459	&	-3926	$\pm$ 	471	&	&	0.11	$\pm$ 	0.02	&	3504	$\pm$ 	372	&	-82	$\pm$ 	20	&	&	…			&	…			&	…			&	&	0.55	$\pm$ 	0.03	&	10418	$\pm$ 	1459	&	-8749	$\pm$ 	1050	&	&	0.45	$\pm$ 	0.03	&	3504	$\pm$ 	372	&	-81	$\pm$ 	20	&	&	…			&	…			&	…\\			
SDSSJ210831.56-063022.5	&	9389	$\pm$	619	&	-0.30	$\pm$	0.01	&	0.42	$\pm$	0.01	&	-5744	$\pm$	365	&	-4982	$\pm$	177	&	-4284	$\pm$	264	&	-3709	$\pm$	334	&	&	0.87	$\pm$ 	0.05	&	13618	$\pm$ 	1907	&	-3250	$\pm$ 	390	&	&	0.13	$\pm$ 	0.01	&	4340	$\pm$ 	461	&	0	$\pm$	10	&	&	…			&	…			&	…			&	&	0.70	$\pm$ 	0.04	&	13618	$\pm$ 	1907	&	-3250	$\pm$ 	390	&	&	0.30	$\pm$ 	0.02	&	4340	$\pm$ 	461	&	0	$\pm$	10	&	&	…			&	…			&	…			\\
SDSSJ212329.46-005052.9	&	7978	$\pm$	524	&	-0.21	$\pm$	0.01	&	0.42	$\pm$	0.01	&	-3752	$\pm$	332	&	-3311	$\pm$	139	&	-2876	$\pm$	231	&	-2551	$\pm$	277	&	&	0.72	$\pm$ 	0.04	&	10217	$\pm$ 	1430	&	-2694	$\pm$ 	323	&	&	0.28	$\pm$ 	0.01	&	5457	$\pm$ 	580	&	0	$\pm$	10	&	&	…			&	…			&	…			&	&	0.63	$\pm$ 	0.04	&	10217	$\pm$ 	1430	&	-4233	$\pm$ 	508	&	&	0.37	$\pm$ 	0.02	&	5457	$\pm$ 	580	&	0	$\pm$	10	&	&	…			&	…			&	…			\\
SDSSJ235808.54+012507.2	&	6504	$\pm$	398	&	-0.25	$\pm$	0.01	&	0.44	$\pm$	0.01	&	-3197	$\pm$	245	&	-2727	$\pm$	114	&	-2312	$\pm$	186	&	-2029	$\pm$	235	&	&	0.80	$\pm$ 	0.05	&	8302	$\pm$ 	1162	&	-2163	$\pm$ 	260	&	&	0.20	$\pm$ 	0.01	&	3468	$\pm$ 	369	&	0	$\pm$	10	&	&	…			&	…			&	…			&	&	0.49	$\pm$ 	0.03	&	9231	$\pm$ 	1292	&	-5323	$\pm$ 	639	&	&	0.51	$\pm$ 	0.03	&	3468	$\pm$ 	369	&	0	$\pm$	10	&	&	…			&	…			&	…			\\

\noalign{\smallskip}

            \hline
            \noalign{\smallskip}
Median & 6934 $\pm$ 2406 & -0.27 $\pm$ 0.10 & 0.39 $\pm$ 0.08 & -3192 $\pm$ 1828 & -2593 $\pm$ 1426 & -2103 $\pm$ 1359 & -1860 $\pm$ 1062 & & 0.72 $\pm$ 0.19 & 10317 $\pm$ 2820 & -2036 $\pm$ 1053 & & 0.28 $\pm$ 0.19 & 3486 $\pm$ 927 & … & & … & … & … & & 0.60 $\pm$ 0.11 & 10217 $\pm$ 2427 & -4249 $\pm$ 2095 & & 0.4 $\pm$ 0.11 & 3504 $\pm$ 1048 & … & & … & … & …  \\
\noalign{\smallskip}

            \hline
            \noalign{\smallskip}
            \multicolumn{32}{c}{Population B}\\
            \noalign{\smallskip}
            \hline
            \noalign{\smallskip}
            
        SDSSJ114358.52+052444.9	&	7078	$\pm$	766	&	-0.30	$\pm$	0.05	&	0.27	$\pm$	0.01	&	-3049	$\pm$	511	&	-2059	$\pm$	230	&	-1375	$\pm$	204	&	-1147	$\pm$	185	&	&	0.60	$\pm$ 	0.06	&	14345	$\pm$ 	1037	&	-1497	$\pm$ 	180	&	&	0.24	$\pm$ 	0.04	&	5055	$\pm$ 	365	&	0	$\pm$	10	&	&	0.16	$\pm$ 	0.02	&	11848	$\pm$ 	2109	&	3822	$\pm$ 	515	&	&	0.57	$\pm$ 	0.06	&	14345	$\pm$ 	1037	&	-3216	$\pm$ 	433	&	&	0.17	$\pm$ 	0.02	&	5055	$\pm$ 	365	&	0	$\pm$	10	&	&	0.26	$\pm$ 	0.03	&	11848	$\pm$ 	2109	&	2540	$\pm$ 	342	\\
SDSSJ115954.33+201921.1	&	5741	$\pm$	585	&	-0.08	$\pm$	0.05	&	0.29	$\pm$	0.01	&	-1671	$\pm$	444	&	-1851	$\pm$	122	&	-1516	$\pm$	166	&	-1253	$\pm$	174	&	&	0.54	$\pm$ 	0.05	&	9933	$\pm$ 	718	&	-1307	$\pm$ 	157	&	&	0.32	$\pm$ 	0.05	&	5716	$\pm$ 	413	&	0	$\pm$	10	&	&	0.14	$\pm$ 	0.01	&	8458	$\pm$ 	1506	&	4397	$\pm$ 	592	&	&	0.36	$\pm$ 	0.04	&	9933	$\pm$ 	718	&	-1307	$\pm$ 	176	&	&	0.07	$\pm$ 	0.01	&	5716	$\pm$ 	413	&	0	$\pm$	10	&	&	0.58	$\pm$ 	0.06	&	8458	$\pm$ 	1506	&	4397	$\pm$ 	592	\\
SDSSJ120147.90+120630.2	&	4995	$\pm$	475	&	-0.08	$\pm$	0.04	&	0.32	$\pm$	0.01	&	-1354	$\pm$	284	&	-1133	$\pm$	104	&	-1061	$\pm$	148	&	-1015	$\pm$	132	&	&	0.50	$\pm$ 	0.05	&	7464	$\pm$ 	540	&	-1454	$\pm$ 	174	&	&	0.35	$\pm$ 	0.05	&	3945	$\pm$ 	285	&	0	$\pm$	10	&	&	0.15	$\pm$ 	0.01	&	6710	$\pm$ 	1194	&	1508	$\pm$ 	203	&	&	0.44	$\pm$ 	0.04	&	7464	$\pm$ 	540	&	-4676	$\pm$ 	630	&	&	0.09	$\pm$ 	0.01	&	3945	$\pm$ 	285	&	0	$\pm$	10	&	&	0.47	$\pm$ 	0.05	&	6710	$\pm$ 	1194	&	1508	$\pm$ 	203	\\
SDSSJ153830.55+085517.0	&	5819	$\pm$	448	&	-0.41	$\pm$	0.02	&	0.35	$\pm$	0.02	&	-2841	$\pm$	303	&	-2108	$\pm$	147	&	-1381	$\pm$	187	&	-961	$\pm$	185	&	&	0.68	$\pm$ 	0.07	&	10024	$\pm$ 	725	&	-992	$\pm$ 	119	&	&	0.32	$\pm$ 	0.05	&	4313	$\pm$ 	312	&	0	$\pm$	10	&	&	…			&	…			&	…			&	&	0.48	$\pm$ 	0.05	&	10024	$\pm$ 	725	&	-992	$\pm$ 	134	&	&	0.52	$\pm$ 	0.08	&	4313	$\pm$ 	312	&	0	$\pm$	10	&	&	…			&	…			&	…			\\

            \noalign{\smallskip}
            \hline
             \noalign{\smallskip}
Median &  5780 $\pm$ 1474 & -0.17 $\pm$ 0.21 & 0.30 $\pm$ 0.06 & -2256 $\pm$ 1563 & -1955 $\pm$ 783 & -1378 $\pm$ 342 & -1081 $\pm$ 252 & & 0.57 $\pm$ 0.15 & 9978 $\pm$ 2162 & -1380 $\pm$ 415 & & 0.30 $\pm$ 0.07 & 4129 $\pm$ 1268 & … & & 0.15 $\pm$ 0.01 & 8458 $\pm$ 2569 & 3822 $\pm$ 1444 & & 0.46 $\pm$ 0.16 & 9978 $\pm$ 2087 & -2261 $\pm$ 2920 & & 0.14 $\pm$ 0.24 & 4129 $\pm$ 1255 & … & & 0.47 $\pm$ 0.16 & 8458 $\pm$ 2569 & 2540 $\pm$ 1444\\
\noalign{\smallskip}

            \hline
         \end{tabular}
}
    
     %\tablefoot{Columns: (1) Source identification. (2) Observation date. (3) Grating. (4) Central wavelength of the filter, in units of $\mu$m.}
   \end{sidewaystable*}
%-----------------------------------------------------------------
\subsubsection{\ion{C}{IV}$\lambda1549$+\ion{He}{II}$\lambda1640$}

\begin{figure}
    \centering
    \includegraphics[width=0.9\columnwidth]{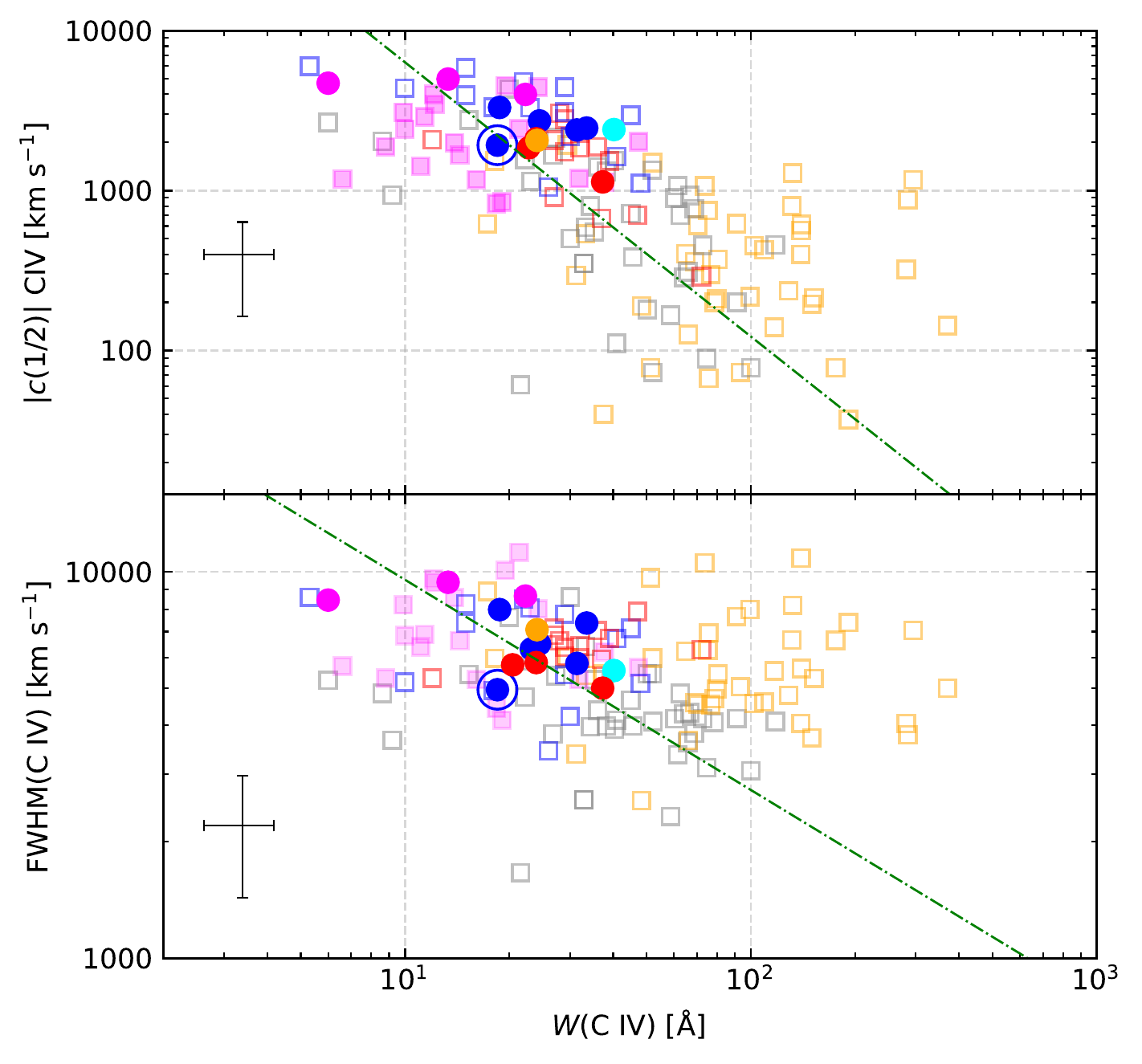}
    \caption{\textit{Top:} Relation between centroid velocity at 1/2 intensity ($c$(1/2)) and rest-frame equivalent width of \civ{}. \textit{Bottom:} FWHM of \civ{} \textit{versus} $W$(\ion{C}{IV}). Colour scheme for our sample as in Fig. \ref{fig:HB_BC_Full}. Filled magenta squares indicate data from \cite{martinezaldama_2018}. Blue and red open squares represent Pop. A and Pop. B HE sources from \citetalias{sulentic_2017}. Low-$z$ FOS data from \citetalias{Sulentic_2007} are represented by grey (Pop. A) and orange (Pop. B) squares. Green point-dashed lines indicate the linear regression obtained through the bisector method including only the Pop. A sources from the four samples.  Error bars refer to 1$\sigma$ level of confidence and were estimated only for our sample.}
    \label{fig:w_c12_civ}
\end{figure}

\begin{figure}
    \centering
    \includegraphics[width=0.8\columnwidth]{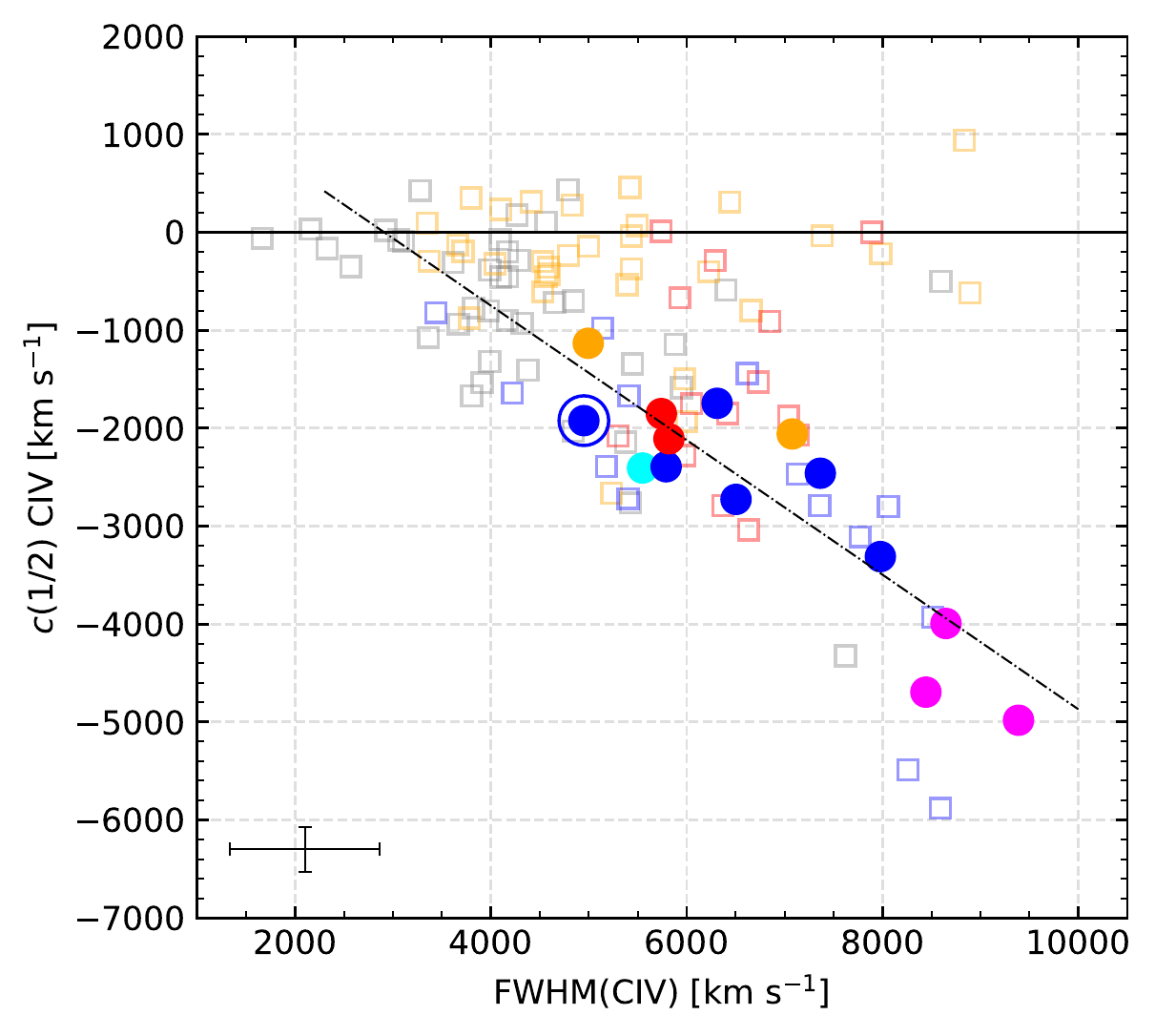}
    \caption{Centroid velocity at 1/2 flux intensity,  $c$(1/2), for the \civ{} emission line \textit{versus} FWHM of the full profile of \civonly{}. Filled circles represent our sample and the colour scheme is the same as in Fig. \ref{fig:HB_BC_Full}. Blue and red open squares are high-redshift sources from \citetalias{sulentic_2017}. Low-$z$ Pop. A and Pop. B are represented by grey and orange open squares, respectively. Dashed line represents the least squares linear regression presented in Eq. \ref{eq:c12fwhm_civ}, including only Pop. A sources. Error bars refer to 1$\sigma$ level of confidence for our sample.}
    \label{fig:civ_c12_fwhm}
\end{figure}

%\par Table \ref{tab:specphot_CIV} reports the spectrophotometric measurements on the \ion{C}{IV}$\lambda1549$+\ion{He}{II}$\lambda1640$ region.
\par Measurements on the \ion{C}{IV}$\lambda1549$+\ion{He}{II}$\lambda1640$ blend are reported in Table \ref{tab:meas_CIV}. As for H$\beta$, we list FWHM (Col. 2), A.I. (Col. 3), kurtosis (Col. 4), and centroids $c$(1/4), $c$(1/2), $c$(3/4) and $c$(9/10) in Cols. 5 to 8 for the \ion{C}{IV}$\lambda1549$ full line profile. Information on individual components is also given for both \ion{C}{IV}$\lambda1549$ and \ion{He}{II}$\lambda1640$. Cols. 9 to 11 give the relative intensity $I/I_{\mathrm{tot}}$, FWHM, and shift for \ion{C}{IV}$\lambda1549$ BLUE; Cols. 12 to 14 for \ion{C}{IV}$\lambda1549$ BC; Cols. 15 to 17 for \ion{C}{IV}$\lambda1549$ VBC; Cols. 18 to 20 for \ion{He}{II}$\lambda1640$ BLUE; Cols. 21 to 23 for \ion{He}{II}$\lambda1640$ BC; and Cols. 24 to 26 for \ion{He}{II}$\lambda1640$ VBC. 

\par We have ten out of 12 Pop. A sources and four out of ten Pop. B quasars with \ion{C}{IV} covered for analysis.   The fits are remarkably good and allow for the ``plateau'' found as an extension of the red wing of \civ{} close to 1575 \AA, which can usually be well reproduced by fitting \ion{C}{IV}$\lambda1549$ and \ion{He}{II}$\lambda1640$ with the same FWHM for blueshifted, broad, and very broad components \citep{fineetal10,Marziani_2010}.  In general, Pop. A objects present a larger FWHM of the full profile, always $\ge$ 4900 km s$^{-1}$, with respect to Pop. B. Pop. A are also those that in general present higher asymmetries and shifts towards the blue.  Apart from SDSSJ212329.46-005052.9, all Pop. A sources present a FWHM(\ion{C}{IV}$_{\rm{BC}}$) $\le 4500$ km s$^{-1}$.  The Pop. A3 SDSSJ210831.56-063022.5 is the source with the strongest blueshift, reaching $c$(1/2) $\approx -4980$ km s$^{-1}$ and with similar values of fractional intensities along the profile (as can be seen in the top plot of Fig. \ref{fig:w_c12_civ}). 
\par In the case of Pop. B sources, we have found that the VBC accounts for only $\sim 15$\% of the full profile. The widest VBC FWHM is seen in SDSSJ114358.52+052444.9, which reaches 11848 km s$^{-1}$ with a shift of 3822 km s$^{-1}$. This same source is the one that also presents  the broadest FWHM of the BLUE (14345 km s$^{-1}$) and larger shift ($-1497$ km s$^{-1}$). In all Pop. B cases, the BLUE component accounts for at least 50\% of the \civ\ full profile, but still a fraction that is lower than in Pop. A ($\sim$ 70 \%). Correspondingly, the Pop. A \civ\ profiles in general present higher blueshifts and asymmetry indexes than the ones measured in Pop. B. 

%\par A similar approach was followed in the analysis of the blueshifted components. In these cases our first guess is that \ion{C}{IV}$\lambda$1549 and \ion{He}{II}$\lambda$1640 blueshifted profiles share the same FWHM with different intensities. However, in the majority of the sources, these components are free to vary also in FWHM. In both Pop. A and Pop. B, the blueshifted component represents the main contribution to the full profile, always $\ge 50$\% (with some cases reaching $\sim 80$\%). Apart from that, the blueshifted component also shows a very wide FWHM and a significant shift towards blue wavelengths in both populations and accounts for the blue shifts seen in the centroid velocities. In the two populations of quasars, the blueshifted components of both \ion{C}{IV}$\lambda$1549 and \ion{He}{II}$\lambda$1640 present very high values of FWHM (in the majority of the cases $\ge10000$ km s$^{-1}$) and blue shifts going from $\sim1000$ to $\sim 5000$ km s$^{-1}$.

\par The top and bottom panels from Fig. \ref{fig:w_c12_civ} show the relation between the rest-frame equivalent widths of \civ{} and the centroid velocities at 1/2 peak intensity ($c$(1/2)) and FWHM respectively). We also include in the figure the bright, high-$z$ xA sample of \cite{martinezaldama_2018} as well as the high-$z$ data from \citetalias{sulentic_2017} and the FOS low-$z$ data from \citetalias{Sulentic_2007}.  Our data are in good agreement with the ones of \cite{martinezaldama_2018} and \citetalias{sulentic_2017}. Our sample together with the high-$L$ comparison samples, presents smaller values of \civ{} equivalent width when compared with the FOS data. Additionally, the high-$z$, high-$L$ sources are those that present the most blueshifted \civ{} profiles among all quasars \citep{shen_2011,Rankine_2020,vietrietal18,vietrietal20}. The Pearson correlation coefficient between $W$ and $c$(1/2) of \civ{}  for the Pop. A sources is -0.64 (with $\rho=2.1\times 10^{-12}$) and their relation is given in Table \ref{tab:linreg}.
%\begin{equation}
%    \log W (\ion{C}{IV})=(-0.583\pm 0.055)\log |c(1/2)|_{\ion{C}{IV}}+(3.217\pm 0.174).
%\end{equation}
There is also a correlation between $W$ and FWHM \civ{} with $r \approx -0.49$, if we restrict the samples to only Pop. A objects (Table \ref{tab:linreg}).  
%In this case, we found a correlation coefficient of -0.49 for the relation between $W$ and FWHM of \civ{}, \textbf{also presented in }.
%\begin{equation}
%    \log W(\ion{C}{IV})=(-1.838\pm 0.189)\log \textrm{FWHM(\ion{C}{IV})}+(8.314+0.710).
%\end{equation}

%If Pop. B objects are considered along with Pop. A sources, this last correlation is wiped out because Pop. B sources roughly show the same width as Pop. A but with much smaller shift especially at low luminosity (see bottom panel of Fig. \ref{fig:w_c12_civ}).
\par Fig. \ref{fig:civ_c12_fwhm} presents the strong correlation between FWHM and $c$(1/2) of the \civonly{} line, with a Pearson $r \approx$ -0.90 for high-$z$\ Pop. A sources:
%\begin{equation}
% c(1/2)_{\mathrm C IV} =(-0.68\pm 0.10)\textrm{FWHM}({\rm C IV})+(2000.00\pm 737.10).   
%\end{equation}
\begin{equation}
\label{eq:c12fwhm_civ}
c(1/2)_{\rm \ion{C}{IV}} =(-0.68\pm 0.10) \textrm{FWHM}({\rm{\ion{C}{IV}}})+(2000\pm 737)  
\end{equation}

This relation is in good agreement with previous results \cite[e.g.][and references therein]{coatman_2016, marziani_2016a, sulentic_2017, Sun_2018, vietrietal18}. As the blueshift of \civonly\ increases, FWHM(\civonly) is likely to increase as well. Both \civonly\ FWHM and  blueshift  parameterised by $c$(1/2)  show a clear increase from spectral types A1 to A3, the latter being those with the highest blueshifts (median value $\approx$ $-4500$ km\,s$^{-1}$) and the widest FWHM (median $\approx$ 9000 km\,s$^{-1}$).

 In the case of \civ{} in Pop. B sources, the shifts at 1/2 flux intensities are also displaced towards blue wavelengths with average value of $c$(1/2) $=-1681$ km s$^{-1}$ and mean  FWHM $\approx 5940$ \kms, somewhat smaller than for Pop. A. 
  However, the $c$(1/2) values for the Pop. B sources from our sample are still higher than the majority of the low-luminosity and low-redshift quasars, which usually present a $c$(1/2) $\le 200$ km s$^{-1}$. Low-luminosity Pop. B spectra show a sort of dichotomy in the  \civ\ shift  distribution: a fraction of sources remains unshifted or with modest shifts to the blue (the orange and grey ``cloud'' of points in Fig. \ref{fig:civ_c12_fwhm}). The FOS low-luminosity and low-redshift sources tend to present smaller $c$(1/2) in the \civ{} emission line when compared with our data. This indicates that the \civ\ blueshifted components seen in higher-luminosity sources are more prominent  than the ones observed in quasars of lower  luminosities. %\chony{Review the points in fig. 8 and previous paragraph}\chonyblue{, could be  some confused squares/colors:  FOS sample is S07 and Bachev 2004.  S17 is high-z high-L UV \textbf{HEMS}.}

%\par The location of our sources in the parameter plane in Fig. \ref{fig:civ_c12_fwhm} is in a good agreement with the related literature based on  the HE high-redshift sources \citep{Marziani_2009,sulentic_2017}.  Our data and the \textbf{HEMS} share the same area on the plot, indicating a common behaviour of high-redshift, high-luminosity \textbf{quasars} in this context. Nevertheless,  

\begin{table}[t!]
      \caption[]{Measurements on the full profiles of  \ion{Si}{IV}$\lambda1397$+\ion{O}{IV}]$\lambda1402$.}
      \label{tab:meas_SiIV}
      \resizebox{\linewidth}{!}{
                 \begin{tabular}{lccccc}
            \hline
            \hline
            \noalign{\smallskip}
             Source & FWHM & A.I. & Kurtosis & $c$(1/2) & $c$(9/10)\\
             & [km s$^{-1}$] & & & [km s$^{-1}$] &  [km s$^{-1}$]\\
             (1) & (2) & (3) & (4) & (5) & (6) \\
            \noalign{\smallskip}
            \hline
            \noalign{\smallskip}
            \multicolumn{6}{c}{Population A}\\
            \noalign{\smallskip}
            \hline
            \noalign{\smallskip}
SDSSJ005700.18+143737.7	&	8659	$\pm$	1202	&	-0.34	$\pm$	0.02	&	0.36	$\pm$	0.03	&	-2110	$\pm$	377	&	-858	$\pm$	443	\\
SDSSJ132012.33+142037.1	&	5128	$\pm$	859	&	-0.26	$\pm$	0.05	&	0.34	$\pm$	0.01	&	-376	$\pm$	249	&	131	$\pm$	293	\\
SDSSJ135831.78+050522.8	&	7438	$\pm$	1658	&	-0.27	$\pm$	0.07	&	0.31	$\pm$	0.01	&	-150	$\pm$	534	&	499	$\pm$	395	\\
Q 1410+096	&	5430	$\pm$	790	&	0.00	$\pm$	0.09	&	0.40	$\pm$	0.01	&	750	$\pm$	157	&	666	$\pm$	335	\\
SDSSJ161458.33+144836.9	&	7370	$\pm$	860	&	-0.26	$\pm$	0.06	&	0.48	$\pm$	0.01	&	-392	$\pm$	227	&	680	$\pm$	844	\\
PKS 2000-330	&	3494	$\pm$	329	&	0.32	$\pm$	0.13	&	0.55	$\pm$	0.02	&	888	$\pm$	63	&	54	$\pm$	112	\\
SDSSJ210831.56-063022.5	&	9253	$\pm$	1880	&	-0.24	$\pm$	0.05	&	0.28	$\pm$	0.03	&	-2613	$\pm$	481	&	-1618	$\pm$	452	\\
SDSSJ212329.46-005052.9	&	6157	$\pm$	892	&	-0.06	$\pm$	0.05	&	0.40	$\pm$	0.01	&	-868	$\pm$	190	&	-635	$\pm$	396	\\
SDSSJ235808.54+012507.2	&	7135	$\pm$	974	&	-0.14	$\pm$	0.03	&	0.40	$\pm$	0.01	&	-533	$\pm$	252	&	-55	$\pm$	443	\\
\noalign{\smallskip}

            \hline
             \noalign{\smallskip}
Median & 7135 $\pm$ 1888 & -0.14 $\pm$ 0.20 & 0.40 $\pm$ 0.04 & -392 $\pm$ 718 & -55 $\pm$ 1134 \\
\noalign{\smallskip}
\hline
            \noalign{\smallskip}
            \multicolumn{6}{c}{Population B}\\
            \noalign{\smallskip}
            \hline
            \noalign{\smallskip}
SDSSJ114358.52+052444.9	&	7070	$\pm$	1027	&	-0.11	$\pm$	0.05	&	0.40	$\pm$	0.01	&	184	$\pm$	242	&	394	$\pm$	439	\\
SDSSJ115954.33+201921.1	&	6813	$\pm$	1372	&	-0.12	$\pm$	0.07	&	0.34	$\pm$	0.01	&	257	$\pm$	396	&	644	$\pm$	378	\\
SDSSJ120147.90+120630.2	&	5000	$\pm$	717	&	-0.18	$\pm$	0.20	&	0.34	$\pm$	0.04	&	667	$\pm$	156	&	609	$\pm$	315	\\
SDSSJ153830.55+085517.0	&	5906	$\pm$	1222	&	-0.18	$\pm$	0.02	&	0.36	$\pm$	0.01	&	159	$\pm$	416	&	504	$\pm$	413	\\
            \noalign{\smallskip}
            \hline
             \noalign{\smallskip}
Median & 6359 $\pm$ 1197 & -0.15 $\pm$ 0.06 & 0.35 $\pm$ 0.03 & 220 $\pm$ 181 & 556 $\pm$ 141\\
\noalign{\smallskip}
\hline
         \end{tabular}
}
    
    % \tablefoot{Columns: (1) Source identification. (2) Observation date. (3) Grating. (4) Central wavelength of the filter, in units of $\mu$m.}
   \end{table}

\subsubsection{\ion{Si}{IV}$\lambda1397$+\ion{O}{IV}]$\lambda1402$}

\par Table \ref{tab:meas_SiIV} presents the measurements on the full profile of \ion{Si}{IV}$\lambda1397$+\ion{O}{IV}]$\lambda1402$. We report values of FWHM (Col. 2), A.I. (Col. 3), kurtosis (Col. 4), $c$(1/2) (Col. 5), and c(9/10) (Col. 6). \ion{Si}{IV}$\lambda1397$+\ion{O}{IV}]$\lambda1402$ BC is set at rest-frame and our initial guess for this component takes into account the results obtained for the \ion{C}{IV}$_{\rm{BC}}$ after fitting the \ion{C}{IV}$\lambda$1549+\ion{He}{II}$\lambda$1640 region. SDSSJ141546.24+112943.4 has very strong and wide absorption lines in this region of the UV spectra, which makes it difficult to perform a reliable fitting (see Fig. \ref{fig:1415_UV} in the Appendix \ref{app_uv}).
\par Regarding the BLUE of \ion{Si}{IV}$\lambda1397$+\ion{O}{IV}]$\lambda1402$, they are very wide (even if apparently not as wide as in \ion{C}{IV}$\lambda$1549+\ion{He}{II}$\lambda$1640), reaching more than 6000 km s$^{-1}$ in all cases, and representing a significant percentage of the full emission line profile. The shifts towards shorter wavelength are also smaller than for \ion{C}{IV}$\lambda$1549, but they are still very high, going from 800 to 2700 km s$^{-1}$, with the extreme amplitude  of SDSSJ135831.78+050522.8 that reaches $\approx 4400$ km s$^{-1}$. 

\par The fluxes of \civ{} and \siv{} presented in Table \ref{tab:specphot_1900A} indicate that on average Pop. A presents higher values of the \civ{}/\siv{} ratio  ($\approx 0.69$) than Pop. B ($\approx 0.49$). This discrepancy may be linked to differences in chemical abundances with Pop. A sources being systematically more metal rich \cite[c.f.][]{sniegowska_2021,punslyetal20}.  However, the issue goes beyond the scope of the present work and will be discussed elsewhere.

%\subsubsection{Error estimates in the UV range}

%\par \textbf{The error evaluation of the UV lines has been performed following the same approach of the optical region. In this case, the FWHM uncertainties are between 10 and 15\% and errors on intensity measurements  are usually $ \lesssim 10\%$ for the strongest emission line components.}
   
%-----------------------------------------------------------------
\section{Discussion}
\label{discussion}
%\subsection{Location in the 4DE1}
%\par Fig. \ref{fig:optical_plane} shows the location of the sources in the optical plane of the 4DE1.  
\par In the previous section, we have shown that   Pop. A and Pop. B sources  may reflect  different contributions of line emitting gas that produces prominent blueshifted features and is most likely associated with an outflow. %such as in the FWHM(H$\beta_{\rm BC}$) \textit{vs.} FWHM(H$\beta_{\rm full}$) relation (Fig. \ref{fig:HB_BC_Full}) as well as in UV emission lines like \civ{} and \aliii{} and \ion{C}{III}]$\lambda1908$. 
To shed further  light on the role of outflows, we now report an interline comparison in both the optical and UV spectral regions. We highlight the effect of the outflowing components on the estimate of physical parameters such as black hole mass and Eddington ratio. 

%the main differences and similarities found between our data and other samples that include both high- and low-redshift, high- and low-luminosity, as well as radio-loud and radio-quiet quasars. The relations between these properties and physical parameters are reported and we also estimate the black hole mass by using H$\beta$, \civ{}, and \aliii{} emission lines and different methods.

\subsection{Defining the outflow}

\subsubsection{H$\beta$ and [\ion{O}{III}]$\lambda\lambda$4959,5007}
\label{disc:hboiii}
\label{section_hb_oiii}

\begin{figure}[t!]
\centering
\includegraphics[width=0.8\columnwidth]{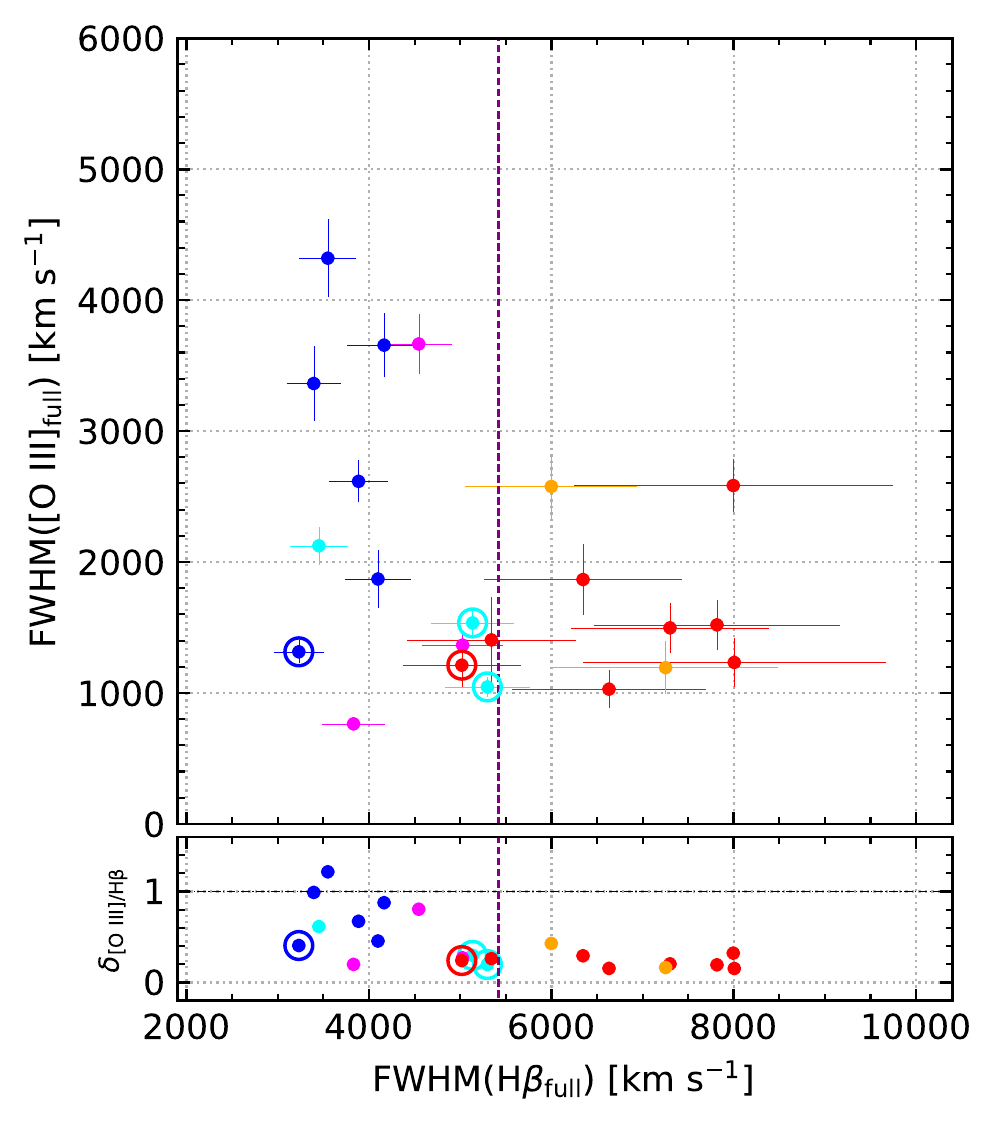}
\caption{\textit{Top panel:} FWHM of the full profile of [\ion{O}{III}]$\lambda$5007 \textit{vs.} FWHM of the full profile of H$\beta$. \textit{Bottom panel:} Ratio $\delta_{\textrm{[O III]/H}\beta}=$ FWHM([\ion{O}{III}]$_{\rm full}$)/FWHM(H$\beta_{\rm full}$) \textit{vs.} FWHM of the full profile of H$\beta$. As in Fig. \ref{fig:HB_BC_Full}, each colour represent one different spectral type, and symbols surrounded by open circles indicate the radio-loud quasars.}
\label{fig:oiii_hb}
\end{figure}

\par The relation between the FWHM of the full profiles of \hb{} and \ion{O}{III}]$\lambda$5007 is shown in Fig. \ref{fig:oiii_hb}. In general, the two populations present a FWHM(H$\beta_{\rm{full}}$) higher than FWHM([\ion{O}{III}]$_{\rm{full}}$), as expected from previous observations \citep[][\citetalias{Marziani_2009}]{Sulentic_2004, Sulentic_2007,zamfir_2010}. At low-z, only in the case of  the   ``blue-outliers'' the \oiii\ profiles appear to be very boxy-shaped and with the FWHM $> 1000$ \kms{}.  Some of these sources are NLSy1s  and the \oiii\ FWHM is becoming comparable to the one of the broad \hb{} profile sources \citep{zamanovetal02, Komossa_2008, Cracco_2016, Komossa_2018, Berton_2021}. At high $z$ and high luminosity the  \oiii{} profiles often appear much broader, as do also the \hb\ broad profiles \citep{Carniani_2015,Fiore_2017,Villar-martin_2020}. Two cases in point from previous works are 2QZJ002830.4-281706 at $z = 2.$\ \citep{canodiazetal12, Carniani_2015} and HE0940-1050 at $z = 3.1$ \citep{Marziani_2017}.  We have four sources that present  \oiii\ FWHM above 3000 \kms, comparable to the \hb\ FWHM, with   one extraordinary case in which FWHM(H$\beta_{\rm{full}}$) is smaller than FWHM([\ion{O}{III}]$_{\rm{full}}$) (SDSSJ135831.78+050522.8, Pop. A1), with FWHM(H$\beta_{\rm{full}}$) $\approx 3550$ km s$^{-1}$ and FWHM([\ion{O}{III}]$_{\rm{full}}$) $\approx 4320$ km s$^{-1}$ (see Fig. \ref{fig:1358_UV}). Although FWHM of \oiii\ $\gtrsim 2000$ \kms\ are frequently observed at high luminosity, some extreme values should be taken with care because the \oiii\ profiles are weak and broad:  it is  difficult to properly define the \oiii{} profiles especially when the broad \hb\ red wing is strong. 

%\paolaQ{[Really necessary?]} The two quasar populations A and B present different trends. Pop. A sources are the ones that present wider \oiii{} FWHM and larger shifts. The mean values of FWHM equal to 2731 km s$^{-1}$ for Pop. A sources and 1589 km s$^{-1}$ for Pop. B. The Pop. B are the ones that present the highest differences between the profiles of H$\beta$ and \oiii{}, with very broad H$\beta$ profile and with narrower and more peaked \oiii{} emission line profiles as observed at low $z$.  This is an expected behaviour for Pop. B sources, consistent with the one observed  in  low-redshift data from \cite{zamfir_2010} and in other related works \citep[e.g.,][\citetalias{Marziani_2009}]{Sulentic_2004,Sulentic_2007, Marziani_2019}. Both the \oiii\ and the \hb\ emission appear to be more "virially dominated" in Pop. B than in Pop. A.

\begin{figure}[t!]
    \centering
    \includegraphics[width=0.8\columnwidth]{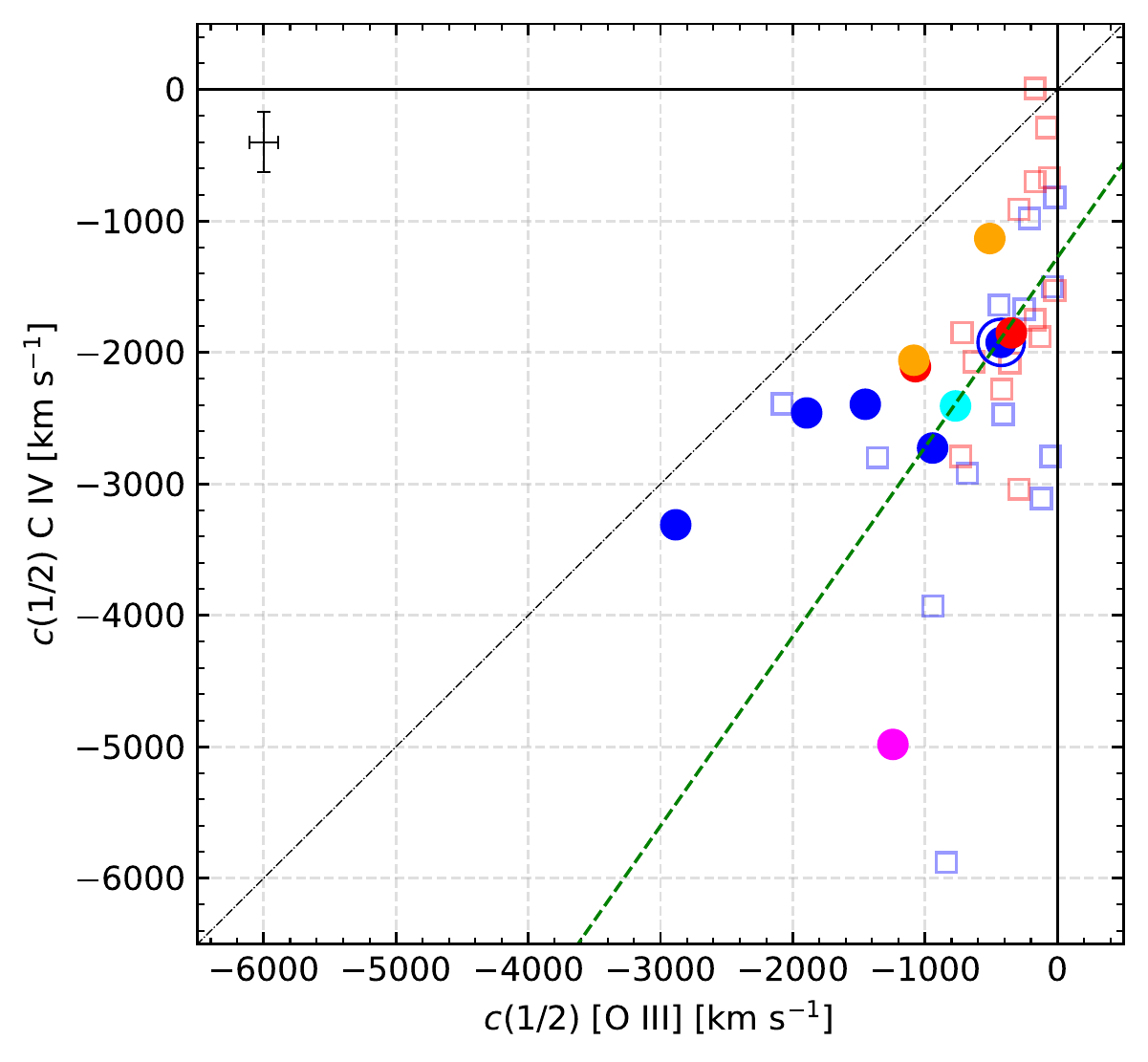}
    \caption{Centroid velocity at 1/2 flux intensity ($c$(1/2)) of \civ{} \textit{vs.} $c$(1/2) of \oiii{}. Blue and red open squares show the HE Pop. A and Pop. B sources studied by \citetalias{Marziani_2009} and \citetalias{sulentic_2017}. Color scheme as in Fig. \ref{fig:HB_BC_Full}. The green line indicates the linear regression between $c$(1/2) of \civ{} and \oiii{} obtained through the orthogonal least square method. The error bars refer to 1$\sigma$ level of confidence for our sample.}
    \label{fig:CIV_oiii}
\end{figure}

\begin{figure}[t!]
    \centering
       \includegraphics[width=0.85\columnwidth]{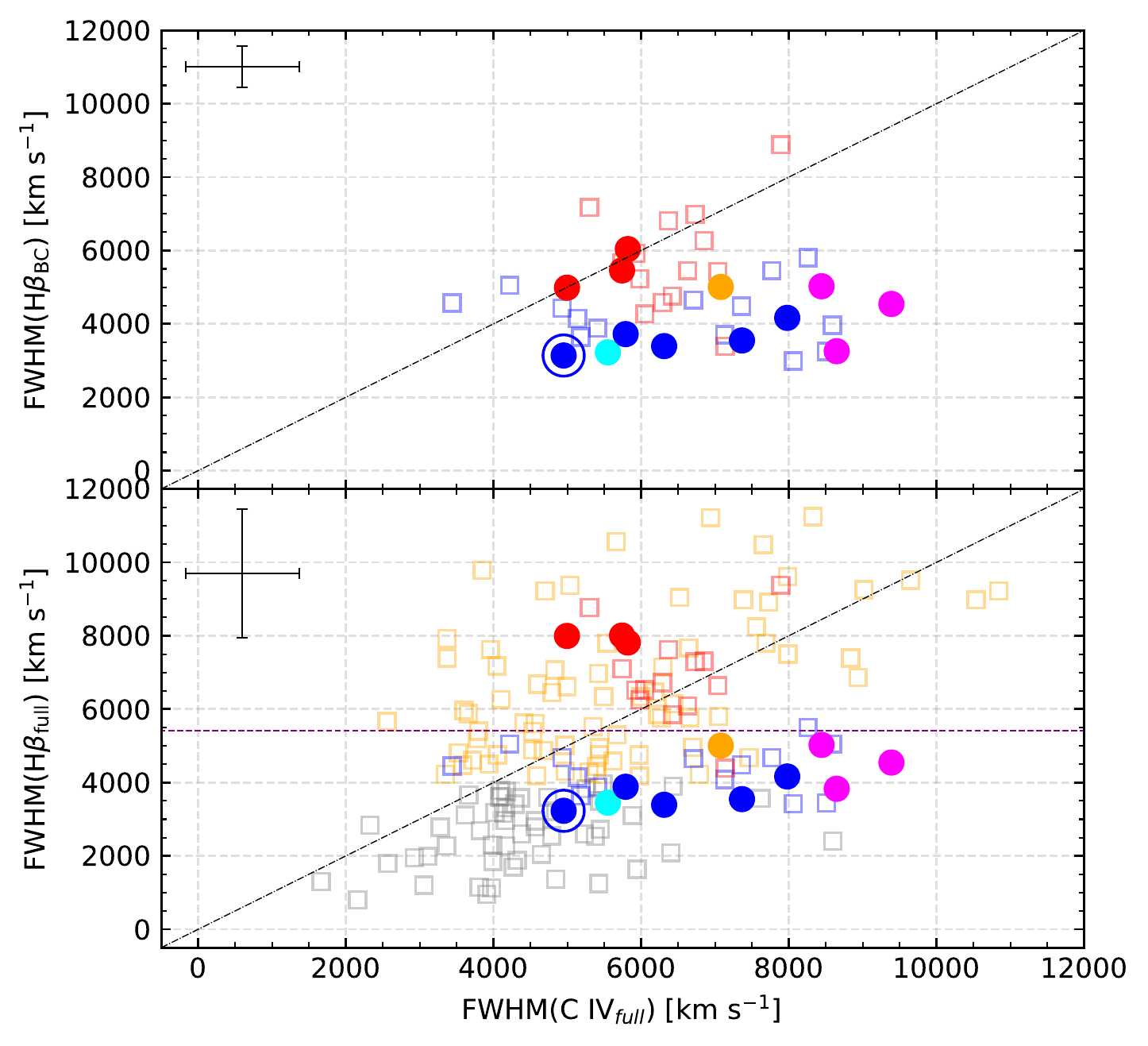}
    \caption{FWHM(H$\beta_{\rm{BC}}$) $vs.$ FWHM(\ion{C}{IV}$_{\rm full}$)   and FWHM(H$\beta_{\mathrm{full}}$) $vs.$ FWHM(\ion{C}{IV}$_{\rm full}$). Sources are identified according to the different bins of spectral types following color scheme from Fig. \ref{fig:HB_BC_Full}. Blue and red open squares represent Pop. A and Pop. B sources from \citetalias{sulentic_2017} and grey and orange open squares indicate Pop. A and Pop. B from \citetalias{Sulentic_2007}, respectively. Horizontal purple line indicates the A/B boundary and the error bars on the top left of the plots refer to 1$\sigma$ level of confidence for our sample.}
    \label{fig:Hb_CIV}
\end{figure}

%On the other hand, Pop. A sources present full \oiii{} profiles that are more in agreement with the full H$\beta$ profile, presenting a broader and less peaked profile.

\subsubsection{\oiii{} \textit{vs.} \ion{C}{IV}$\lambda1549$   }
\label{oiiiciv}
\par Fig. \ref{fig:CIV_oiii} shows the relation between   $c$(1/2) for \oiii{} and \civ{}. The HE comparison sample analysed in \citetalias{Marziani_2009} and \citetalias{sulentic_2017} is included. The sources that present strong shifts in the \oiii{} emission line profiles will also present it in \civ{},  in agreement with the HE data. Both lines show a correlation between their widths and shifts (Fig. \ref{fig:oiii_c12_fwhm} for \oiii\ and Fig. \ref{fig:civ_c12_fwhm} for \civ) indicating that the broadening is mainly associated with a blueshifted component that is increasing in prominence with increasing shift. The least squares linear regression of Fig. \ref{fig:CIV_oiii} (considering both HEMS and our sample) is given by:

%\begin{equation}
%    c(1/2)_{\rm \ion{C}{IV}}= (3.23 \pm 1.28)c(1/2)_{\rm [\ion{O}{III}]}+(-45.21 \pm 661.61).
%\end{equation}

\begin{equation}
    c(1/2)_{\rm \ion{C}{IV}}= (1.50 \pm 0.31)\  c(1/2)_{\rm [\ion{O}{III}]}+(-1239\pm 276),
\end{equation}

\noindent in a good agreement with the study of \citet{Coatman_2019}. The trend (actually, a weak, marginally significant correlation at a $3\sigma$ confidence level) of Fig. \ref{fig:CIV_oiii} raises the issue of the relation between \oiii\ and \civ\ outflows. There is evidence suggesting that the semi-broad component of \oiii\ $W$\ remains almost constant with luminosity \citep[][Section 4.3]{Marziani_2016}, overwhelming the narrow, core component. The narrow component is however mainly associated with the NLR that may extend up to tens of kpc \citep{bennertetal02,bennertetal06}. The dispersion in the relation with more sources around 0 shift in \oiii\ might be explained by a  narrow component whose strength may depend on the past AGN evolution \citep{Storchi_Bergmann_2018}. It is therefore not surprising that, even if a large \civ\ shift is measured, the \oiii\ profile may be unshifted or show only a modest blueshift.  However, the presence of a shift correlation and the relatively large \oiii/\civ\ FWHM ratio support  a physical connection between an inner outflow on scales of a few hundreds gravitational radii where the  BLUE \civ\  component is emitted, and an outflow at the outer edge of the BLR, beyond $10^4 - 10^5$\ gravitational radii \citep{zamanovetal02} where the \oiii\ semibroad component likely originates.  The nature of this connection remains unclear.  It is beyond the scope of the present paper and might be investigated elsewhere.

\subsubsection{\ion{C}{IV}$\lambda1549$ \textit{versus} \hb{}}
\label{hbciv}

\par  Top panel of Fig. \ref{fig:Hb_CIV} represents the relation between   FWHM(H$\beta_{\rm BC}$) and FWHM(C IV$_{\rm full}$), including the \citetalias{sulentic_2017}  sample. Our data is in agreement with this comparison sample in terms of a clear A/B separation.  Pop. A sources seem to present a trend from A1 to A3 in the FWHM(C IV$_{\rm full}$) where the A3 sources show the largest values.   The FWHM(C IV$_{\rm full}$) is larger than the FWHM(H$\beta_{\rm BC}$), and   the widths of the two lines are not correlated. There is a degeneracy between \civ\ and \hb: to a single value of \hb\ FWHM corresponds a wide range of \civ\ FWHM \citepalias[c.f. ][\citealt{Capellupo_2015,mejiarestrepo_2016, coatman_2017}]{Sulentic_2007}. The systematically broader  profile  can be interpreted as a consequence of the dominance of the outflow component in the \civ\ line profile \citep{Marziani_2019}.    

The bottom panel of Fig. \ref{fig:Hb_CIV} shows the FWHM(C IV$_{\rm full}$) \textit{vs.} FWHM(H$\beta_{\rm full}$) where, apart from including \citetalias{sulentic_2017} data, we also display the low-$z$ sample from \citetalias{Sulentic_2007} for comparison. In this plot, the separation between Pop. A and Pop. B is even clearer: Pop. A sources show broader \civ, while those of Pop. B have values consistent with, or narrower, than \hb.

%Regarding the Pop. B sources,  for our three Pop. B1  the relation between FWHM(C IV$_{full}$) and FWHM(H$\beta_{\rm BC}$) is almost one to one, but that 
%FWHM(C IV$_{\rm full}$) and FWHM(H$\beta_{\rm full}$) are weakly correlated with a Pearson correlation coefficient of 0.49, including not only our sample but also \citetalias{sulentic_2017}, \citetalias{Sulentic_2007}.   
%However, the only radio-loud source present in this plot (PKS 2000-330) is the Pop. A one with smaller value of \civ{} FWHM.
%The outlying case is the Pop. A3 SDSSJ210831.56-063022.5, which presents a huge shift in \civ{} ($\sim$ -5000 km s$^{-1}$) and a still large shift of -1243 km s$^{-1}$ in the \oiii{} profile.  

\begin{figure}[t!]
    \centering
    \includegraphics[width=0.8\columnwidth]{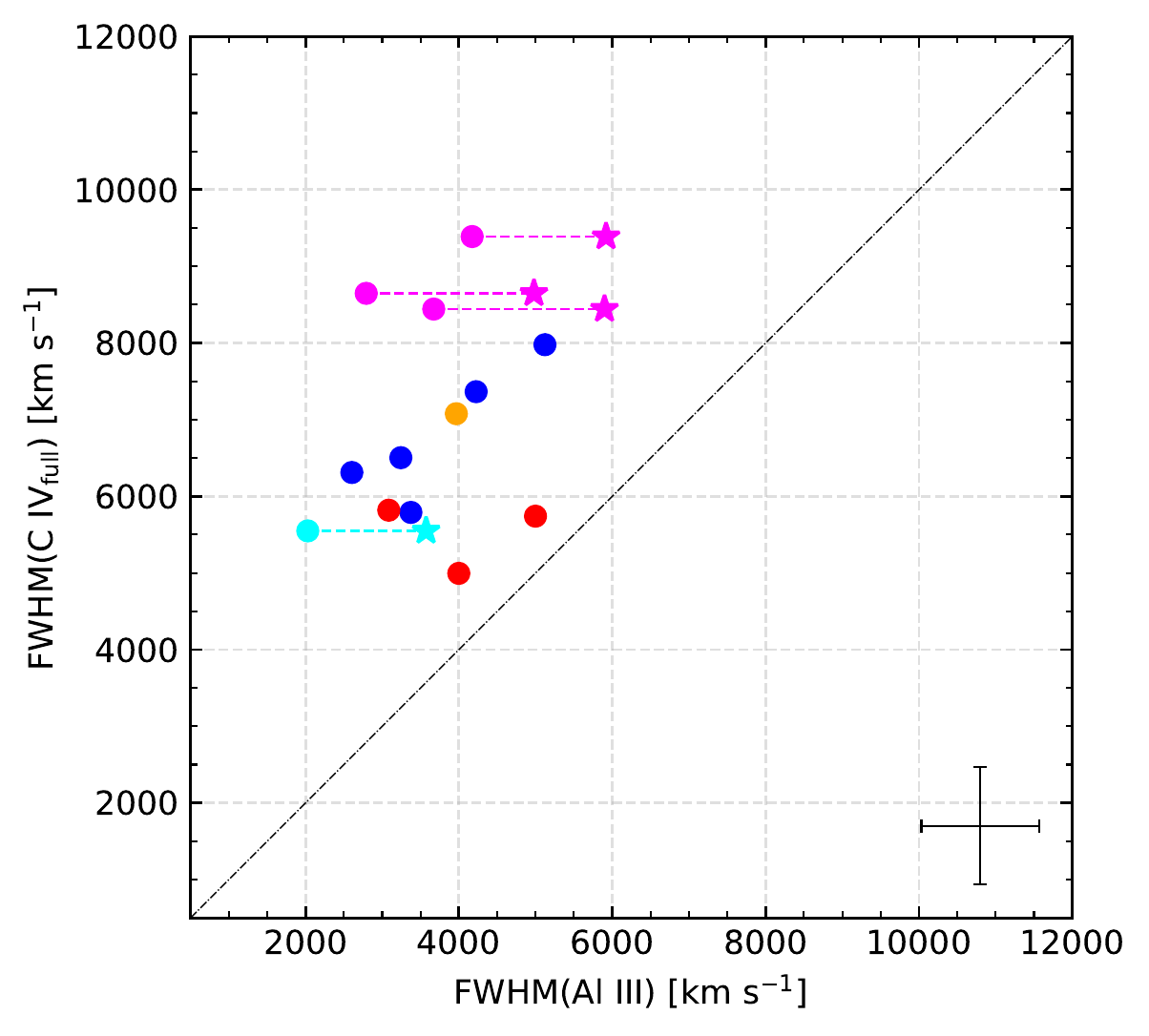}
    \caption{\ion{C}{IV}$_{\rm full}$ relation with \aliii{}. Colour scheme as in Fig. \ref{fig:HB_BC_Full}. The data points joined by a  dashed line are those for which a significant blueshifted emission component has been isolated. The left data points refer to the \aliii\ BC only and the right ones (star symbol) to the full profile FWHM (BC+BLUE). Error bars refer to 1$\sigma$ level of confidence.}
    \label{fig:civ_aliii}
\end{figure}

\subsubsection{\ion{C}{IV}$\lambda1549$ \textit{versus} \aliii{}}
\label{civaliii}

\par Fig. \ref{fig:civ_aliii} shows FWHM of the full profiles of \civ{} \textit{vs.} FWHM of \aliii{}. The FWHM(\ion{C}{IV}$_{\rm full}$) presents significantly higher values than FWHM(\ion{Al}{III}).  This indicates that the relation between \civ{} and \aliii{} is similar to  the one of \civ{} and \hb{}. In both cases the \civ\ FWHM tends to be larger: always significantly so for Pop. A, and larger or consistent for Pop. B. The cyan and magenta dashed lines connect the FWHM estimated from the \aliii\ BC with and without a BLUE component: without the BLUE (stars), the \aliii\ FWHM becomes closer to,  but remains significantly smaller than the FWHM of the \civ\ full profile.  
%\chony{Explain here the meaning of magenta stars}

\subsection{Virial black hole mass estimates}
\label{disc:mass}

%\subsubsection{1900$\AA$ blend %\textit{versus} \hb{}}
%\label{disc:hbaliii}

\par
\begin{figure*}[t!]
    \centering
    \includegraphics[scale=0.65]{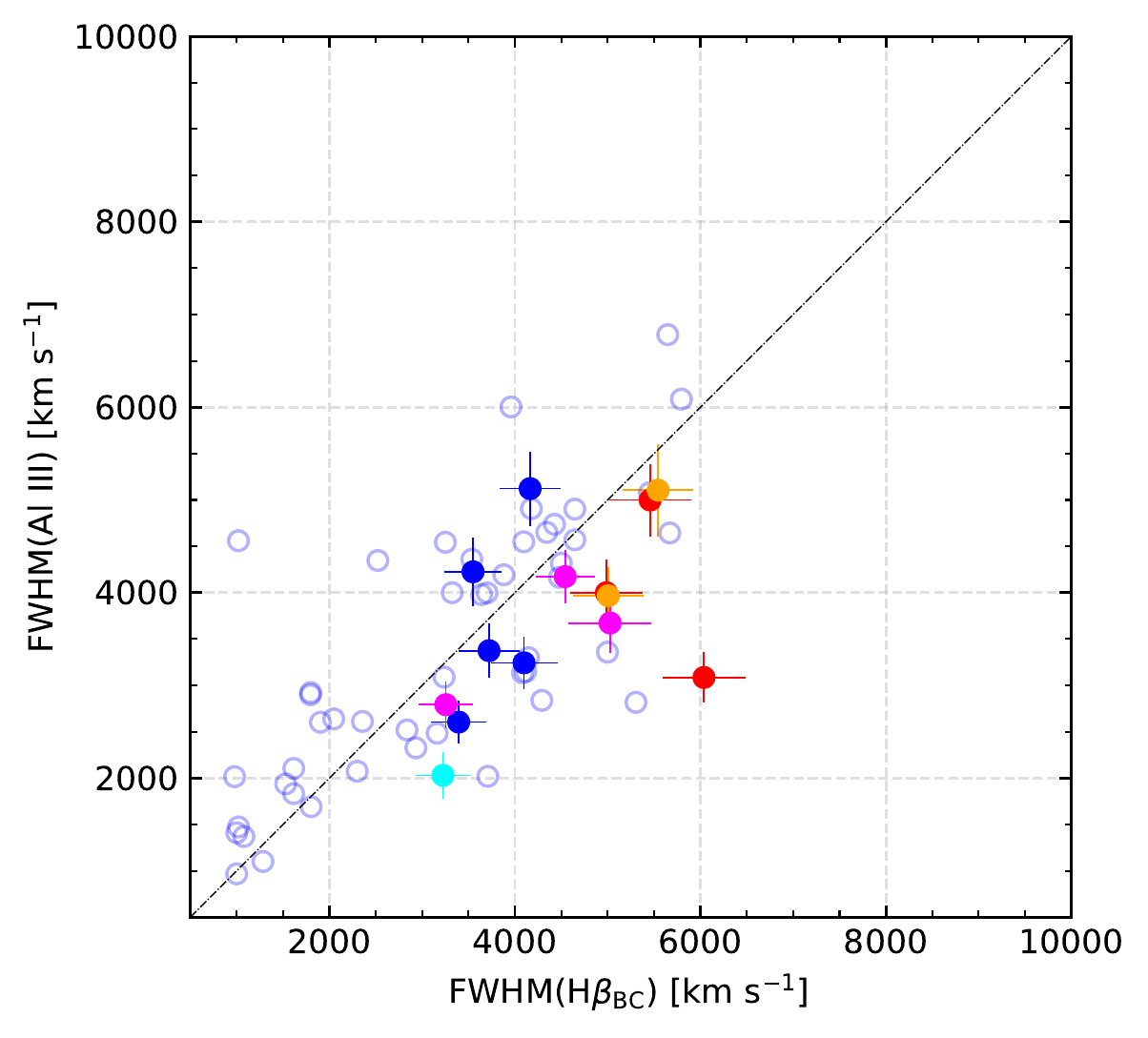}
    \includegraphics[scale=0.65]{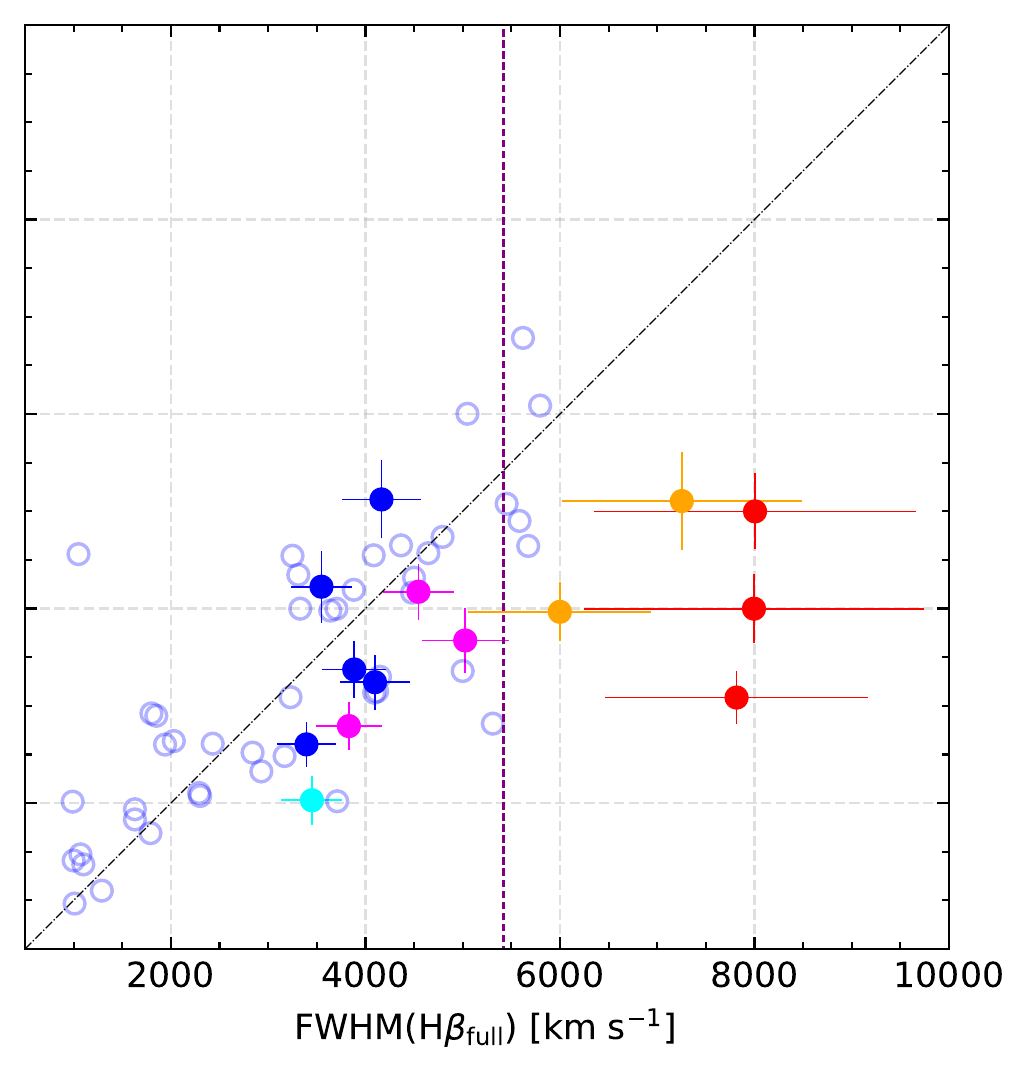}
    \\
     
    \caption{ FWHM(\ion{Al}{III}) $vs.$ FWHM(H$\beta_{\rm{BC}}$) (left panel) and FWHM(\ion{Al}{III}) $vs.$ FWHM(H$\beta_{\mathrm{full}}$) (right panel).  Sources are identified according to the different bins of spectral types following colour scheme from Fig. \ref{fig:HB_BC_Full}. Blue open circles show the data from M22 Pop. A sources for a comparison. Vertical purple  line indicates the A/B boundary for our sample.}
    \label{fig:Hb_AlIII}
\end{figure*}

 A comparison between the FWHM of \ion{Al}{III}$\lambda$1860 and H$\beta$ (BC and full) profiles is shown in  Fig. \ref{fig:Hb_AlIII}. The figure includes the data from \citet[][hereafter M22]{Marziani_2022apj} that involve exclusively Pop. A sources. The FWHM of H$\beta_{\rm{BC}}$\ is consistent with the FWHM of \ion{Al}{III}$\lambda$1860 for Pop. A: their average  ratio (FWHM \ion{Al}{III} over FWHM H$\beta$) in our sample is $\approx 0.87 \pm 0.18$. %The \hb{} BC in some sources are overestimated sweeping the BLUE component. A good example  is provided by SDSSJ210524.49+000407.3 that has a FWHM$({\rm H\beta})\approx 5000$ \kms\ and its \hb\ profile is almost completely represented by a very broad BC only. 
%If we assume that the \hb\ profiles may share similarities with the UV emission line profiles, we could expect some contribution of BLUE components on the \hb\ and the \aliii\ profiles: \civ\  profiles present  asymmetries in the blue part of the profile. In some cases, we do see relatively significant blueward asymmetries on the \hb{} profile but a much weaker or no BLUE contribution in the \aliii\ profile.}
%On average, the \aliii{} profile has a value of 3739 km s$^{-1}$, while for \hb{} this value is 4919 km s$^{-1}$. 
%The only three sources in which FWHM(\ion{Al}{III}) $>$ FWHM(H$\beta_{\rm{BC}}$) are SDSSJ135831.78+050522.8 (Pop. A2), SDSSJ212329.46-005052.9 (Pop. A2), and SDSSJ141546.24+112943.4 (Pop. B2). In the case of SDSSJ141546.24+112943.4 there are absorption lines at the \aliii{} wavelength. The absorptions generate an additional source of uncertainty on the measurement that may explain why FWHM(\ion{Al}{III}) is broader than FWHM(H$\beta_{\rm{BC}}$).
The right panel of Fig. \ref{fig:Hb_AlIII} shows a comparison between  FWHM(\ion{Al}{III}) \textit{vs.} FWHM of full H$\beta$. The purple dotted vertical line indicates the A/B boundary as in Fig. \ref{fig:optical_plane}. The location of our Pop. A objects remains in agreement with the results from M22, and with those for H$\beta_{\rm{BC}}$. There is  a clear deviation from the 1:1 line in the location of  Pop. B sources: for the 5 Population B sources of our sample the average ratio is just $0.57 \pm 0.12$ (c.f. \citealt{Marziani_2017a}). %\chonyblue{I think we must remove this ref., no data from ref, and too many Marzianis... one is more than enough}. 
These results provide evidence that Pop. B \aliii\ is systematically narrower than \hb\ full profile, most likely because of the strong VBC contribution. The ratio FWHM \aliii\ over \hbbc\  is in fair agreement $\approx 0.79 \pm 0.16$ for  Pop. B,   and  therefore the H$\beta$ VBC should not be included on the black hole mass estimation. The 20 \%\ difference between the FWHM \hbbc\ and FWHM \aliii\ for Pop. B sources is marginally significant and should be confirmed by larger samples.  

\begin{figure*}[ht!]
    \centering
    \includegraphics[width=0.81\columnwidth]{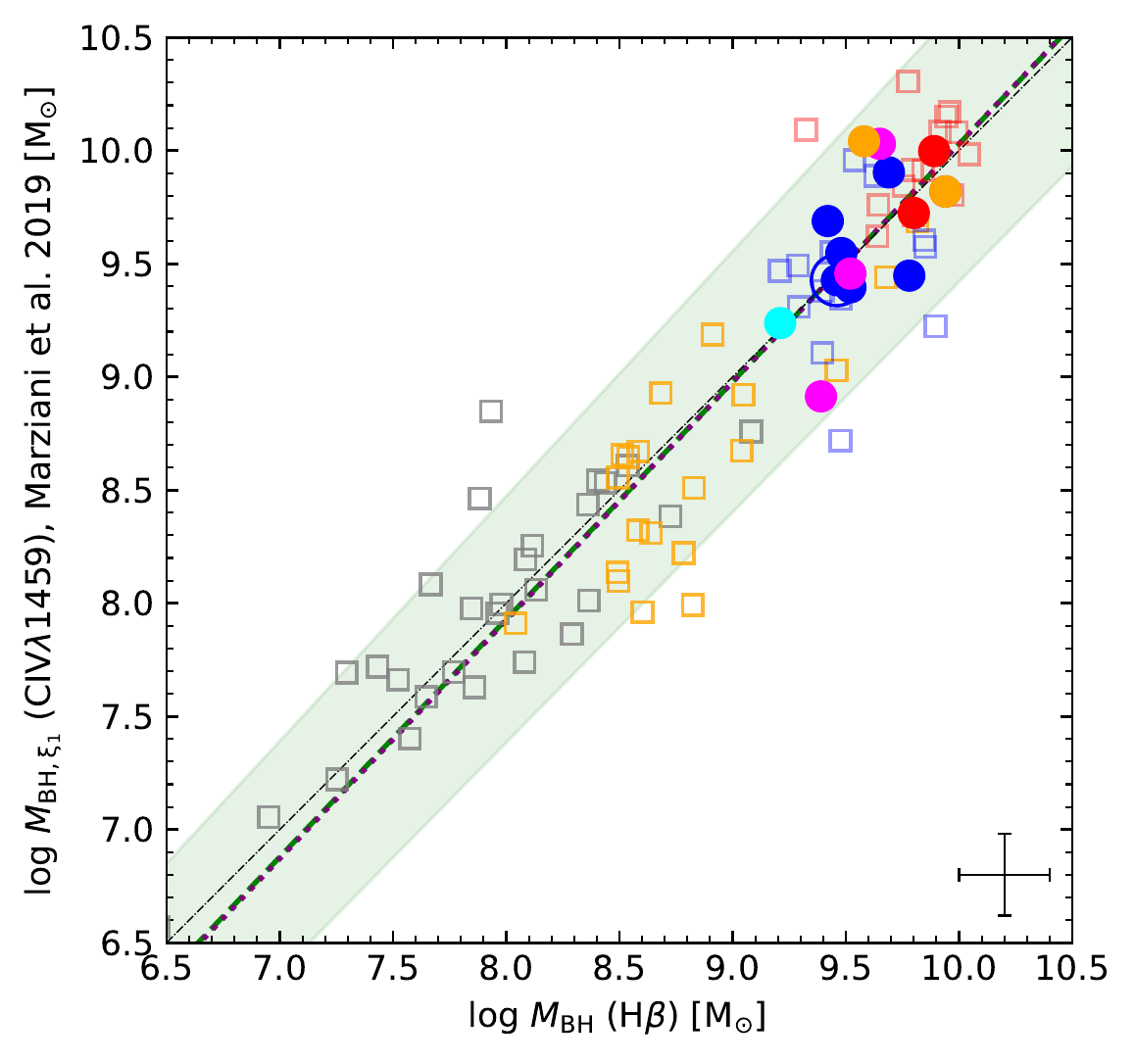}
    \includegraphics[width=0.81\columnwidth]{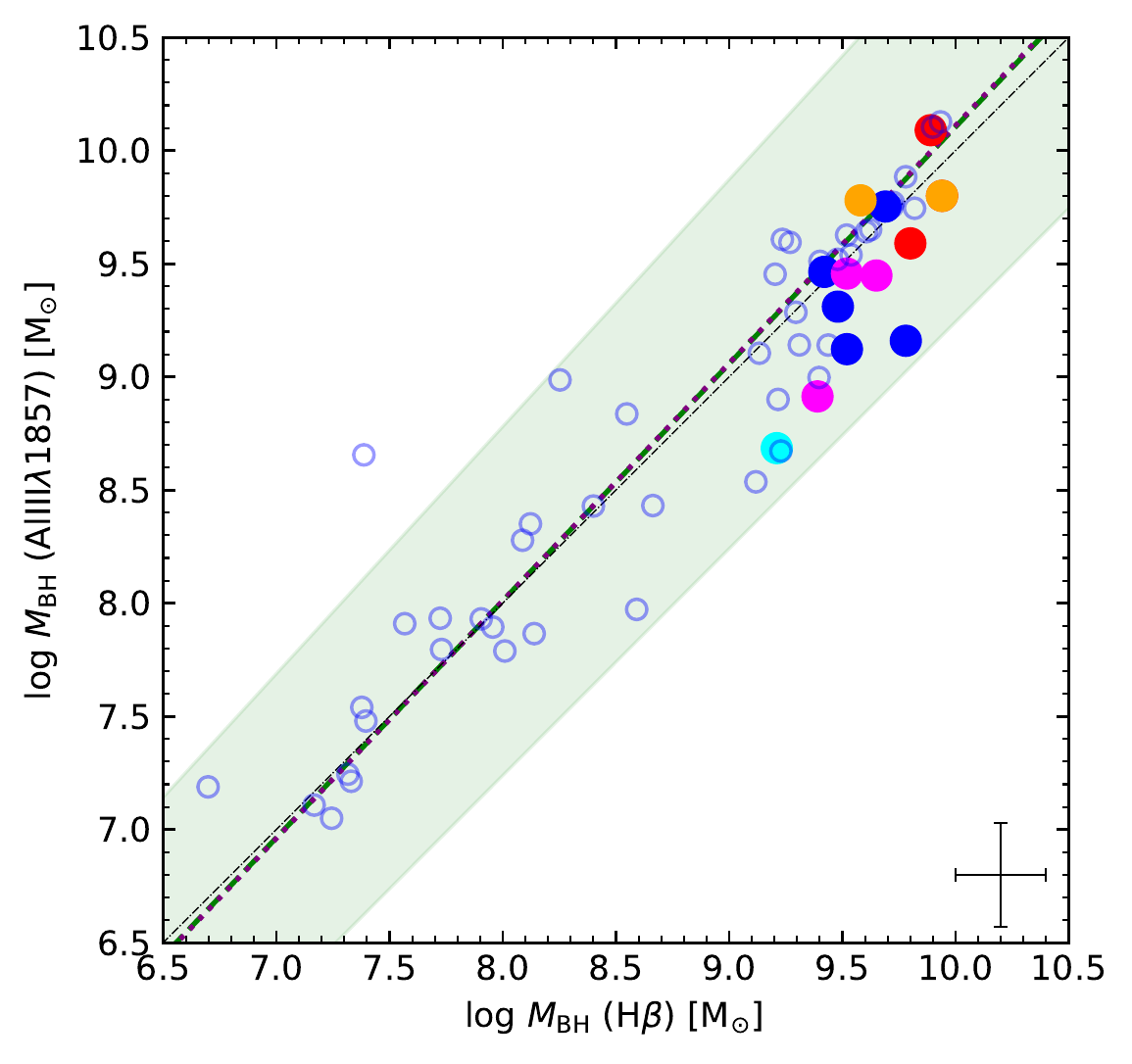}
    \caption{$M_{\rm BH}$(H$\beta$) compared to $M_{\rm BH}$(\ion{C}{IV}) (left panel) and to $M_{\rm BH}$(\ion{Al}{III}) (right panel). Open squares on the left plot represent the low- and high-redshift sources analysed by \cite{Marziani_2019} (blue and red for high-$z$ and grey and orange for low-$z$ Pop. A and Pop. B, respectively). Blue dots on the right panel are for the M22 data. Green and purple dashed lines indicate the linear regressions using bisector and orthogonal methods, respectively. Colour scheme as in Fig. \ref{fig:HB_BC_Full}.}
    \label{fig:mbh_Hb_CIV}
\end{figure*}

%\subsection{$M_{BH}$ estimates using different optical and UV emission lines}

\par The black hole masses are estimated through different scaling relations using emission lines such as those based on the FWHM of H$\beta$, \ion{Al}{III}$\lambda1860$,  and \ion{C}{IV}$\lambda1549$. The scaling law of \cite{Vestergaard_2006} was used to estimate the mass \mbh\ with  FWHM \hbbc. We  utilize Equation (18) from \citet[][$\sigma=0.33$]{Marziani_2019} to estimate $M_{\textrm{BH}}$ with the \ion{C}{IV}$\lambda1549$ FWHM.
%following the equation below:
%\begin{equation}
%\begin{split}
%    \log M_{\textrm{BH}}(\textrm{H}\beta)=\log \left\{ \left[ \frac{\textrm{FWHM}(\textrm{H}\beta)}{1000 \textrm{ km s}^{-1}}\right]^2 \left[   \frac{\lambda L_{\lambda}(5100\AA)}{10^{44} \textrm{ erg s}^{-1}}\right]^{0.50} \right\}\\
%    + (6.91 \pm 0.02).
%\end{split}
%\end{equation}

%\par We can also use \cite{Vestergaard_2006} to estimate $M_{\rm BH}$ using the \ion{C}{IV}$\lambda1549$ emission line, from the following equation:
%\begin{equation}
%    \begin{split}
 %       \log M_{\rm BH}(\ion{C}{IV})= \log \left\{ \left[ \frac{\textrm{FWHM} (\ion{C}{IV})}{1000 \textrm{ km s}^{-1}}\right]^2 \left[\frac{\lambda L_{\lambda}(1350\AA)}{10^{44} \textrm{ erg s}^{-1}}\right]^{0.53} \right\}\\
 %       + (6.73 \pm 0.01).
 %   \end{split}
 %   \label{eq:civ_vestergaard}
%\end{equation}

%\begin{equation}
%\begin{split}
%    \log M_{\textrm{BH}}(\ion{C}{IV})\approx 0.63\log \left[\frac{\lambda L_{\lambda}(1450\AA{})}{10^{44} \textrm{ km s}^{-1}}\right]\\
    %+2\log (\xi_{\ion{C}{IV}}\textrm{FWHM(\ion{C}{IV})}) +0.525,
%\end{split}    
%\label{eq:civ_marziani}
%\end{equation}
%in which $\xi_{\ion{C}{IV}}$ is a bias correction factor (see \citealt{Marziani_2019} for a detailed description) needed to account for the broadening of the \civ\ profile by the outflow component. 

\par The left panel of Fig. \ref{fig:mbh_Hb_CIV} presents a comparison between the $M_{\textrm{BH}}$ estimates using the scaling law for \ion{C}{IV}$\lambda1549$\ from \citet[][]{Marziani_2019} and the H$\beta$\ scaling law of \citet{Vestergaard_2006}. The open squares in this plot indicate the data from \citet{Marziani_2019} which include high- and low-$z$ sources. Our sample agrees with the previous estimates and corresponds to the extreme cases, with the largest $M_{\textrm{BH}}$\ reaching $10^{10}$ M$_\odot$. %We can also highlight that the Pop. B sources are the ones that present the highest $M_{\textrm{BH}}$ for both \hb{} and \civ{} emission lines. However, this is a selection effect. 

\par The plot of $M_{\textrm{BH, Al III}}$ \textit{vs.} $M_{\textrm{BH, H$\beta$}}$ (right panel of Fig. \ref{fig:mbh_Hb_CIV}). relies on the $M_{\rm BH}$   scaling law from M22.
%\begin{equation}
%\label{eq:alii}
%\begin{split}
%    \log M_{\rm BH} (\ion{Al}{III})\approx 0.578\log \left[\frac{\lambda L_{\lambda}(1700\AA)}{10^{44} \textrm{erg s}^{-1}} \right]\\
 %   +2\log(\rm FWHM(\ion{Al}{III}))+0.490.
%\end{split}
%\end{equation}
The $M_{\rm BH, H\beta}$ and $M_{\rm BH, \ion{Al}{III}}$ values obtained for our sources are compatible within the confidence interval of the M22 scaling law for \aliii{}. We caution that  the relation given in this equation has been derived for Pop. A sources and, consequently, it may lead to larger uncertainties for Pop. B quasars. For these sources we apply a correction $\xi=\rm{FWHM}_{\rm{H\beta,BC}}/\rm{FWHM}_{\rm{\ion{Al}{III}}}=1.35\pm 0.10$ to the FWHM(\ion{Al}{III}) \citep[][who assumed  FWHM  H$\beta_{\rm{BC}}$ as the reference virial broadening estimator]{Marziani_2017}.
 %However, the mass estimates based on \aliii\ $M_{\rm BH, \ion{Al}{III}}$ reach somewhat smaller values than $M_{\rm BH, H\beta}$. The origin of this discrepancy is not clear, also because it is easy to see that the \mbh\ estimated using \civ\ and \hb\ are almost equivalent.

%However, the mass estimates based on \aliii\ $M_{\rm BH}$\ reach somewhat smaller values than  $M_{\rm BH, H\beta}$, with an offset by a factor $\approx$ 1.5. \chonyblue{Instead of remarking that there is a factor 1.5 between the masses of Hb and AlIII (that can be due to a blue comp in Hb, and larger uncertainties in masses CIV), we should first enhance that the values obtained with AlIII and Hb are fully compatible within the confidence interval of the M22 scaling law for AlIII. We are not in orders of magnitude of difference, just a factor of 1.5. It may be also that the scaling law of M22 is influenced by many NLSy1 low-z sources.} The origin of this discrepancy is not clear, also because it is easy to see that the $M_{\rm BH}$ estimated using \civ{} and \hb{} are almost equivalent. %\chony{remove CIII paragraph, we do not include those values}. \chony{and paper AlIII is not yet published}  \mbh\ values  also found when considering the \ciii{} emission line, following a \ciii\ scaling law given by M22. The median difference between the estimates of \mbh{} using \ciii{} and \aliii{} is -0.10, with a standard deviation of 0.17. These values are in good agreement, with \ciii{} \mbh\ $\approx$ 20\% less than \aliii\ \mbh{}. 

\par The equations that describe the relations between $M_{\rm BH}$ estimations using $\xi$-corrected \civ\ and \aliii\ are listed in Table \ref{tab:masses}. Col. 1 lists the method used for the linear regression (orthogonal and bisector); Cols. 2 and 3 present the linear and angular coefficients together with the respective standard deviations; rms error and Pearson correlation coefficient $r$ are reported in Cols. 4 and 5, respectively.

%\par The left panel of Fig. \ref{fig:mbh_CIV_CIV} shows the comparison between the $M_{\textrm{BH}}$ estimated through the \civ{}  FWHM to which the $\xi$\ correction has been applied     versus the masses derived from the original scaling law of \citetalias{Vestergaard_2006} without any correction because of the excess broadening of the \civ\ profile.  The right panel analogously compares the \mbh\ obtained from \hb\ to the ones obtained with uncorrected \civ\ scaling law. In both cases there is a larger discrepancy in the slope of the best fitting relation with respect to the 1:1 expectation. In the case of the \hb\ vs. \civ\ comparison the dispersion is also much increased (see Table \ref{tab:masses}). 

The original \civ\ \citetalias{Vestergaard_2006} relation  lacks  a correction because of the bias introduced by non-virial broadening (i.e., by the blueshifted component; \citealt{Brotherton_2015,coatman_2017,Marziani_2019}). If \mbh\ estimated with the uncorrected \citetalias{Vestergaard_2006} relations are used, there is a significant deviation in the 1:1 relation between H$\beta$ and \ion{C}{IV}$\lambda1549$ $M_{\rm{BH}}$ estimates and a large scatter (rms $\approx 0.42$).  We therefore utilise the scaling laws  based on \aliii\ and $\xi$-corrected \civ\ that may provide less biased estimators with respect to \hb-based \mbh\ scaling law.

%\par The average $M_{\rm BH, H\beta}$ is 9.47 for the full sample using the \cite{Vestergaard_2006} relation. For Pop. A sources this value is 9.39 while for Pop. B it is 9.57. In the case of the estimates based on \aliii{}, we found $M_{\rm BH}=9.38$ for the full sample, and 9.28 for Pop. A and  9.58 for Pop. B. Similarly, using the \cite{Marziani_2019} relation for \civ{} we obtain an average value of 9.59 for the complete sample and for Pop. A and Pop. B the values are 9.49 and 9.83, respectively. On the other hand, the averaged values for $M_{\rm BH, C IV}$ using the \cite{Vestergaard_2006} relation are higher: 9.72 for the full sample, and 9.69 for Pop. A and 9.77 for Pop. B. We have also derived a weighted average for our sample based on the uncertainties found in the full profiles of \civ{} and \hb{}. For \hb{}, we found weighted averages of 9.43 and 9.58 for Pop. A and Pop. B, respectively. Regarding the \civ{} estimates based on the \cite{Marziani_2019} relation, we obtain $M_{\rm BH}=9.51$ for Pop. A and $M_{\rm BH}=9.85$ for Pop. B.

\begin{table}[t!]
    \centering
     \caption{Linear relations between the different $M_{\rm BH}$ estimates using orthogonal and bisector methods. %We also list RMS error and Pearson's correlation coefficient.
     }
     \resizebox{\linewidth}{!}{
    \begin{tabular}{lcccr}
    \hline
    \hline
    \noalign{\smallskip}
      Method   & $a\pm \delta a$ & $b \pm \delta b$ & RMS & CC \\
      (1) & (2) & (3) & (4) & (5)\\
      \noalign{\smallskip}
    \hline
    \noalign{\smallskip}
    \multicolumn{5}{c}{$M_{\rm BH, C IV, M19}=a+b*M_{\rm BH, H\beta, VP06}$} \\
    \noalign{\smallskip}
    \hline
    \noalign{\smallskip}
      %LS   & $-0.142 \pm 0.301$  & $1.016 \pm 0.034$ & \\
      Orthogonal   &  $-0.495\pm 0.306$ & $1.053\pm 0.034$ &\multirow{2}{*}{0.315} & \multirow{2}{*}{0.949}\\
      Bisector   &  $-0.471\pm 0.289 $ & $1.050\pm 0.032$ & & \\\noalign{\smallskip}
         \hline
         \noalign{\smallskip}
     \multicolumn{5}{c}{$M_{\rm BH, AlIII, M22}=a+b*M_{\rm BH, H\beta, VP06}$} \\
     \noalign{\smallskip}
    \hline
    \noalign{\smallskip}
      %LS   &  $0.201\pm 0.463$  & $0.977\pm 0.050$ & \multirow{3}{*}{0.319} & \multirow{3}{*}{0.949}\\
      Orthogonal   & $-0.389\pm 0.431$ & $1.050 \pm 0.047$ & \multirow{2}{*}{0.320} & \multirow{2}{*}{0.948}\\
      Bisector   & $-0.366\pm 0.411$ & $1.047\pm 0.045$ & &\\
      \noalign{\smallskip}
          \hline
 %    \multicolumn{5}{c}{$M_{\rm BH, CIV, M19}=a+b*M_{\rm %BH, CIV, VP06}$} \\
 %   \hline
 %     %LS   & $-0.317\pm 0.338$ & $1.037\pm 0.039$ & %\multirow{3}{*}{0.341} & \multirow{3}{*}{0.941}\\
 %     Orthogonal   & $-0.948\pm 0.327$ & $1.109\pm 0.037$ & %\multirow{2}{*}{0.341} & \multirow{2}{*}{0.941}\\
 %     Bisector   &  $-0.887\pm 0.307$ &    $1.102\pm 0.035$ %& &\\
 %         \hline
 %    \multicolumn{5}{c}{$M_{\rm BH, H\beta, %VP06}=a+b*M_{\rm BH, CIV, VP06}$} \\
 %   \hline
 %     %LS   & $0.804\pm 0.448$ & $0.909\pm 0.050$ & \\
 %     Orthogonal   & $-0.204\pm 0.403$ &    $1.023\pm %0.045$ & \multirow{3}{*}{0.424} & %\multirow{3}{*}{0.890}\\
 %     Bisector   & $-0.180\pm 0.358$ & $1.021\pm 0.040$ & %&\\
         \hline
    \end{tabular}
   }
    \label{tab:masses}
\end{table}

%\begin{figure*}[t!]
%    \centering
%    \includegraphics[width=\columnwidth]{figs/mbh_civ_civ.pdf}
%    \includegraphics[width=\columnwidth]{figs/mbh_hb_civ_vestergaard.pdf}
%    \caption{\textit{Left panel:} Comparison between the $M_{\rm %BH}$(\ion{C}{IV}) estimated using \cite{Vestergaard_2006} ($x$ axis) %and \cite{Marziani_2019} ($y$ axis). \textit{Right panel:} Comparison %between the estimation of $M_{\rm BH}$ from \cite{Vestergaard_2006} %using both \hb{} and \civ{} emission lines. Cyan (low-$z$) and blue %(high-$z$) squares represent the Pop. A and the orange (low-$z$) and %red (high-$z$) ones show the Pop. B quasars from %\cite{Marziani_2019}. Green and purple lines indicate the linear %regressions using least squares and bisector methods, respectively. %The different spectral types on our sample are shown following the %same color scheme as in Fig. \ref{fig:HB_BC_Full}.} 
%    \label{fig:mbh_CIV_CIV}
%\end{figure*}     

\begin{table}[t!]
    \centering
     \caption{Weighted averaged masses, luminosity at 5100 \r{A}, and Eddington ratio ($L/L_{\rm Edd}$), in logarithmic scale.}
     \resizebox{\linewidth}{!}{
    \begin{tabular}{lccccr}
    \hline
    \hline
    \noalign{\smallskip}
      Source & $M_{\rm BH, H\beta}$  & $M_{\rm BH}$ & $L$(1450\r{A}) & $L$(5100\r{A}) & $L/L_{\rm Edd}$ \\
      (1) & (2) & (3) & (4) & (5) & (6) \\
      \noalign{\smallskip}
    \hline
    \noalign{\smallskip}
    \multicolumn{6}{c}{Population A}\\
    \noalign{\smallskip}
    \hline
    \noalign{\smallskip}
SDSSJ005700.18+143737.7	&	9.39	$\pm$	1.41	&	9.37	$\pm$	0.27	&	46.65	$\pm$	5.60	&	46.41	$\pm$	5.57	&	0.10	$\pm$	0.01	\\
SDSSJ132012.33+142037.1	&	9.21	$\pm$	1.38	&	9.19	$\pm$	0.14	&	46.74	$\pm$	5.61	&	46.42	$\pm$	5.57	&	0.03	$\pm$	0.01	\\
SDSSJ135831.78+050522.8	&	9.42	$\pm$	1.41	&	9.56	$\pm$	0.11	&	46.97	$\pm$	5.64	&	46.67	$\pm$	5.60	&	0.07	$\pm$	0.01	\\
Q 1410+096	&	9.52	$\pm$	1.43	&	9.42	$\pm$	0.20	&	47.11	$\pm$	5.65	&	46.92	$\pm$	5.63	&	0.32	$\pm$	0.04	\\
B1422+231$^{(a)}$	&	9.96	$\pm$	1.49	&	-			&	-			&	47.07	$\pm$	5.65	&	-0.07	$\pm$	0.01	\\
SDSSJ161458.33+144836.9	&	9.48	$\pm$	1.42	&	9.49	$\pm$	0.10	&	47.05	$\pm$	5.65	&	46.69	$\pm$	5.60	&	0.04	$\pm$	0.01	\\
PKS 1937-101$^{(a)}$	&	10.14	$\pm$	1.52	&	-			&	-			&	47.36	$\pm$	5.68	&	0.05	$\pm$	0.01	\\
PKS 2000-330	&	9.46	$\pm$	1.42	&	9.21	$\pm$	0.04$^{(c)}$	&	46.58::	&	46.94	$\pm$	5.63	&	0.30	$\pm$	0.04	\\
SDSSJ210524.47+000407.3	&	9.84	$\pm$	1.48	&	9.83	$\pm$	0.28$^{(b)}$	&	47.16	$\pm$	5.66	&	46.89	$\pm$	5.63	&	-0.13	$\pm$	0.02	\\
SDSSJ210831.56-063022.5	&	9.52	$\pm$	1.43	&	9.63	$\pm$	0.12	&	46.98	$\pm$	5.64	&	46.43	$\pm$	5.57	&	-0.26	$\pm$	0.03	\\
SDSSJ212329.46-005052.9	&	9.69	$\pm$	1.45	&	9.77	$\pm$	0.08	&	47.19	$\pm$	5.66	&	46.92	$\pm$	5.63	&	0.05	$\pm$	0.01	\\
SDSSJ235808.54+012507.2	&	9.78	$\pm$	1.47	&	9.57	$\pm$	0.31	&	46.85	$\pm$	5.62	&	47.11	$\pm$	5.65	&	0.15	$\pm$	0.02	\\

\noalign{\smallskip}
\hline
\noalign{\smallskip}
\multicolumn{6}{c}{Population B}\\
\noalign{\smallskip}
\hline
\noalign{\smallskip}
HE 0001-2340$^{(a)}$	&	9.63	$\pm$	1.44	&	-			&	-			&	46.48	$\pm$	5.58	&	-0.33	$\pm$	0.04	\\
$[\rm HB89]$ 0029+073$^{(a)}$	&	9.71	$\pm$	1.46	&	-			&	-			&	46.81	$\pm$	5.62	&	-0.07	$\pm$	0.01	\\
CTQ 0408$^{(a)}$	&	10.20	$\pm$	1.53	&	-			&	-			&	47.26	$\pm$	5.67	&	-0.12	$\pm$	0.01	\\
H 0055-2659$^{(a)}$	&	9.69	$\pm$	1.45	&	-			&	-			&	46.75	$\pm$	5.61	&	-0.12	$\pm$	0.01	\\
SDSSJ114358.52+052444.9	&	9.58	$\pm$	1.44	&	9.75	$\pm$	0.25	&	47.16	$\pm$	5.66	&	46.39	$\pm$	5.57	&	-0.05	$\pm$	0.01	\\
SDSSJ115954.33+201921.1	&	9.89	$\pm$	1.48	&	9.90	$\pm$	0.05	&	47.36	$\pm$	5.68	&	46.84	$\pm$	5.62	&	-0.23	$\pm$	0.03	\\
SDSSJ120147.90+120630.2	&	9.94	$\pm$	1.49	&	9.89	$\pm$	0.2	&	47.19	$\pm$	5.66	&	47.09	$\pm$	5.65	&	0.03	$\pm$	0.01	\\
SDSSJ141546.24+112943.4	&	9.80	$\pm$	1.47	&	9.78	$\pm$	0.07$^{(b)}$	&	47.08	$\pm$	5.65	&	46.64	$\pm$	5.60	&	-0.33	$\pm$	0.04	\\
SDSSJ153830.55+085517.0	&	9.80	$\pm$	1.47	&	9.76	$\pm$	0.24	&	47.23	$\pm$	5.67	&	46.50	$\pm$	5.58	&	-0.47	$\pm$	0.06	\\
PKS 2126-15$^{(a)}$	&	9.78	$\pm$	1.47	&	-			&	-			&	47.11	$\pm$	5.65	&	0.16	$\pm$	0.02	\\

\noalign{\smallskip}
         \hline
\end{tabular}
}
    \label{tab:mass_lum}
\tablefoot{$^{(a)}$ The \civ{} and the 1900 \r{A} blend regions were not fitted in these cases. $^{(b)}$ We compute the weighted averaged mass between the \hb{} and \aliii{} estimates only. $^{(c)}$ We compute the weighted averaged mass between the \hb{} and \civ{} estimates only.}
\end{table}

\par Table  \ref{tab:mass_lum} lists the individual values of \hb{} black hole masses (Col. 2), the weighted averages of $M_{\rm BH}$ (Col. 3), the luminosity at 1450\r{A} and 5100\r{A} (Cols. 4 and 5, respectively), as well as the Eddington ratio of each source (Col.6). We have considered the mass estimates using \hb{}, \aliii{}, and \civ{} to determine the weighted average of \mbh\ and used the FWHM errors as weights. \hb{} \mbh\ are also reported since they are available for all sources of our sample and needed to compare our data to previous samples. We have adopted a bolometric correction of 10 for the optical range in accordance with the value assumed in the previous works we consider for comparison \citepalias[e.g., \citealt{Richards_2006};][]{Sulentic_2007,sulentic_2017, Marziani_2019}, although lower bolometric corrections are expected in the $L_{\rm bol} > 10^{47}$ erg s$^{-1}$ luminosity range \citep{Marconi_2004, Netzer_2019}, which may introduce a bias towards higher accretion rates. The bolometric correction adopted in the UV range is 3.5 \citep{Elvis_1994}. 

\subsection{Dependence on accretion parameters}

\begin{figure*}[ht!]
    \centering
    \includegraphics[width=0.95\columnwidth]{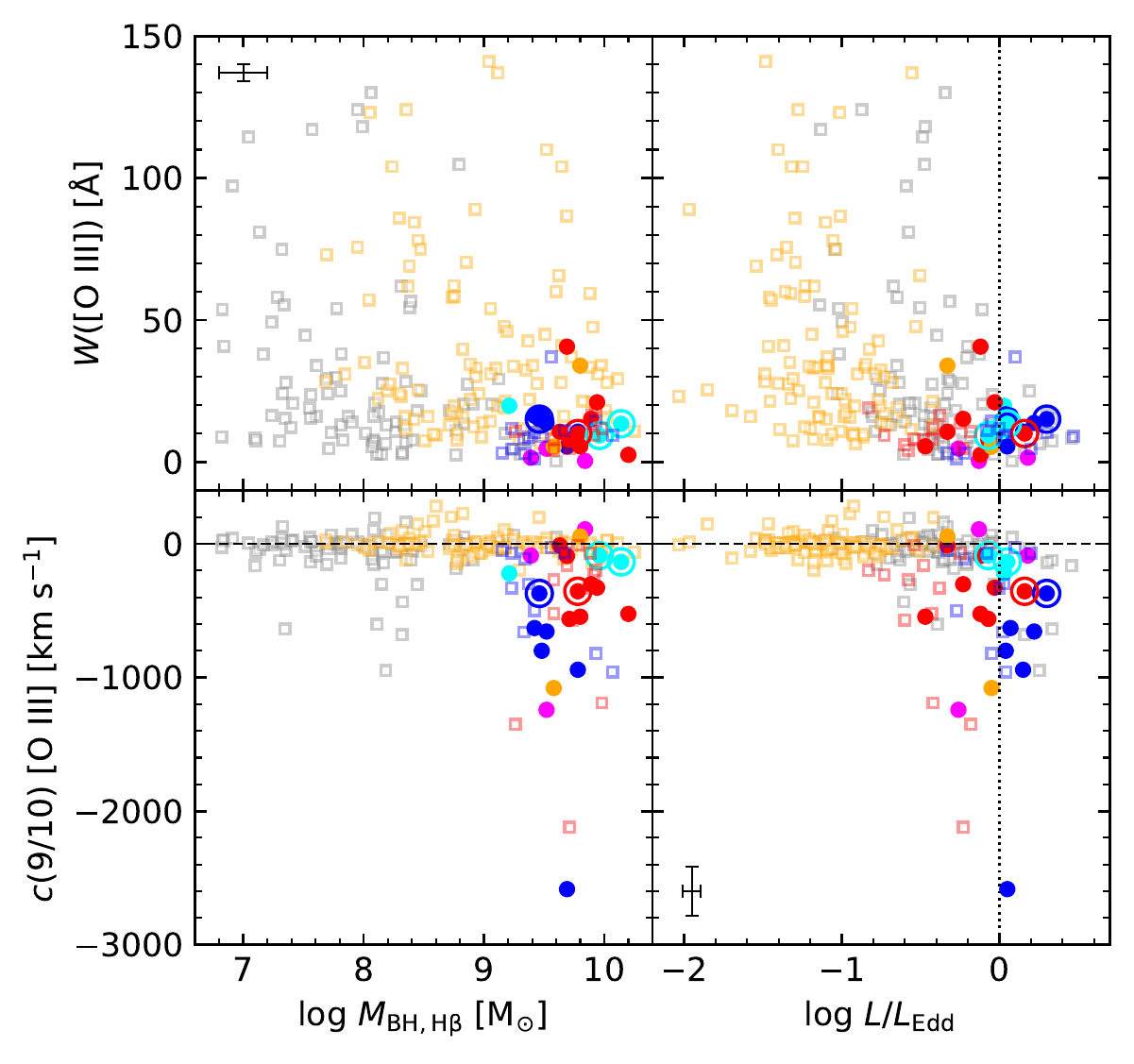}
    \includegraphics[width=0.805\columnwidth]{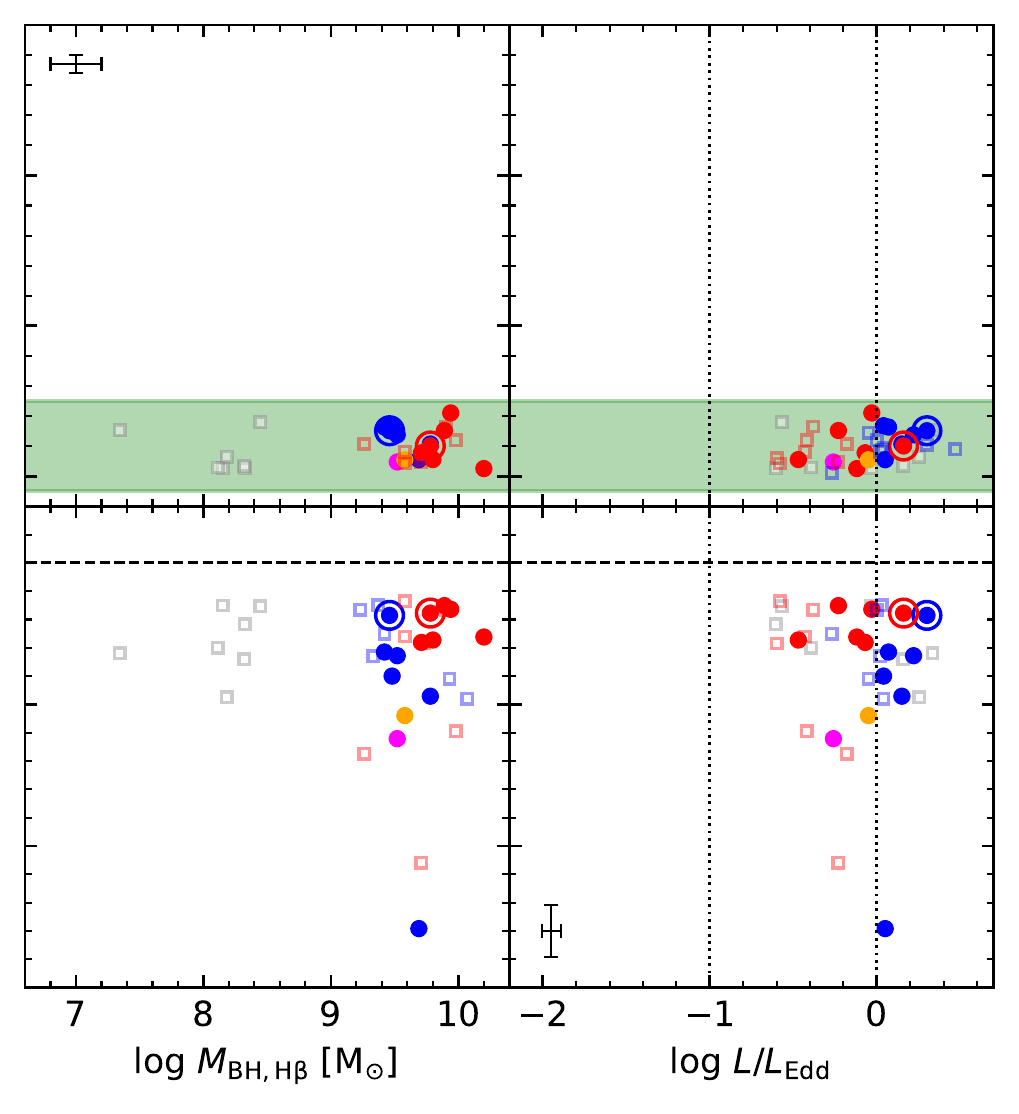}
    \caption{\oiii{} dependence on physical parameters. \textit{Left panel:} Our full sample in comparison with the full \citetalias{Marziani_2003} (grey for Pop. A and orange for Pop. B) and HE (blue for Pop. A and red for Pop. B) samples. \textit{Right panel:} Only sources with $c$(9/10) $< -250$ \kms{}. Our data follow the same colour scheme as in Fig. \ref{fig:HB_BC_Full}.}
    \label{fig:physical_parameters_oiii}
\end{figure*}
\subsubsection{\oiii{}}

\par Fig. \ref{fig:physical_parameters_oiii} shows the dependence of $W$([\ion{O}{III}]) and $c$(9/10) [\ion{O}{III}] on M$_{\rm BH,H\beta}$ and L/L$_{\rm Edd}$. Our sample presents $W$([\ion{O}{III}]) $< 25$\r{A} in the majority of the cases, with only two outliers (SDSSJ141546.24+112943.4 with $\approx$ 34\r{A} and H 0055-2659 with $\approx$ 41\r{A}). The $W$([\ion{O}{III}]) of our sample is relatively small when compared with low-$z$ sources. However, our data show more frequently blueshifts in $c$(9/10). %\textcolor{green}{Significant shifts are found for \lledd\ $\gtrsim 0.1$\ but are otherwise observed over a broad range of \mbh.}  %The most extreme cases 
%indicating \oiii{} profiles that are much more shifted to the blue than the ones at low-$z$. 

\par The right panels of Fig. \ref{fig:physical_parameters_oiii} consider only sources that present $c$(9/10) $<-250$ \kms{}. In this case, all the sources fit within  $W(\rm [\ion{O}{III}])\lesssim 25$\r{A} and are found in a wide range of \mbh{} as for the  distribution of the full samples. Sources with $c$(9/10) $<-250$ \kms{} have  an average Eddington ratio of -0.09:  at high Eddington ratios the \oiii{} profiles tend to be strongly affected or dominated by the blueshifted SBC.  In addition, the \oiii{} $W$ remains roughly constant over a wide range of masses (and luminosity), indicating that the luminosity of the blueshifted SBC is proportional to the continuum luminosity. This result provides evidence that the outflow traced by the blueshifted \oiii{} is directly related to the active nucleus, as proportionality between line and continuum luminosity is a classical proof of photoionisation \citep[][]{osterbrockshuder82}.    

%The outlying point in Fig. \ref{fig:physical_parameters_oiii} is  one  special case which apparently presents a huge blueshift on the profile peak (SDSSJ212329.46-005052.9, with a 9/10 intensity shift of -2885 km s$^{-1}$). 
\subsubsection{\civ{}}

\par
\begin{figure*}[t!]
    \centering

    \includegraphics[width=1\linewidth]{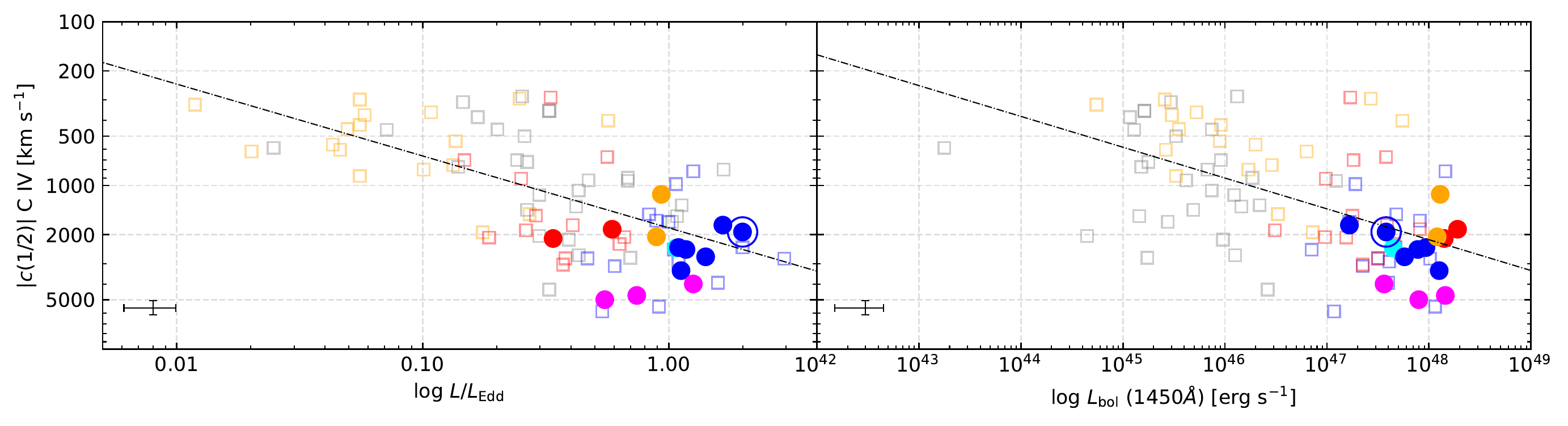}
    \caption{\textit{Left panel:} Relation between \ion{C}{IV} $c$(1/2) and $L/L_{\rm{Edd}}$. \textit{Right panel:} \ion{C}{IV} $c$(1/2) \textit{vs.} $L_{\rm bol}$(1450\r{A}). Grey and orange squares represent low-$z$ Pop. A and Pop. B quasars respectively, while blue and red open squares indicate HE Pop. A and Pop. B sources from \citetalias{sulentic_2017}. Color scheme as Fig. \ref{fig:HB_BC_Full}. Linear regressions are shown by the dashed lines. The errors refer only to our sample.}
    \label{fig:civ_comparison}
\end{figure*}

 In Fig. \ref{fig:civ_comparison}, we provide an analysis of the dependence of \civ{} shift measured by $c$(1/2) on luminosity and $L/L_{\rm Edd}$.  We include the data analysed by \citet{Marziani_2019} along with the present sample data. The same  trend appears to be followed by both ours and \cite{Marziani_2019} samples. Sources that present the largest values of \civ\ $c$(1/2) are Population A, with two xAs (SDSSJ210524.47+000407.3 and SDSSJ210831.56-063022.5) reaching $c$(1/2) $\approx$ --5000 \kms. %\chony{ is c(1/2)$\approx$ --5000, and the sources are J2105 and J2108.} \chonyblue{In the figure is represented c(1/2). In c(1/4) some blueshifts achieve larger values as -4950 for J0057, -5060 for J2105 or -5744 for J2108}   

%\par Following \citetalias{sulentic_2017}, we present Fig. \ref{fig:civ_comparison} in order to perform an analysis of the relation between \civ{} c(1/2), $L/L_{\rm{Edd}}$, and $\lambda$L$_{\lambda}$(1450\r{A}). 

As in \citetalias{sulentic_2017}, the left plot of Fig. \ref{fig:civ_comparison} indicate a clear relation (with a Pearson correlation coefficient of 0.59) between $c$(1/2) of \civ{} and $L/L_{\rm Edd}$, given by the following equation for the absolute value of the centroid shift at 1/2:
\begin{equation}
    \log |c(1/2)|_{\rm \ion{C}{IV}}=(0.43 \pm 0.06)\log L/L_{\rm Edd}+(3.25 \pm 0.04).
\end{equation}
\par A similar relation can also be derived between the absolute value of \civ\ $c$(1/2)  and $L_{\rm bol}$(1450\r{A}) (right panel of Fig. \ref{fig:civ_comparison}):
\begin{equation}
 \log |c(1/2)|_{\rm \ion{C}{IV}}=(0.18 \pm 0.03)\log L_{\rm bol}+(-5.70\pm 1.46).   
\end{equation}
The Pearson correlation coefficient in this case is 0.54. These relations are consistent  with those of \citetalias{sulentic_2017} who found slope $\approx 0.5$\ and $\approx 0.2$\ for the dependence on Eddington ratio and luminosity, respectively. Since the \citetalias{sulentic_2017} data were included in the linear regression, we can say that the new data confirm the slope difference that suggests a major role of the \lledd\ for governing the shift amplitude, and a secondary effect of luminosity. 

\begin{figure}
    \centering
    \includegraphics[width=0.49\columnwidth]{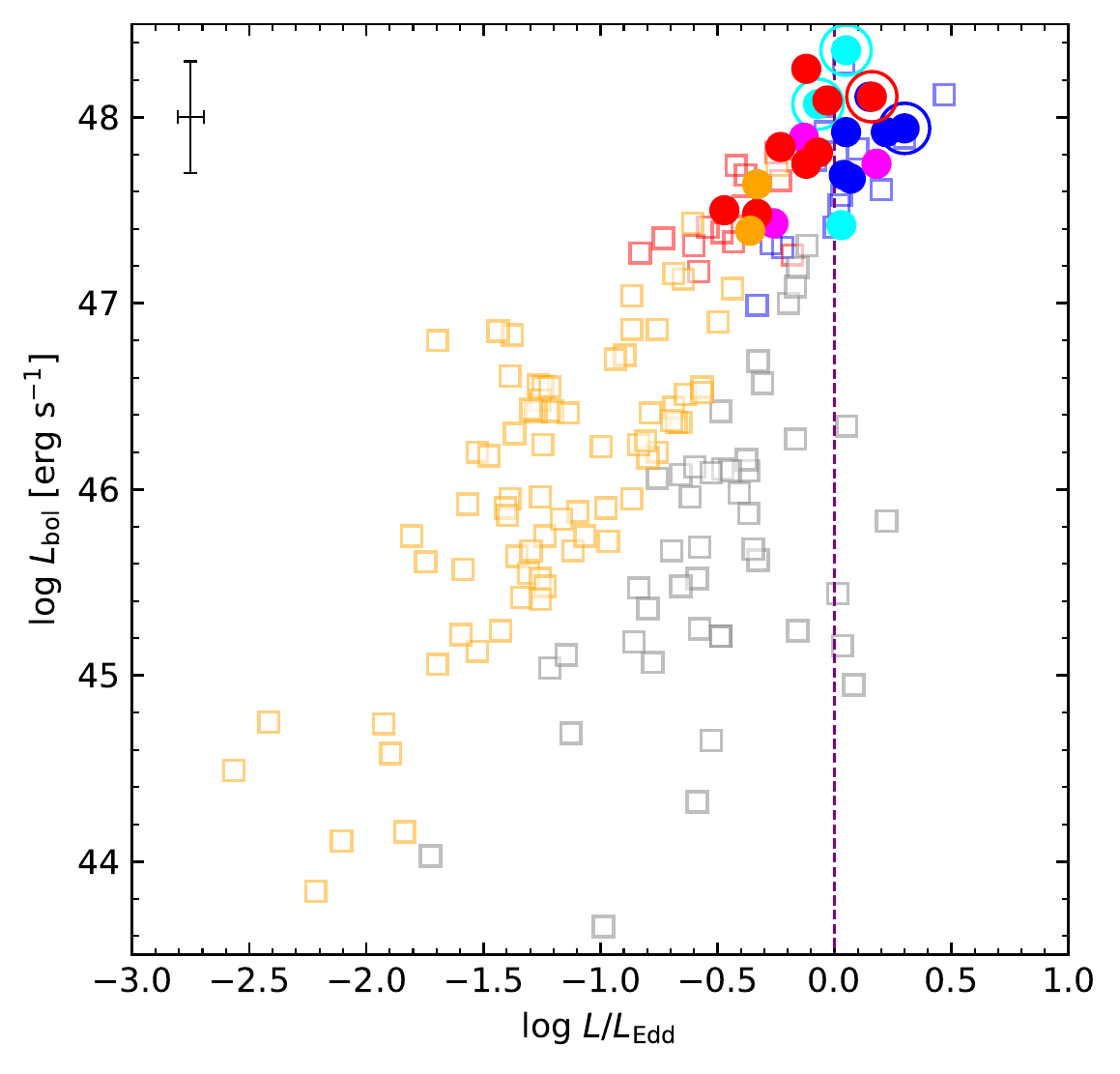}
   \includegraphics[width=0.50\columnwidth]{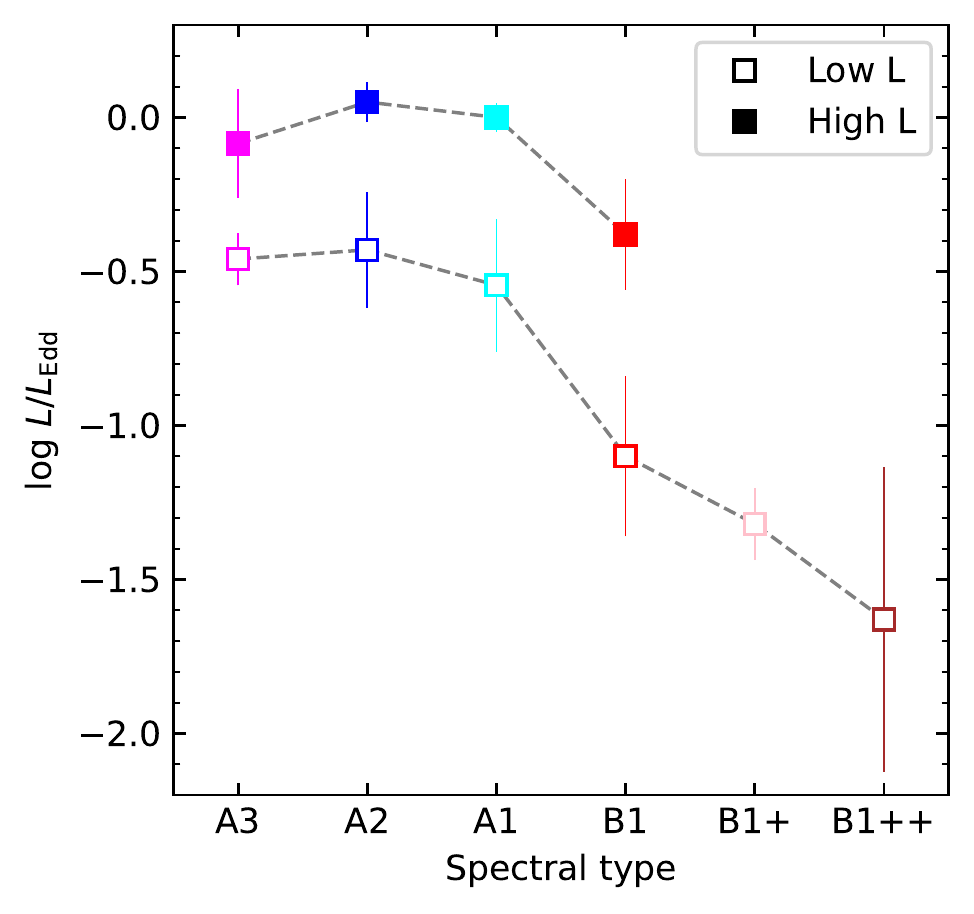}
    \caption{\textit{Left plot:} Bolometric luminosity $L_{\rm bol}$ \textit{versus} Eddington ratio ($L/L_{\rm Edd}$). Squared symbols identify the samples of \citetalias{Sulentic_2007} (low-$z$) and of \citetalias{sulentic_2017} (high-$z$ sources). Vertical purple line indicates  $L/L_{\rm Edd} = 1$. Colour scheme as in Fig. \ref{fig:HB_BC_Full}. \textit{Right plot:} Median values of the Eddington ratio separated into high-luminosity ($\log L>47$ erg s$^{-1}$) and low luminosity for each spectral type. We have included our own sources together with the comparison samples. Error bars indicate the semi-interquartile range.}
    \label{fig:lum_ledd}
\end{figure}

\subsection{Redefining optical properties of high-$z$ sources}
\label{disc:eddratio}

%\par The Eddington ratio for each source is listed in Col. 5 of Table \ref{tab:mass_lum} together with \hb{} black hole masses (Col. 2), the individual weight averaged M$_{BH}$ (Col. 3), and the luminosity at 5100 \r{A} (Col. 4). We have considered the masses estimates using \hb{}, \aliii{}, and \civ{} to determine the weighted averaged black hole masses and used the FWHM errors as weights.

\par Fig. \ref{fig:lum_ledd} shows the relation between the Eddington ratio and the bolometric luminosity. The distinction between Pop. A and Pop. B is very clear in both low- and high-redshift samples including our sample, and the data analysed by \citetalias{Sulentic_2007} and \citetalias{sulentic_2017}. The systematically higher Eddington ratio of the high-luminosity samples may be in part the result of a Malmquist type bias \citep{sulentic_2014}, of the assumption of a constant bolometric correction over a wide range in luminosity,  of orientation effects, and of intrinsic evolution of the Eddington ratio \citep[e.g.,][]{cavalierevittorini00,volonterietal03,shankaretal09}.  For instance, the radio-loud sources from our sample seem to be the more extreme cases of their respective population, reaching $L/L_{\rm Edd} \sim 1$.  They are core-dominated and presumably beamed in the direction of the observer. The line profiles appear composite, in the sense that they show  a narrower core superimposed on a VBC. The width of the narrower core may be lowered if the emitting regions are constrained within a flattened geometry that is seen pole-on. This effect blurs the distinction between Pop. A and B, letting sources that are intrinsically Pop. B enter into the domain of Pop. A in the MS optical plane (Fig. \ref{fig:optical_plane}; \citealt{marziani_2001,Zamfir_2008}).  

\par The right panel of Fig. \ref{fig:lum_ledd} presents the median values of the Eddington ratio for each individual spectral type, distinguishing between higher ($L > 10^{47}$ erg s$^{-1}$) and lower ($L < 10^{47}$ erg s$^{-1}$) luminosity sources \citepalias[][together with our sample]{Sulentic_2007, Marziani_2009,sulentic_2017}. The Eddington ratio is consistently higher in Pop. A sources rather than in Pop. B sources for both cases of luminosity and  redshift separation.   
Low-redshift, low-luminosity sources  present very small $L/L_{\rm Edd}$ values for Pop. B. The more extreme case is for low-redshift Pop. B1++, with a $L/L_{\rm Edd} \approx -1.65$. On the converse Pop. A sources have more similar values at low and high luminosity. There is no notable difference between high/low-luminosity and high/low-redshift distribution on the spectral type and log L/L$_{\rm{Edd}}$ plane, which may indicate that the more luminous sources present the highest Eddington ratios because of a  Malmquist-type bias \citep{sulentic_2014} rather than  evolutionary effects on \lledd{}.
%, and especially if the sources are distinguished on the basis of redshift. This may indicate that we are seeing the contribution of an intrinsic evolution in the Eddington ratio. If \lledd\ scales $\propto (1+z)^3$\ \citep{cavalierevittorini00}, then from $z \approx 0.7$\ to $z \approx 1.5$\ we might well have a ten-fold \lledd\ increase at higher $z$.  

%The increase in Eddington ratio at high $L$\ could be due in part to the bias introduced by a fixed (or almost fixed) flux limit \citep{sulentic_2014}, and in part to an intrinsic redshift evolution \citep[e.g.,][]{cavalierevittorini00,volonterietal03,shankaretal09}. In this respect, the RL sources of the sample show extremely high values if compared with the distribution of Eddington ratios in low-$z$ samples. 

Of the four quasars that are  radio-loud, B1422+231, PKS 1937-101, and PKS 2000-330 are classified as Pop. A while PKS 2126-15 is a Pop. B source. From four sources   we can only make  some preliminary considerations that will be properly discussed in a companion paper involving mainly radio-loud objects. The basic result is that the four RLs do not have outstanding properties with respect to the RQ samples considered in this paper.  They share the trends  between  \hb{} and \oiii{}  RQ profile parameters. A remarkable aspect is that the four RL AGN studied in this paper show high Eddington ratio,  a fact that is never observed in powerful RL sources at low-$z$ \citep{sikoraetal07,Zamfir_2008} that are more frequently classified as population B sources. At low-$z$, low-power, ``jetted'' sources for which high Eddington ratio has been estimated to be $\sim$ 1 are known to exist \citep[the so-called RL NLSy1s; ][]{komossaetal06,bertonetal15,foschinietal15}. Even if \lledd\ might be overestimated, the implication is that there is apparently no empirical inconsistency between powerful jetted emission, and high accretion rate.

%\par  On the other hand, the \hb{} full profile presents a FWHM$\sim 5400$ km s$^{-1}$ for our radio-loud sources, with the exception of PKS 2000-330 which present a FWHM of 3138 km s$^{-1}$. In addition, the four sources present a |c(1/2)|$_{\rm H\beta}$ $\sim 0$ km s$^{-1}$ and very small values of |c(1/2)|$_{\rm[O III]}$.
%\par Regarding the bolometric luminosity, the radio-loud sources are between the ones that present the highest values. Also, when considering only Pop. B sources, PKS 2126-15 is the one with the highest value (L/L$_{\rm Edd}=-0.05$).
%\par 

\section{Conclusions} \label{conclusions}

We presented a sample of 22 high-redshift and high-luminosity quasars (12 Pop. A; 10 Pop. B) observed with the VLT/ISAAC spectrograph to cover the \hb\ spectral range shifted into the near infrared.  A dedicated analysis has been performed on the most prominent emission features in the new optical spectra (\hb, \oiii, \feii), as well as in survey or published UV spectra (\aliii, \ciii, \civ, \siv, \ion{Fe}{\sc iii}, \feiii), to measure several parameters related to the emitting gas physical conditions and dynamics. 

\par The main conclusions are as follows. 
   \begin{enumerate}
      \item We confirm that the full profile of \hb{} can be in general well represented by only a Lorentzian BC for Pop. A quasars, while for the Pop. B ones two Gaussians (BC and VBC) provide a satisfactory fit, with the broader, redshifted VBC accounting for $\approx 50$\% of the emission line. Also, some Pop. A H$\beta$ profiles from our sample present outflowing components that seem to be related to those observed in the \oiii\ lines; 
      \item Compared to low-$z$ and low-luminosity samples, our data presents a displacement in the MS in the direction of higher values of FWHM of the H$\beta$ full profile and $R_{\rm{\ion{Fe}{II}}}$. The relation between profile parameters and the \mbh\ and \lledd\ differences between Pop. A and B are also confirmed, in the form seen in previous high-luminosity samples;
      %(Fig. \ref{fig:HB_BC_Full}). 
      %\item In general, Pop. A are the ones that present the highest L/L$_{\rm Edd}$ when compared to Pop. B, with the highest Eddington ratio found for Pop. A2 sources. In addition, the Pop. B are the ones which present the highest black hole masses;
      \item The \oiii\ profile is broad and { not reproducible with a simple Gaussian}. All \oiii\ profiles of our sample are blueshifted by more than 250 \kms. Differently from the low $z$ range, high-$z$ Pop. B \oiii{} profiles present significant contribution of blueshifted semi-broad components, even though they are still less strong than the ones found in Pop. A sources. Nevertheless, both population seem to share similar asymmetry indexes, which indicates that their profiles present comparable shapes;
      \item In some cases \oiii{} FWHM reach values comparable to those of \hb{}. This effect is found especially in sources with high \lledd{}, and may be compared with the case of  highly-accreting NLSy1s at low $z$. Line widths appear extraordinarily large (thousands \kms) because of the high black hole mass values involved.  
       \oiii\ correlations between FWHM and shift, and also between $W$ and shift  mirror the correlations seen in \civ;
       \item {The \oiii\ lines should be avoided for accurate redshift estimates in high-$L$, high-$z$ quasars, as in most of sources the [\ion{O}{III}] lines are dominated by a blueshifted SBC and no rest-frame narrow component could be clearly identified. The average shift of the [\ion{O}{III}] profile peak is $\sim  -520$ km s$^{-1}$, which may lead to a considerable systematic blueshift  of $\delta z \sim 0.0017$ on the redshift estimation; }
       \item Several \civ\ correlations (most notably the one between shift and width) and the \civ\ parameters' dependence on Eddington ratio and mass (or luminosity) are confirmed (including the ones of \citetalias{sulentic_2017}).  Outflow-dominated profiles tend to have low-$W$, large FWHM and shifts. The three parameters  are definitely related;   
      % \item \aliii\ Also the difference between \civ\ and \hb\ profile parameters.
           % \item The relations found by \citetalias{sulentic_2017} between \civ{} $c$(1/2), L/L$_{\rm Edd}$, and $\lambda$L$_{\lambda}$(1450\r{A})  in high-redshift sources have been reconfirmed with our data; 
      
      \item We exploit the \citetalias{Vestergaard_2006} scaling law as reference for \mbh\ estimates, and we verified that the scaling law based on \aliii\ is a reliable estimator with respect to black hole masses derived from \hb. The \civ{} scaling law requires a FWHM correction;
           % and on a \textbf{\textit{corrected}} FWHM \civ\  measurements are \textbf{reliable} estimators with respect to black hole masses derived from \hb; 
      \item \oiii{} and \civ{} seem to follow a very similar, strong  correlation between  FWHM and shift, measured by the $c$(1/2).  
      The \oiii\ and \civ\  outflow velocities are related, as suggested by the correlation between their shift amplitudes shown in Fig. \ref{fig:CIV_oiii};
      
      \item The radio loud sources in the high-$z$ range seem to be radiating at high values of Eddington ratio (with \lledd\ $\sim 1$), at variance with low-$z$ where the most powerful jetted sources belong to Population B and radiate at modest Eddington ratio. 
   \end{enumerate}
 
 The sample studied in this paper is the first part of a sample involving roughly an equal number of RQ and of RI and RL sources. The data in this paper provide a  RQ comparison sample for the ISAAC spectra of RL sources that will be presented in a future paper.

\begin{acknowledgements}
      The authors thank the anonymous referee for her/his valuable suggestions that helped us to significantly improve the present paper. A.D.M. acknowledges the support of the INPhINIT fellowship from ``la Caixa'' Foundation (ID 100010434). The fellowship code is LCF/BQ/DI19/11730018. A.D.M., A.d.O., and J.P. acknowledge financial support from the Spanish State Agency MCIN/AEI/10.13039/501100011033 through research grants PID2019–106027GB–C41 and  PID2019–106027GB–C43 and the Centre of Excellence Severo Ochoa award to the Instituto de Astrofísica de Andalucía under contract SEV–2017–0709. P.M. acknowledges the support of the IAA-CSIC for a visit in November 2021. A.D.M. is thankful for the kind hospitality at the Padova Astronomical Observatory. This research has made use of the NASA/IPAC Extragalactic Database (NED) which is operated by the Jet Propulsion Laboratory, California Institute of Technology, under contract with the National Aeronautics and Space Administration. Funding for the Sloan Digital Sky Survey IV has been provided by the Alfred P. Sloan Foundation, the U.S. Department of Energy Office of Science, and the Participating Institutions. SDSS-IV acknowledges support and resources from the Center for High Performance Computing  at the University of Utah. The SDSS website is www.sdss.org. SDSS-IV is managed by the Astrophysical Research Consortium for the Participating Institutions of the SDSS Collaboration including the Brazilian Participation Group, the Carnegie Institution for Science, Carnegie Mellon University, Center for Astrophysics | Harvard \& Smithsonian, the Chilean Participation Group, the French Participation Group, Instituto de Astrof\'isica de Canarias, The Johns Hopkins University, Kavli Institute for the Physics and Mathematics of the Universe (IPMU) / University of Tokyo, the Korean Participation Group, Lawrence Berkeley National Laboratory, Leibniz Institut f\"ur Astrophysik Potsdam (AIP),  Max-Planck-Institut f\"ur Astronomie (MPIA Heidelberg), Max-Planck-Institut f\"ur Astrophysik (MPA Garching), Max-Planck-Institut f\"ur Extraterrestrische Physik (MPE), National Astronomical Observatories of China, New Mexico State University, New York University, University of Notre Dame, Observat\'ario Nacional / MCTI, The Ohio State University, Pennsylvania State University, Shanghai Astronomical Observatory, United Kingdom Participation Group, Universidad Nacional Aut\'onoma de M\'exico, University of Arizona, University of Colorado Boulder, University of Oxford, University of Portsmouth, University of Utah, University of Virginia, University of Washington, University of Wisconsin, Vanderbilt University, and Yale University.
\end{acknowledgements}
  \bibliographystyle{aa} % style aa.bst
   \bibliography{bib} % your references Yourfile.bib

\begin{appendix} %First appendix

\onecolumn
\section{Multicomponent fits in the optical and UV ranges and individual notes}
\label{app_uv}

\par In this appendix we present the new VLT/ISAAC infrared spectra and the  multi-component fittings for the optical spectral range, including the fits of \hb{}+\oiii{}. An additional analysis of the UV spectra is also presented for the objects with available UV spectra, from the literature or from SDSS. The UV analysis includes fittings for \siv{}+\oiv{}, \civ{}+\heiiuv{} as well as for the 1900 \r{A} blend. We also include  notes for some of the individual objects. 

%\par Optical continua are shown in grey dashed lines and UV continua are represented by orange, pink and purple dashed lines (for \siv{}, \civ{}, and 1900\r{A} blend, respectively). \ion{Fe}{II} contributions are in light green. The color scheme for each component in both optical and UV ranges is as follows: blue for BLUE and SBC, black for BC, red for VBC, orange for NC, and brown for absorption lines. The modelled full profile is represented in pink dashed lines. 

%--------------------------------------------
\subsection{HE 0001-2340}
\label{HE 0001-2340}

\begin{figure}[h!]
    \centering
    \includegraphics[width=0.69\linewidth]{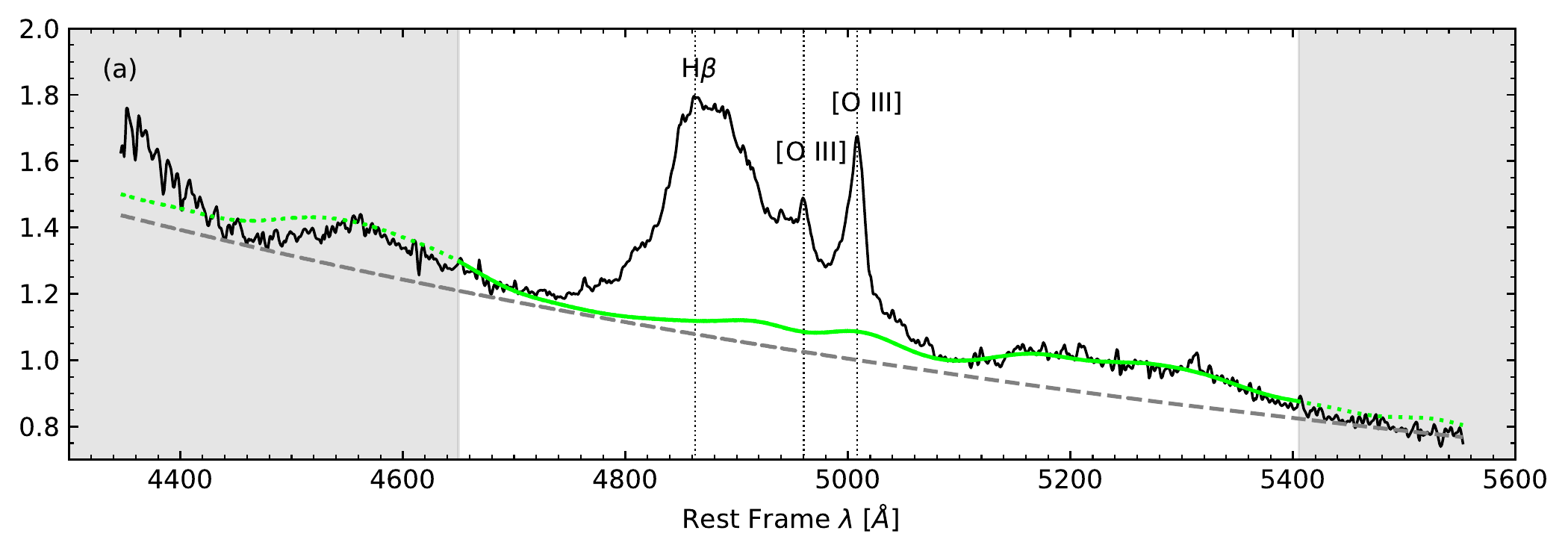}
    \includegraphics[width=0.30\linewidth]{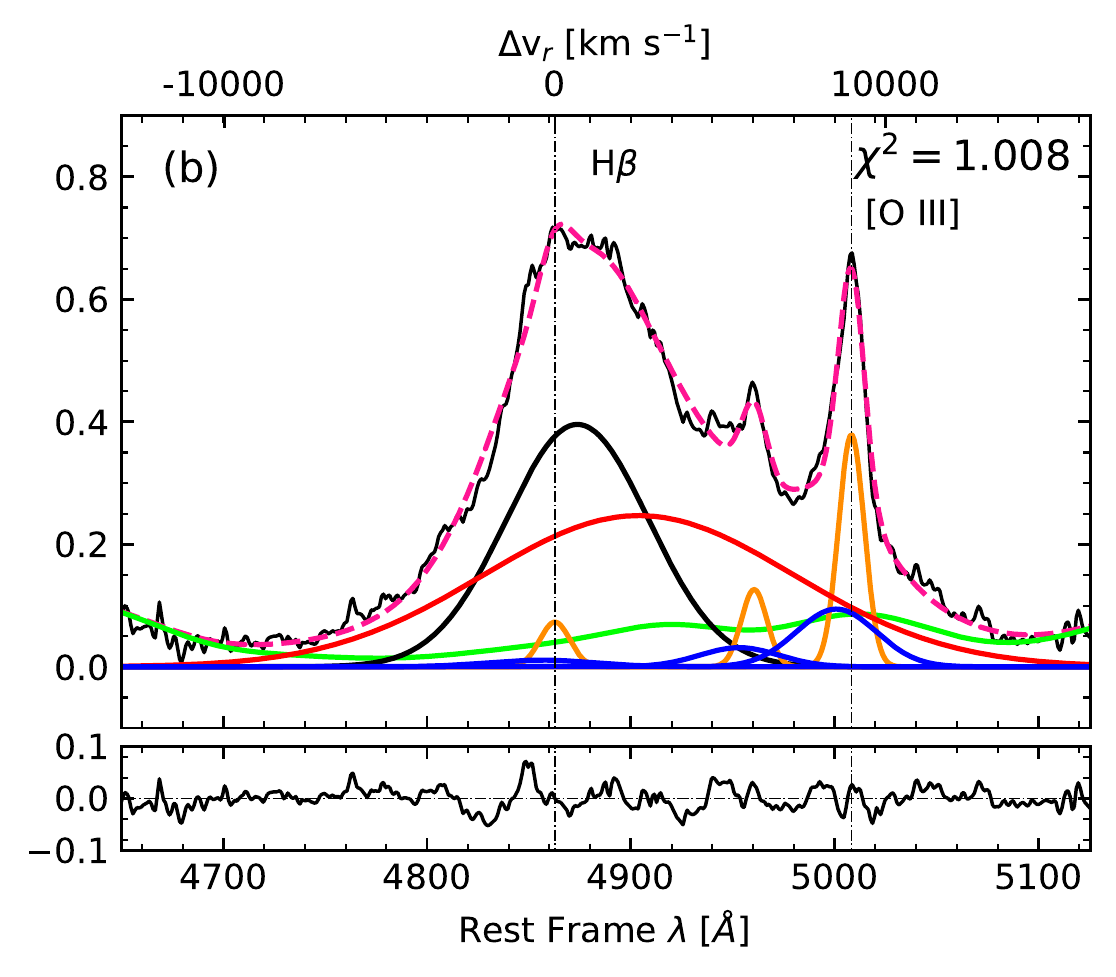}
    \caption{HE 0001-2340. \textit{(a)} Rest-frame  spectrum covering the \hb\ spectral range obtained with VLT/ISAAC. The spectrum is normalised by the continuum at 5100\r{A} (the flux values are available at Table \ref{tab:specphot_HB}). The grey dashed line traces a power law that defines the continuum level as obtained with the \textsc{specfit} multicomponent analysis. The green line shows the \ion{Fe}{II} contribution. For this object \ion{Fe}{II} was fitted only with the red \ion{Fe}{II} blend due to the presence of atmospheric absorption in the blue. The  vertical dotted lines indicate the rest-frame of the main emission lines in the \hb\ spectral range and the grey-shaded area indicate the regions that were \textit{not} considered in the fittings. The white area indicates the region used to anchor both the continuum and the \ion{Fe}{II} template. \textit{(b)} Result of the fitting after continuum subtraction (upper panel) and the residuals (bottom panel) for the \hb\ region. Pink dashed line shows the final fit. Broad \hb\ component (BC) is represented by a black line meanwhile red line show the VBC. Orange lines represent narrow components and the blue ones correspond to the blueshifted components. The region in which the \ion{Fe}{II} template was fitted is represented by the solid green line. Dotted green line indicate the expected \ion{Fe}{II} contribution for the other parts of the spectra.}
    %\textit{Bottom:} Contour map of radio emission obtained from NVSS. 
    \label{fig:0001_HB}
\end{figure}

%\par The \ion{Fe}{II} templates for HE 0001-2340 were adjusted based on the blue part of the spectrum.
%--------------------------------------------
\subsection{$[\rm HB89]$ 0029+073}
\label{0029+073}
\begin{figure}[h!]
    \centering
    \includegraphics[width=0.69\linewidth]{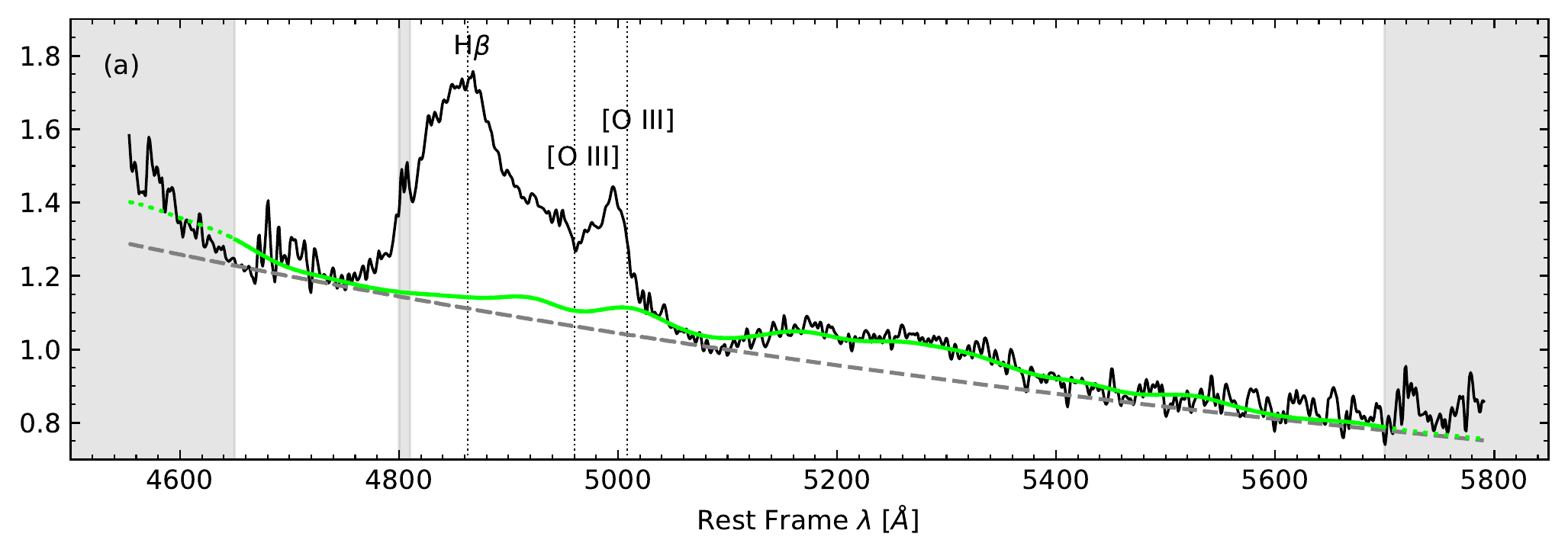}
    \includegraphics[width=0.305\linewidth]{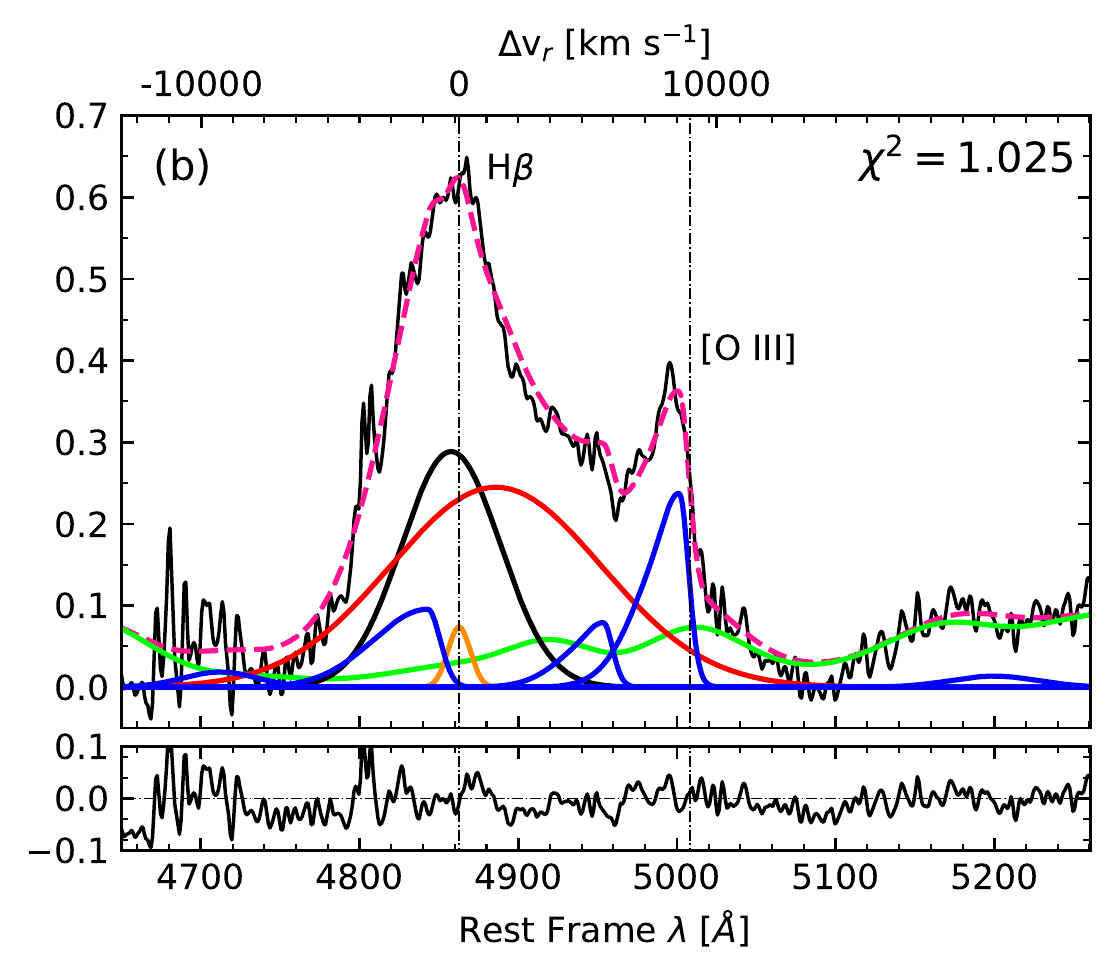}
    \caption{$[\rm HB89]$ 0029+073. Colors and lines as Figure \ref{fig:0001_HB}.}
    \label{fig:0029_HB}
\end{figure}

\par \cite{Jaussen_1995} list this source as a gravitational lens candidate. $[\rm HB89]$ 0029+073 is another case in which the \ion{Fe}{II} contribution was measured at $\lambda \ge$ 5000\r{A}, and rescaled to obtain the intensity of \ion{Fe}{II}$\lambda4570$ blend. 

\newpage
%--------------------------------------------
\subsection{CTQ 0408}
\label{CTG0408}
\begin{figure}[h!]
    \centering
    \includegraphics[width=0.69\linewidth]{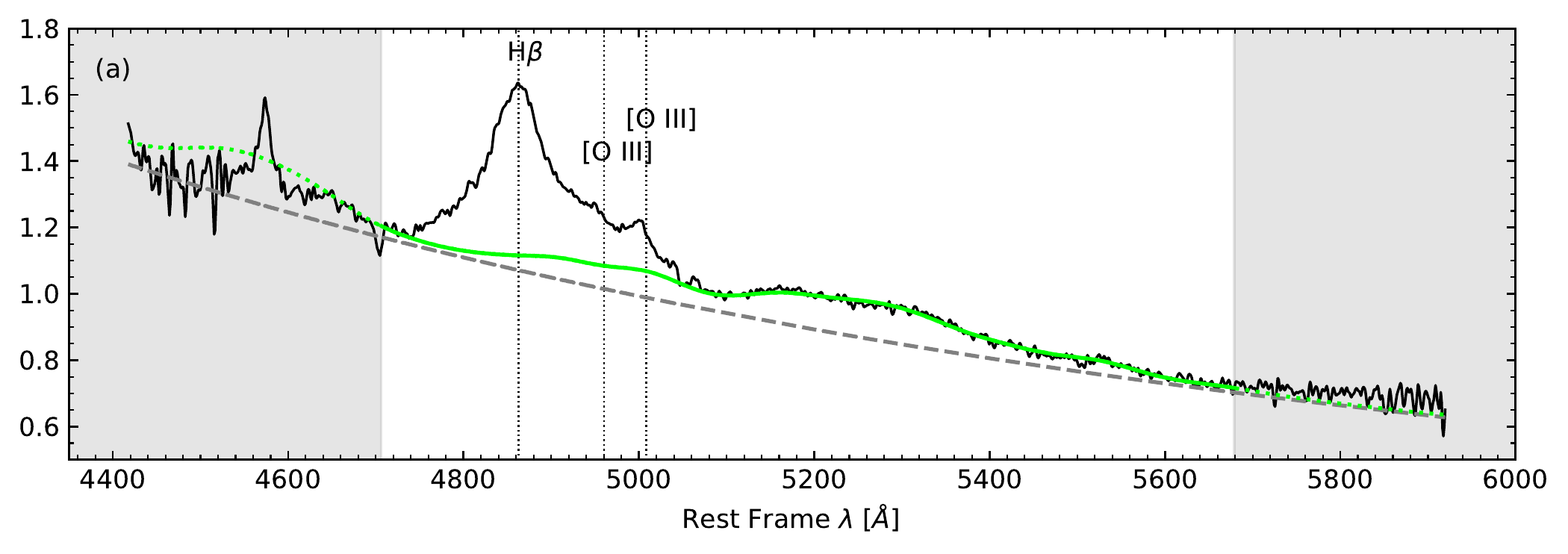}
    \includegraphics[width=0.30\linewidth]{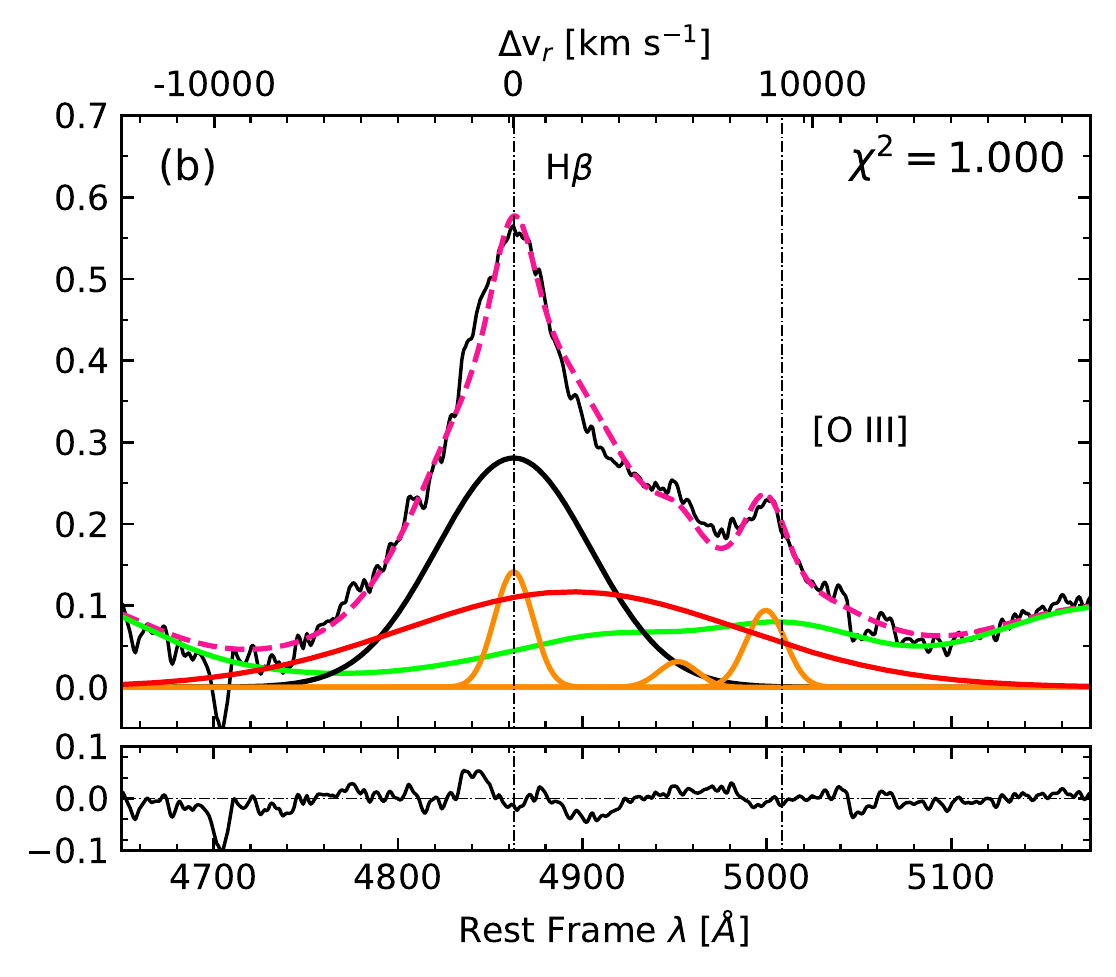}
    \caption{CTQ 0408. Same as Figure \ref{fig:0001_HB}.}
    \label{fig:1801_HB}
\end{figure}

\par As in the case of $[\rm HB89]$ 0029+073, the \ion{Fe}{II} of CTQ 0408 was fit on  the red side of \hb\ with $\lambda \ge$ 5000\r{A}. %once it is difficult to analyse the blue part due to its proximity to the border of the spectrum. 
\par Radio information about this source was found only in the SUMSS catalogue \citep{Mauch_2003}. There is no radio map from NVSS nor FIRST for this object, because it is out of the respective survey fields.

\newpage

%--------------------------------------------
\subsection{SDSSJ005700.18+143737.7}
\label{SDSSJ0057}
\begin{figure}[h!]
    \centering
    \includegraphics[width=0.68\linewidth]{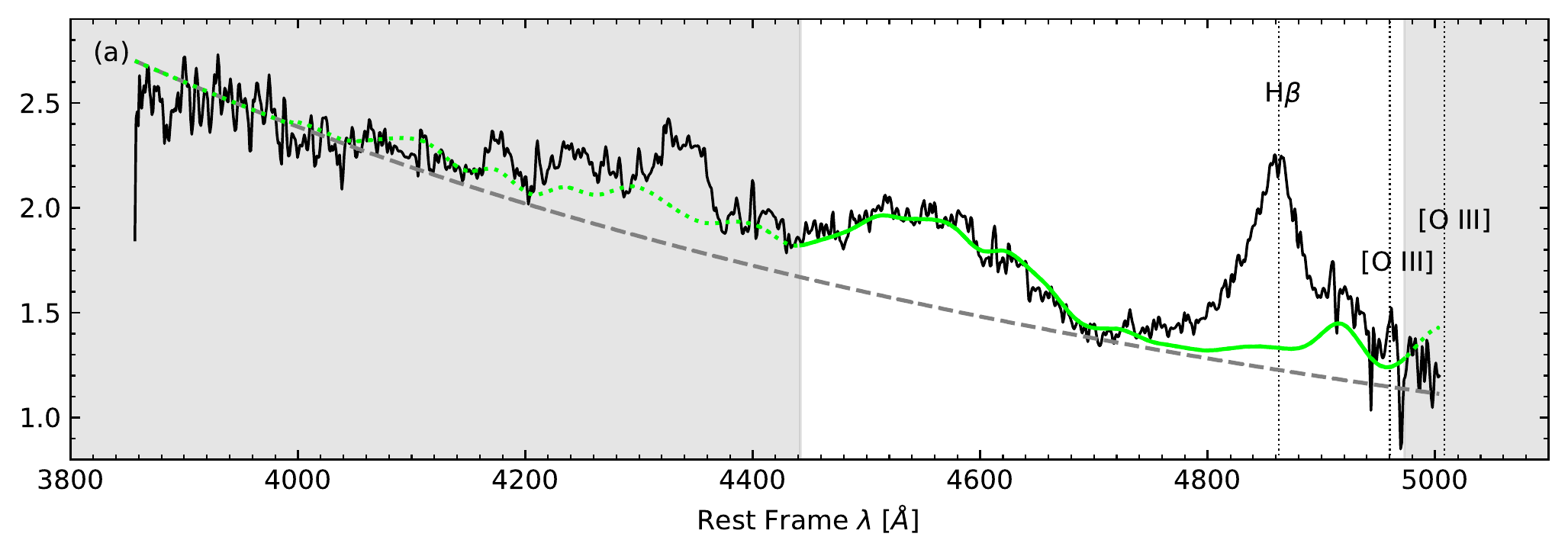}
    \includegraphics[width=0.315\linewidth]{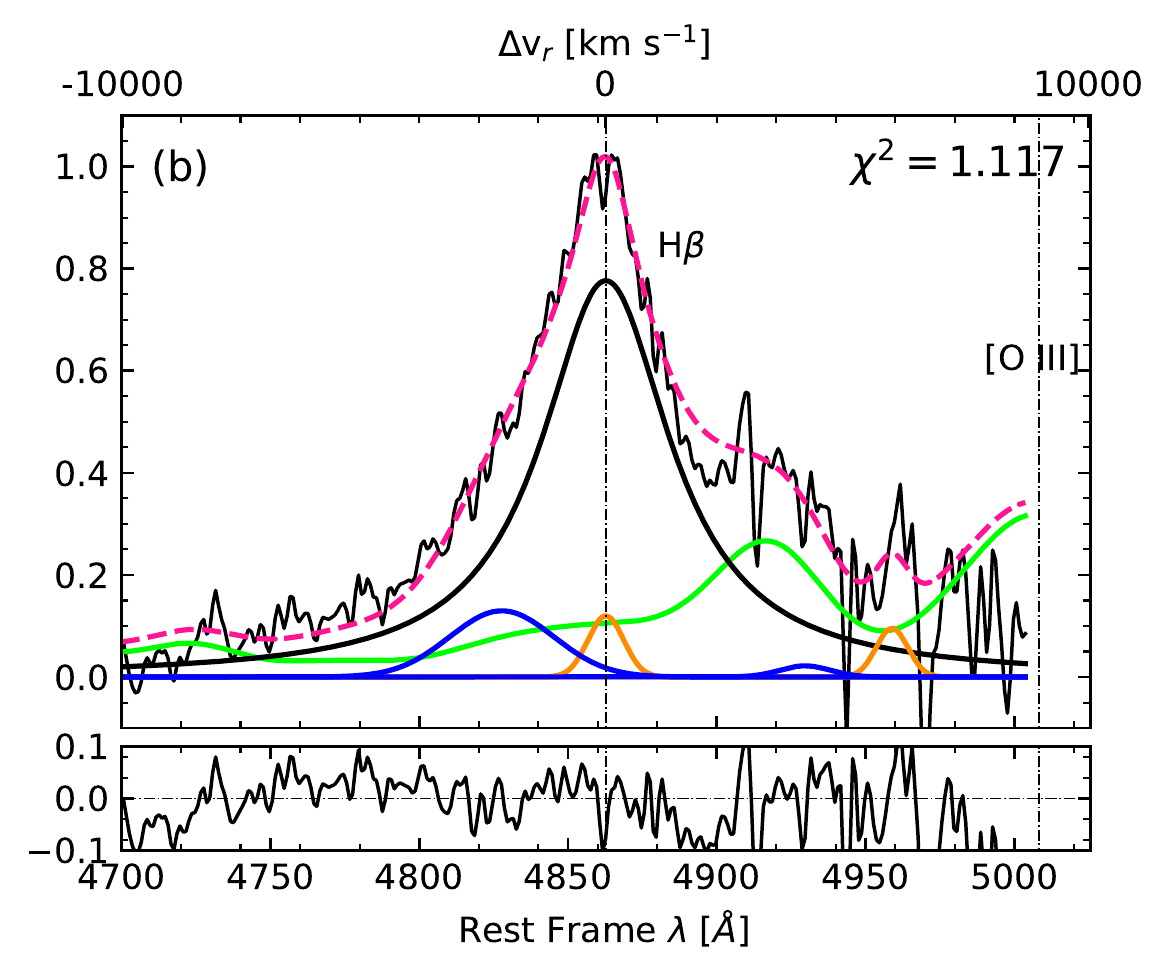}
    \\
    \centering
    \includegraphics[width=\linewidth]{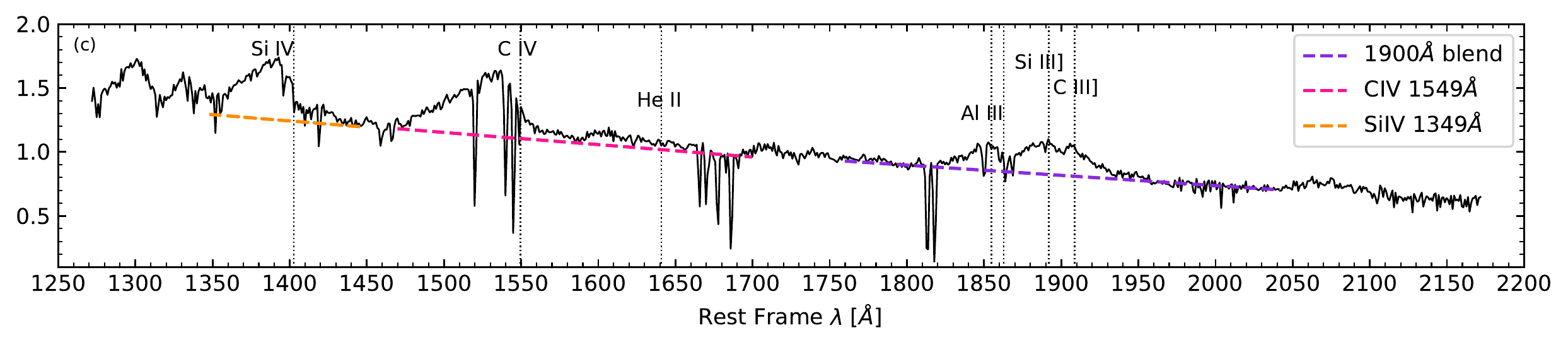}
    \includegraphics[width=0.32\linewidth]{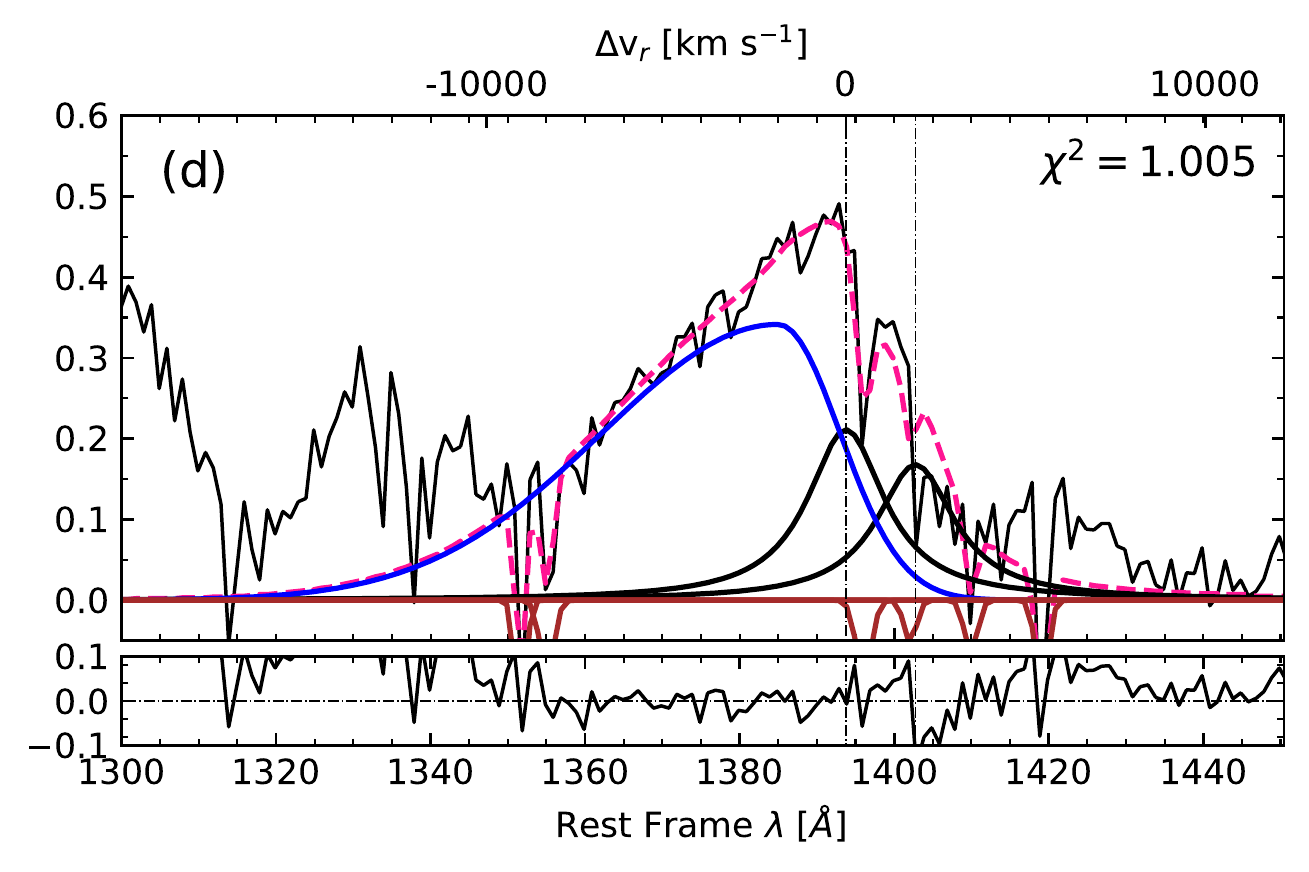}
    \includegraphics[width=0.325\linewidth]{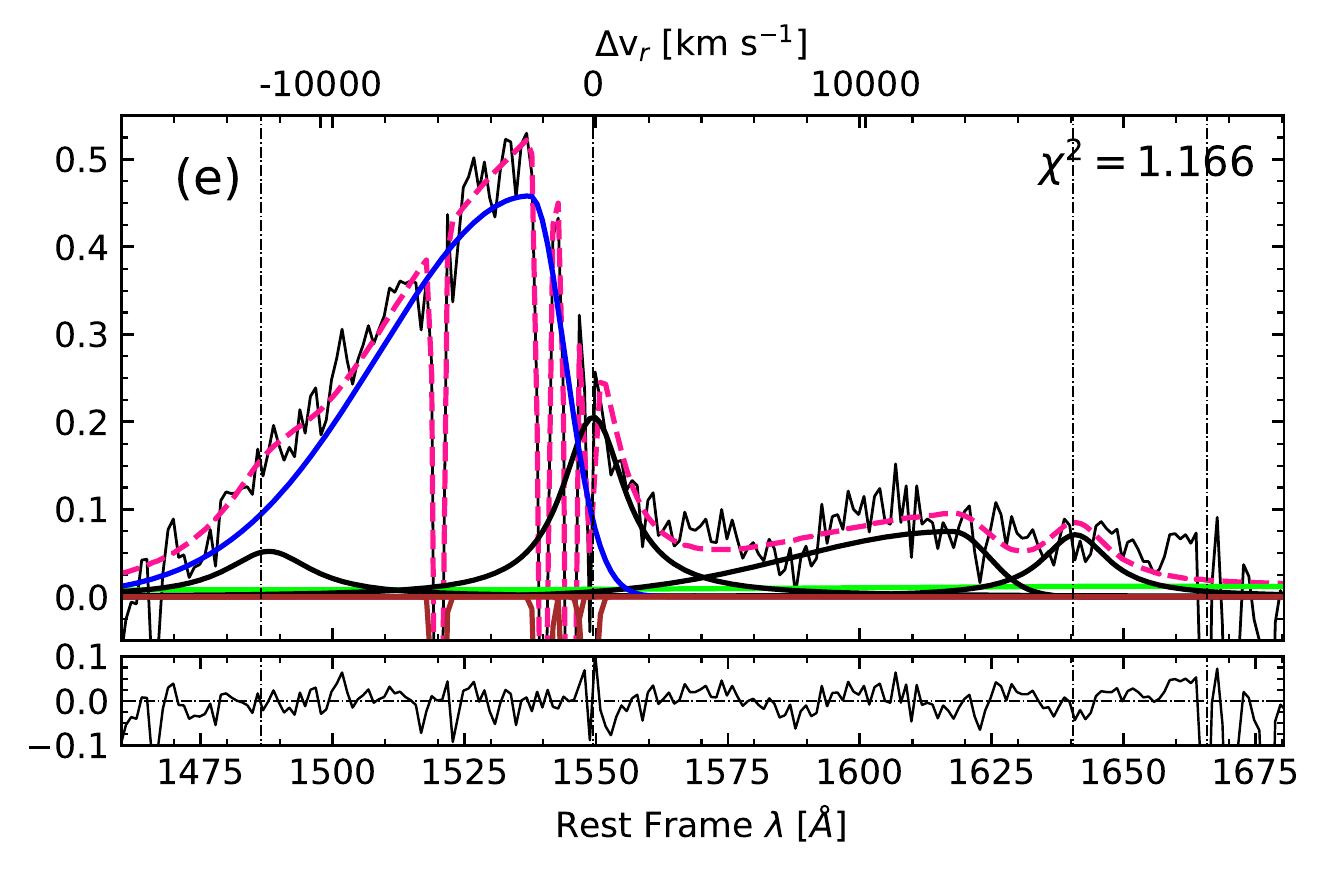}
    \includegraphics[width=0.34\linewidth]{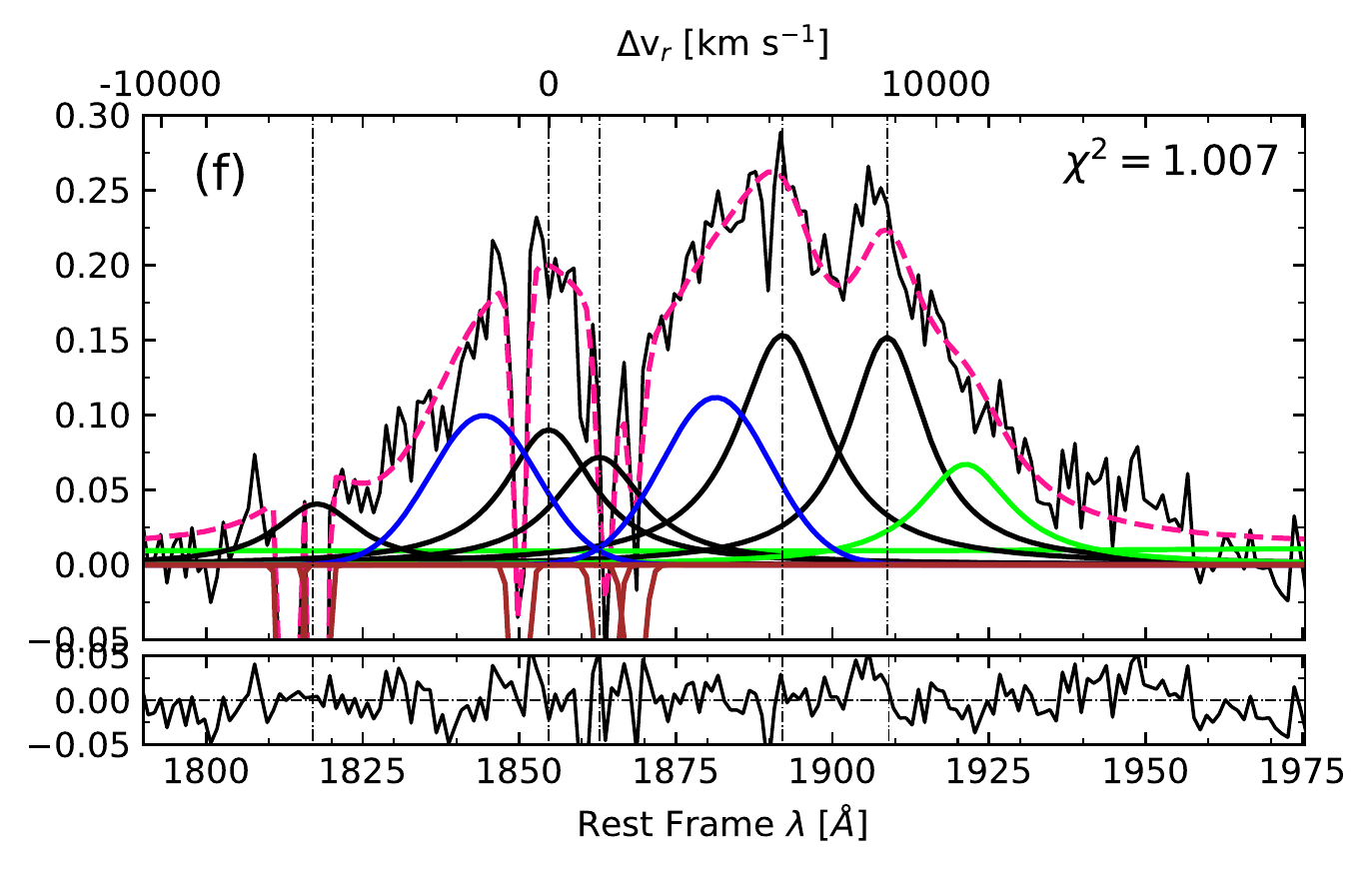}
    \raggedleft
    \caption{SDSSJ005700.18+143737.7. \textit{Top panels:} Same as Fig. \ref{fig:0001_HB}. \textit{(c)} continuum-normalised UV spectrum from SDSS DR-16 with adopted continuum marked in colour; \textit{Bottom panels:} fits for (d) \siv{}+\oiv{}, (e) \civ{}+\heiiuv{}, and (f) the 1900 \r{A} blend. Pink dashed lines show the final fitting.  Broad components are represented by black lines, while blueshifted components are in blue. Green line represents the additional \ion{Fe}{III}$\lambda$1914 line in the red side of \ion{C}{III}], observed in extreme Pop. A sources. Brown lines represent the absorptions seen in the spectrum and modelled as negative-flux Gaussians. }
    \label{fig:0057_UV}
\end{figure}

\par %As can be seen in Fig. \ref{fig:0057_UV}, 
H$\beta$+\oiii{} falls at the red border of the observed spectrum. Consequently, the [\ion{O}{III}] measurements   should be treated as highly uncertain and special marks have been included in the corresponding tables in the paper. The measurements and the analysis on the [\ion{O}{III}] profile were performed with the [\ion{O}{III}]$\lambda$4959 instead of [\ion{O}{III}]$\lambda$5007. With respect to the UV spectrum, the object  presents strong absorption lines likely associated with intervening absorbers throughout the three UV regions of interest. In the analysis we also included Gaussians profiles for the absorption lines in the fittings. It was necessary to include two blueshifted components for \aliii{} and \ion{Si}{III}]$\lambda$1892, otherwise both \aliii{} and \ion{Si}{III}]$\lambda$1892 BC are shifted by more than 1000 km s$^{-1}$.%Regarding the 1900\r{A} blend, we provide two different fittings in order to show the necessity of including blueshifted components for \aliii{} and \ion{Si}{III}]$\lambda$1892.  

\clearpage
%--------------------------------------------
\subsection{H 0055-2659}
\label{H0055}
\begin{figure}[h!]
    \centering
    \includegraphics[width=0.685\linewidth]{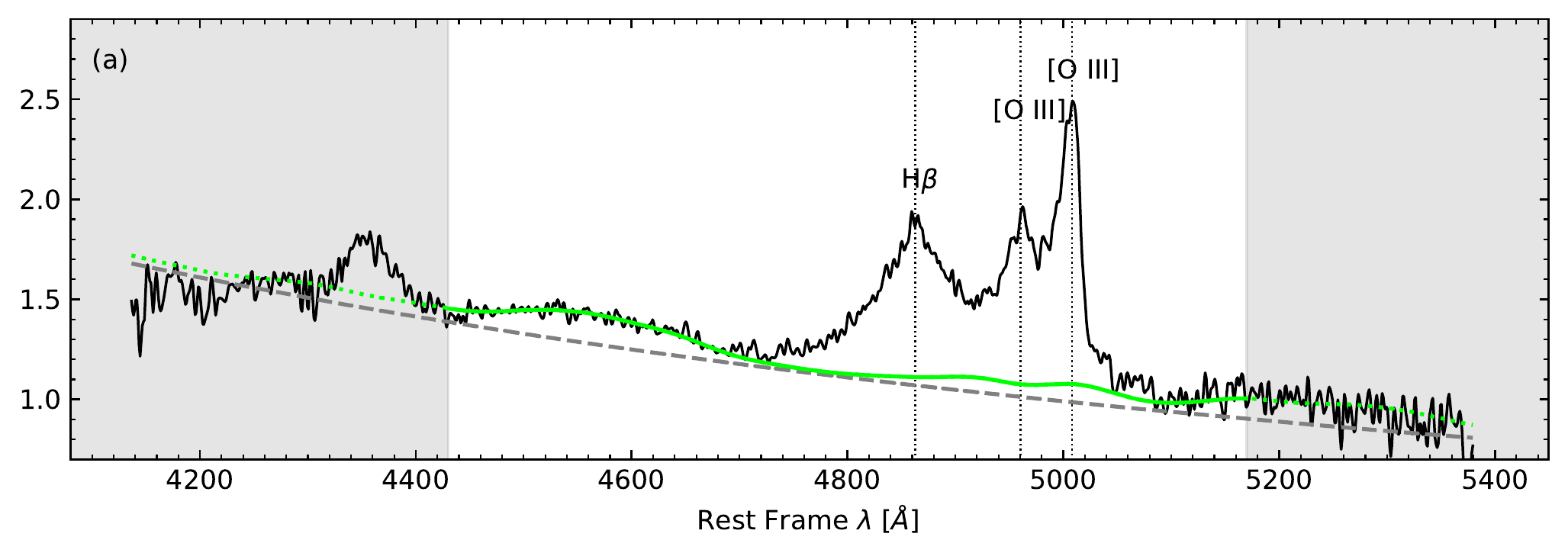}
    \includegraphics[width=0.31\linewidth]{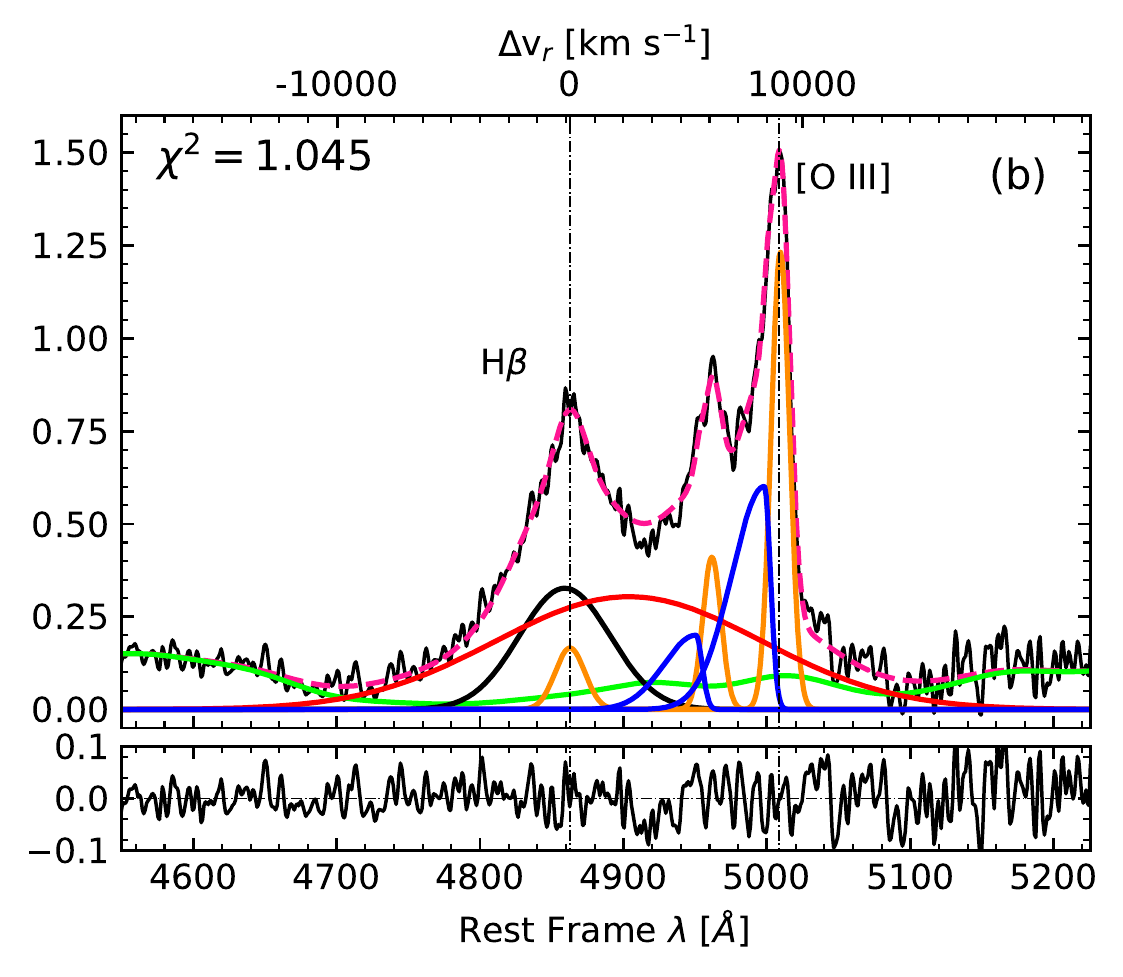}
    \caption{H 0055-2659. Same as Figure \ref{fig:0001_HB}.}
    \label{fig:0055_HB}
\end{figure}

\par The VLT rest-frame optical spectrum of H 0055-2659 presents a very flat profile and small \ion{Fe}{II} multiplets contributions.

\newpage
%--------------------------------------------
\subsection{SDSSJ114358.52+052444.9}
\label{SDSSJ1143}

\begin{figure}[h!]
    \centering
    \includegraphics[width=0.685\linewidth]{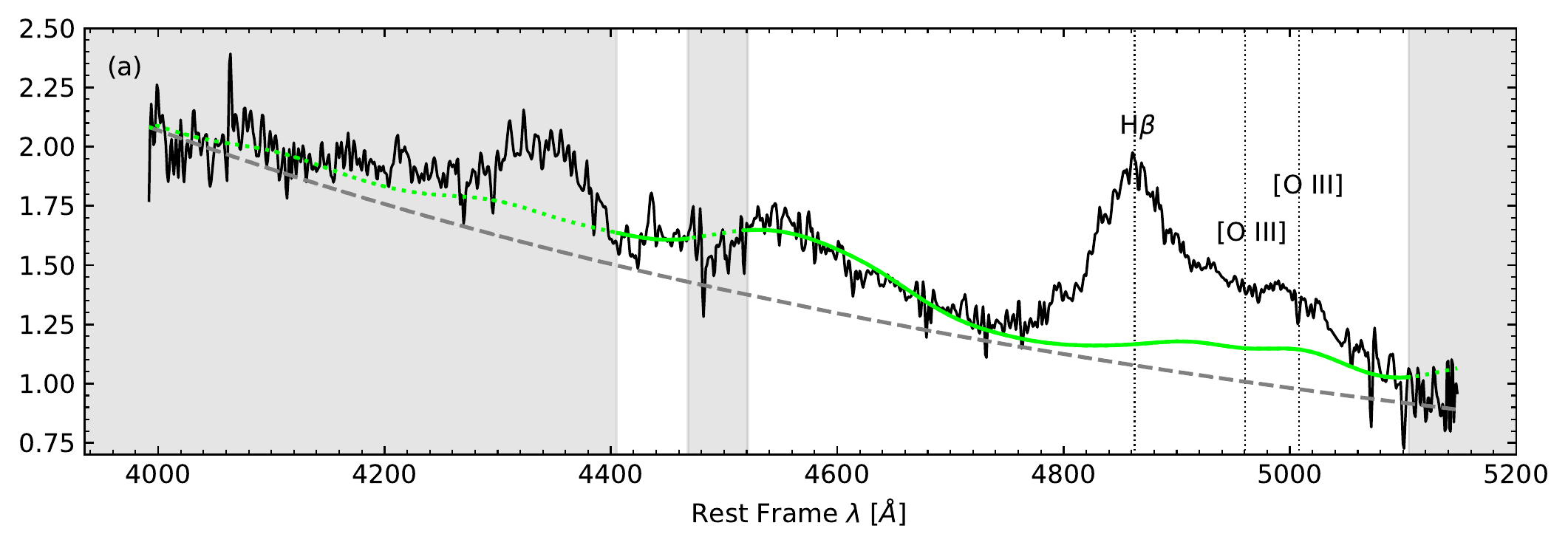}
    \includegraphics[width=0.30\linewidth]{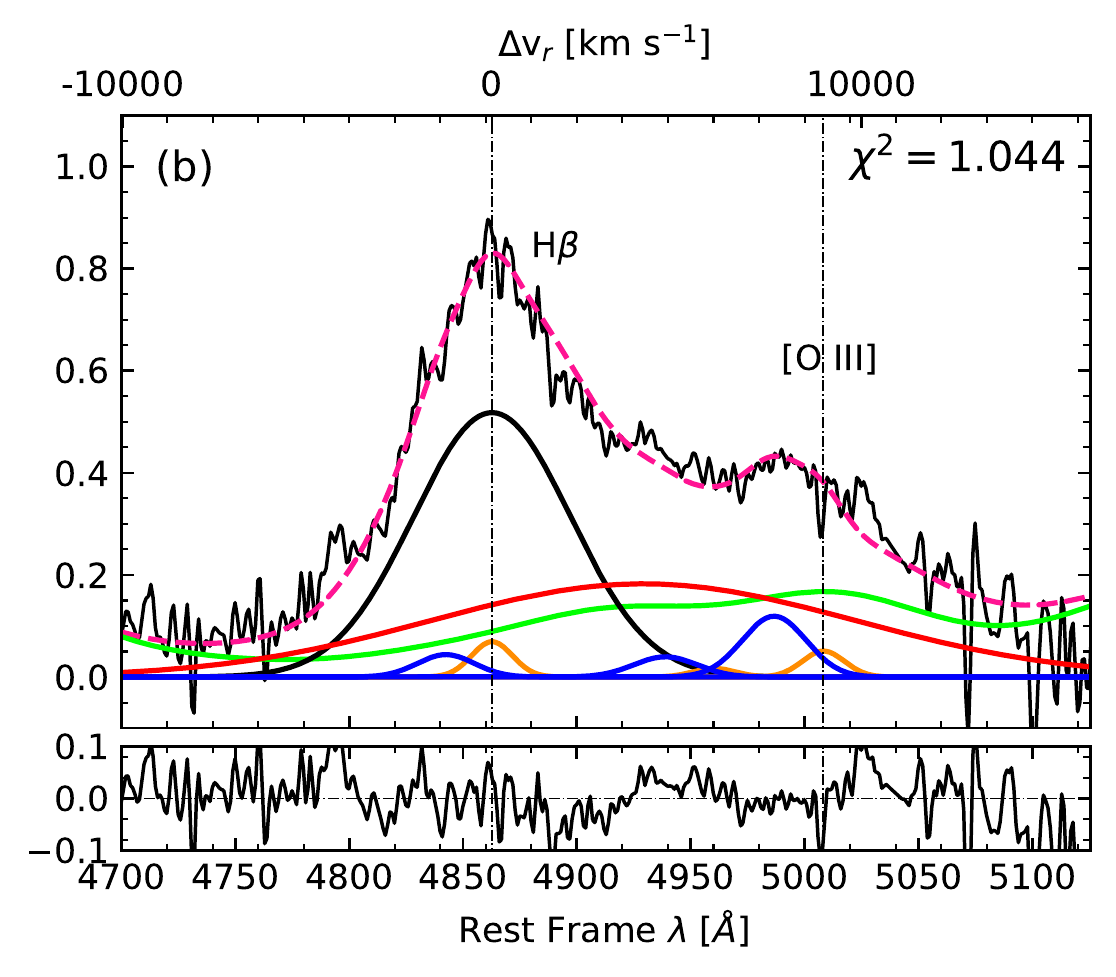}
    \\
    \centering
    \includegraphics[width=\linewidth]{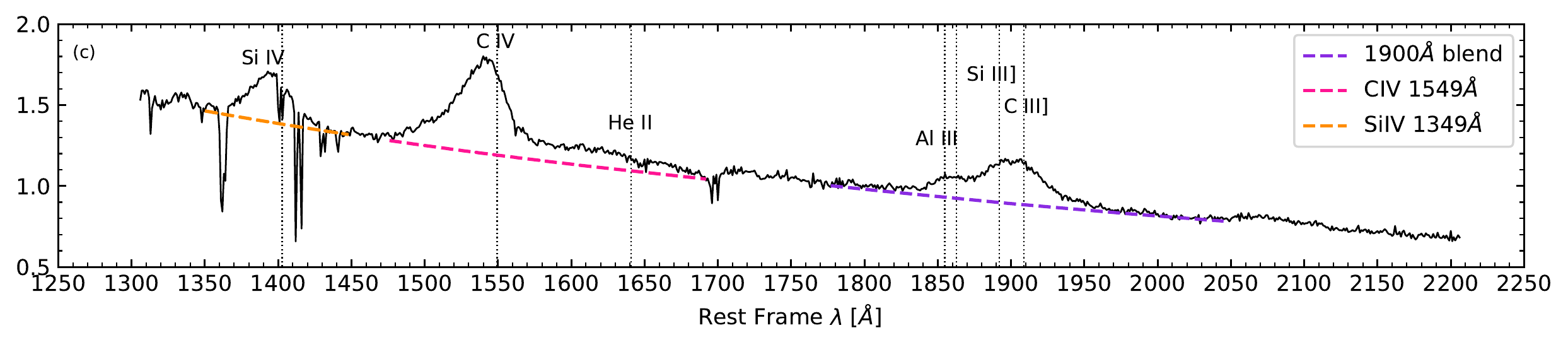}
    \includegraphics[width=0.32\linewidth]{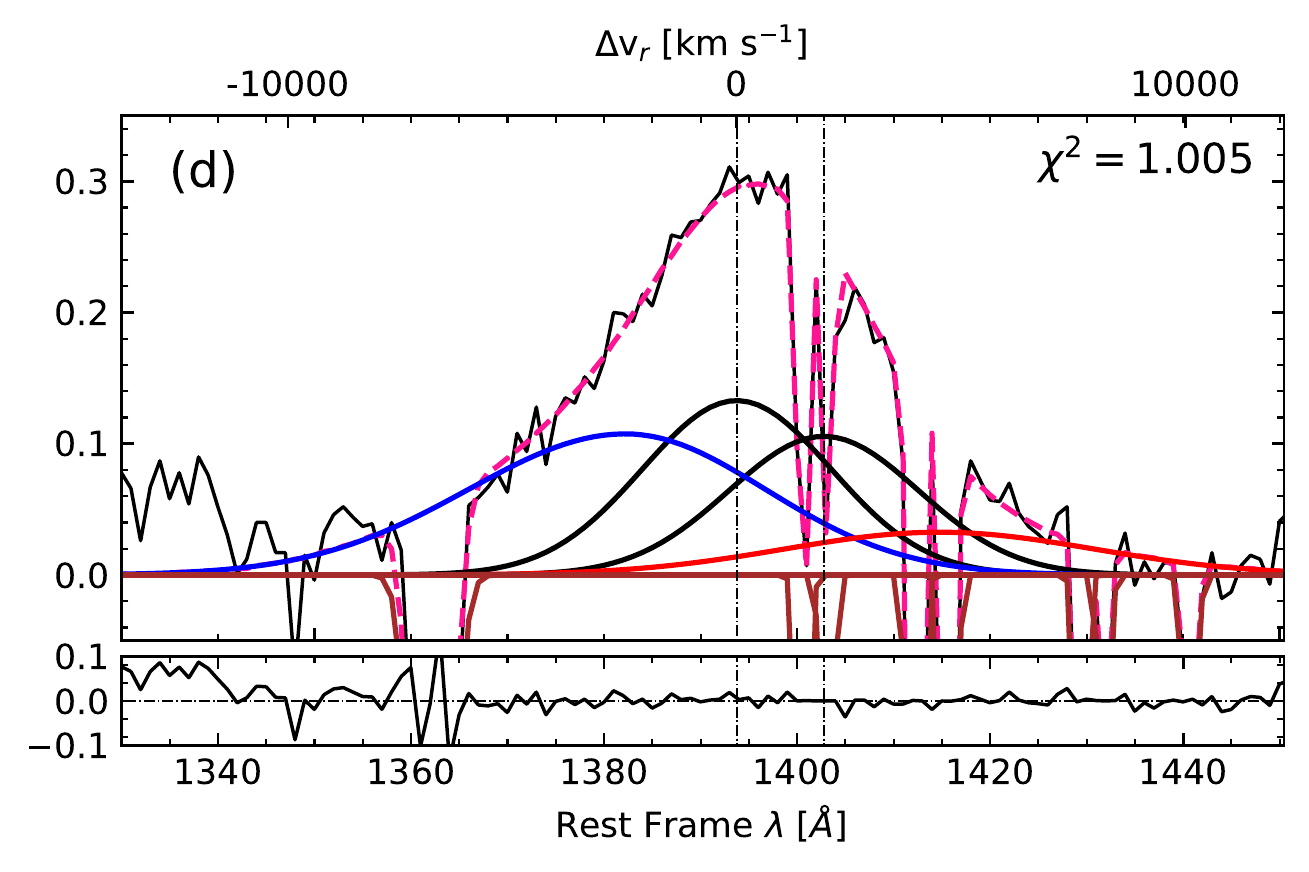}
    \includegraphics[width=0.32\linewidth]{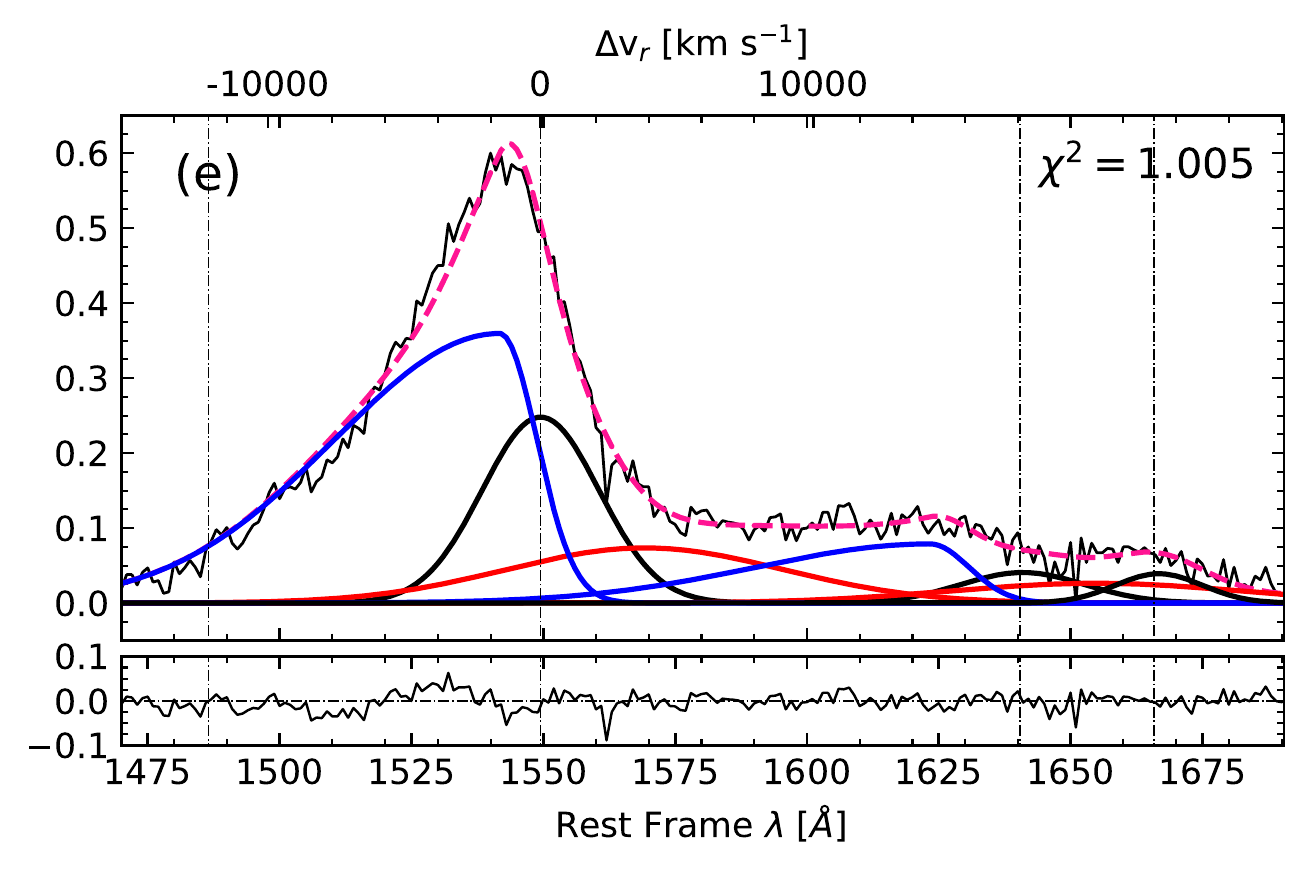}
    \includegraphics[width=0.34\linewidth]{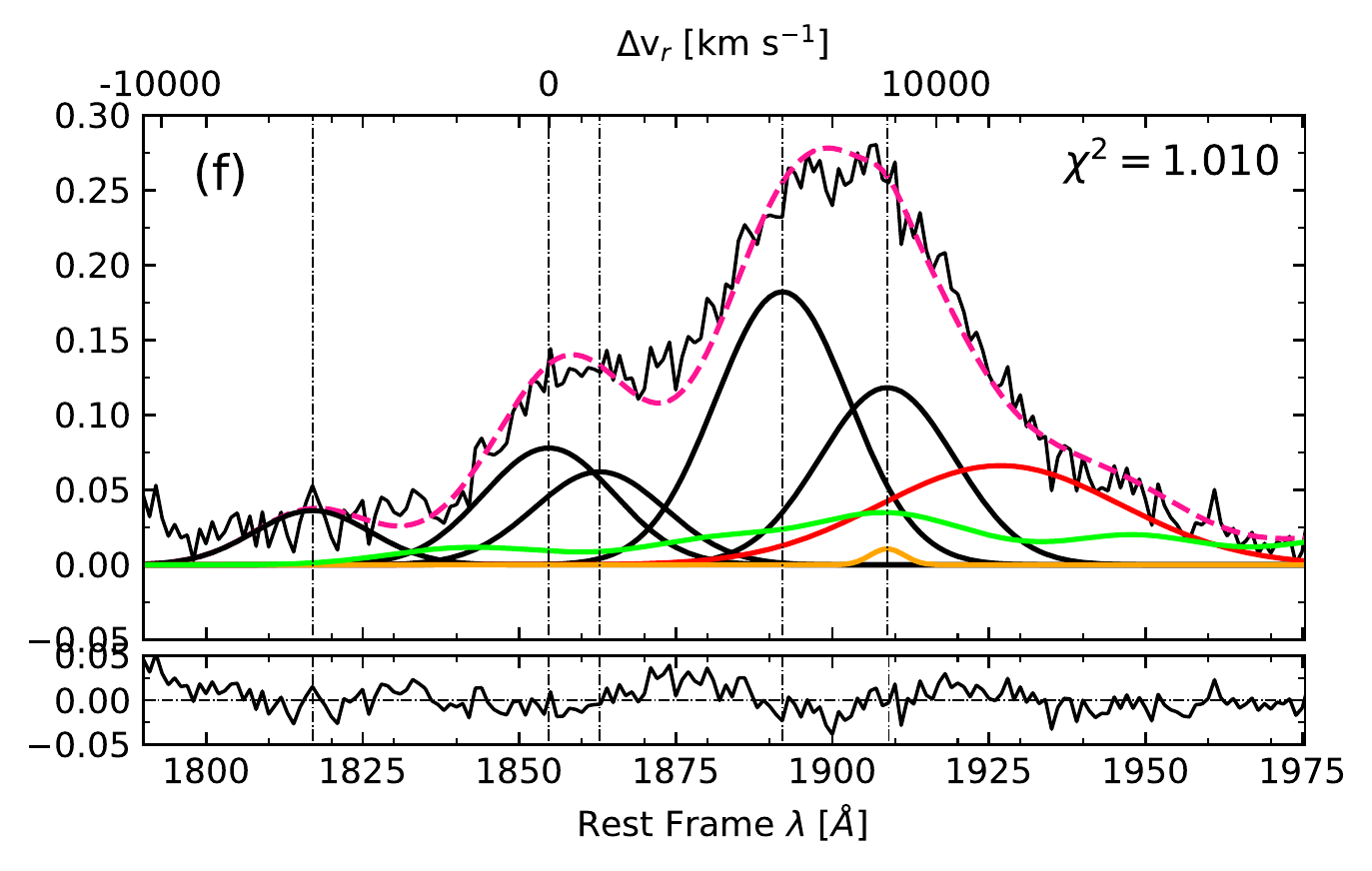}
    \caption{SDSSJ114358.52+052444.9. \textit{Top panels:} same as Fig. \ref{fig:0001_HB} \textit{Middle and bottom panels:} same as Fig. \ref{fig:0057_UV}. Pink dashed lines show the final fitting. Broad components are represented by black lines, while blueshifted components are in blue. Brown lines represent the absorption lines seen in the spectrum.}
    \label{fig:1143_UV}
\end{figure}

\clearpage

%--------------------------------------------
\subsection{SDSSJ115954.33+201921.1}
\label{SDSSJ1159}

\begin{figure}[h!]
    \centering
    \includegraphics[width=0.68\linewidth]{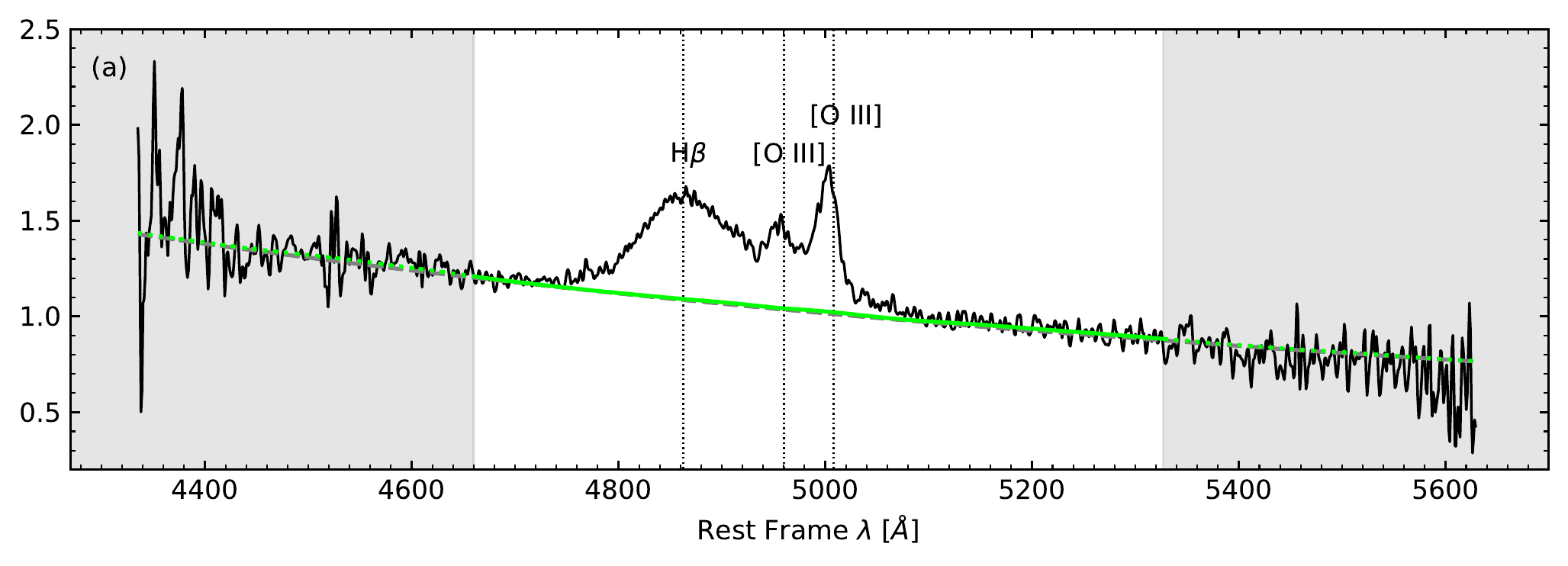}
    \includegraphics[width=0.31\linewidth]{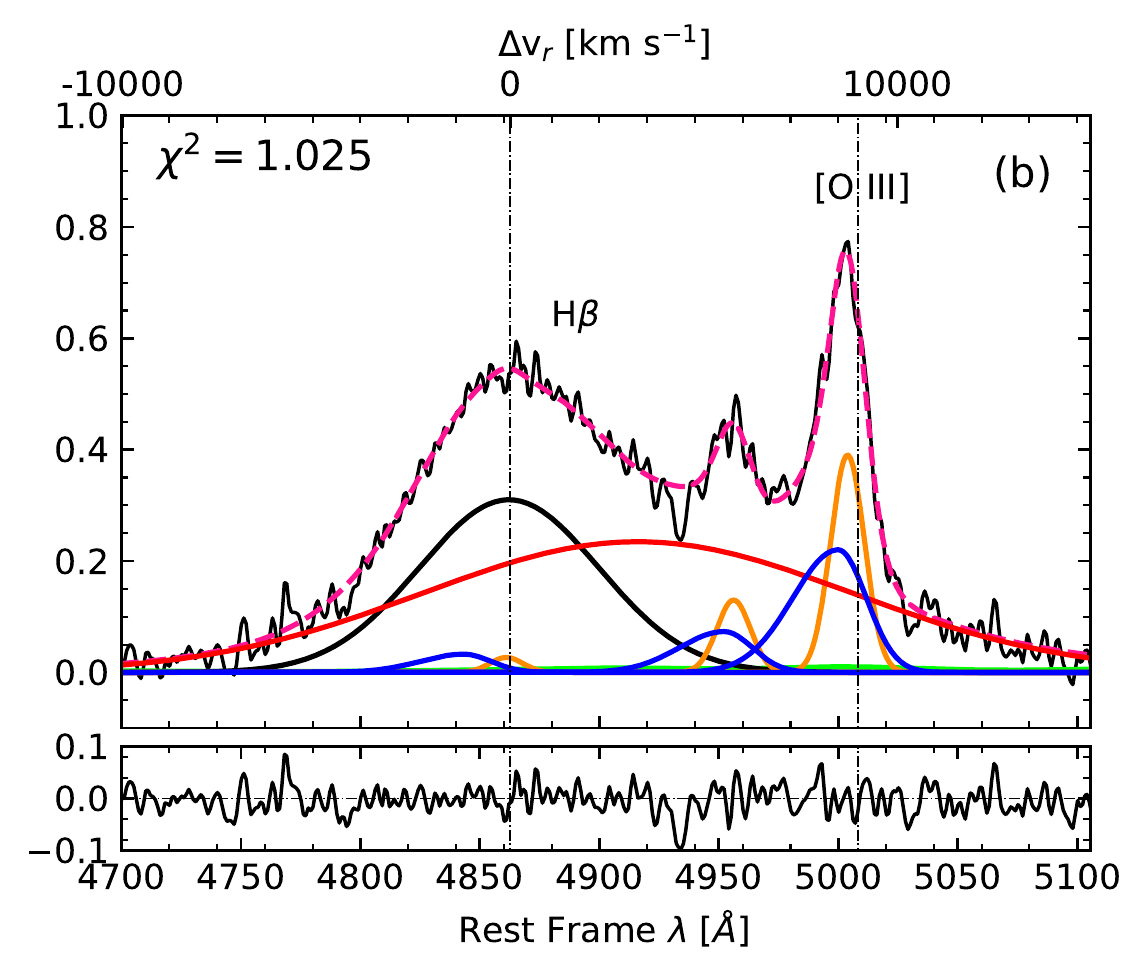}
    \\
    \centering
    \includegraphics[width=\linewidth]{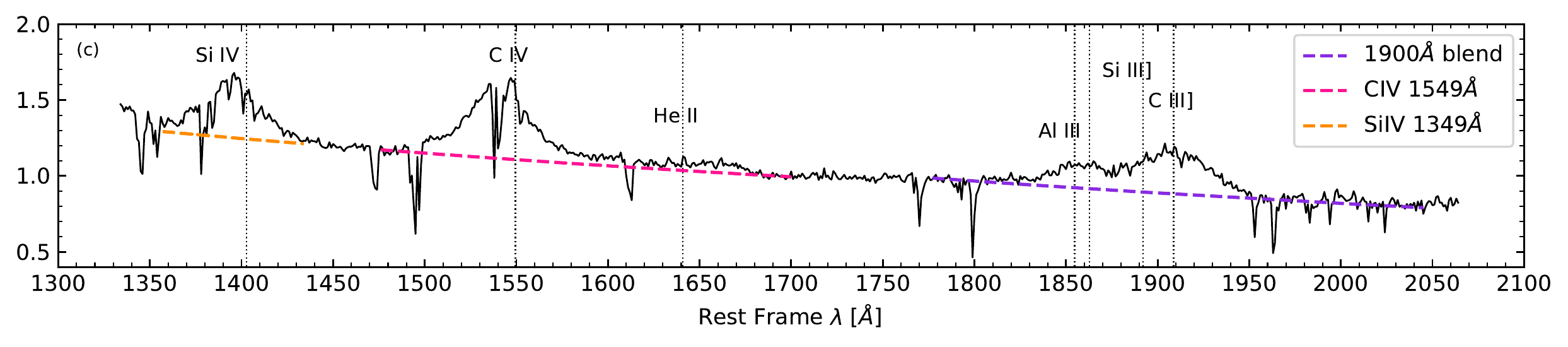}
    \includegraphics[width=0.32\linewidth]{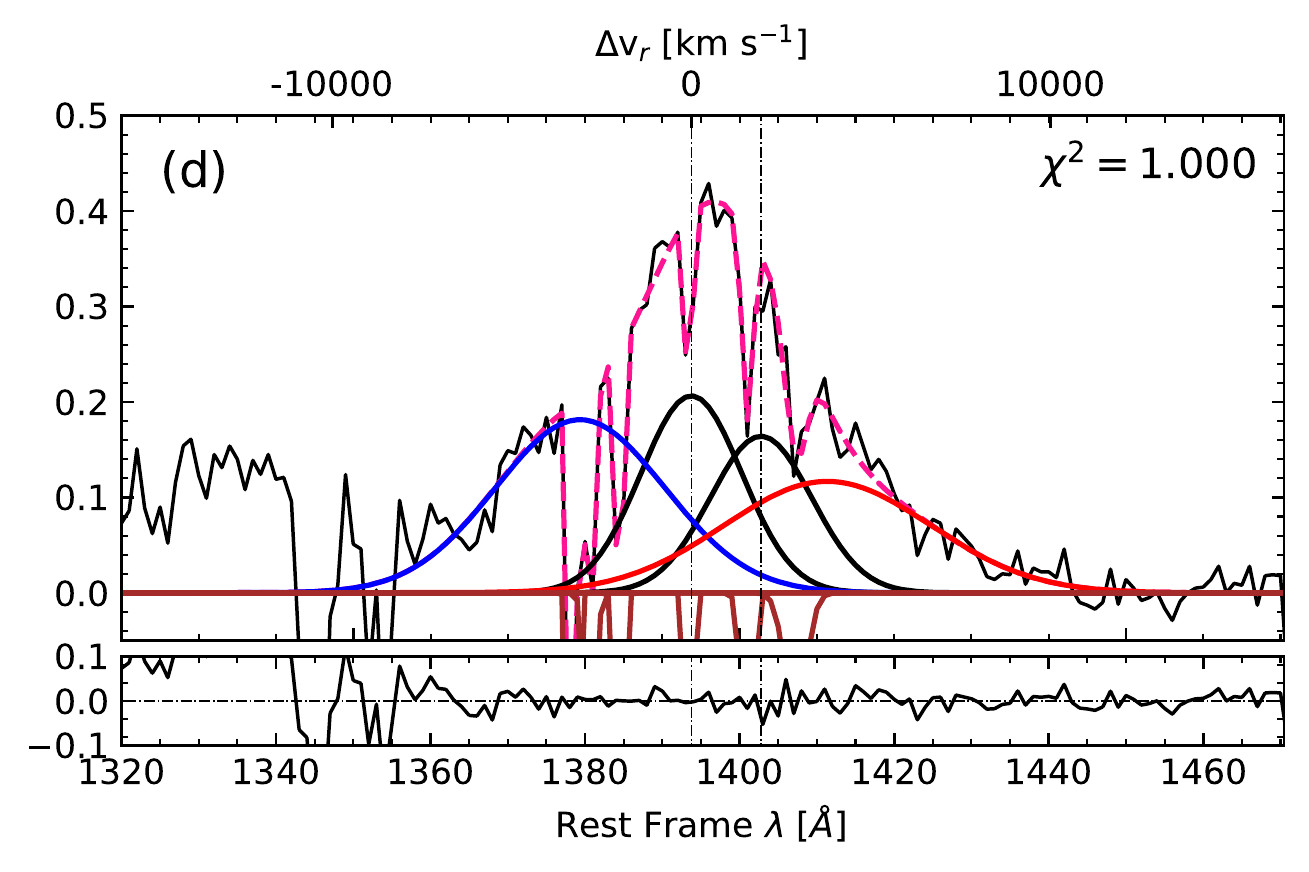}
    \includegraphics[width=0.32\linewidth]{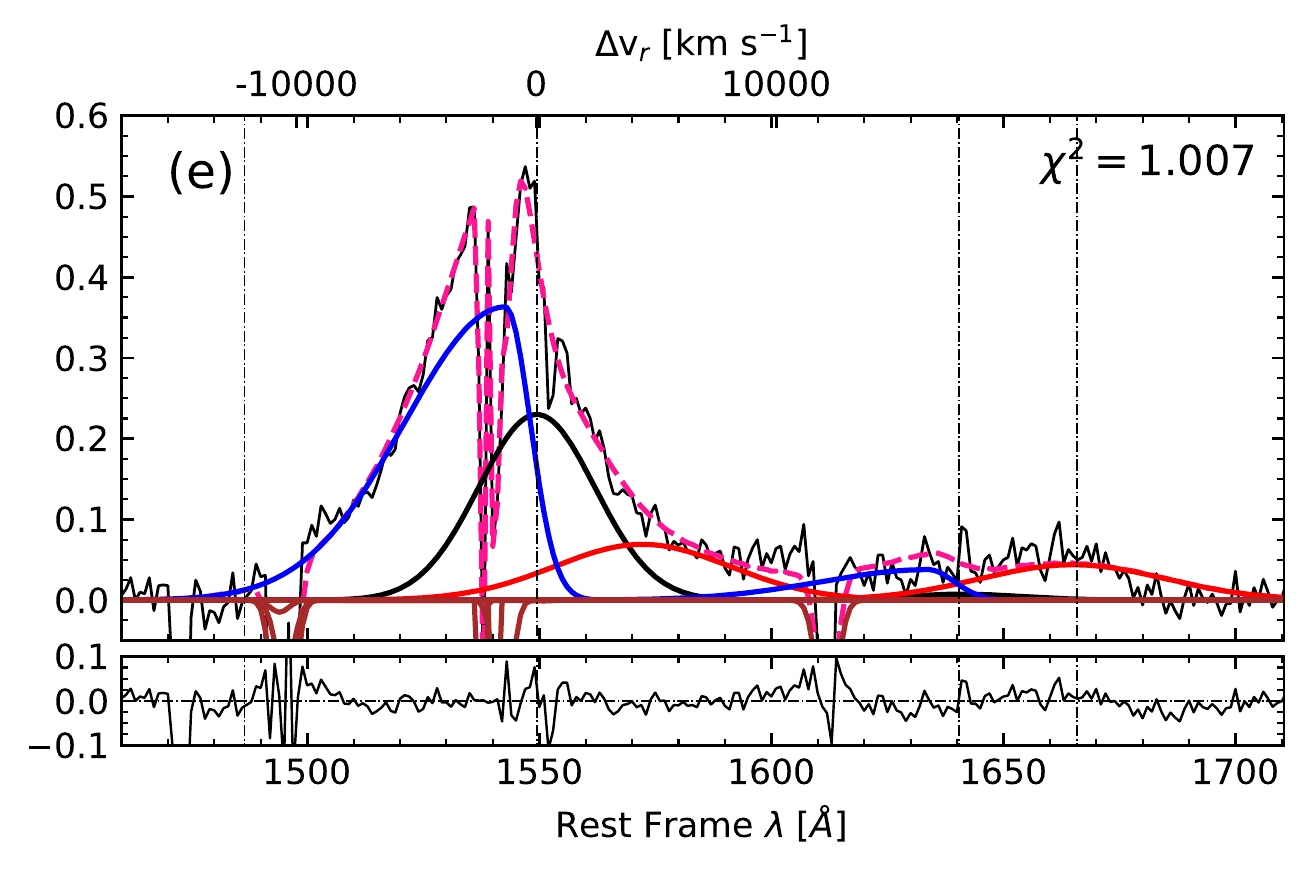}
    \includegraphics[width=0.34\linewidth]{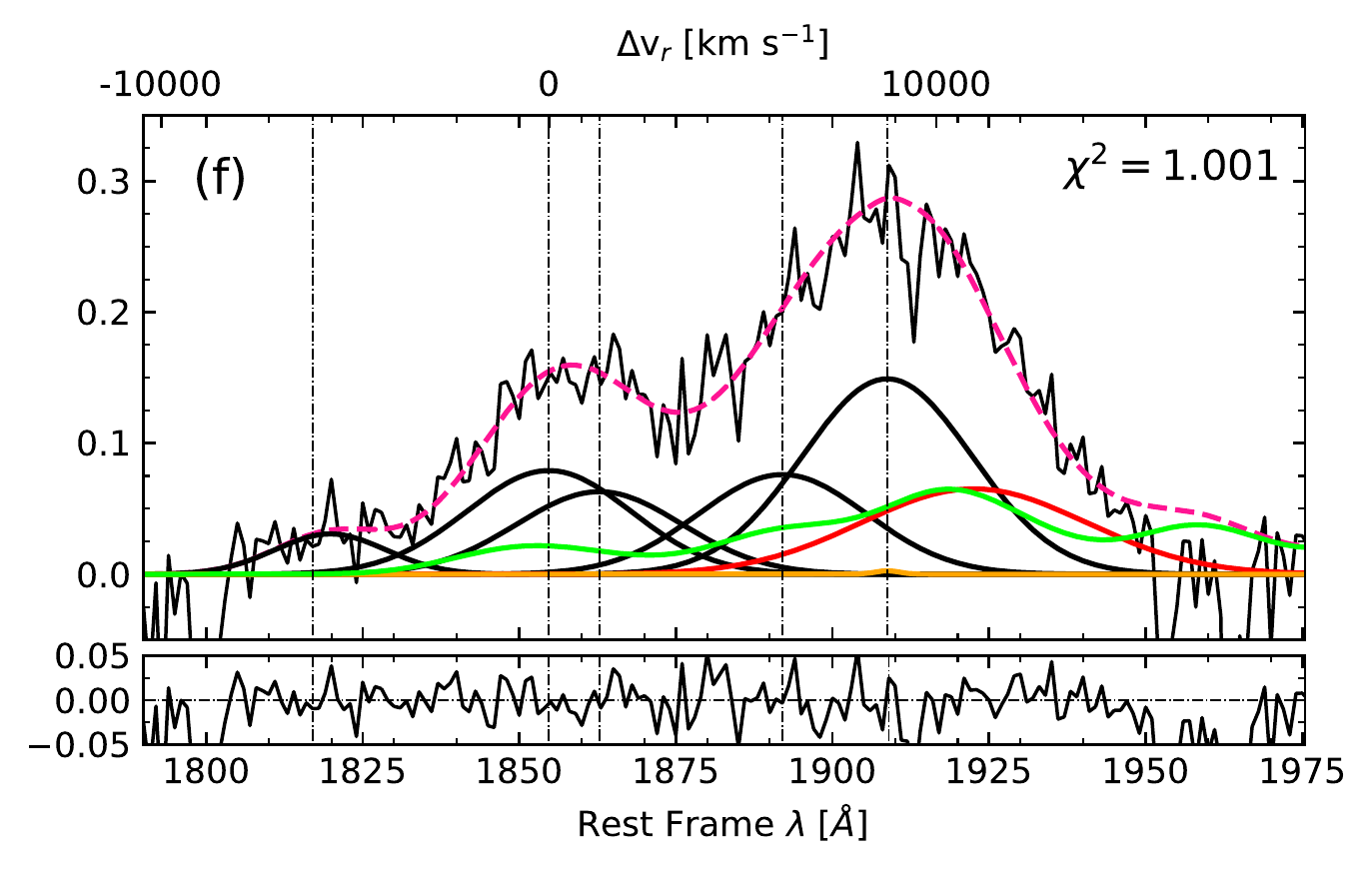}
    \caption{SDSSJ115954.33+201921.1. Same as Figure \ref{fig:1143_UV}.}
    \label{fig:1159_UV}
\end{figure}

\par In the optical range, this source presents a very flat spectrum and almost no \ion{Fe}{II} contributions.

\clearpage

%--------------------------------------------
\subsection{SDSSJ120147.90+120630.2}
\label{SDSSJ1201} 

\begin{figure}[h!]
    \centering
    \includegraphics[width=0.68\linewidth]{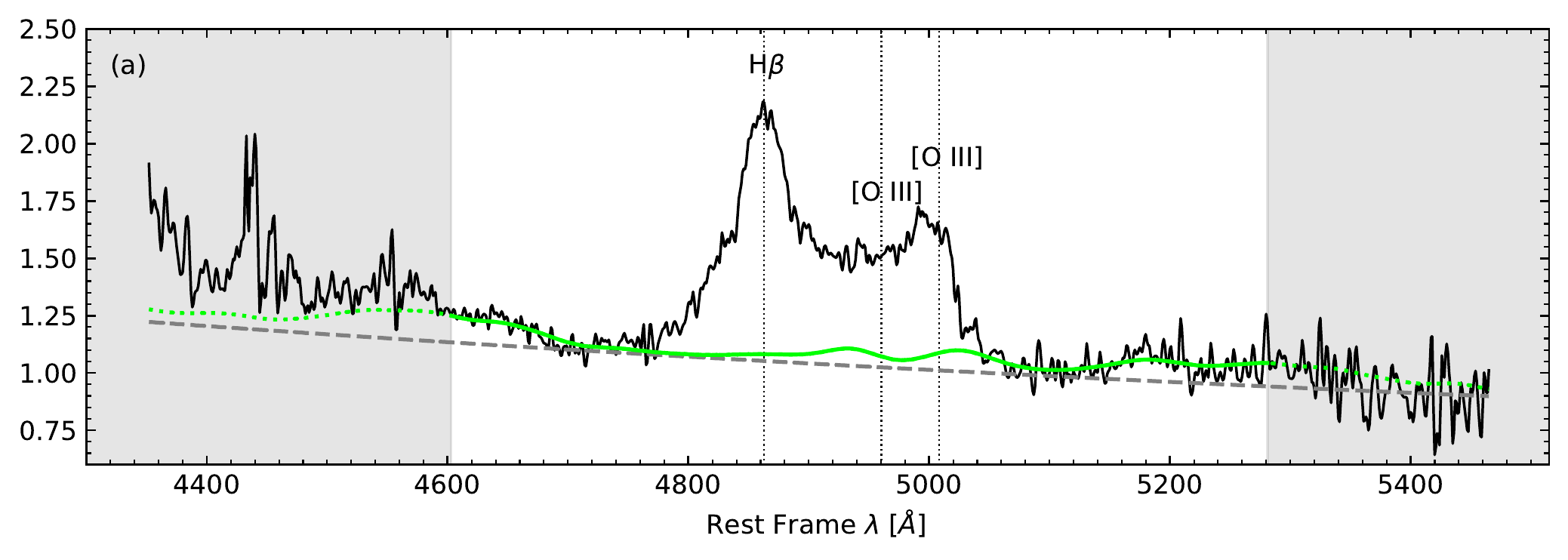}
    \includegraphics[width=0.31\linewidth]{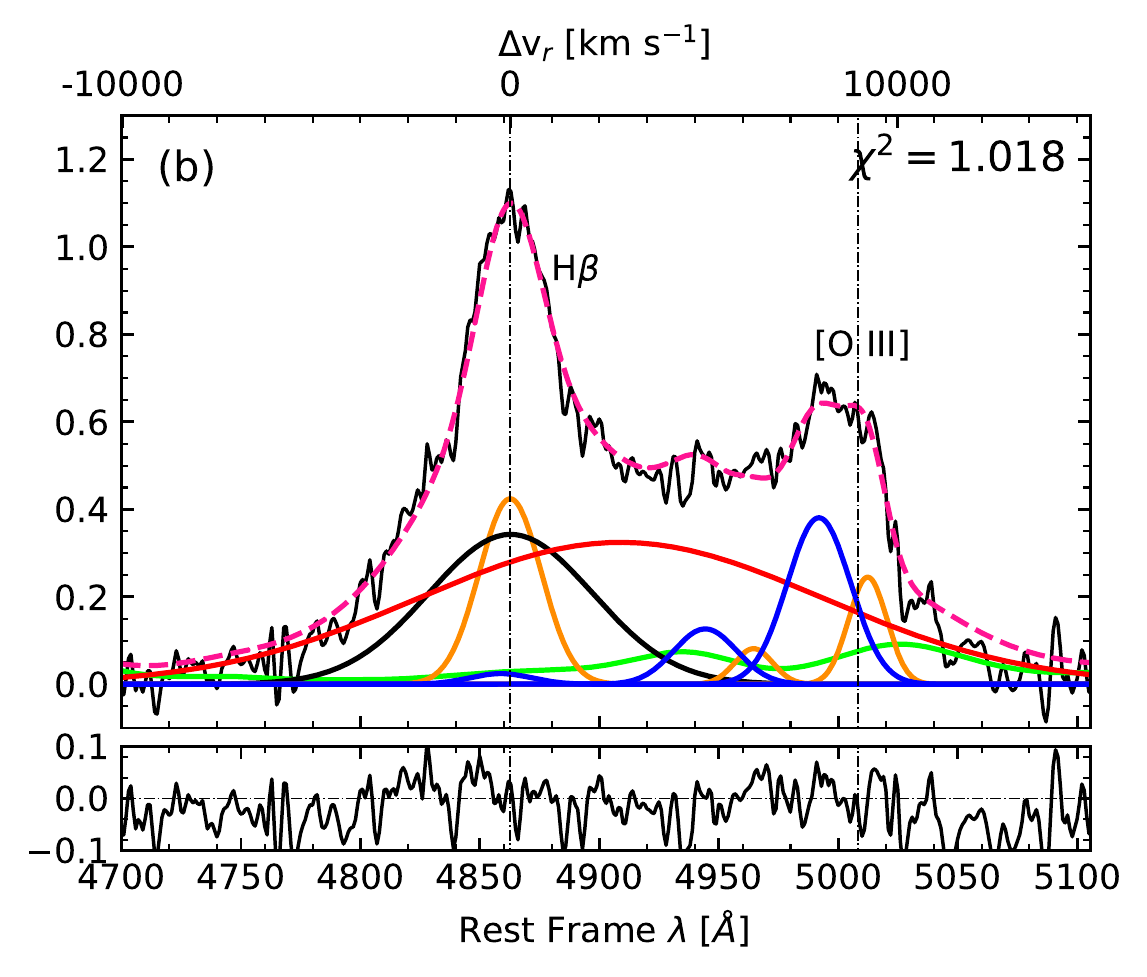}
    \\
    \centering
    \includegraphics[width=\linewidth]{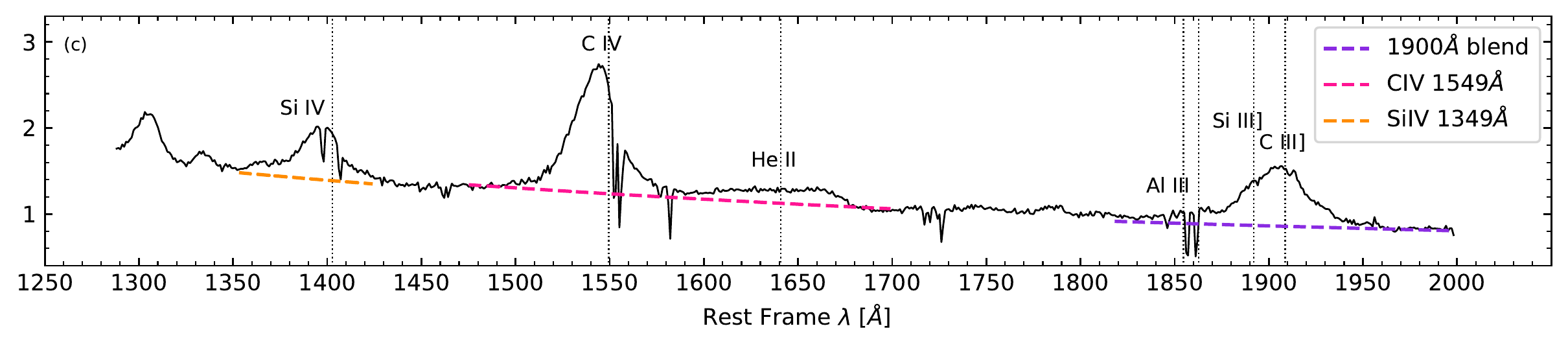}
    \includegraphics[width=0.325\linewidth]{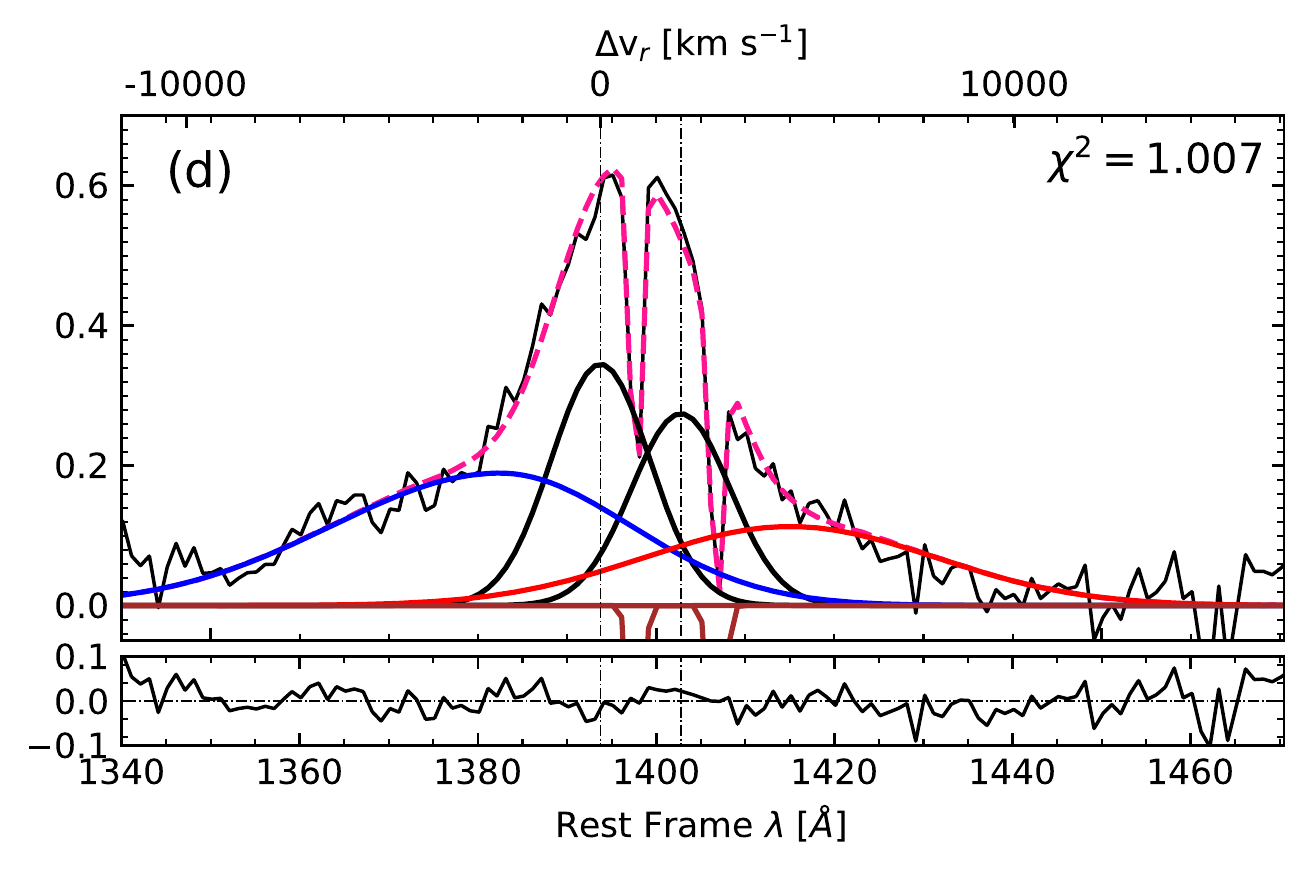}
    \includegraphics[width=0.325\linewidth]{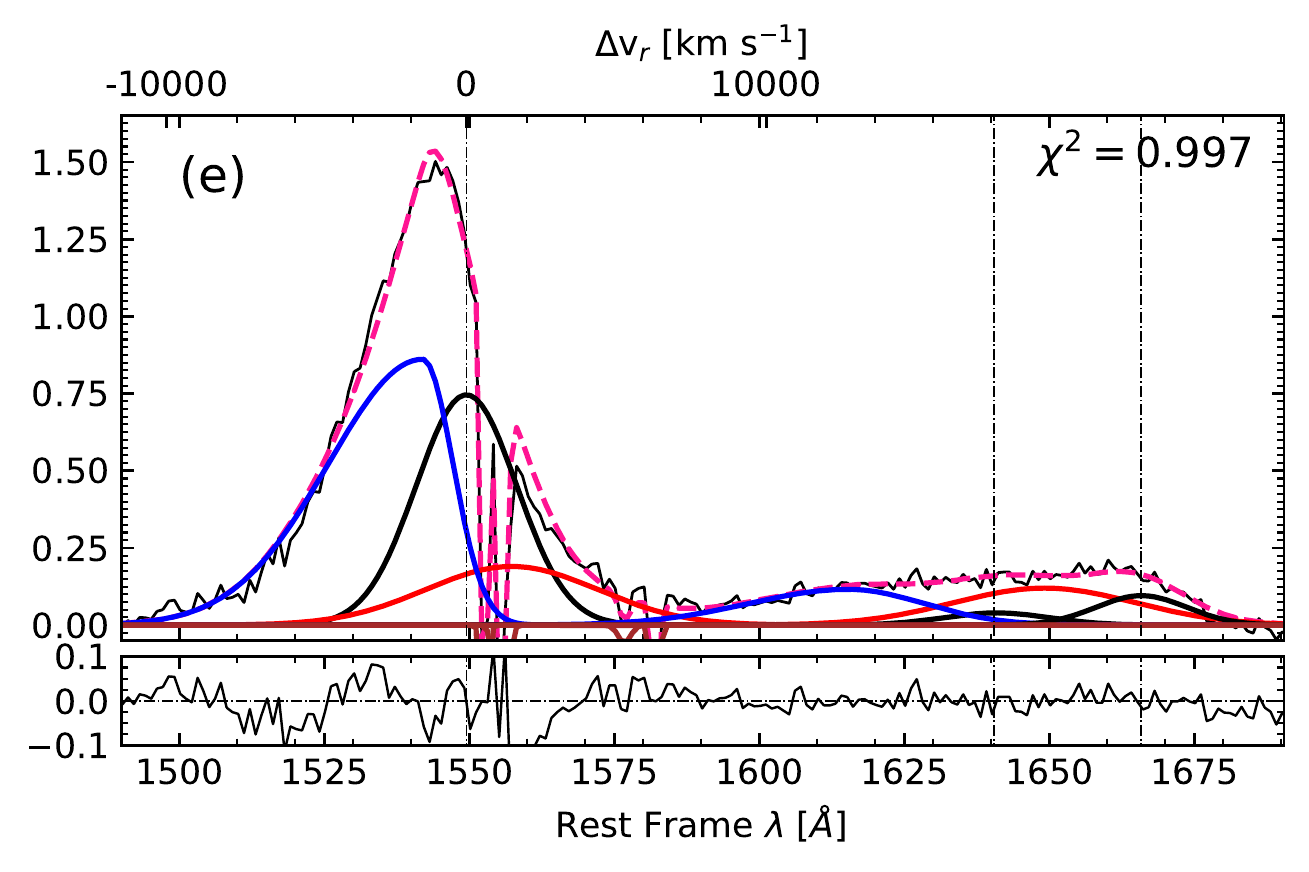}
    \includegraphics[width=0.34\linewidth]{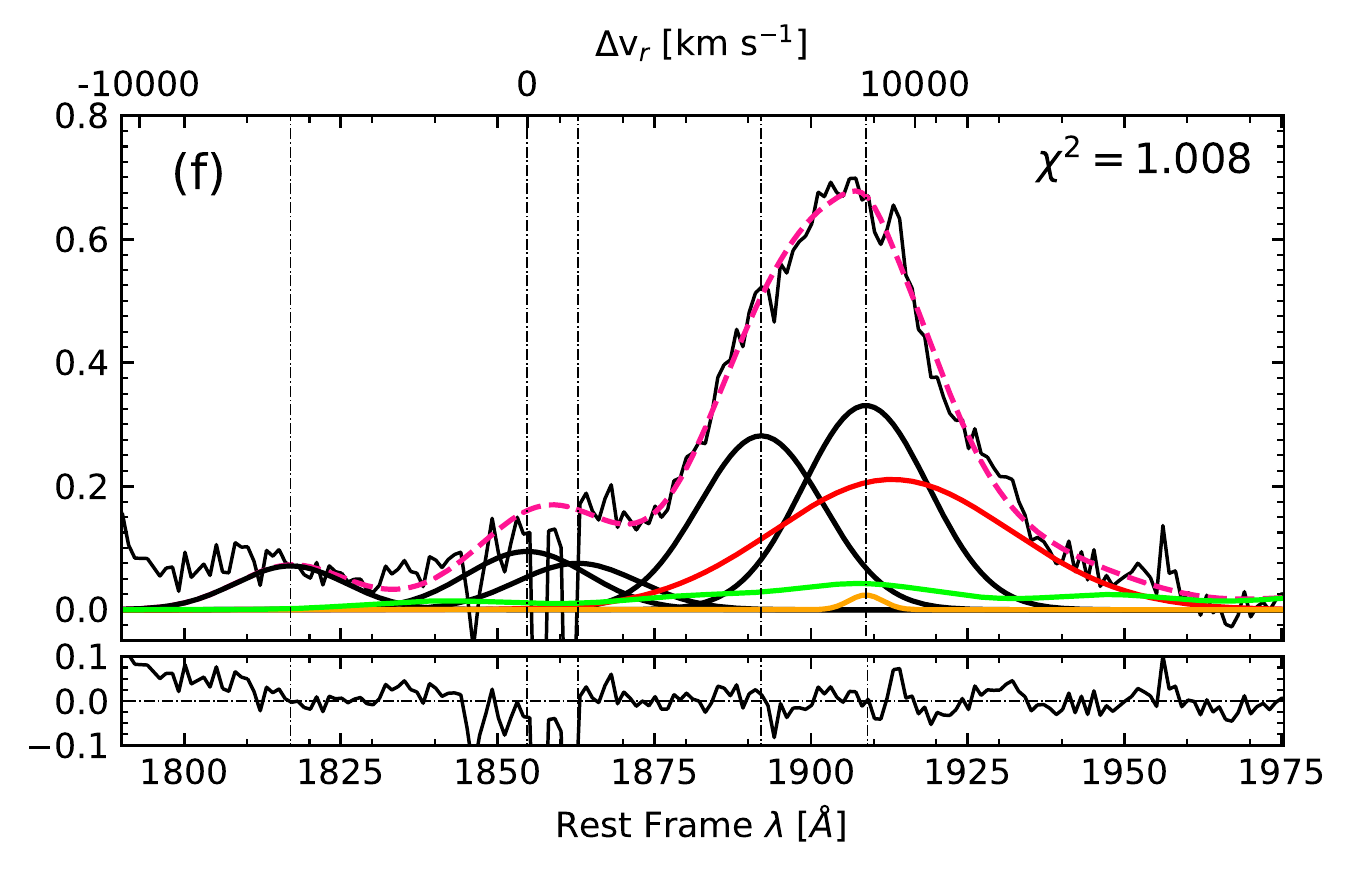}
    \caption{SDSSJ120147.90+120630.2. Same as Figure \ref{fig:1143_UV}.}
    \label{fig:1201_UV}
\end{figure}

\par This quasar has observations of the H$\beta$ region at the Large Binocular Telescope (LBT), as reported in \citet{bischettietal17}. We stress that their spectrum is very similar to the one of ISAAC/VLT.
\clearpage

%--------------------------------------------
\subsection{SDSSJ132012.33+142037.1}
\label{SDSSJ1320}

\begin{figure}[h!]
    \centering
    \includegraphics[width=0.685\linewidth]{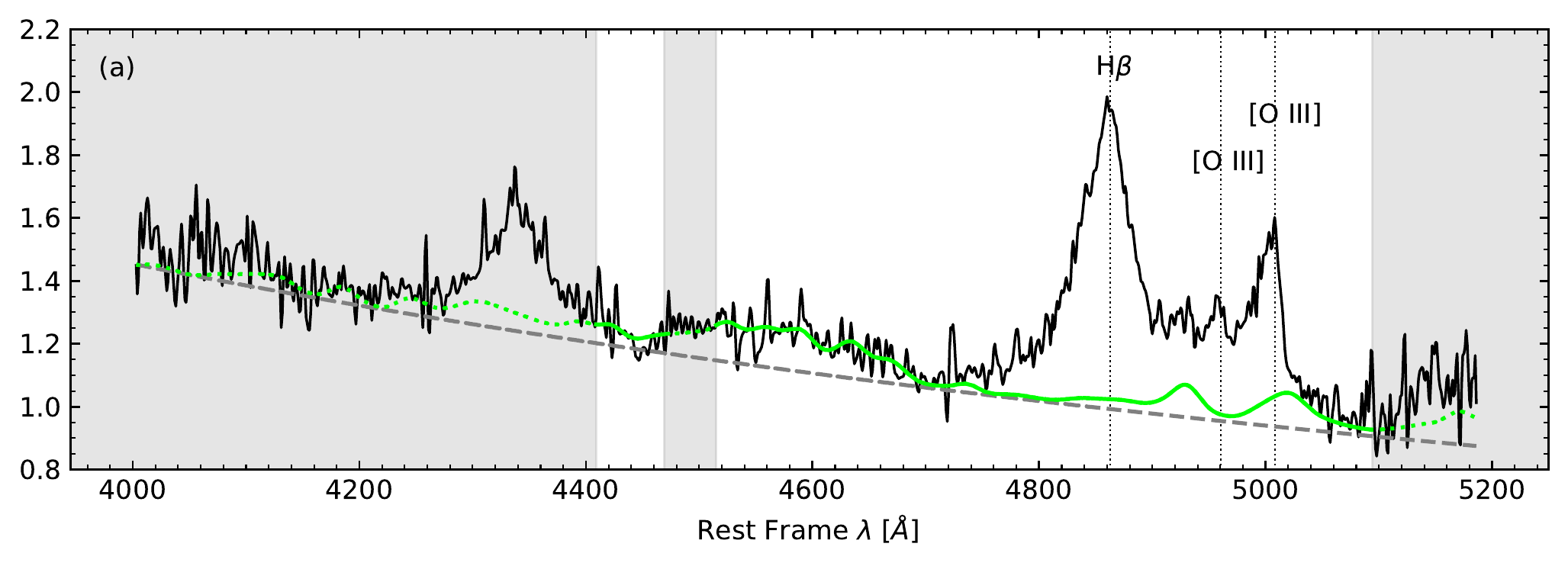}
    \includegraphics[width=0.31\linewidth]{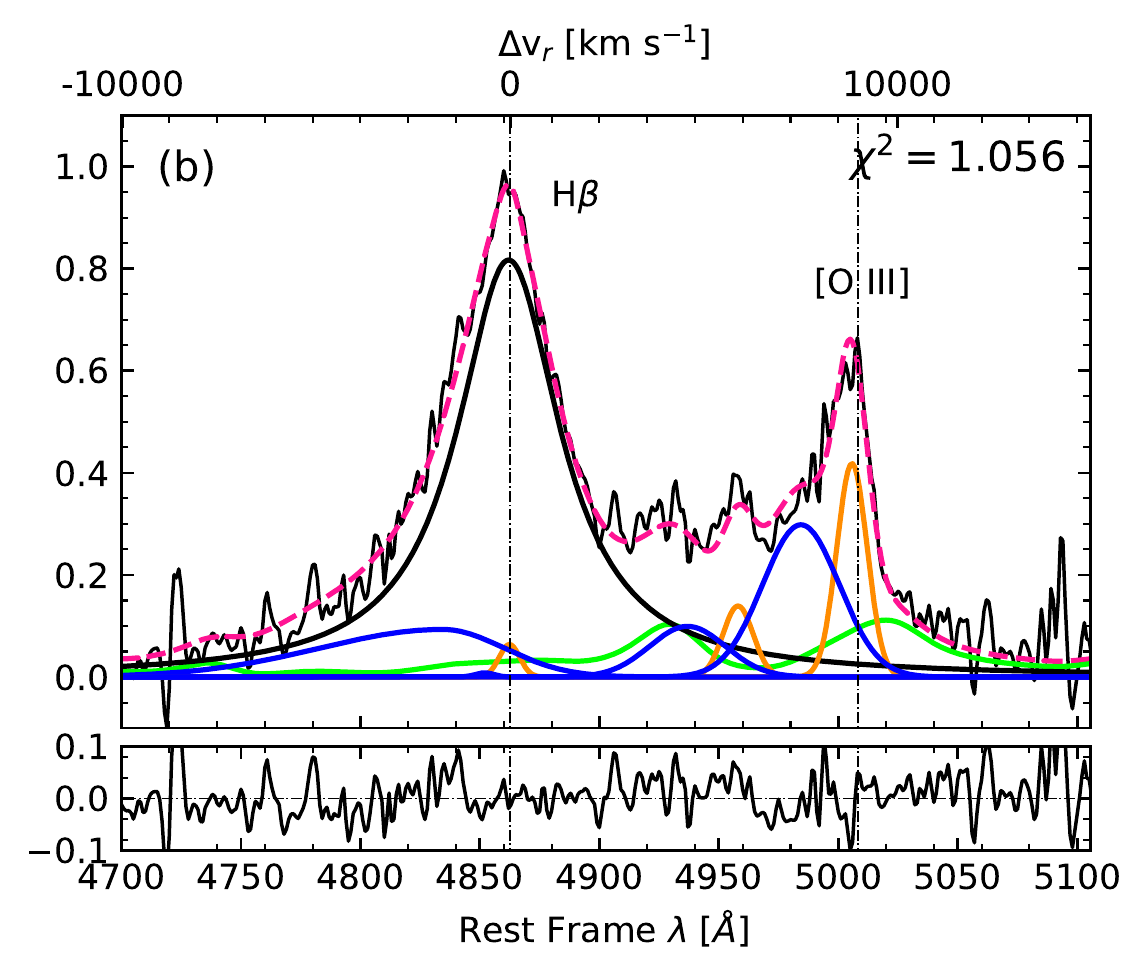}
     \\
    \centering
    \includegraphics[width=\linewidth]{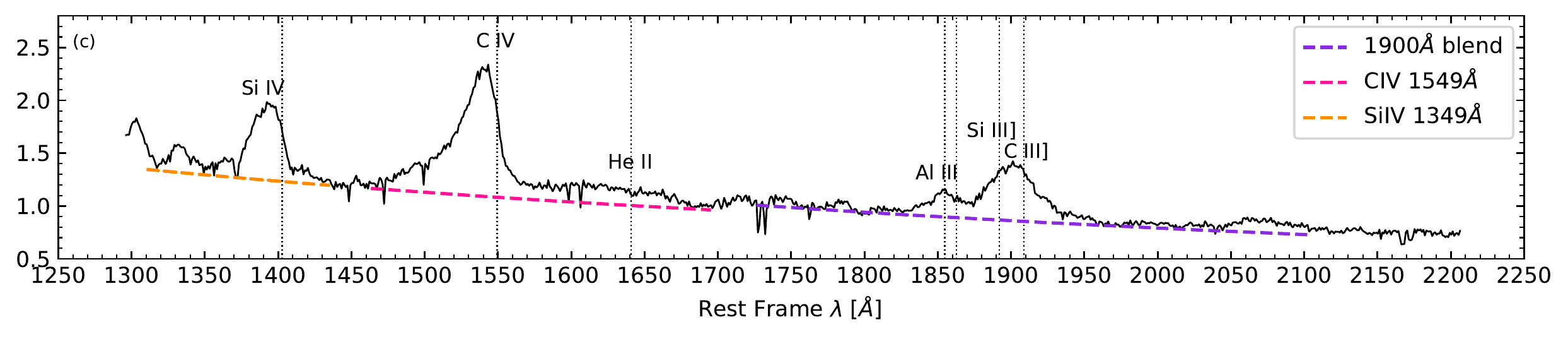}
    \includegraphics[width=0.33\linewidth]{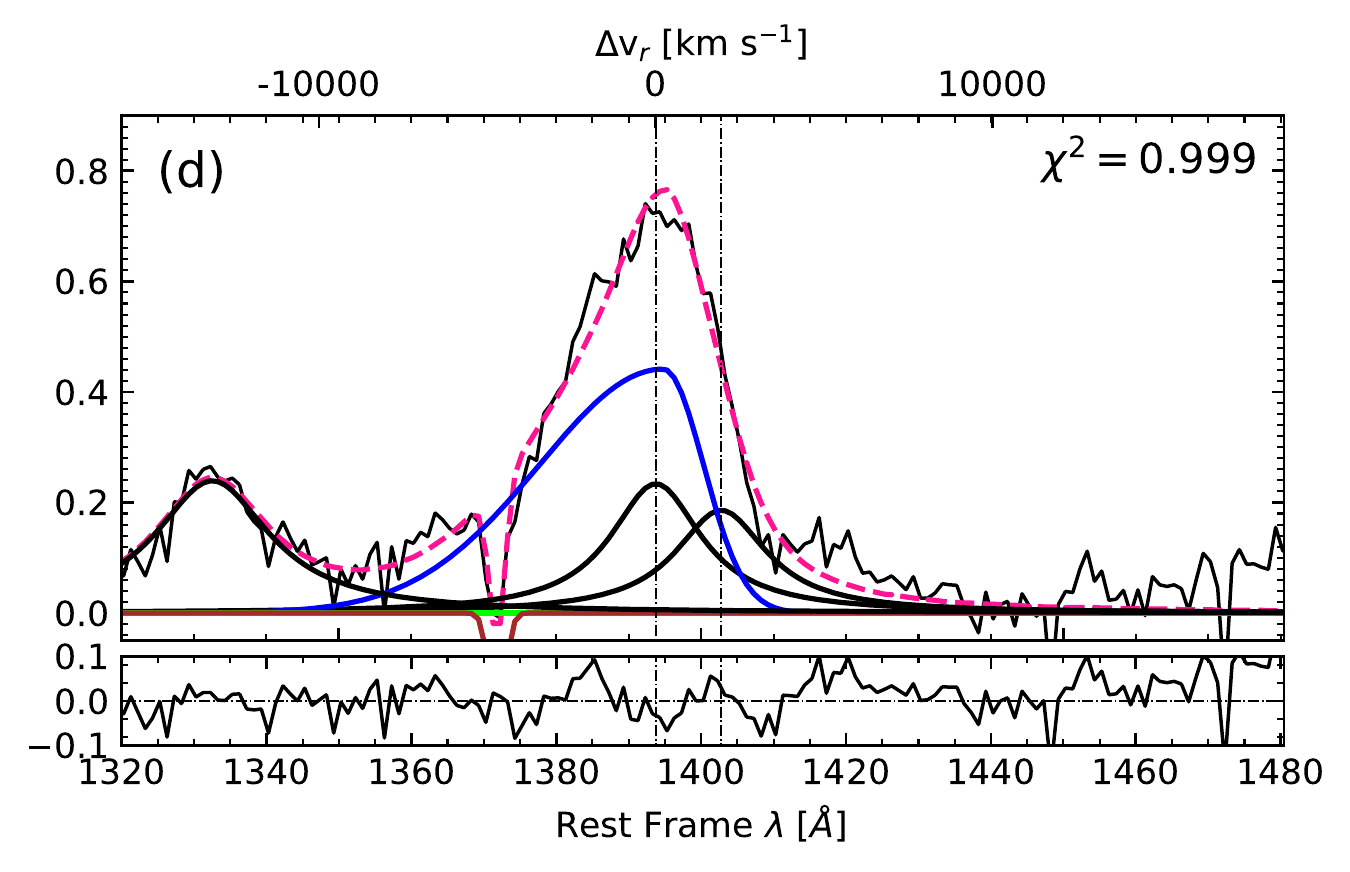}
    \includegraphics[width=0.325\linewidth]{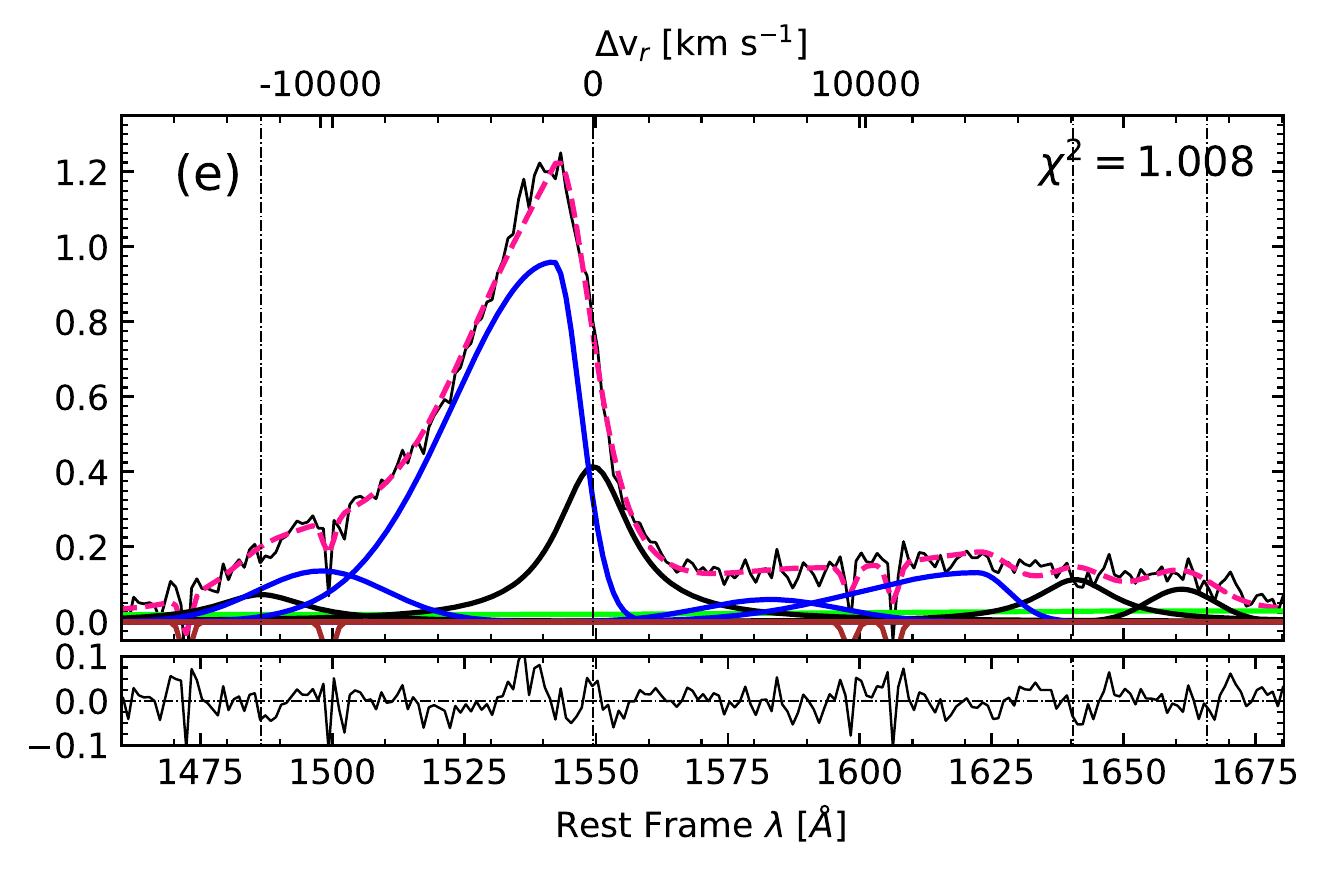}
    \includegraphics[width=0.32\linewidth]{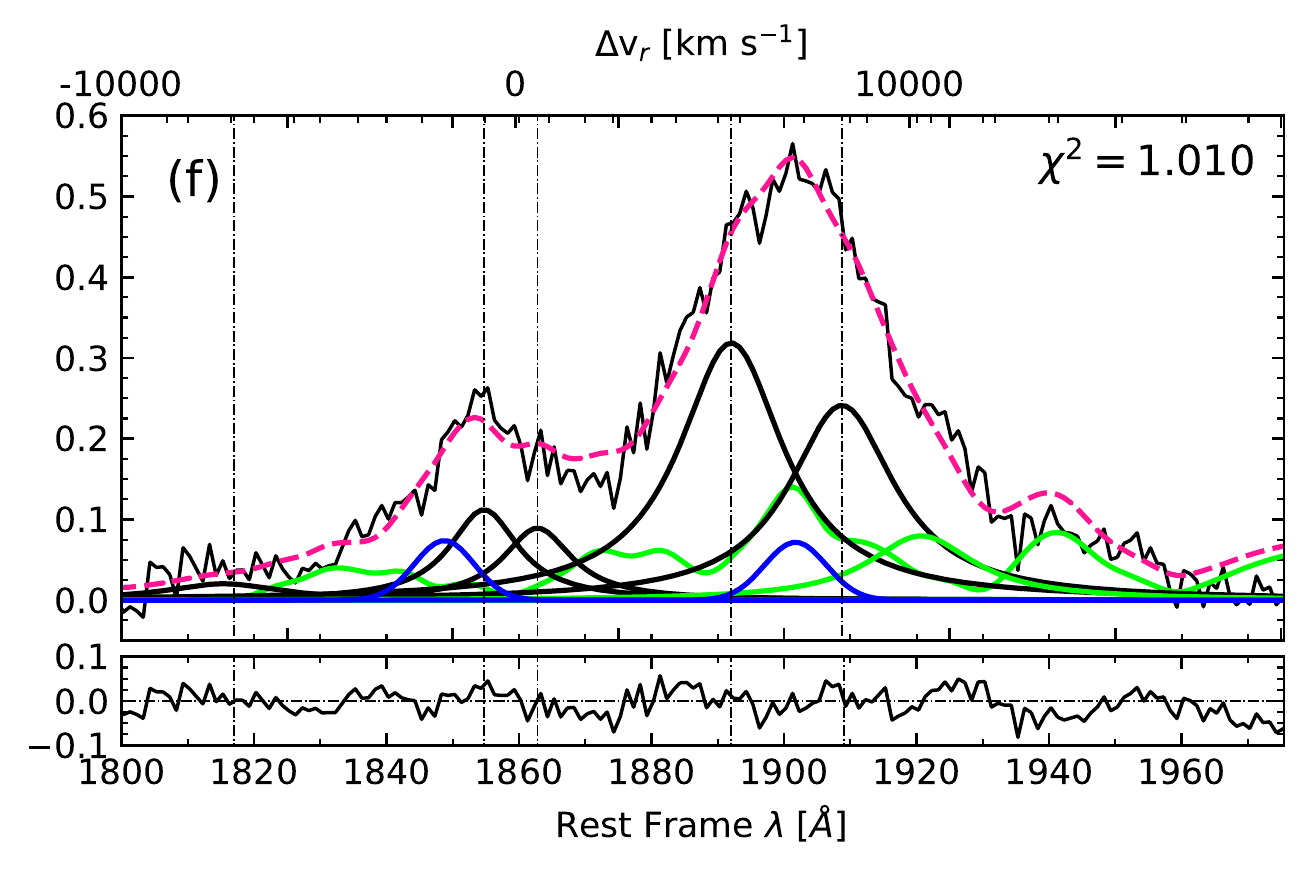}
    \caption{SDSSJ132012.33+142037.1. Same as Figure \ref{fig:1143_UV}.}
    \label{fig:1320_UV}
\end{figure}

\clearpage
%--------------------------------------------
\subsection{SDSSJ135831.78+050522.8}
\label{SDSSJ1358}

\begin{figure}[h!]
    \centering
    \includegraphics[width=0.68\linewidth]{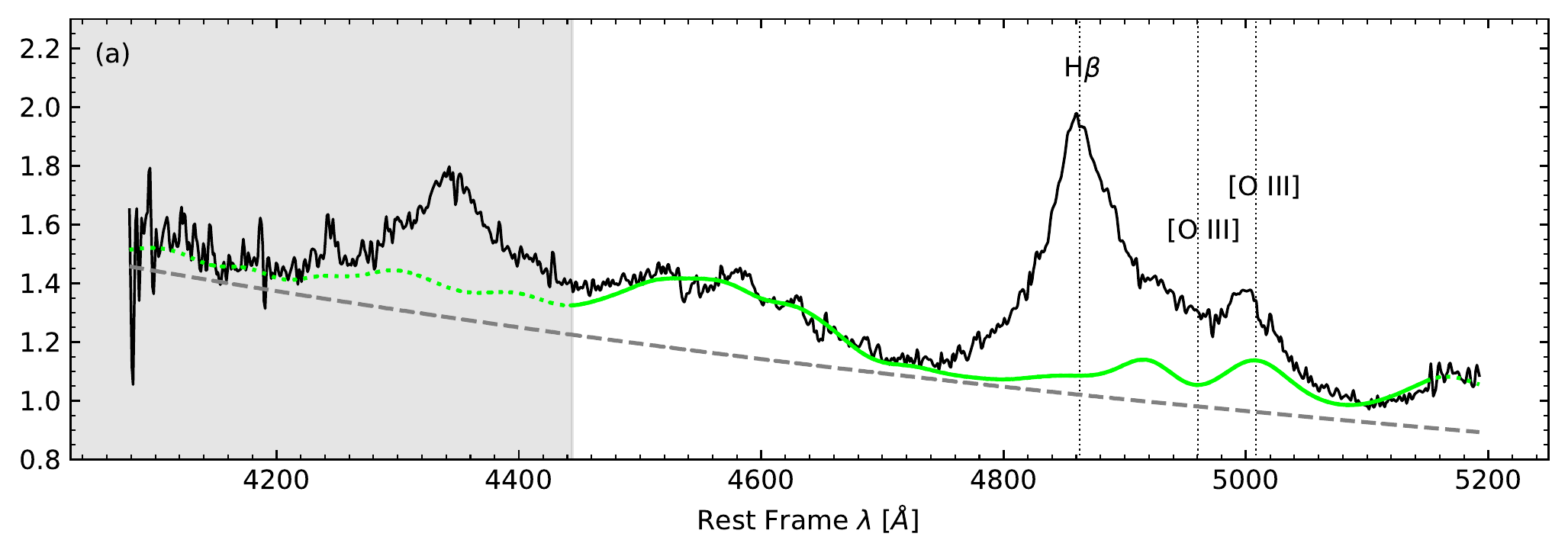}
    \includegraphics[width=0.315\linewidth]{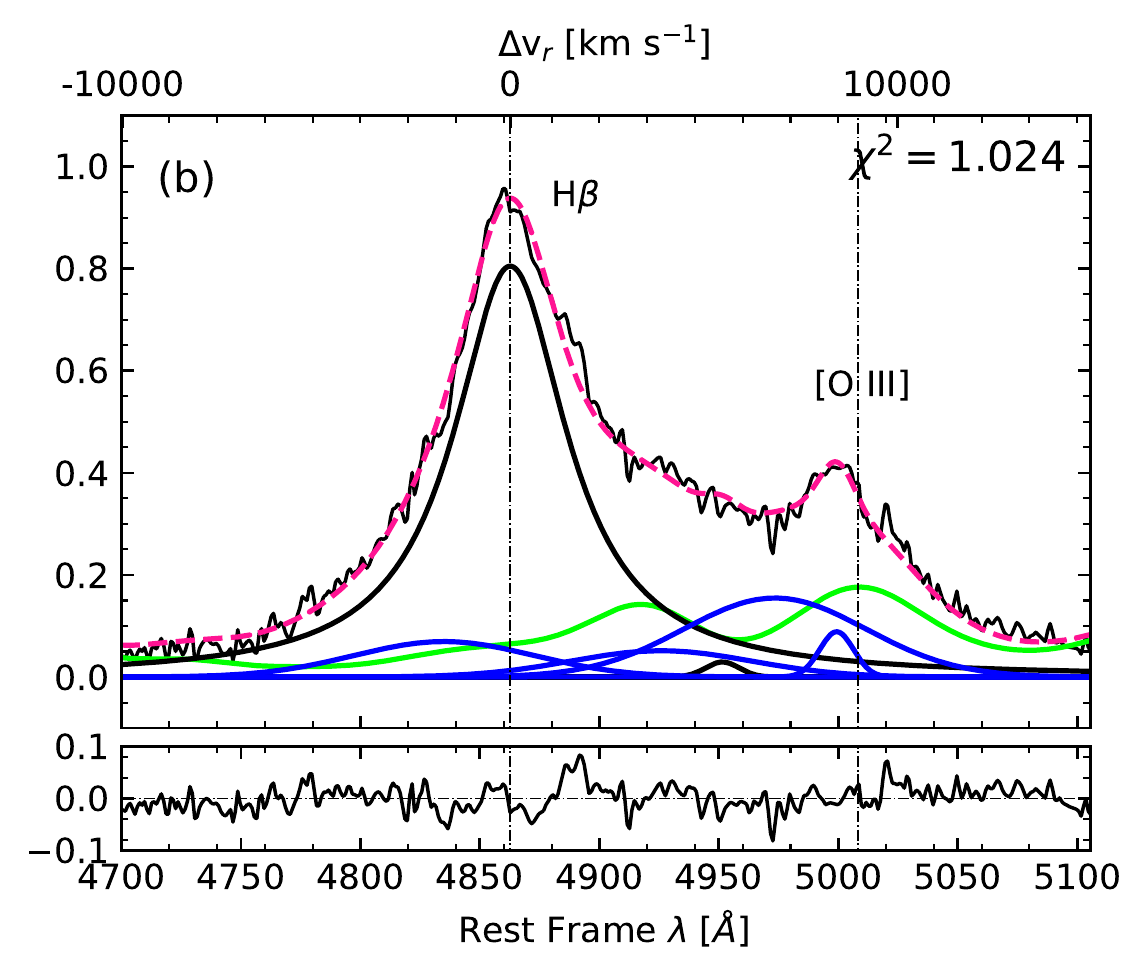}
     \\
    \centering
    \includegraphics[width=\linewidth]{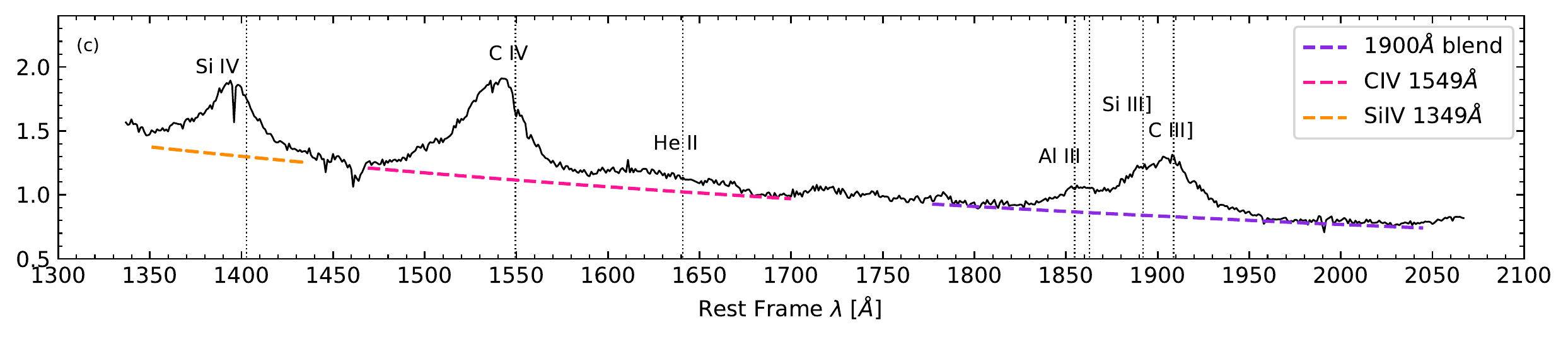}
    \includegraphics[width=0.325\linewidth]{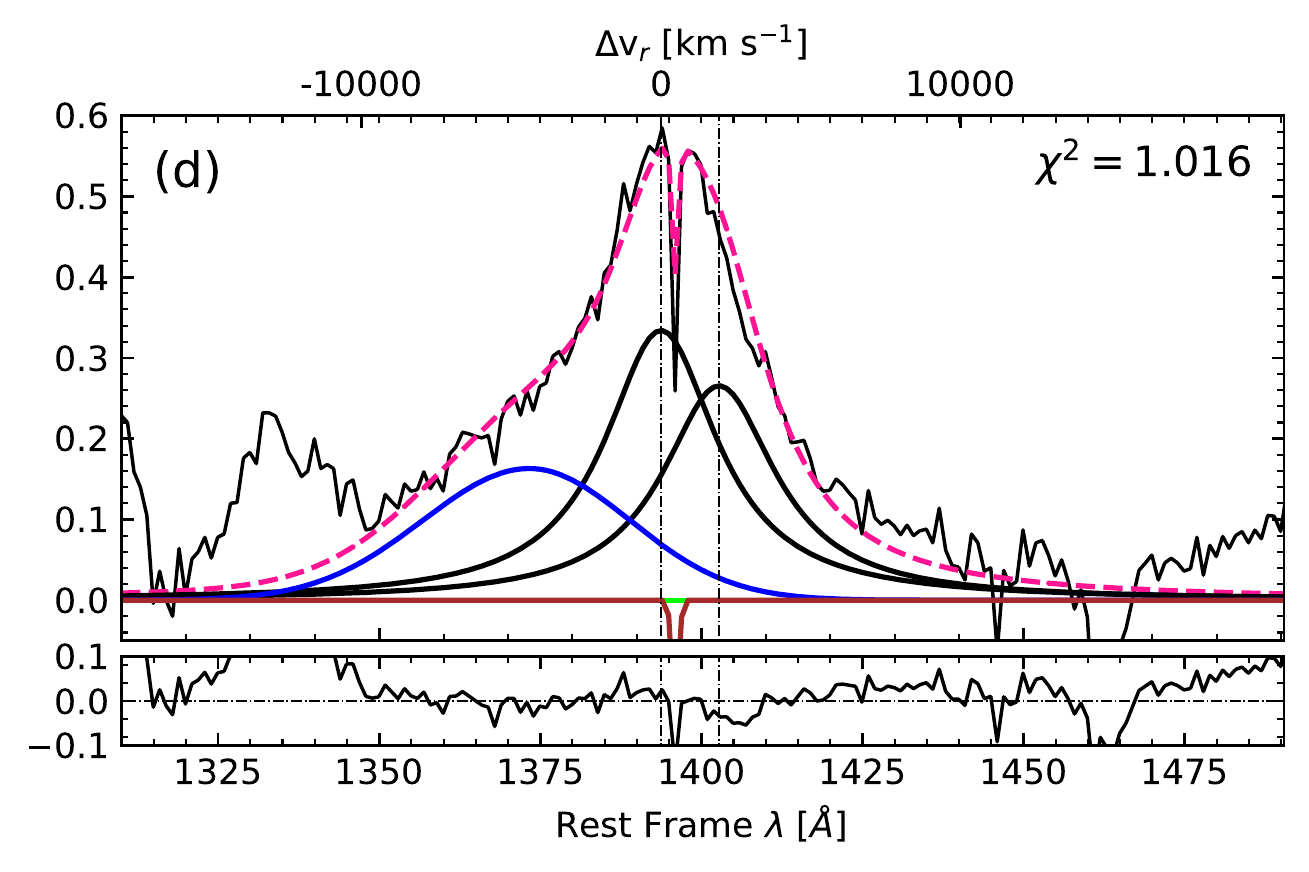}
    \includegraphics[width=0.33\linewidth]{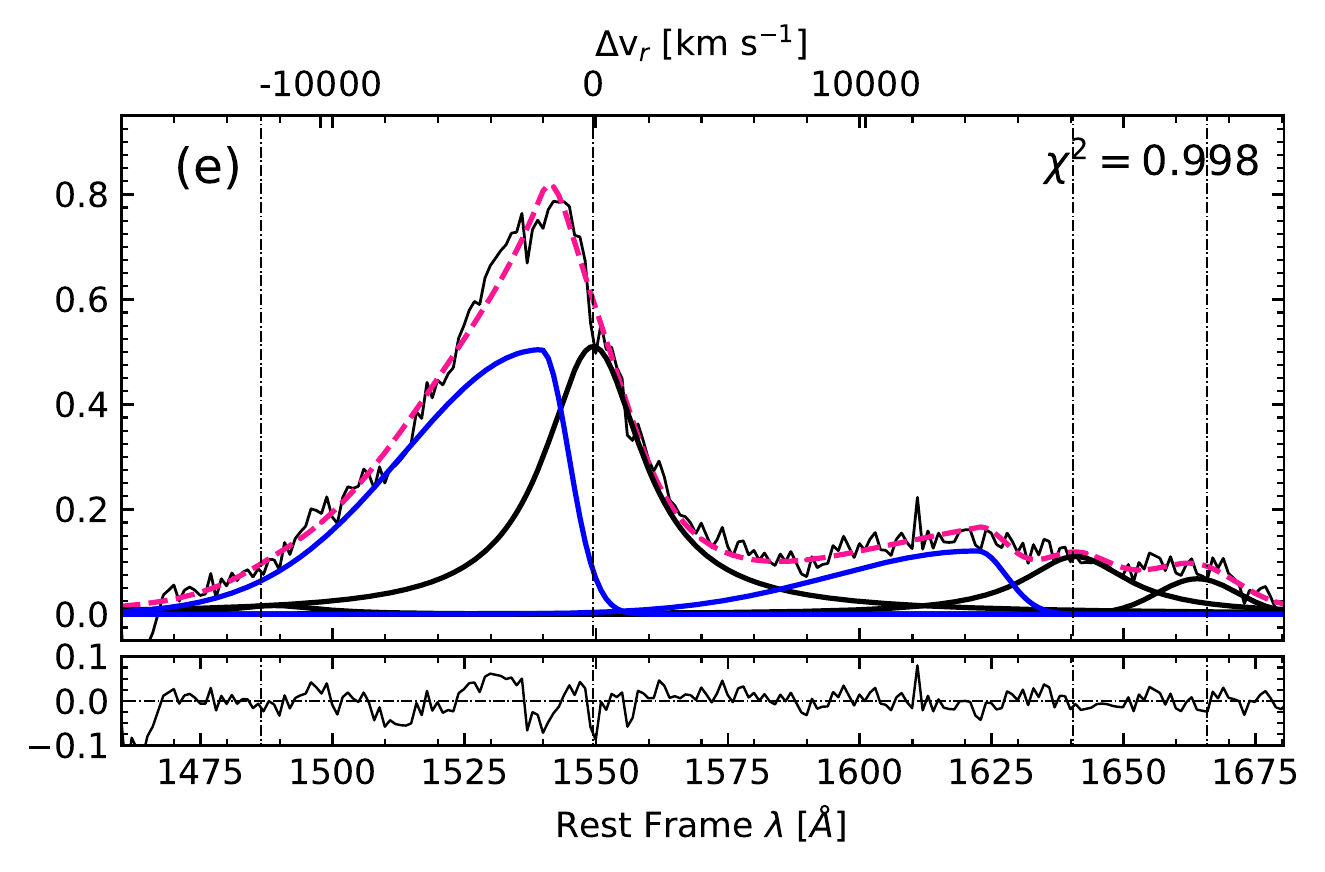}
    \includegraphics[width=0.335\linewidth]{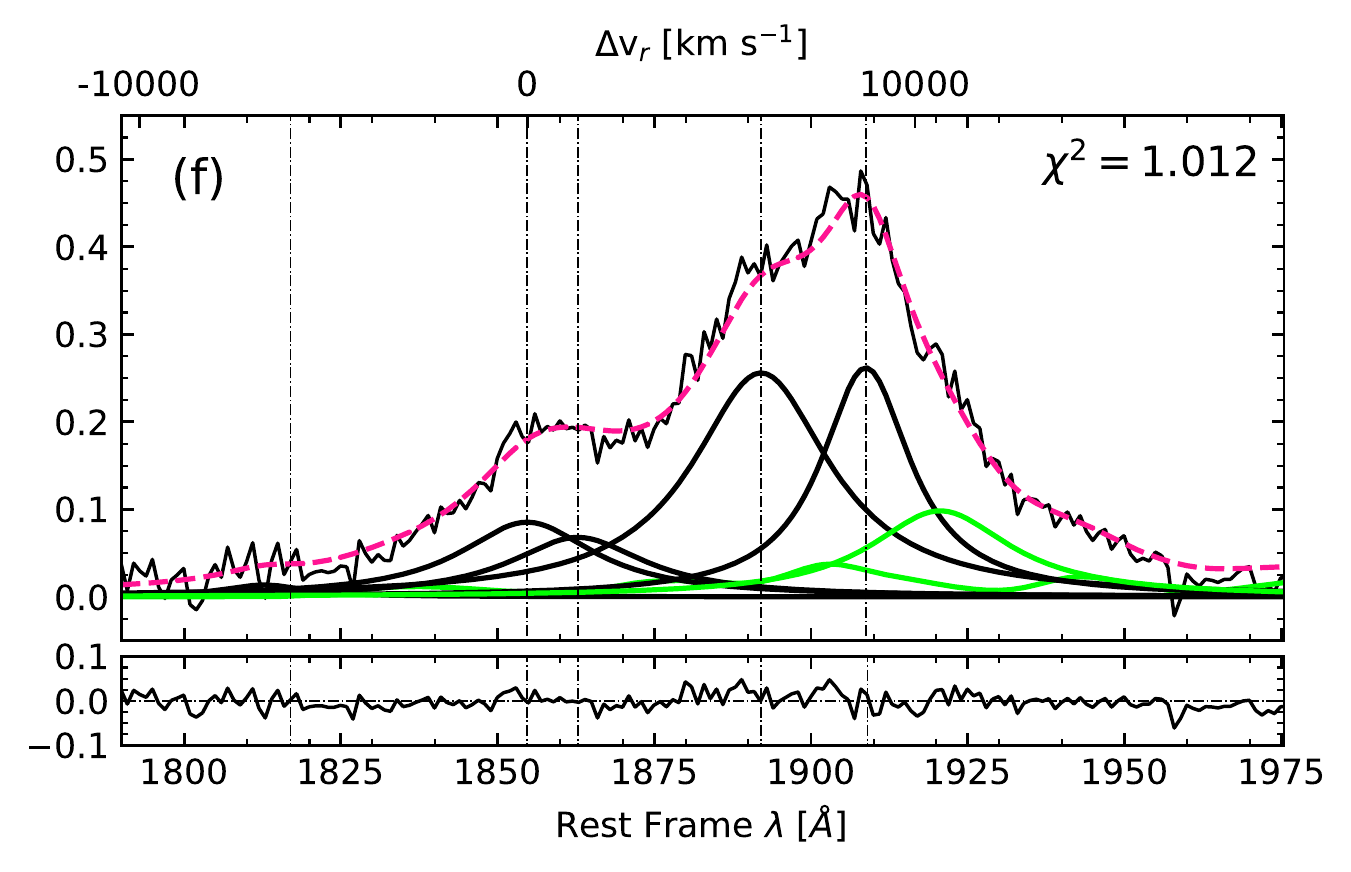}
    \caption{SDSSJ135831.78+050522.8. Same as Figure \ref{fig:1143_UV}.}
    \label{fig:1358_UV}
\end{figure}

\clearpage
%--------------------------------------------
\subsection{Q 1410+096}
\label{Q1410}

\begin{figure}[h!]
    \centering
    \includegraphics[width=0.685\columnwidth]{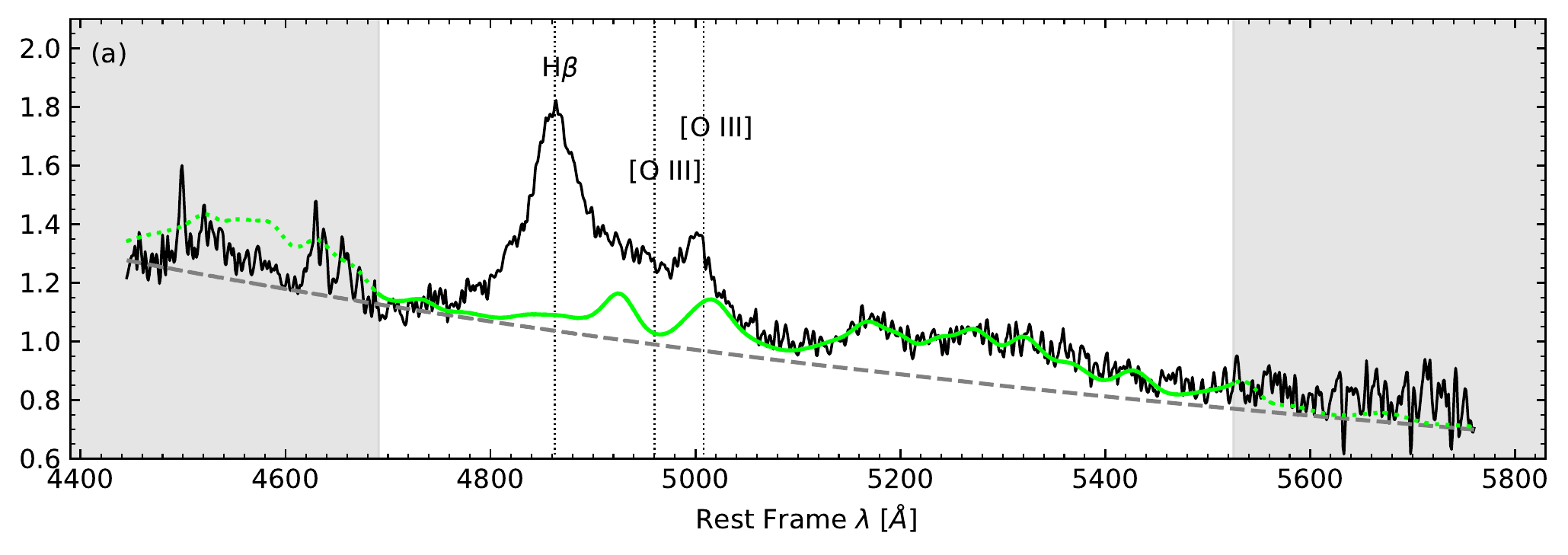}
    \includegraphics[width=0.31\columnwidth]{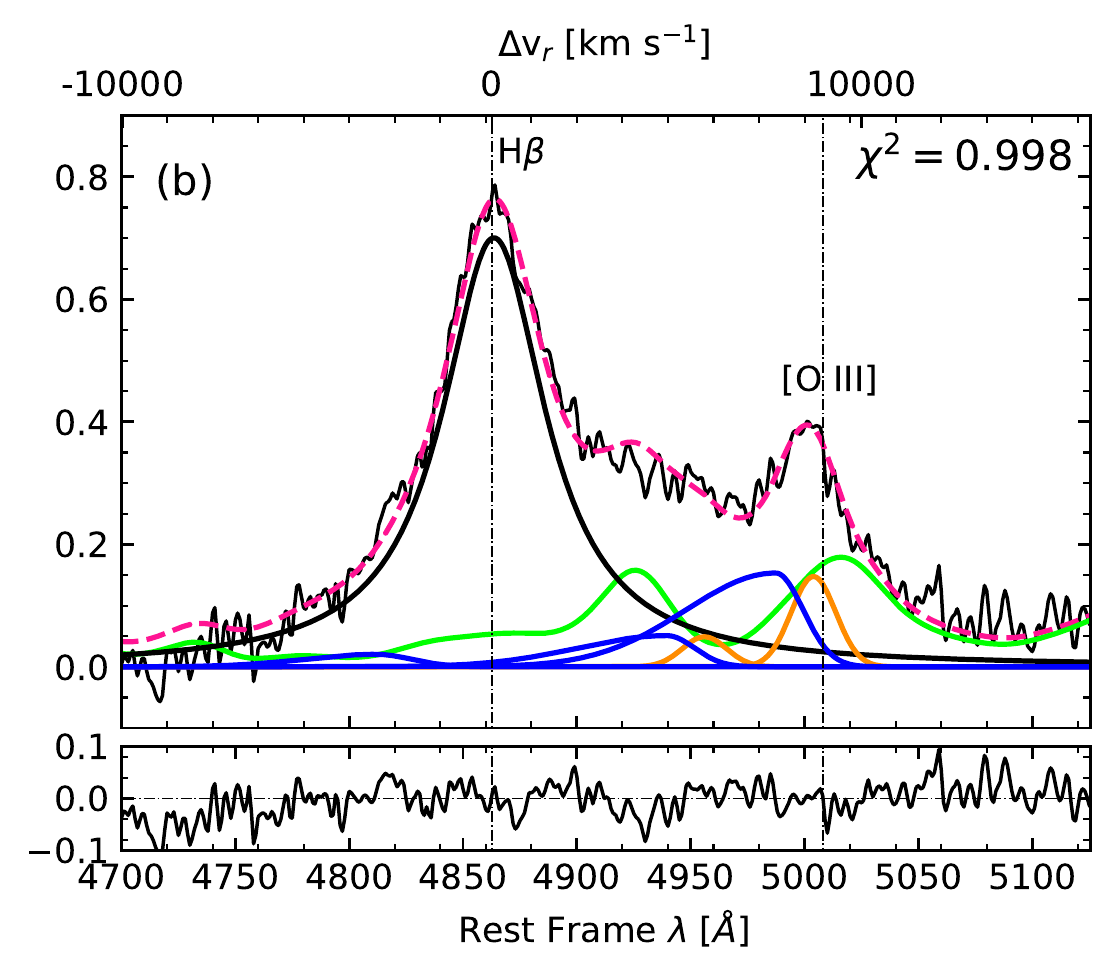}
     \\
    \raggedright
    \includegraphics[width=\linewidth]{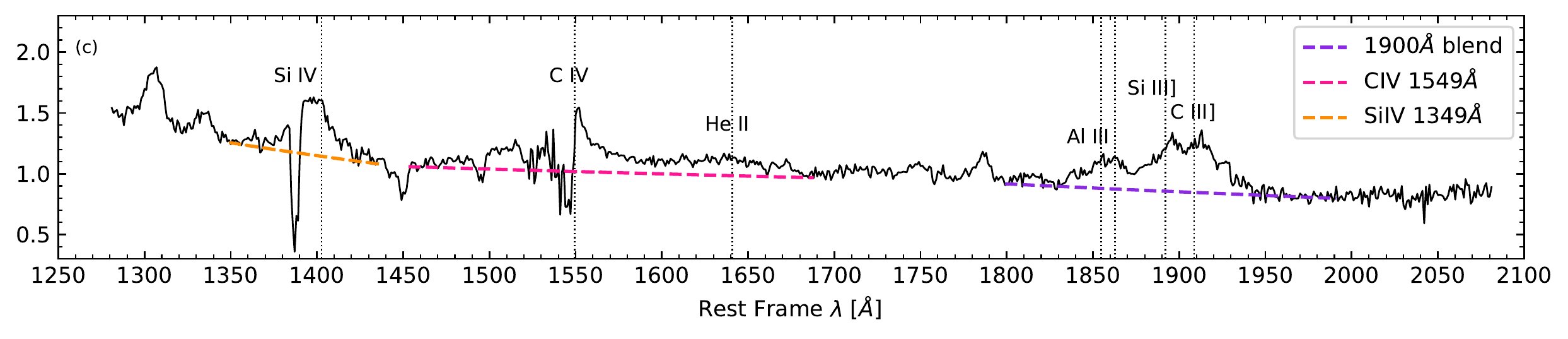}
    \\
    \includegraphics[width=0.325\linewidth]{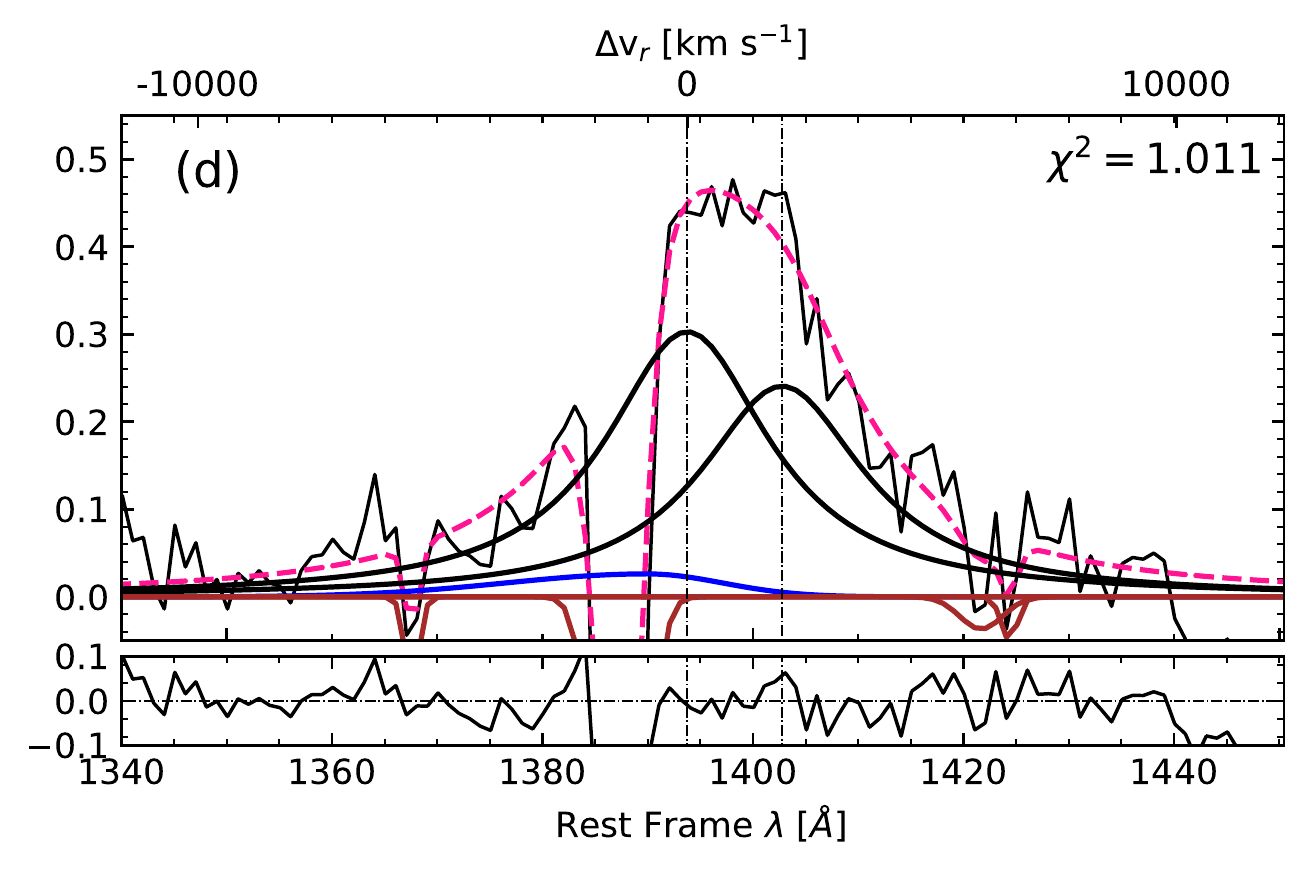}
    \includegraphics[width=0.335\linewidth]{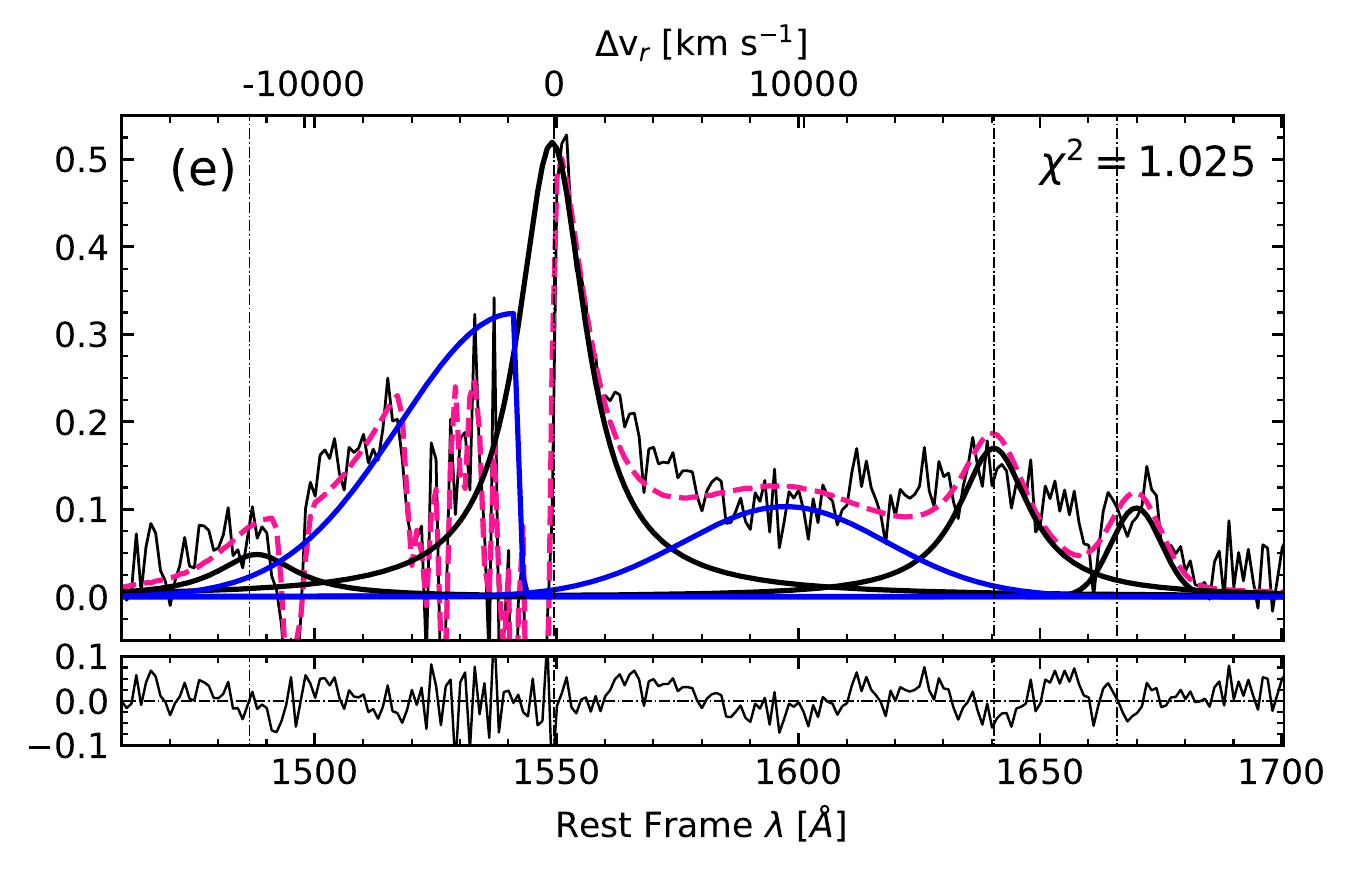}
    \includegraphics[width=0.325\linewidth]{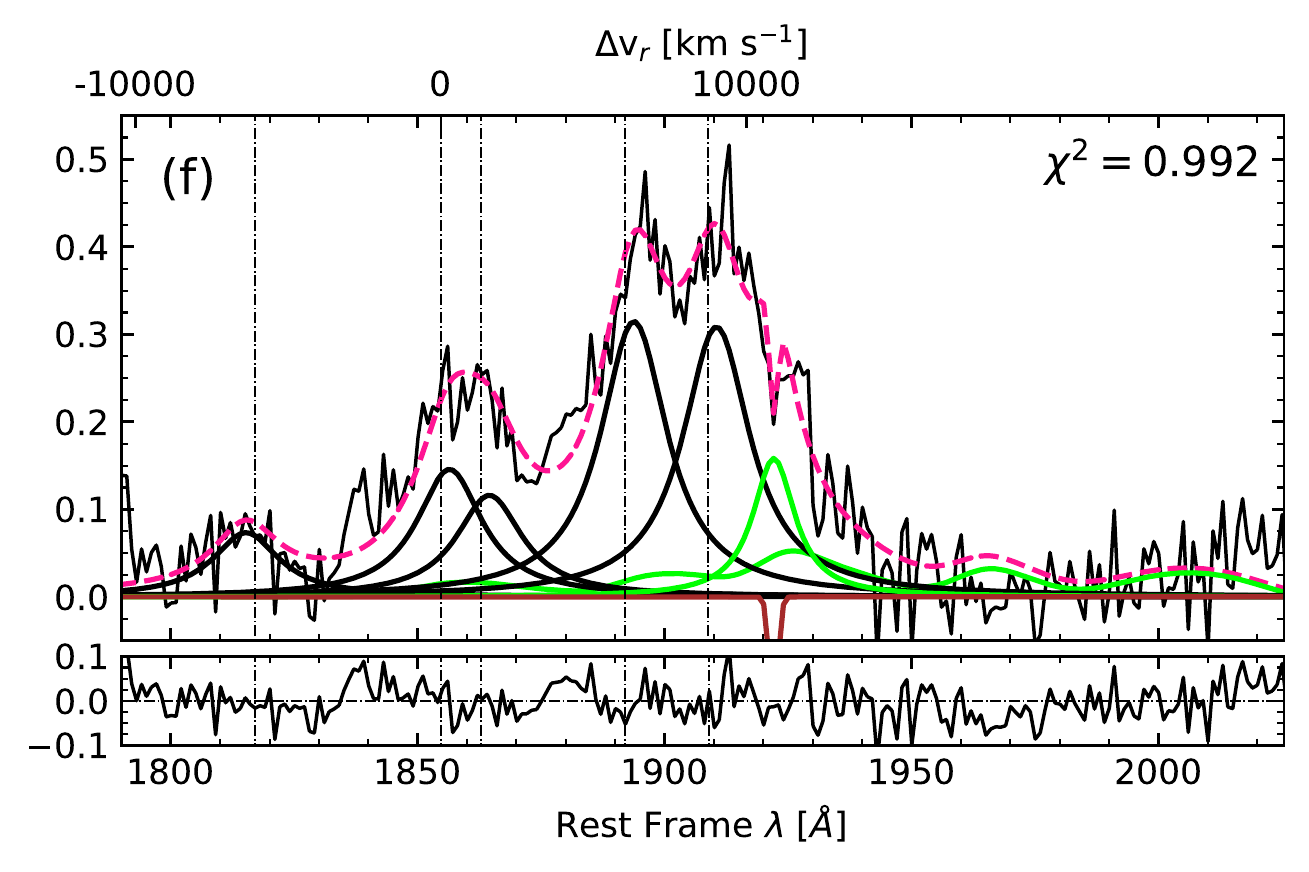}
    \caption{Q 1410+096. Same as Figure \ref{fig:1143_UV}.}
    \label{fig:1410_UV}
\end{figure}

\par This source has been identified as a BAL quasar by \cite{Allen_2011} and the absorption lines are seen in the regions of the \siv{} and \civ{} emission lines. In order to account for the  multiplet contributions in the optical region, we performed a fitting of this \ion{Fe}{II} emission on the red side of \hb, since the blue part of the FeII$\lambda$4570  region  is close to the border of the spectrum and evidently affected by background subtraction or telluric absorption correction. 

For this quasar, we perform the fitting of the \ion{C}{IV}$\lambda$1549 line based on the \ion{He}{II}$\lambda$1640 profile. The \aliii\ BC of 1410+096 is  narrow   when compared to \hb, which may indicate that \hb\ might be affected by blueshifted emission. An excess (not included in the fit) is also visible on the blue side of \aliii\ (bottom rightmost panel of Fig. \ref{fig:1410_UV}).

This source has also been analysed in detail in \cite{alice_2021}. 
\clearpage

%--------------------------------------------
\subsection{SDSSJ141546.24+112943.4}
\label{SDSSJ1415}

\begin{figure}[h!]
    \centering
    \includegraphics[width=0.685\linewidth]{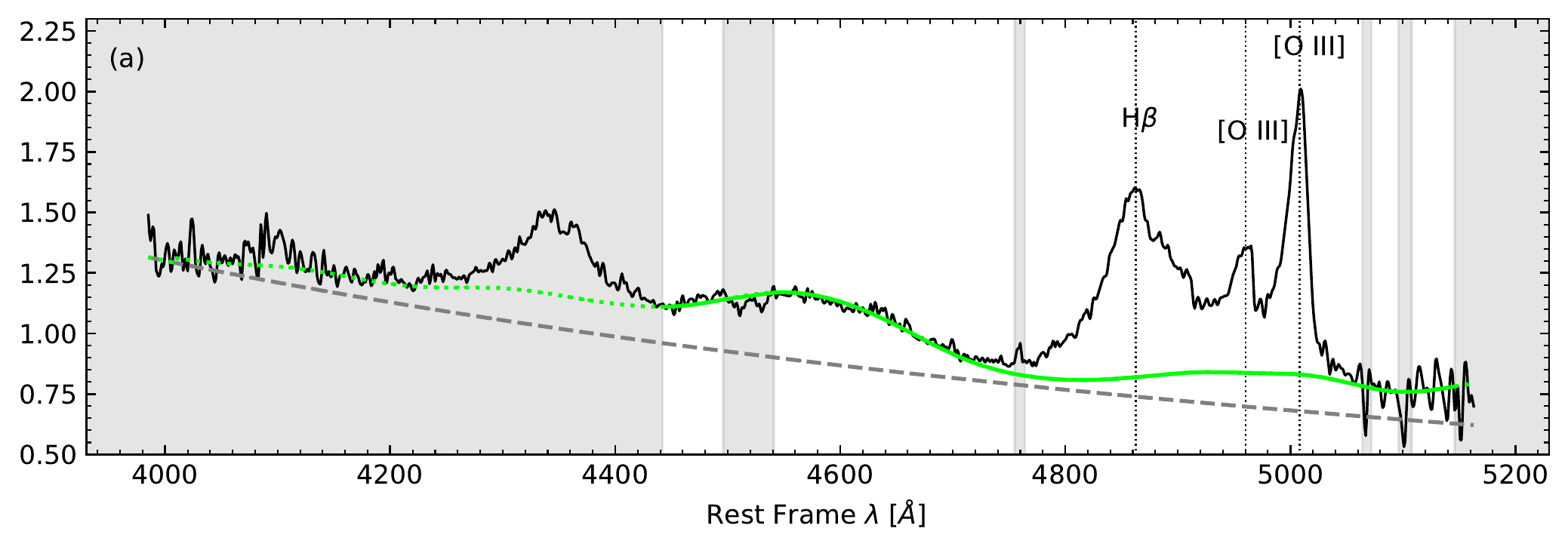}
    \includegraphics[width=0.30\linewidth]{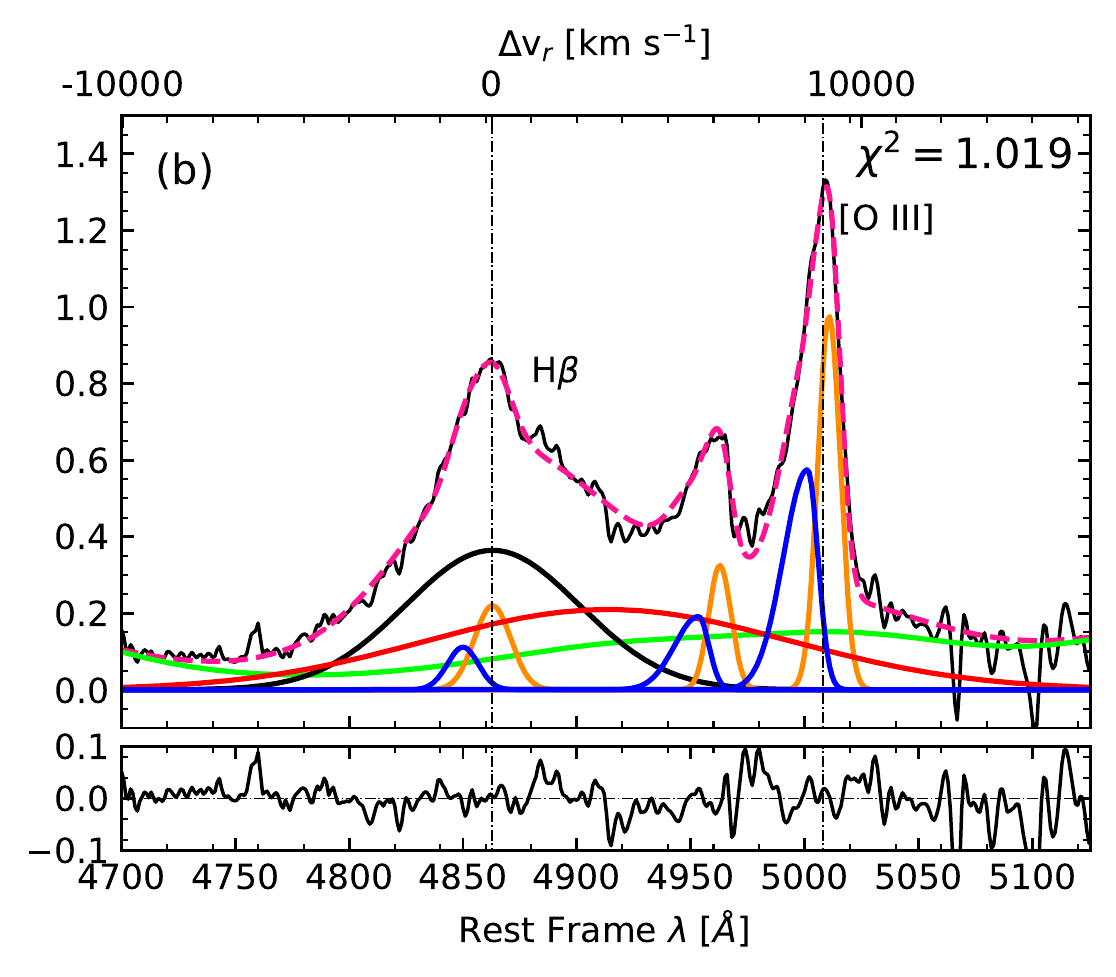}
     \\
    \centering
    \includegraphics[width=\linewidth]{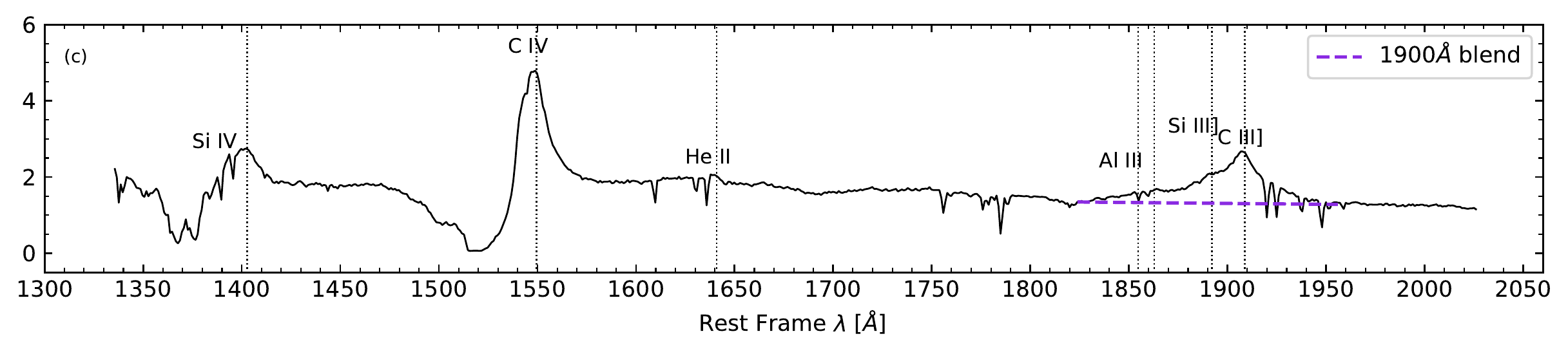}
    \raggedleft
    \includegraphics[width=0.33\linewidth]{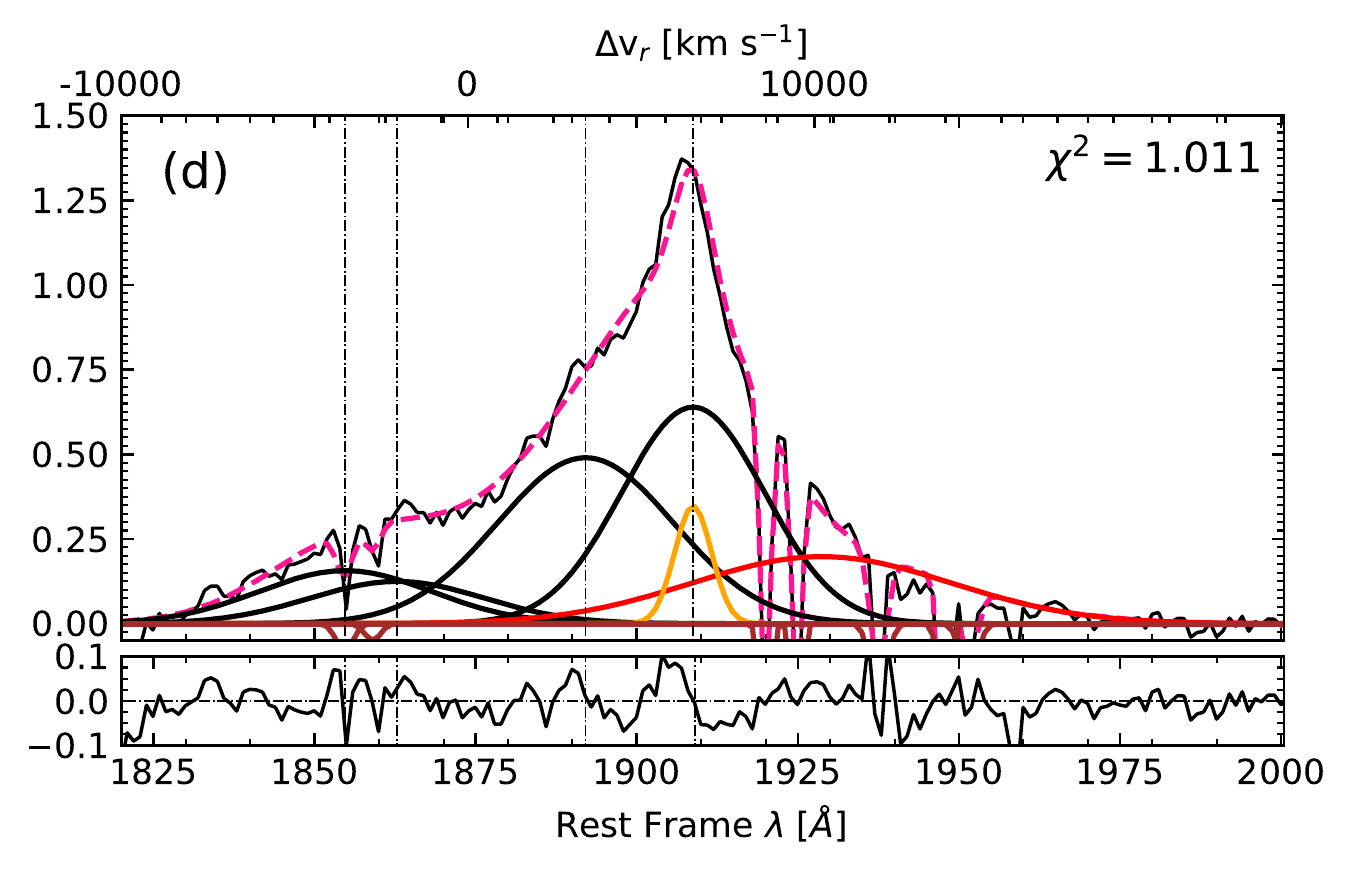}
    \\
    \centering
    \caption{SDSSJ141546.24+112943.4. \textit{Top panels:} Colours and lines as in Fig. \ref{HE 0001-2340}. \textit{(c)} UV spectrum. \textit{(d)} Fittings for the 1900 \r{A} blend. Pink dashed lines show the final fitting. Broad components are represented by black lines, while narrow and very broad components are in orange and red, respectively. Brown lines represent the absorption seen in the spectrum.}
    \label{fig:1415_UV}
\end{figure}

\par This object is known as a weak microlensing candidate in the literature \citep{sluse_2012, takahashi_2014}. According to \cite{welling_2014}, the quasar is a gravitationally lensed object split into 4 images separated by $\sim$ 1''. SDSSJ141546.24+112943.4  is a modest core-dominated radio-loud source with a logarithmic radio loudness parameter of 1.09.  

As can be seen in the UV spectra presented in Fig. \ref{fig:1415_UV}, the source is a BAL quasar, with broad absorption lines especially in the \civ{} and \siv{} regions. %\cite{welling_2014} analysed the \civ{} broad absorption line variability of this object with data from SDSS and FIRST. 
\cite{Hazard_1984} also report broad absorption lines in \aliii{} that we can see on the blue side of the blend at $\lambda \approx 1820$ \AA.  % In this case we perform the fitting of the \ion{C}{IV}$\lambda$1549 line based on the \ion{He}{II}$\lambda$1640 profile.\paolaQ{Not shown?}

\clearpage
%--------------------------------------------
\subsection{B1422+231}
\label{B1422}

\begin{figure}[h!]
    \centering
    \includegraphics[width=0.69\linewidth]{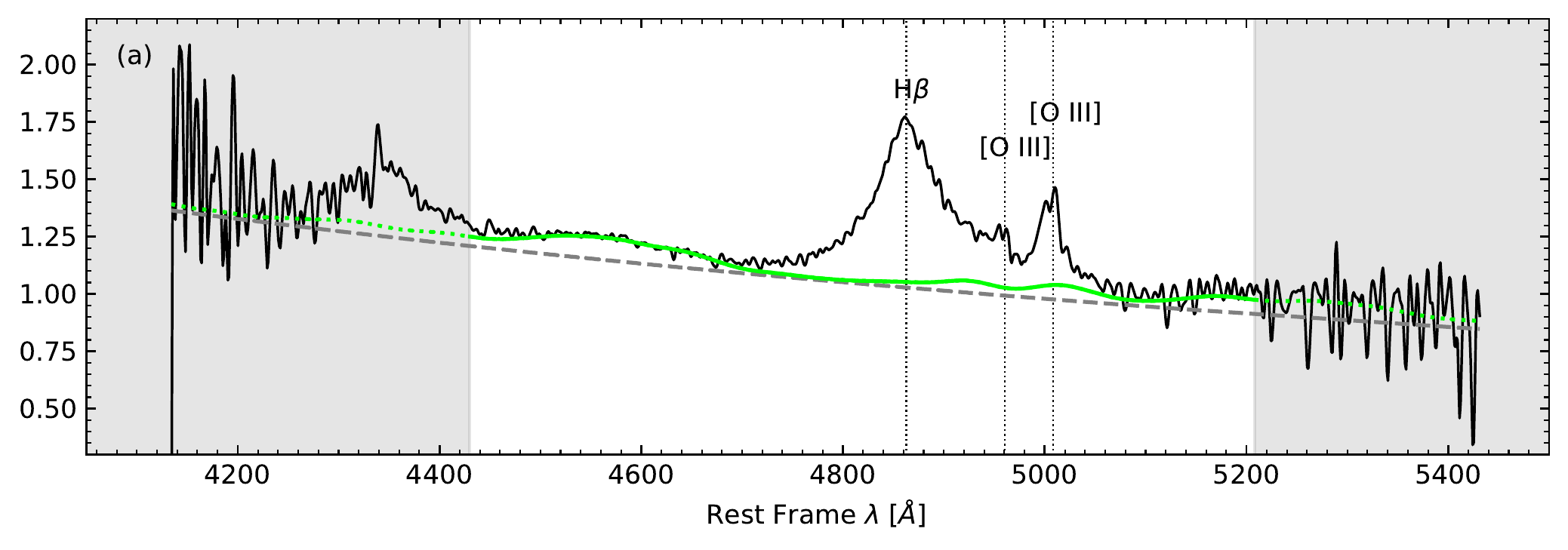}
    \includegraphics[width=0.305\linewidth]{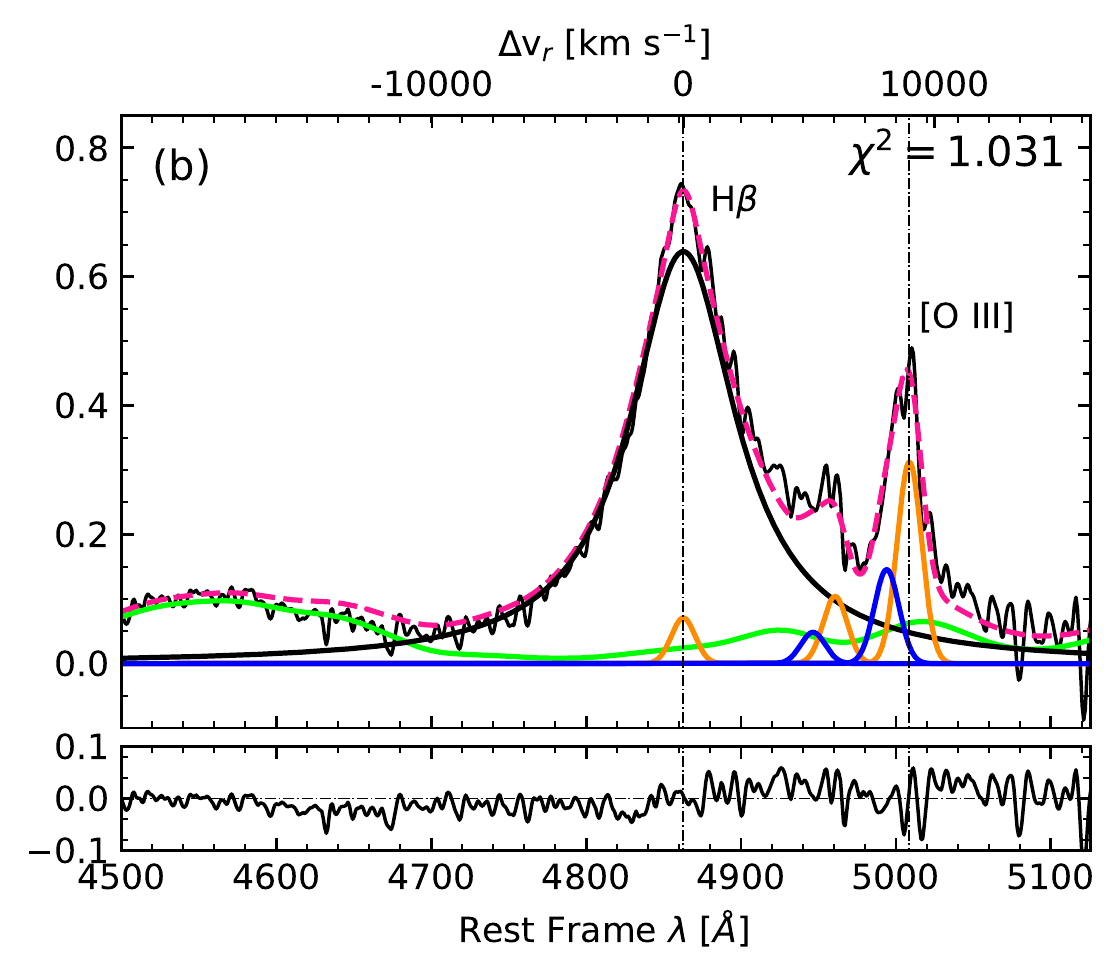}
    \caption{B1422+231. Colours and lines as in Figure \ref{fig:0001_HB}.}
    \label{fig:1422_HB}
\end{figure}

\par The source has been studied in detail for the first time by \cite{patnaik_1992}. It consists of a gravitationally lensed system with a diameter of 1.3 arcsec and four non-resolved components from VLA observations. This quasar is lensed into four images by a galaxy with $z=0.34$ \citep{Tonry_1998}. \cite{Dadina_2016} used XMM-Newton to study the matter inflow at the centre of this source and found that its X-ray spectrum is quite similar to the one of a typical Seyfert galaxy. Several UV spectra were found in literature \citep{Tonry_1998, Assef_2011, kundic_1997, odowd_2017}, but are not available for public usage or do not correspond to the regions we want to analyze. No SDSS spectrum was found.  It has also been analysed in the infrared by \cite{lawrence_1992} who conclude that the source presents similar structures in both radio and infrared. This is one of the radio-loud sources from our sample.

\newpage
%--------------------------------------------
\subsection{SDSSJ153830.55+085517.0}
\label{SDSSJ1538}

\begin{figure}[h!]
    \centering
    \includegraphics[width=0.68\linewidth]{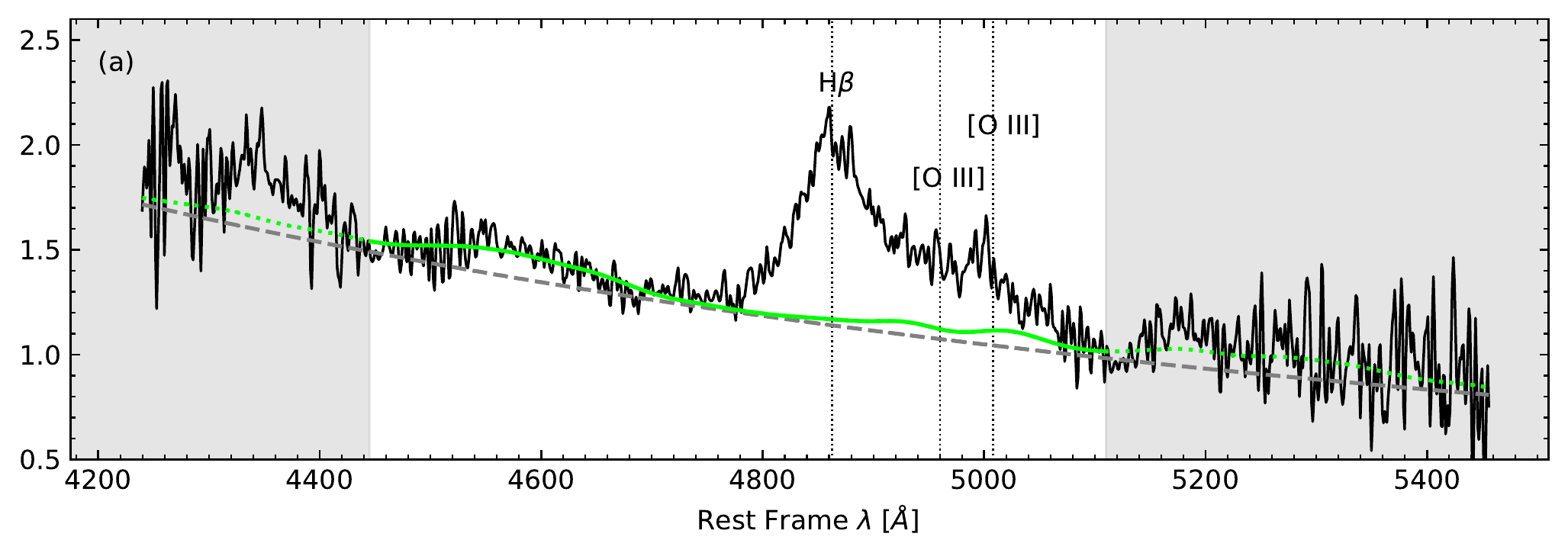}
    \includegraphics[width=0.305\linewidth]{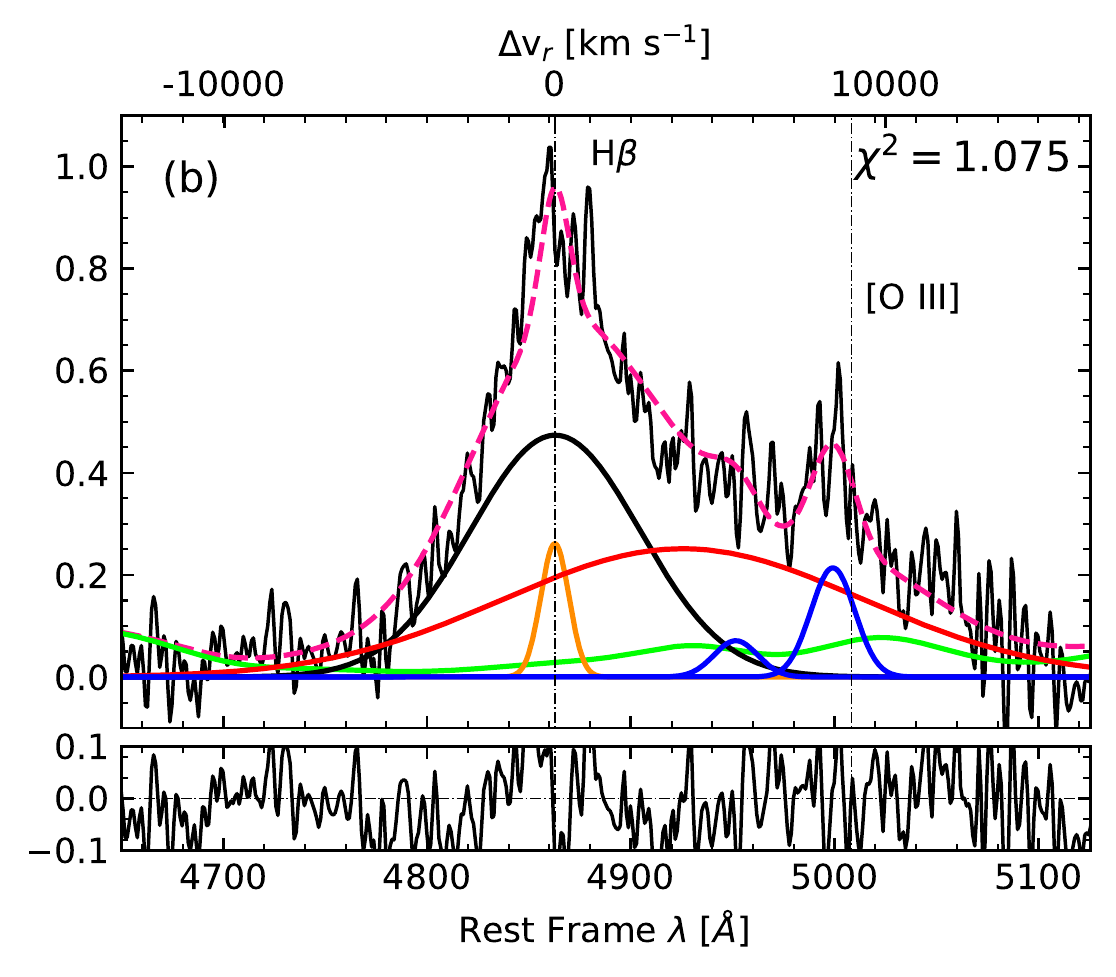}
     \\
    \centering
    \includegraphics[width=\linewidth]{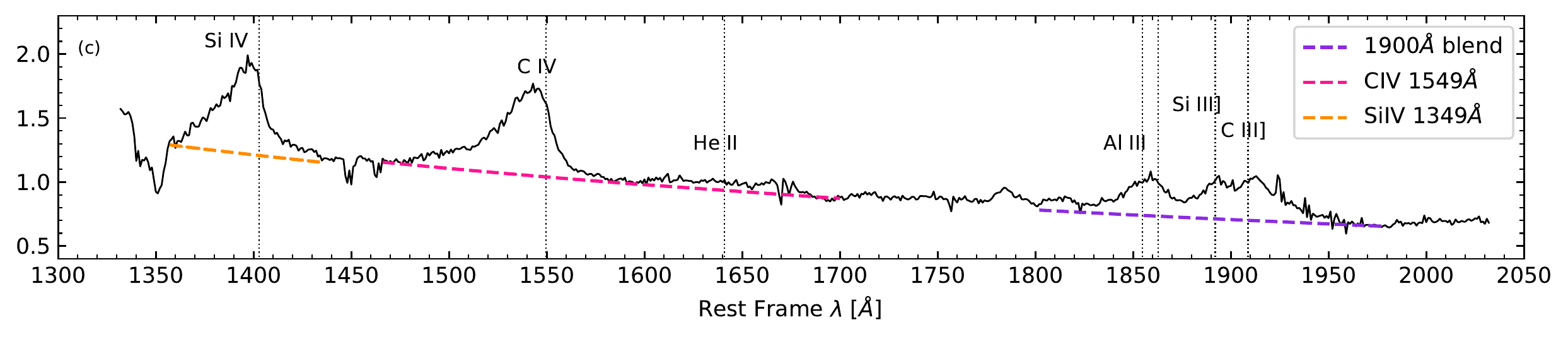}
   \includegraphics[width=0.325\linewidth]{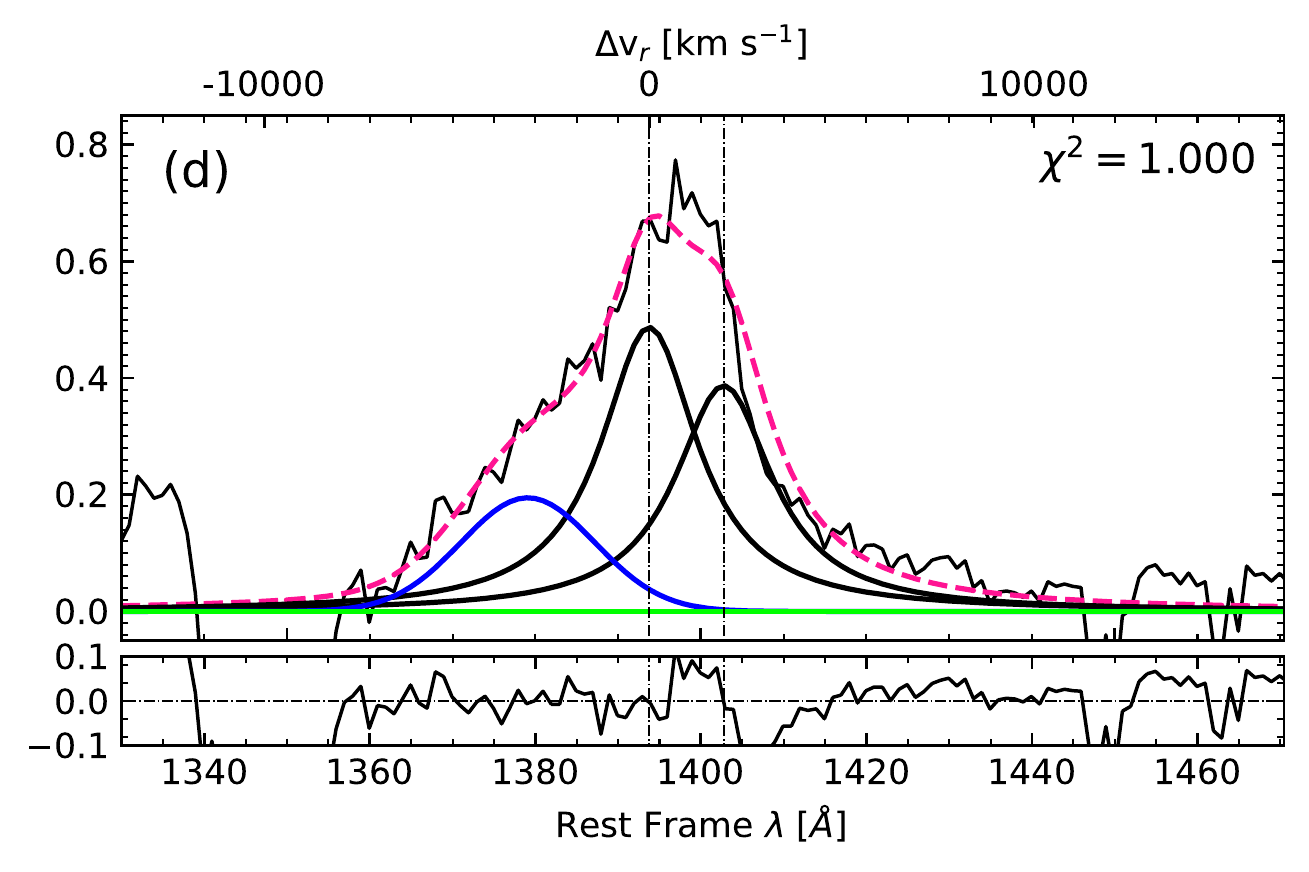}
    \includegraphics[width=0.33\linewidth]{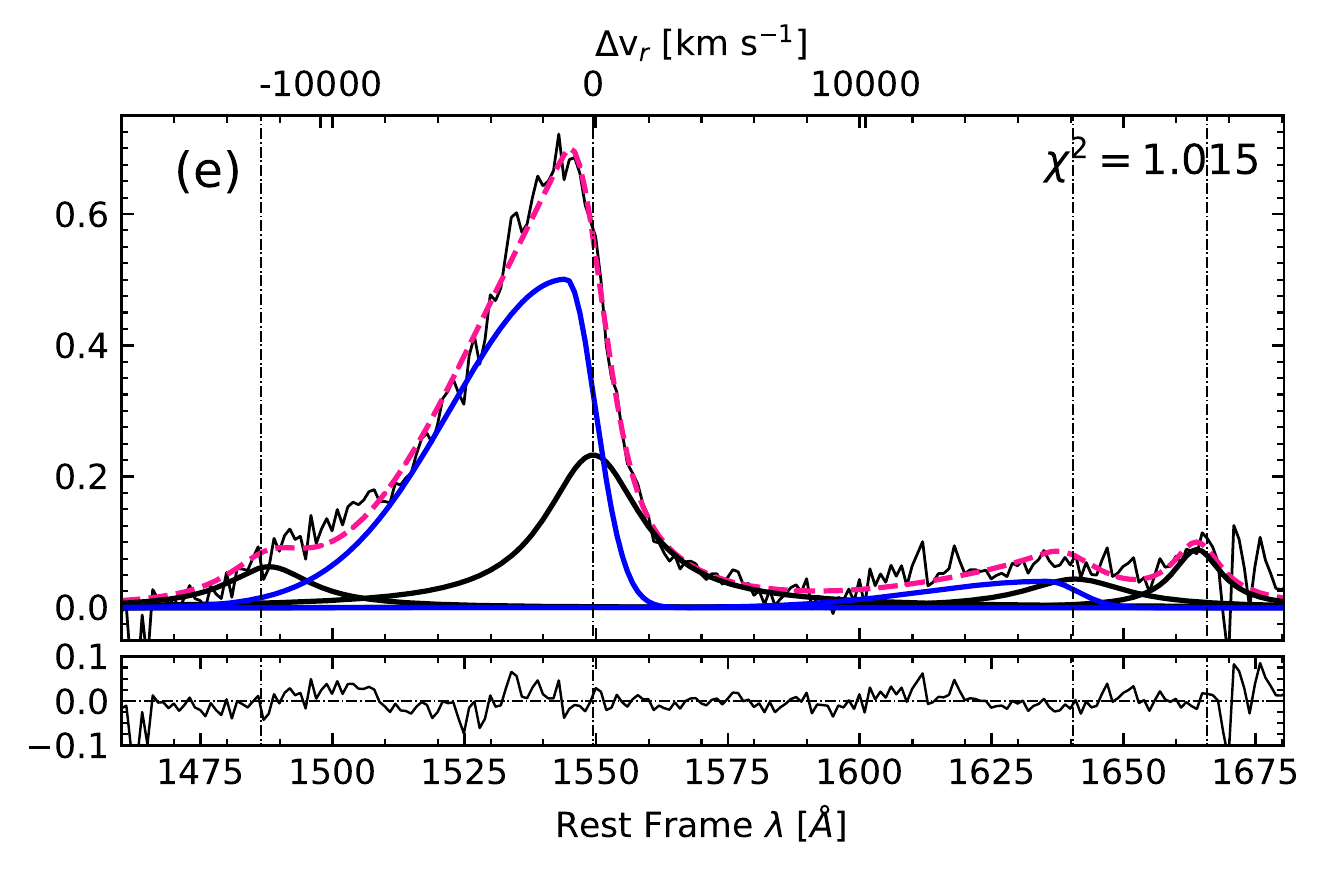}
    \includegraphics[width=0.33\linewidth]{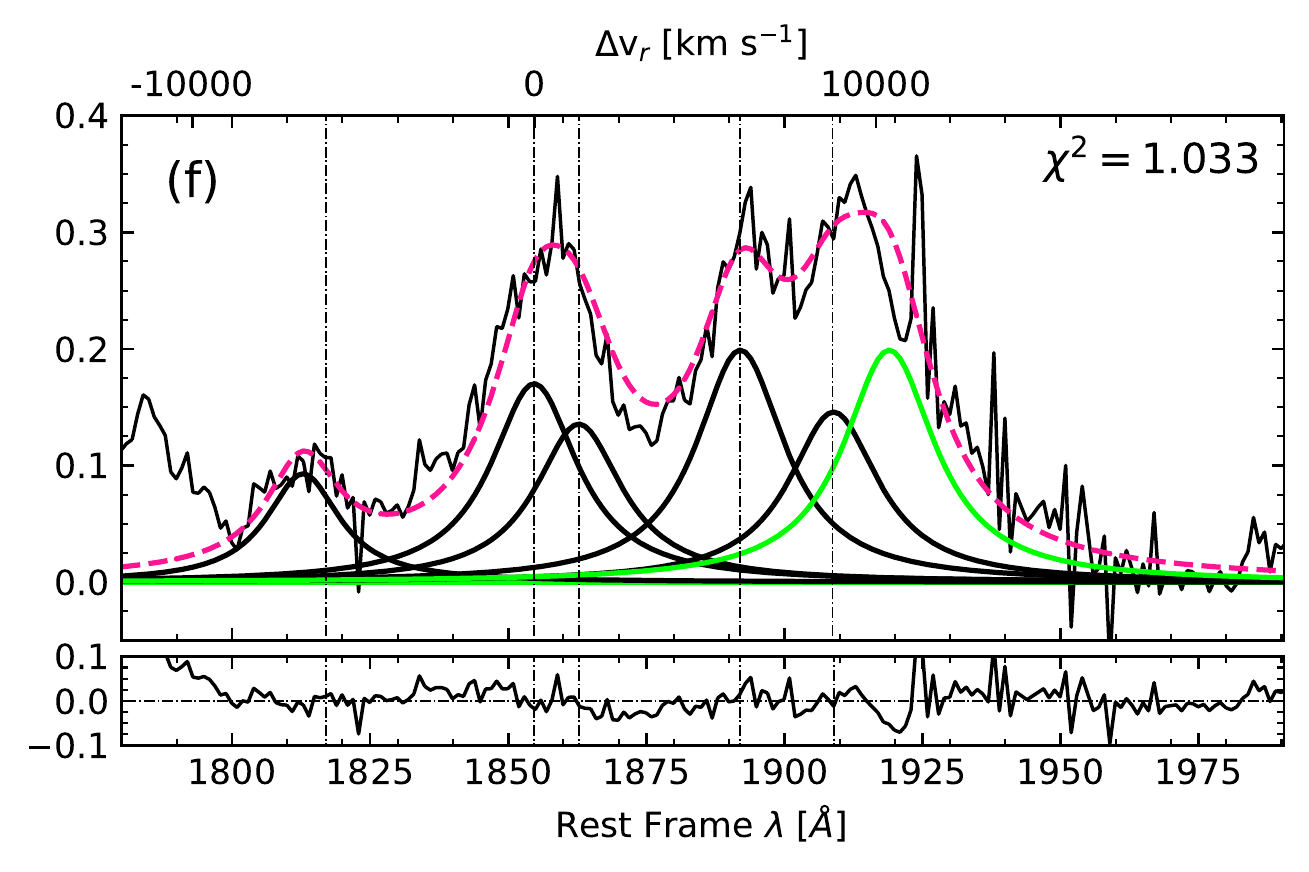}
    \caption{SDSSJ153830.55+085517.0. Same as Figure \ref{fig:1143_UV}.}
    \label{fig:1538_UV}
\end{figure}

\par When compared with the other sources of the sample, this source presents one of the spectra with lowest S/N.  
%which increases the uncertainties, for example when determining \oiii{}.
The low S/N might be due to the presence of a star that is located very close to the field of the quasar significantly increasing the background of the spectrum. This also may explain why the UV and the optical spectra of SDSSJ153830.55+085517.0 differ so much: while the optical spectrum shows a Pop. B-like profile, the UV is clearly the expected one for a Pop. A3 quasar. If real, this would be the first case of discordant population classification from the UV and from the optical spectrum. SDSSJ153830.55+085517.0 had the 1900\r{A} blend fit as Pop. A since it can not reproduce any Pop. B model and its UV spectrum presents all the characteristics of a Pop. A quasar. SDSSJ153830.55+085517.0 has previous observations in the $K$-band at LBT \citep{vietrietal18} and at UKIRT by \citet{Dix_2020}. In both cases our NIR spectrum is compatible with those shown for the H$\beta$ region by these authors. As we remarked before, the UV spectrum obtained with BOSS shows a blend at 1900\r{A} characteristic of the extreme Pop. A. An additional old UV SDSS spectrum exists but it has no information on the 1900\r{A} blend. A new UV spectrum would be necessary to disentangle its classification and possible peculiarity. Additionally, \citet{Bruni_2019} classifies this source as a high-velocity BAL.

\clearpage

%--------------------------------------------
\subsection{SDSSJ161458.33+144836.9}
\label{SDSSJ1614}

\begin{figure}[h!]
    \centering
    \includegraphics[width=0.69\linewidth]{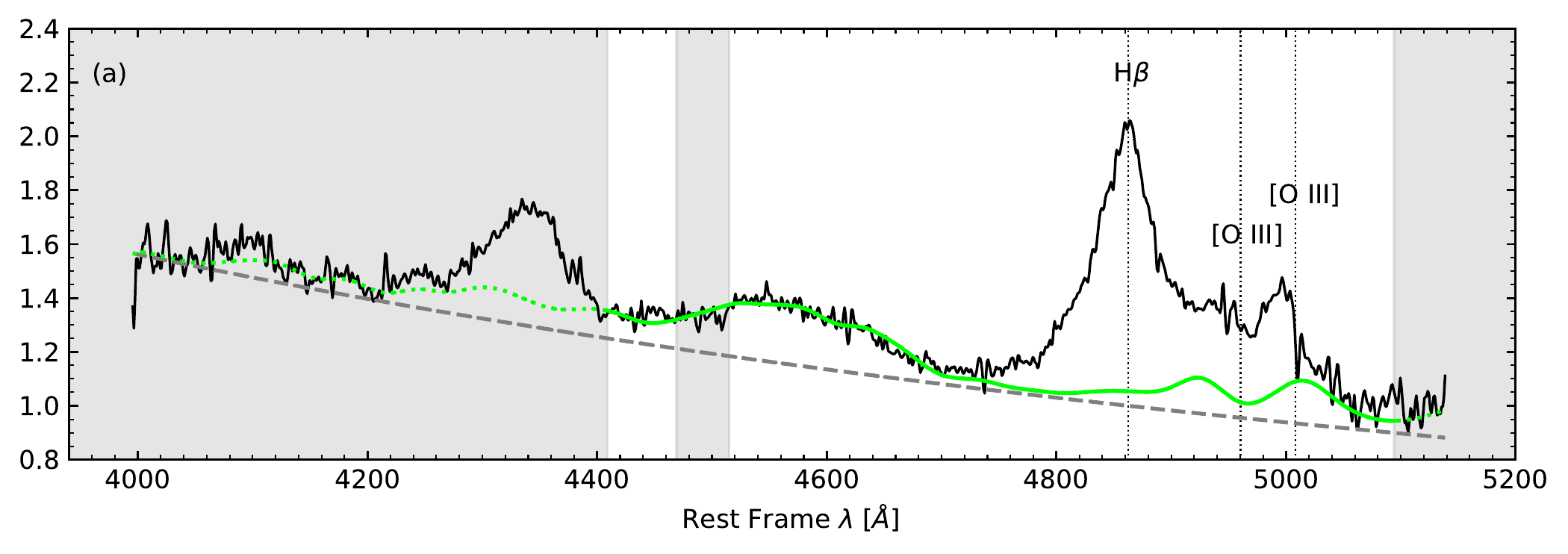}
    \includegraphics[width=0.305\linewidth]{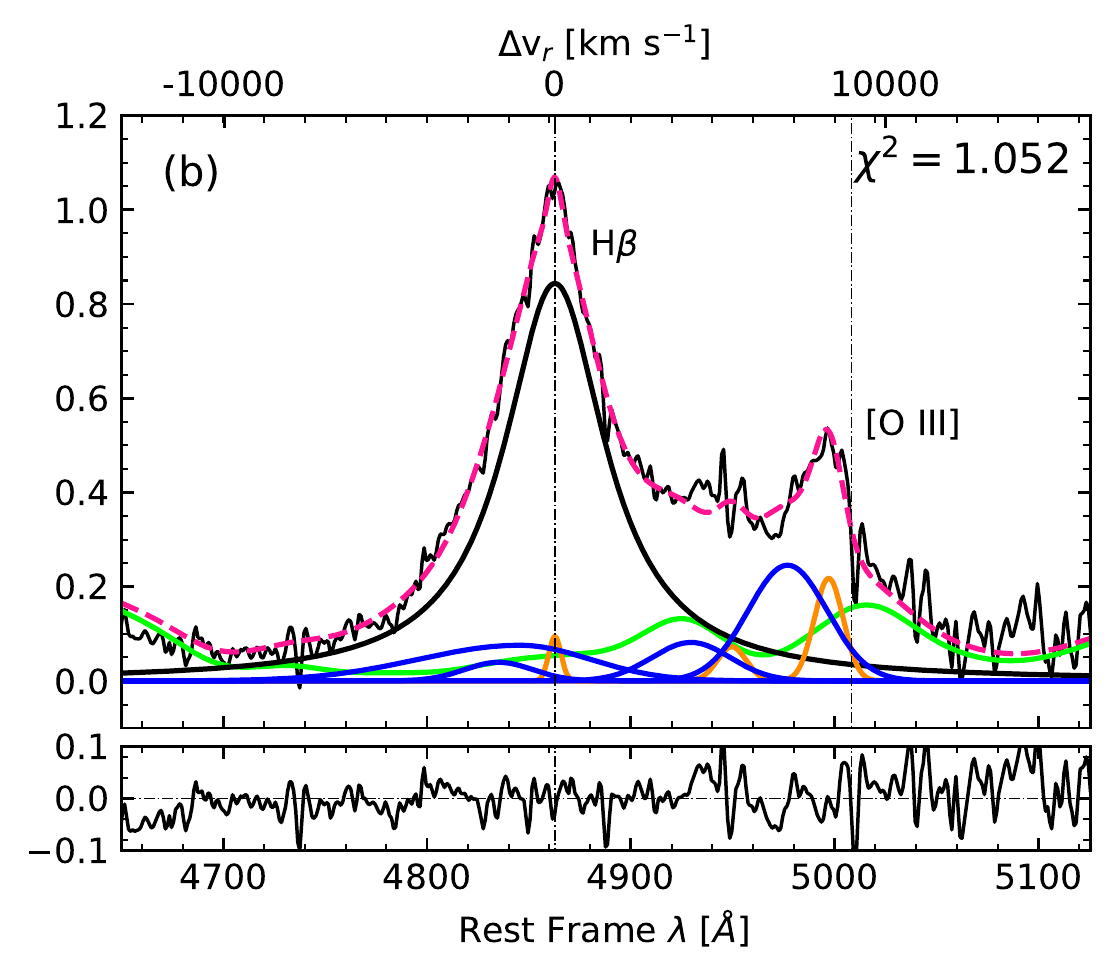}
    \\
    \centering
    \includegraphics[width=\linewidth]{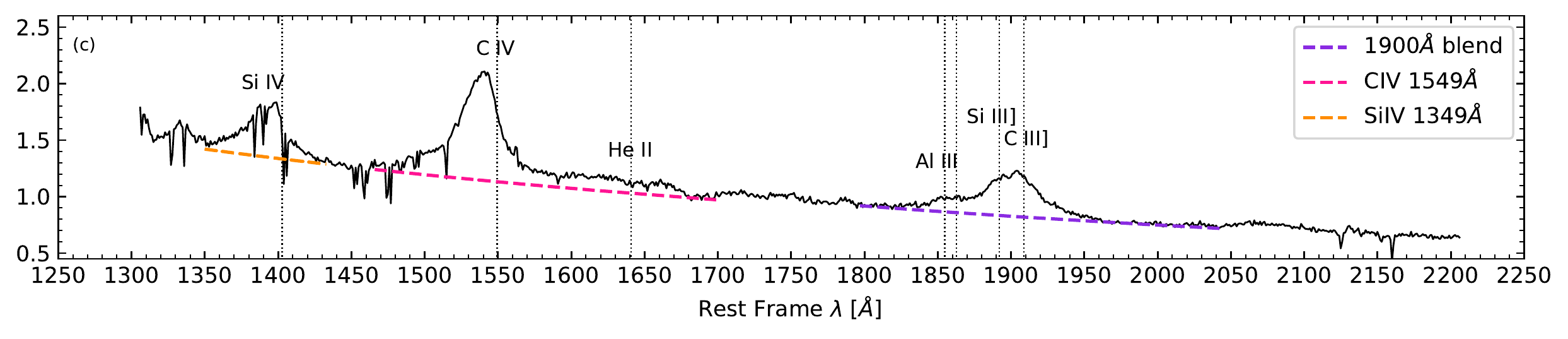}
    \includegraphics[width=0.33\linewidth]{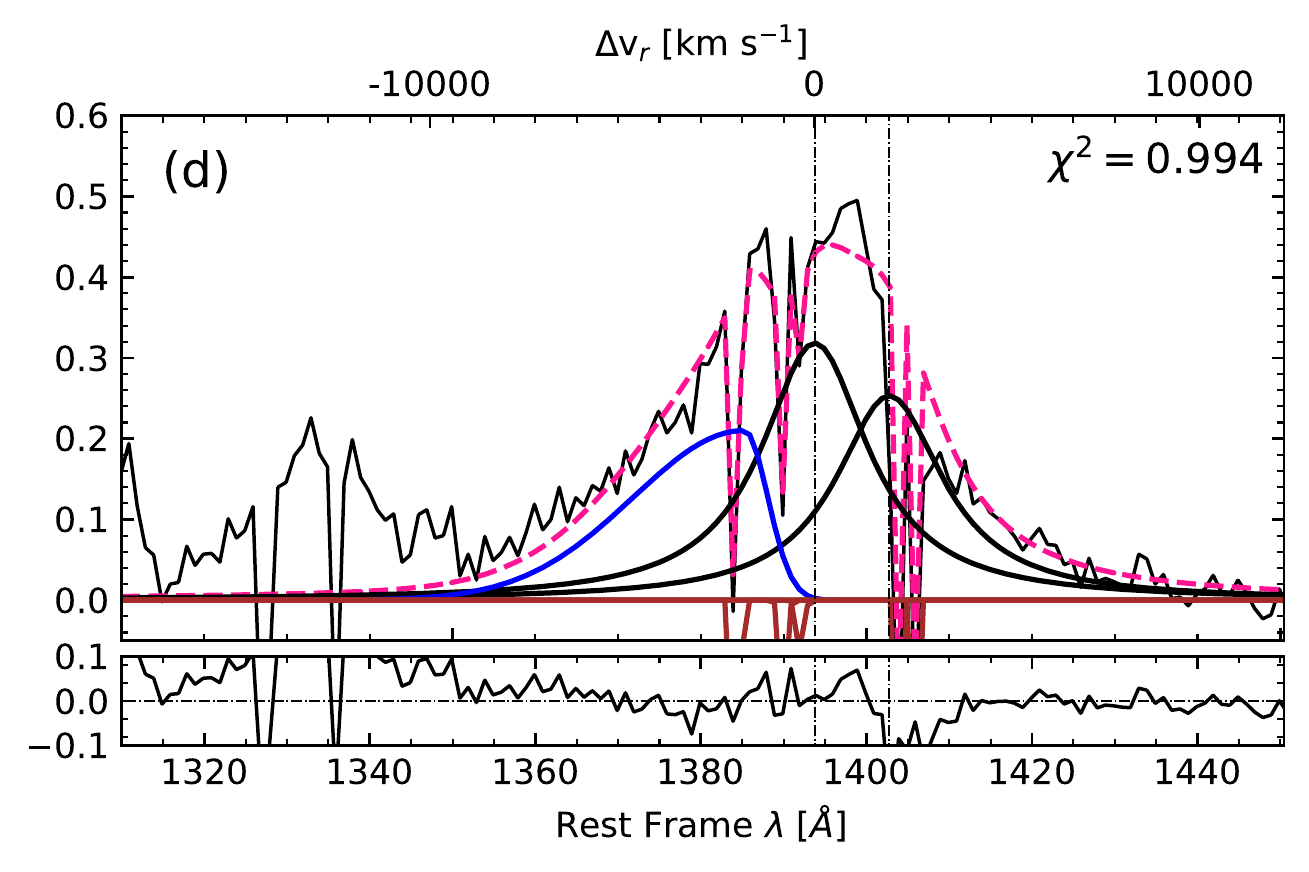}
    \includegraphics[width=0.33\linewidth]{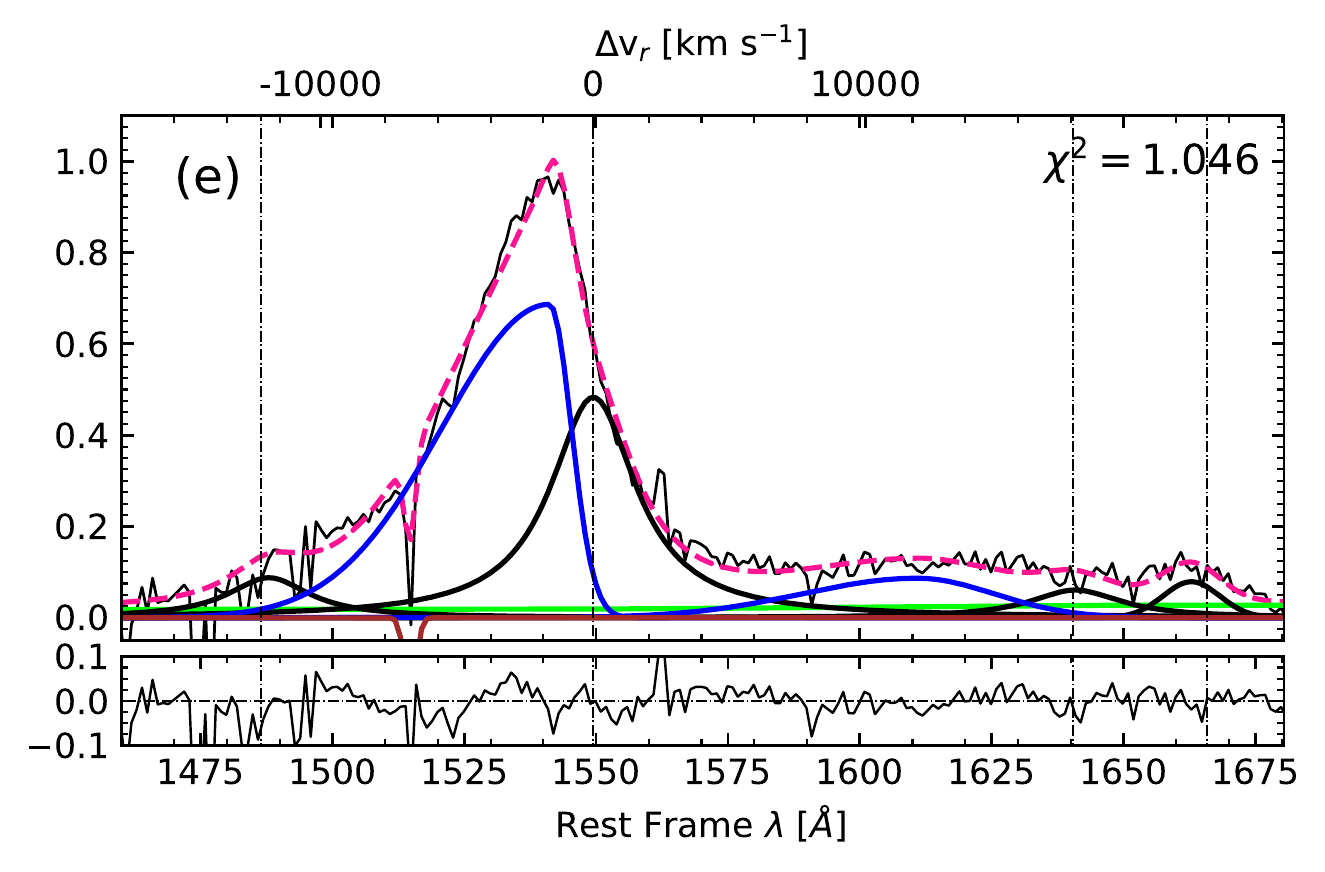}
    \includegraphics[width=0.33\linewidth]{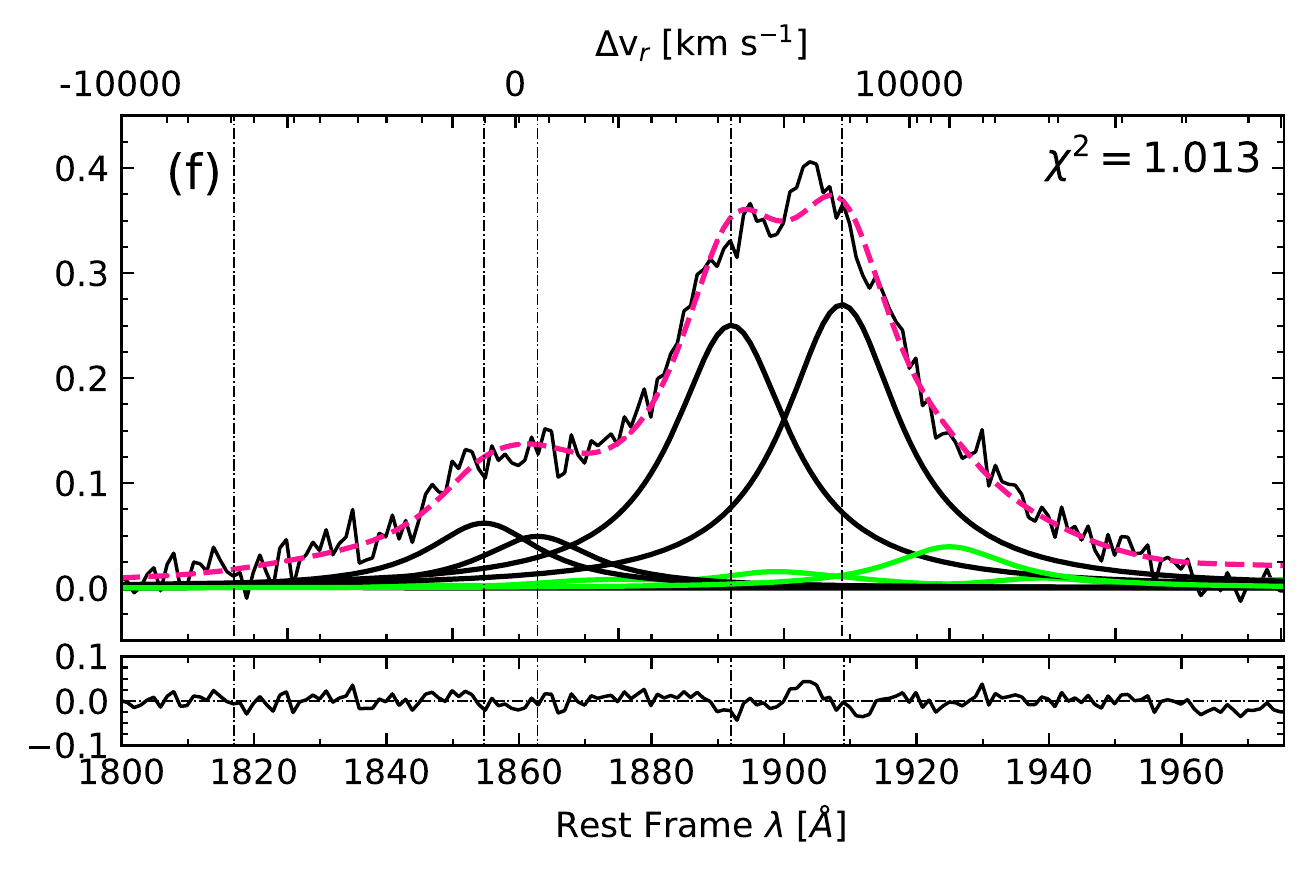}
    \caption{SDSSJ161458.33+144836.9. Same as Figure \ref{fig:1143_UV}.}
    \label{fig:1614_UV}
\end{figure}

\clearpage

%--------------------------------------------
\subsection{PKS 1937-101}
\label{PKS1937}

\begin{figure}[h!]
    \centering
    \includegraphics[width=0.69\linewidth]{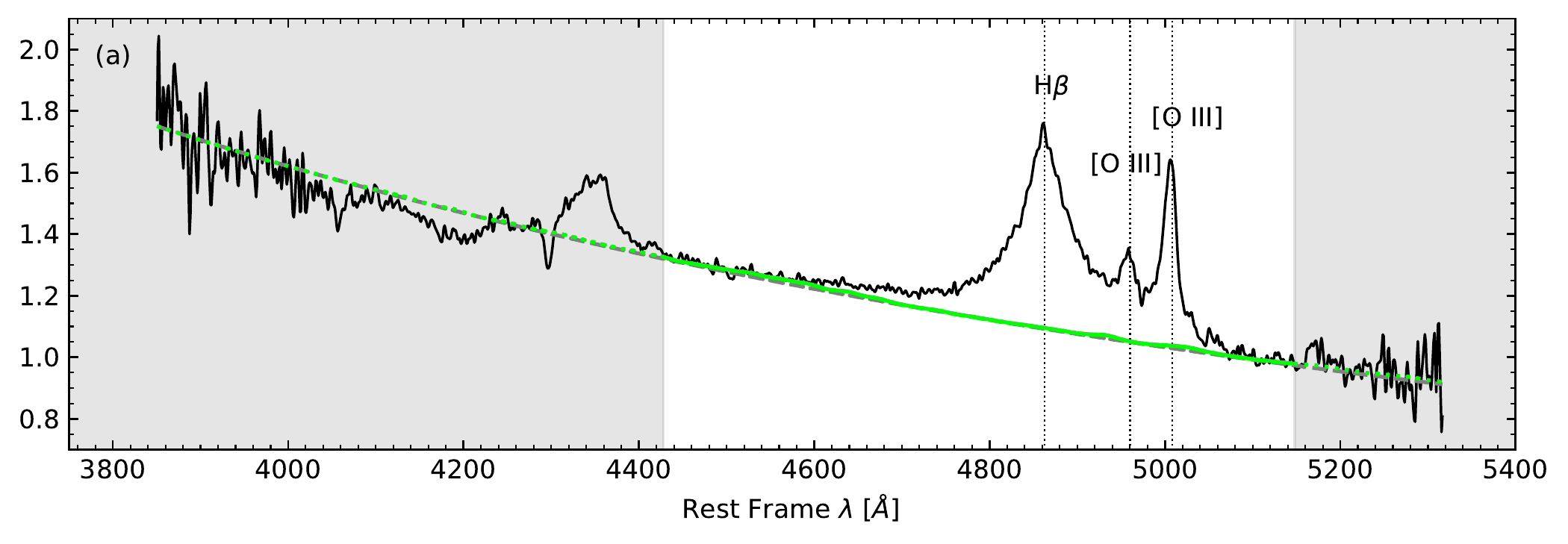}
    \includegraphics[width=0.30\linewidth]{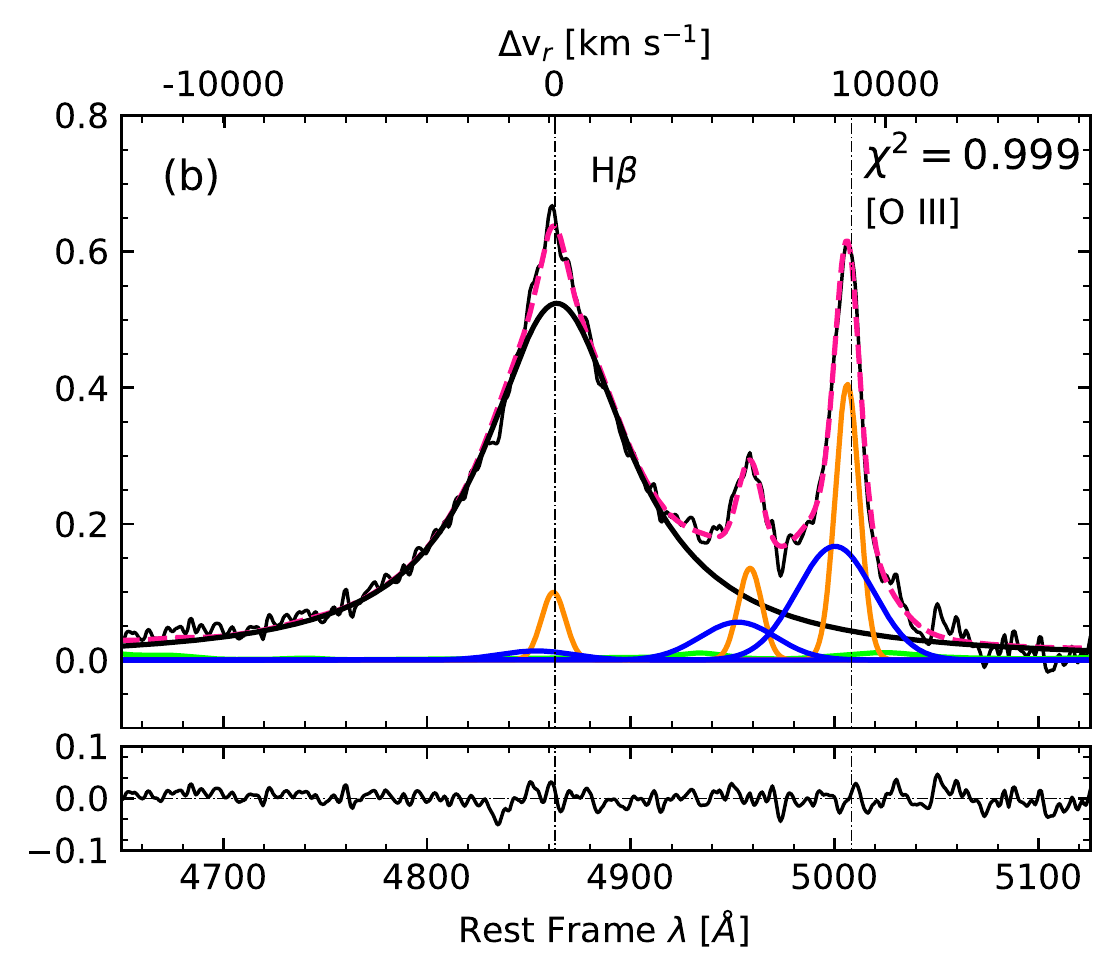}
    \caption{PKS 1937-101. Colours and lines as Figure \ref{fig:0001_HB}.}
    \label{fig:1937_HB}
\end{figure}

\par This  object shows a very faint \ion{Fe}{II} emission, as is typical of A1 spectra. A UV spectrum from the MOJAVE/2CM atlas for this source is presented in \cite{torrealba_2012}. 
It is also one of the strongest radio-loud sources we analyse in this work. Data from the Parkes 2.7 GHz Survey are shown in \cite{savage_1990}. PKS 1937-101 is seen as a compact radio source after analysis of VLBI data \citep{Lee_2016}. 

\cite{brinkmann_1995} report PKS 1937-101 as one of the most distant objects detected with the ROSAT survey \citep{voges_1993}.

\newpage
%--------------------------------------------
\subsection{PKS 2000-330}
\label{PKS2000}

\begin{figure}[h!]
    \centering
    \includegraphics[width=0.68\linewidth]{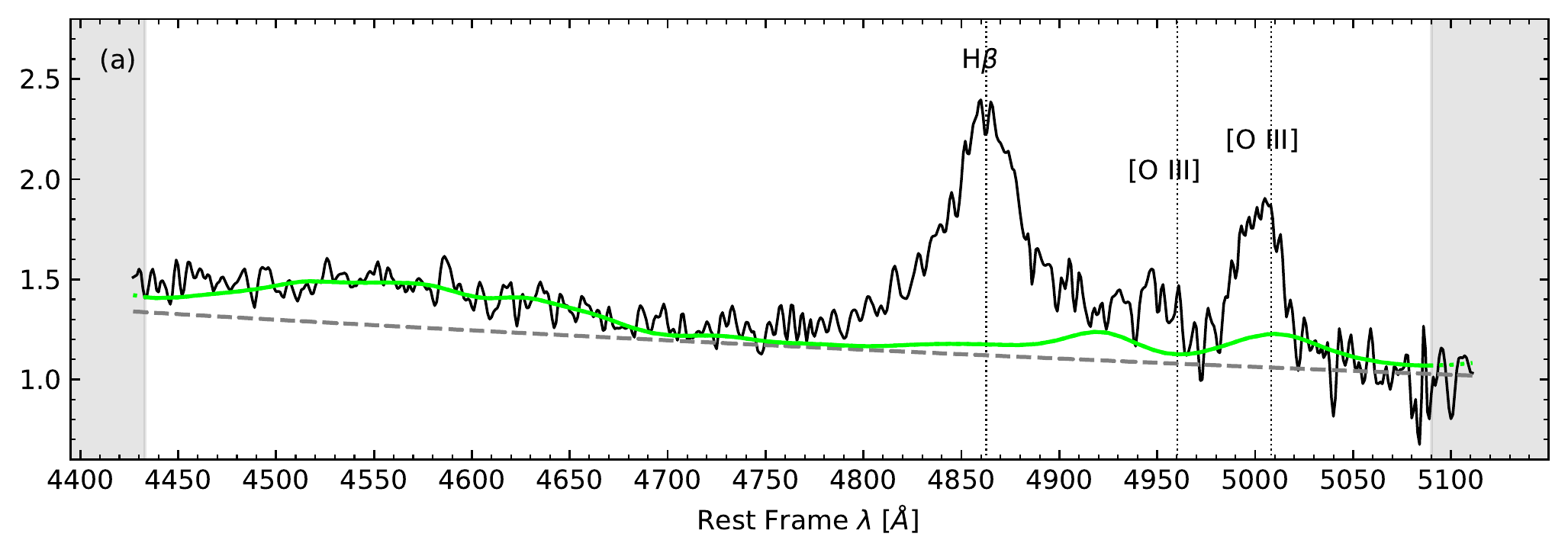}
    \includegraphics[width=0.30\linewidth]{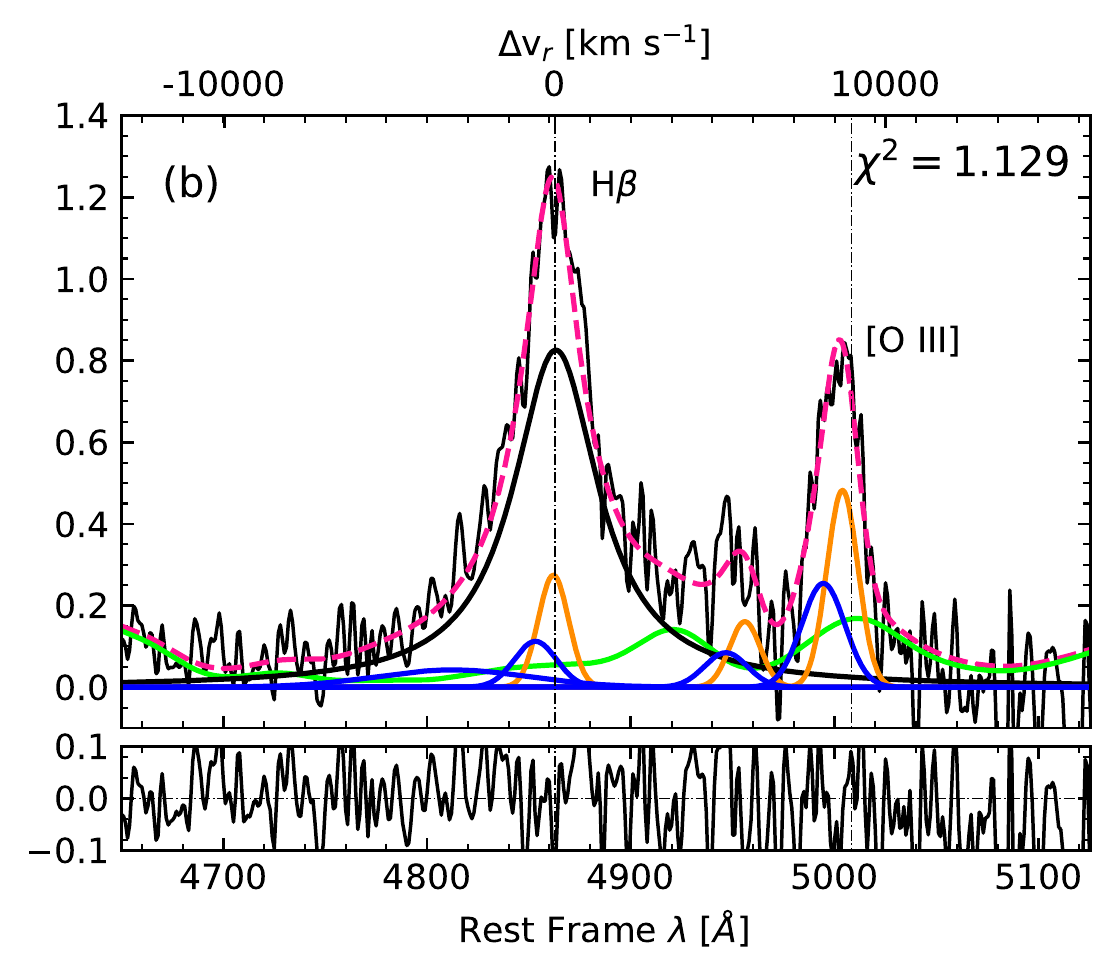}
    \\
    \raggedright
    \includegraphics[width=\linewidth]{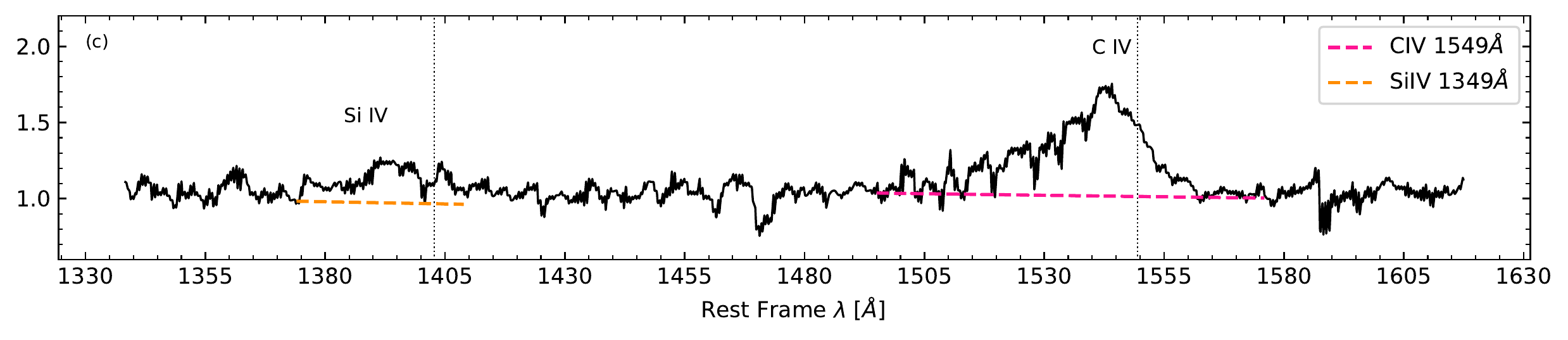}
    \includegraphics[width=0.333\linewidth]{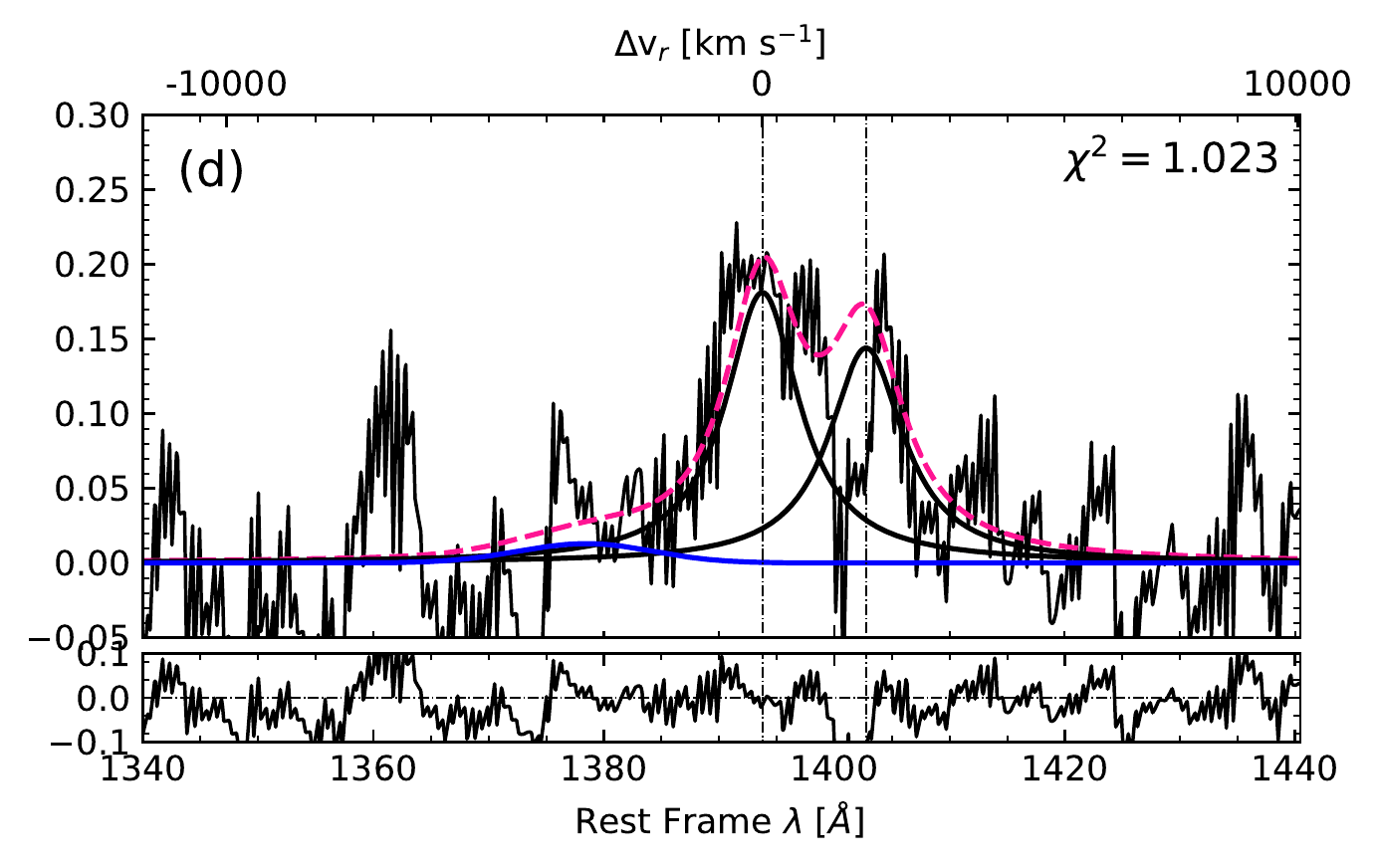}
    \includegraphics[width=0.32\linewidth]{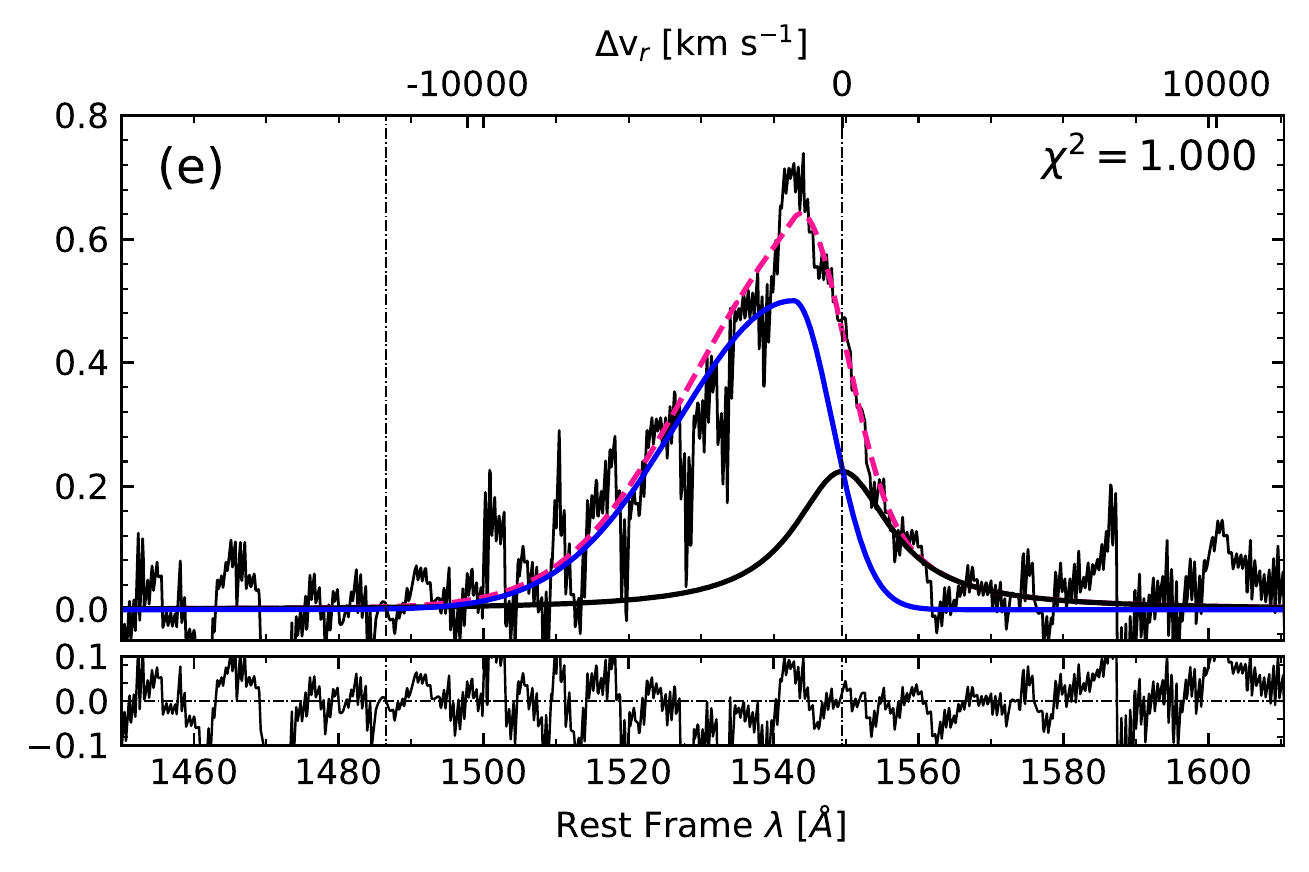}
    \\
    \centering
    \caption{PKS 2000-330. \textit{Top panels:} Same as Figure \ref{fig:1143_UV}. \textit{(c)} UV spectrum. \textit{Bottom panels:} fittings for (d) \siv{} and (e) \civ\. The UV spectrum was digitised from \cite{barthel_1990}.}
    \label{fig:2000_HB}
\end{figure}

\par The UV spectrum used in the fittings was obtained through the digitisation of the spectrum from \cite{barthel_1990}. However, since this spectrum presents a small wavelength range that does not account either for the 1900 \r{A} blend or for \ion{He}{II}$\lambda$1640, we analyse only the \siv{} and \civ{} emission lines. A detailed comparison between this source and Q 1410+096 is performed in \cite{alice_2021}.

This quasar is the strongest radio-loud source from the present sample \citep{savage_1990}. The position of this source is not within the area covered by the FIRST catalogue. 

\newpage
%--------------------------------------------
\subsection{SDSSJ210524.49+000407.3}
\label{SDSSJ2105}

\begin{figure}[h!]
    \centering
    \includegraphics[width=0.68\linewidth]{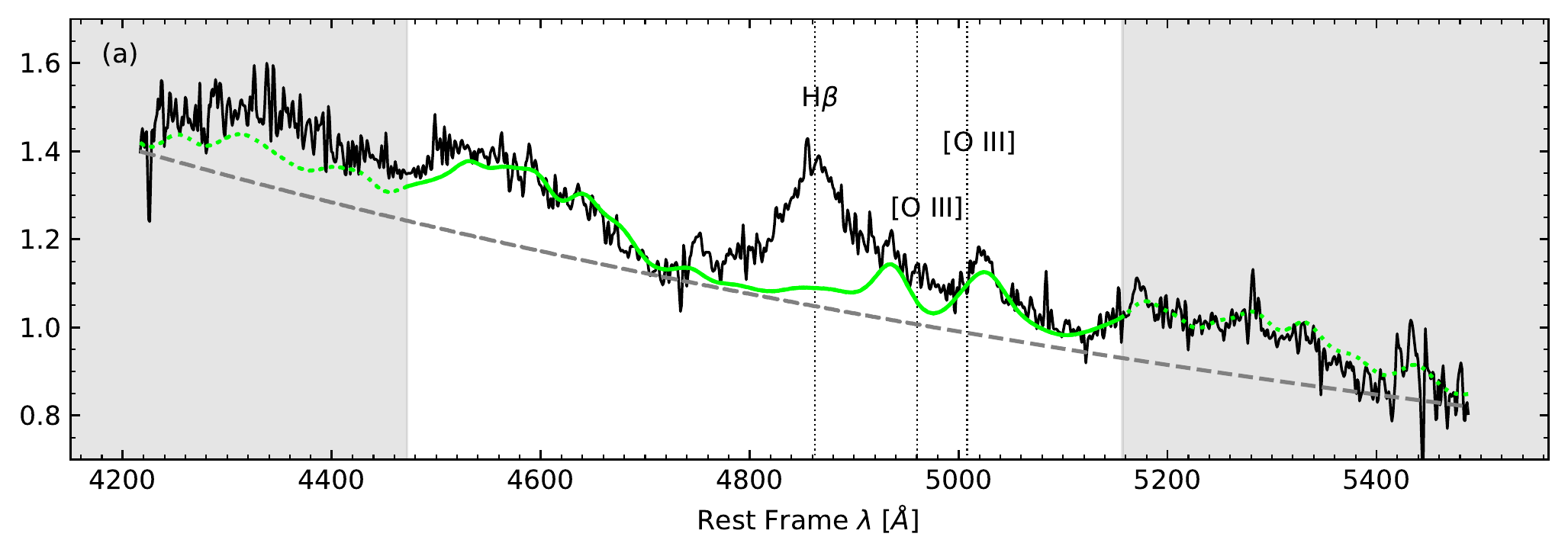}
    \includegraphics[width=0.305\linewidth]{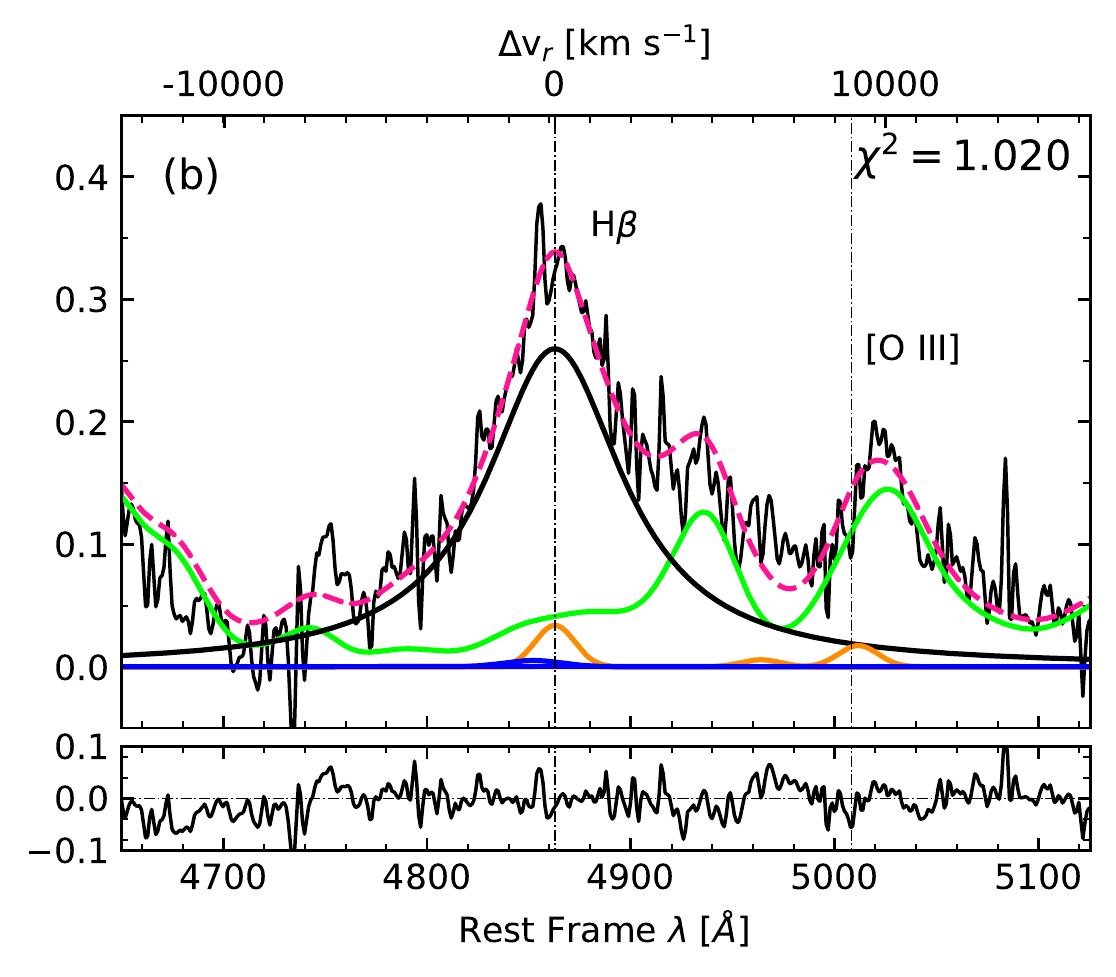}
    \\
    \centering
    \includegraphics[width=\linewidth]{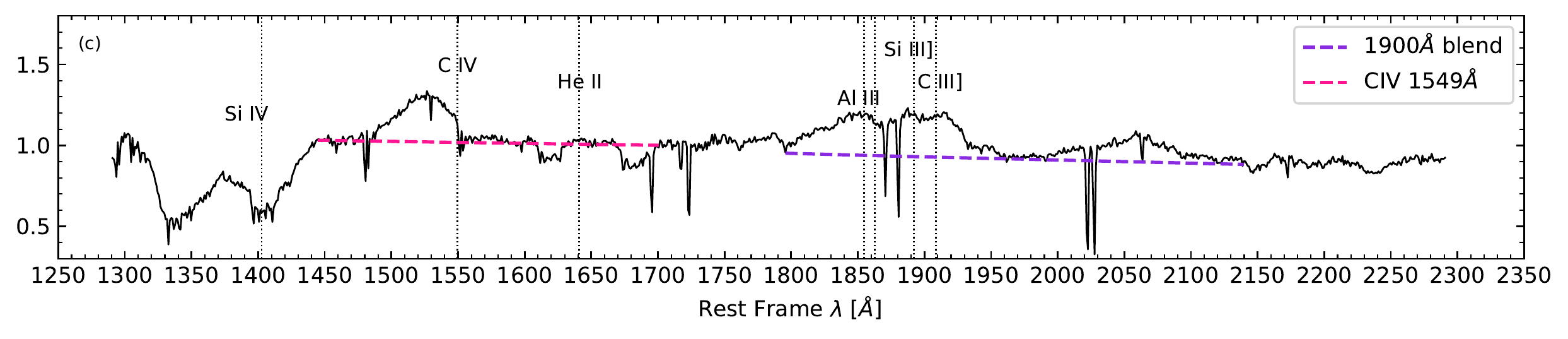}
    \raggedleft
    \includegraphics[width=0.33\linewidth]{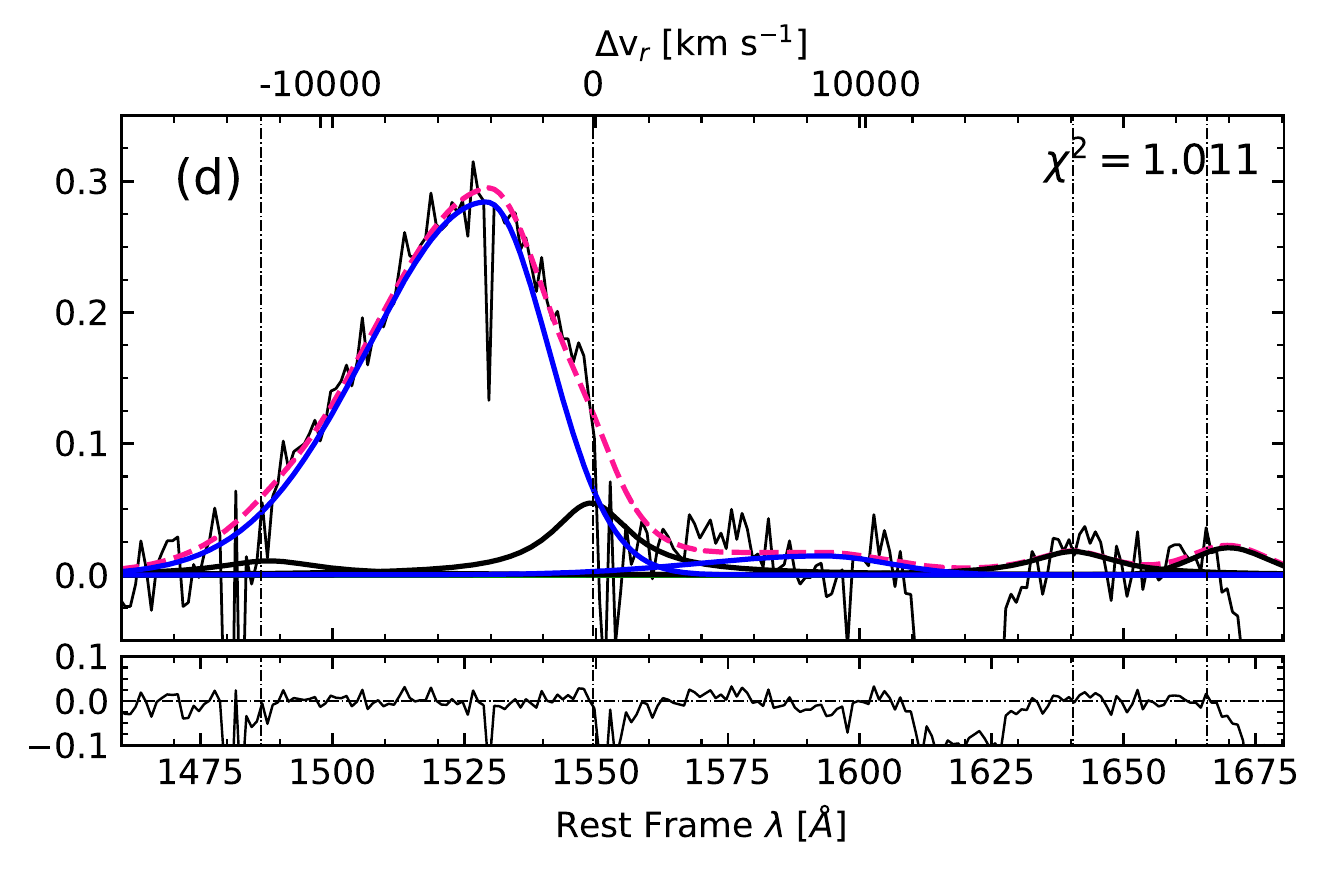}
    \includegraphics[width=0.33\linewidth]{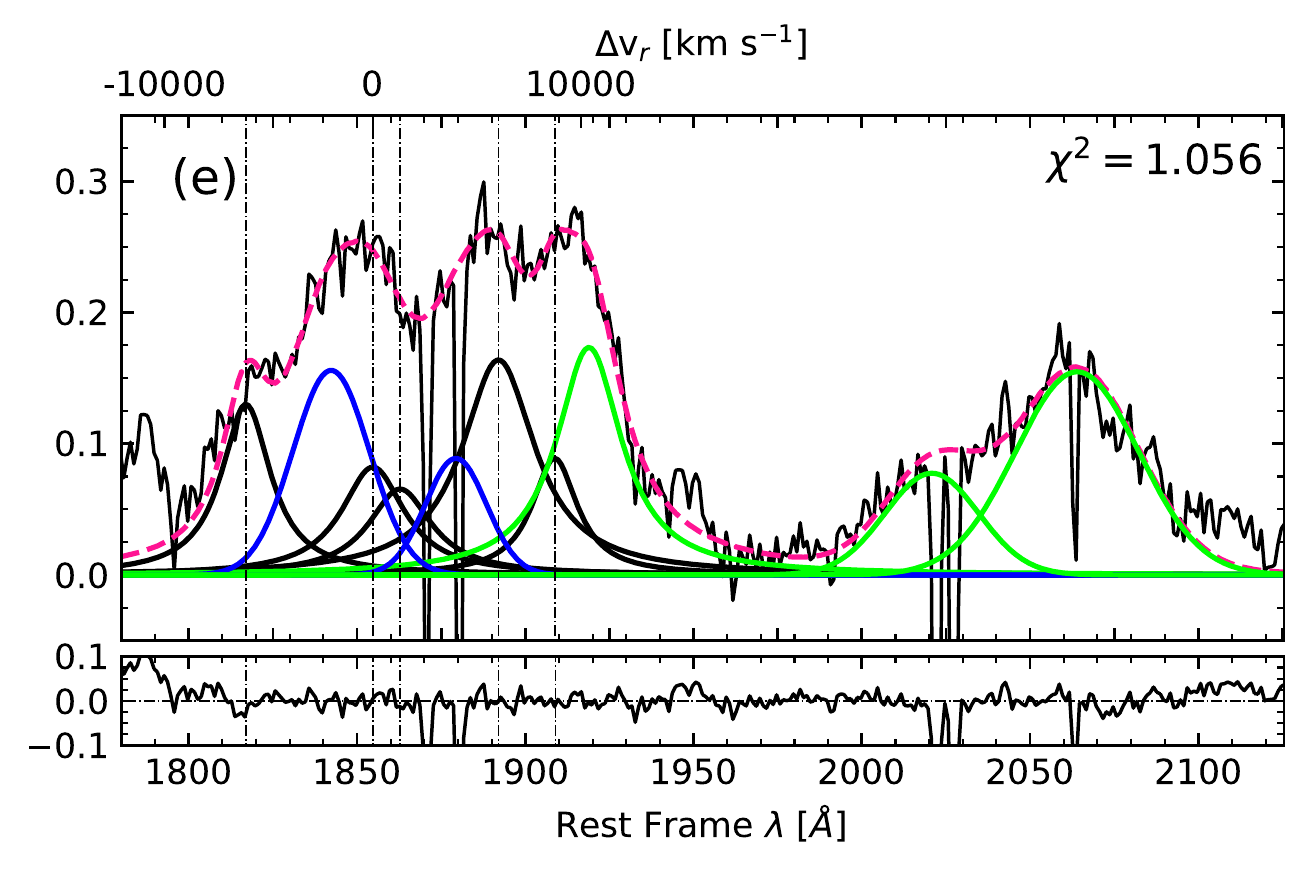}
    \\
    \centering
    \raggedleft
    \caption{SDSSJ210524.49+000407.3. \textit{Top panels:} Colours and lines as in Figure \ref{fig:0001_HB}. \textit{(c)} SDSS UV spectrum. \textit{Bottom panels:} fittings for (d) \civ{} and  (e) the 1900 \r{A} blend.}
    \label{fig:2105_UV}
\end{figure}

\par This source presents a $W$([\ion{O}{III}]) $\approx$ 0.45\AA. The low equivalent width makes it difficult to discern what corresponds to \oiii{} or to Fe II. 

The UV region of this source is very unwieldy because of very broad absorption lines shortwards of \ion{C}{IV}$\lambda1459$. For the 1900 $\AA$ blend we present two different fittings, one including blueshifted components for \aliii{} and \ion{Si}{III}]$\lambda 1892$ and another without these lines. The motivation here is that if we do not include blueshifted components then the BC of the lines present a displacement of $- 1640$ km s$^{-1}$, which can be seen as an indicative of some outflowing gas. 

\clearpage

%--------------------------------------------
\subsection{SDSSJ210831.56-063022.5}
\label{SDSSJ2108}

\begin{figure}[h!]
    \centering
    \includegraphics[width=0.68\linewidth]{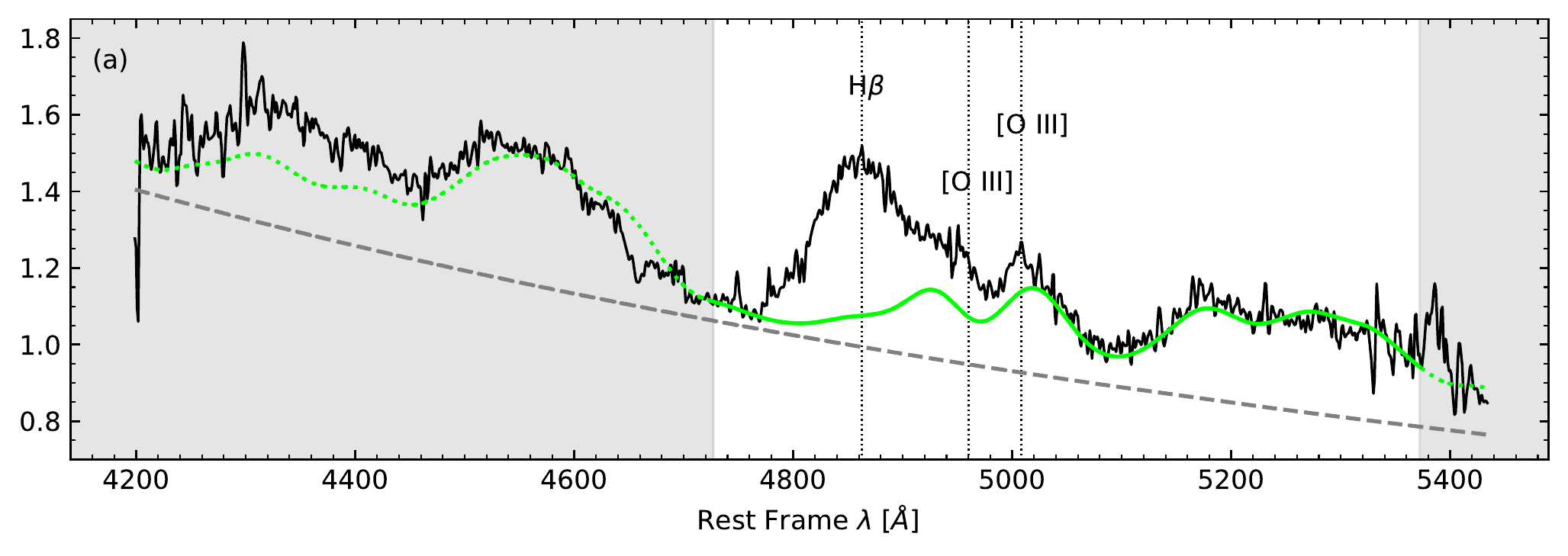}
    \includegraphics[width=0.305\linewidth]{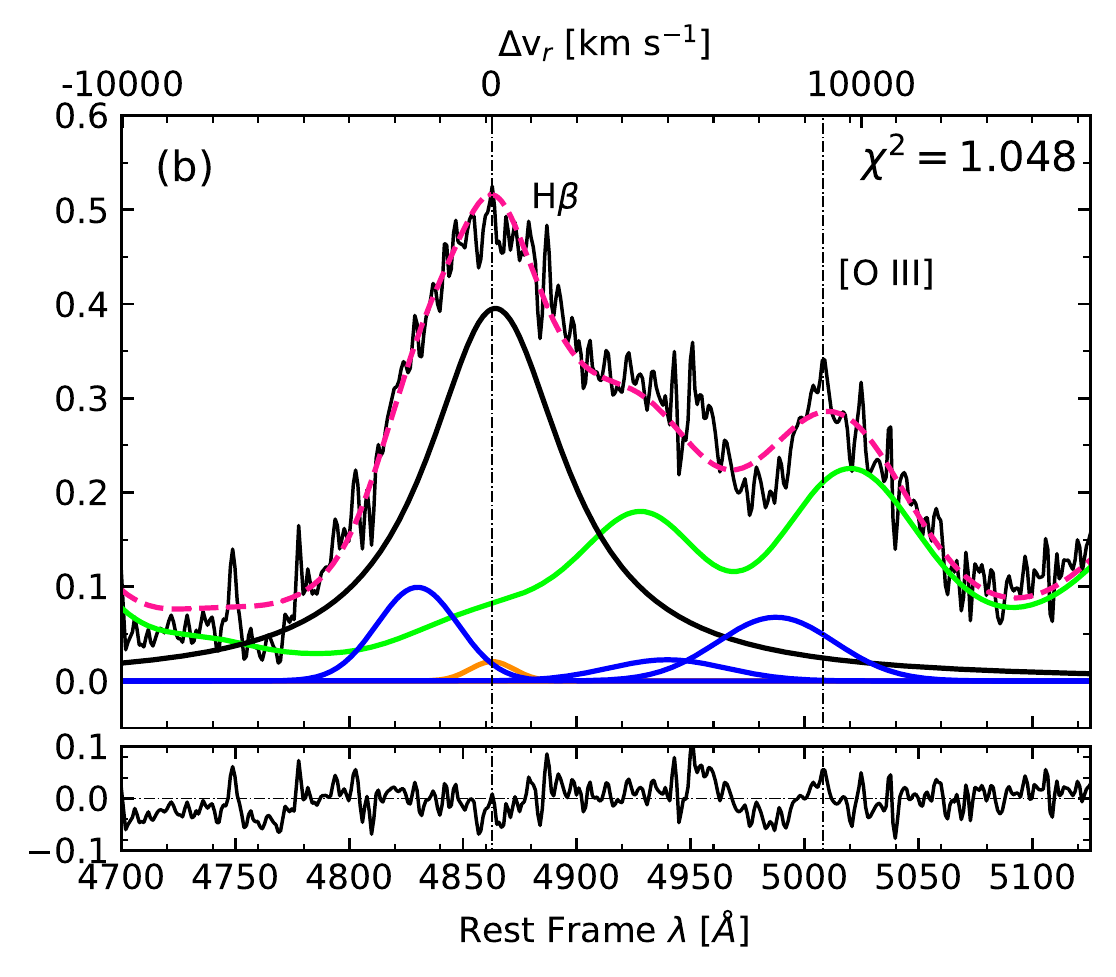}
    \\
    \centering
    \includegraphics[width=\linewidth]{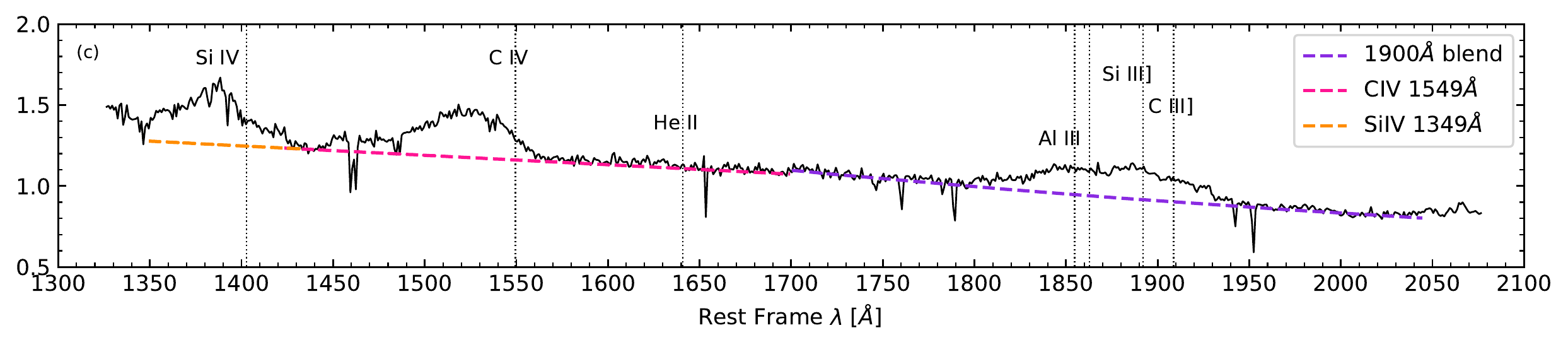}
    \includegraphics[width=0.325\linewidth]{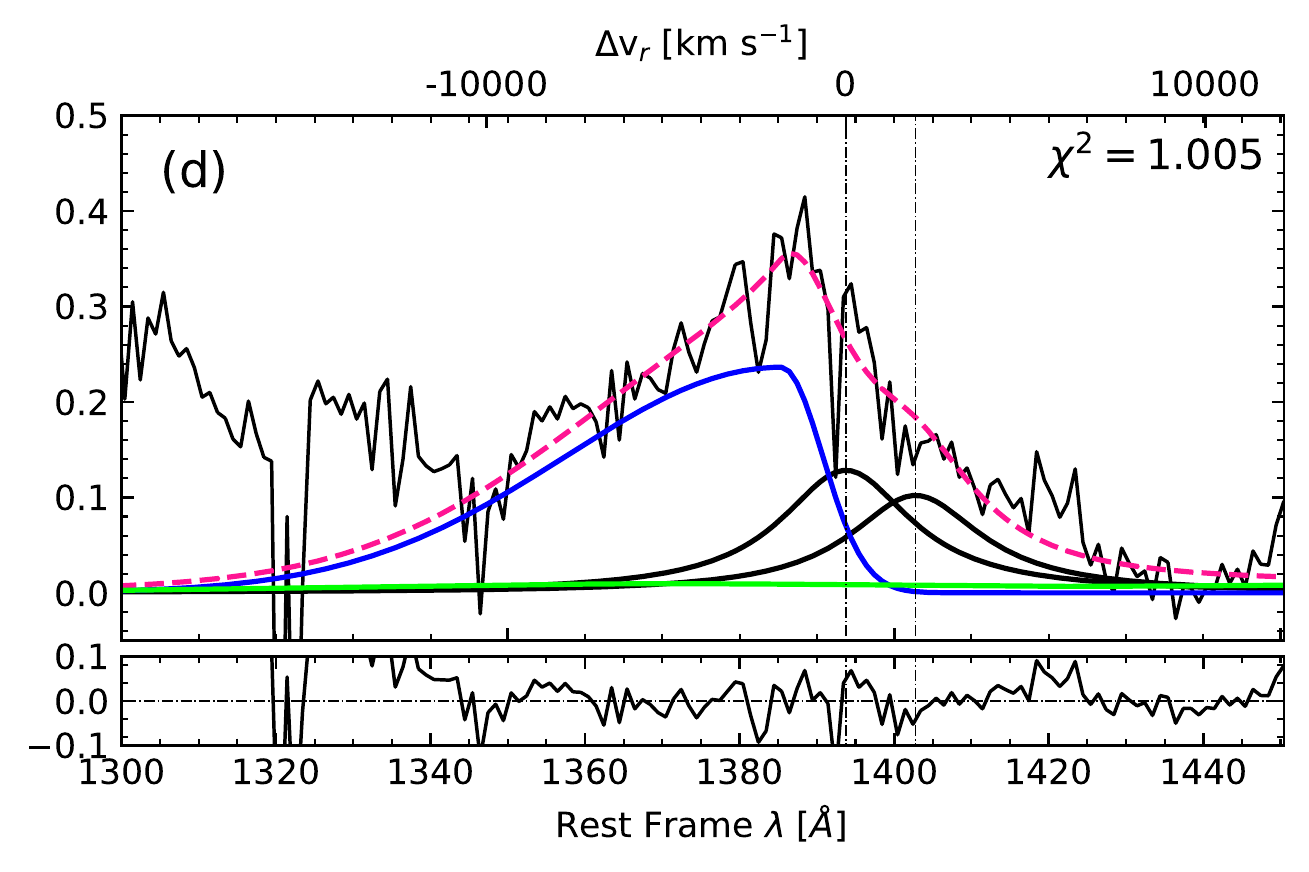}
    \includegraphics[width=0.325\linewidth]{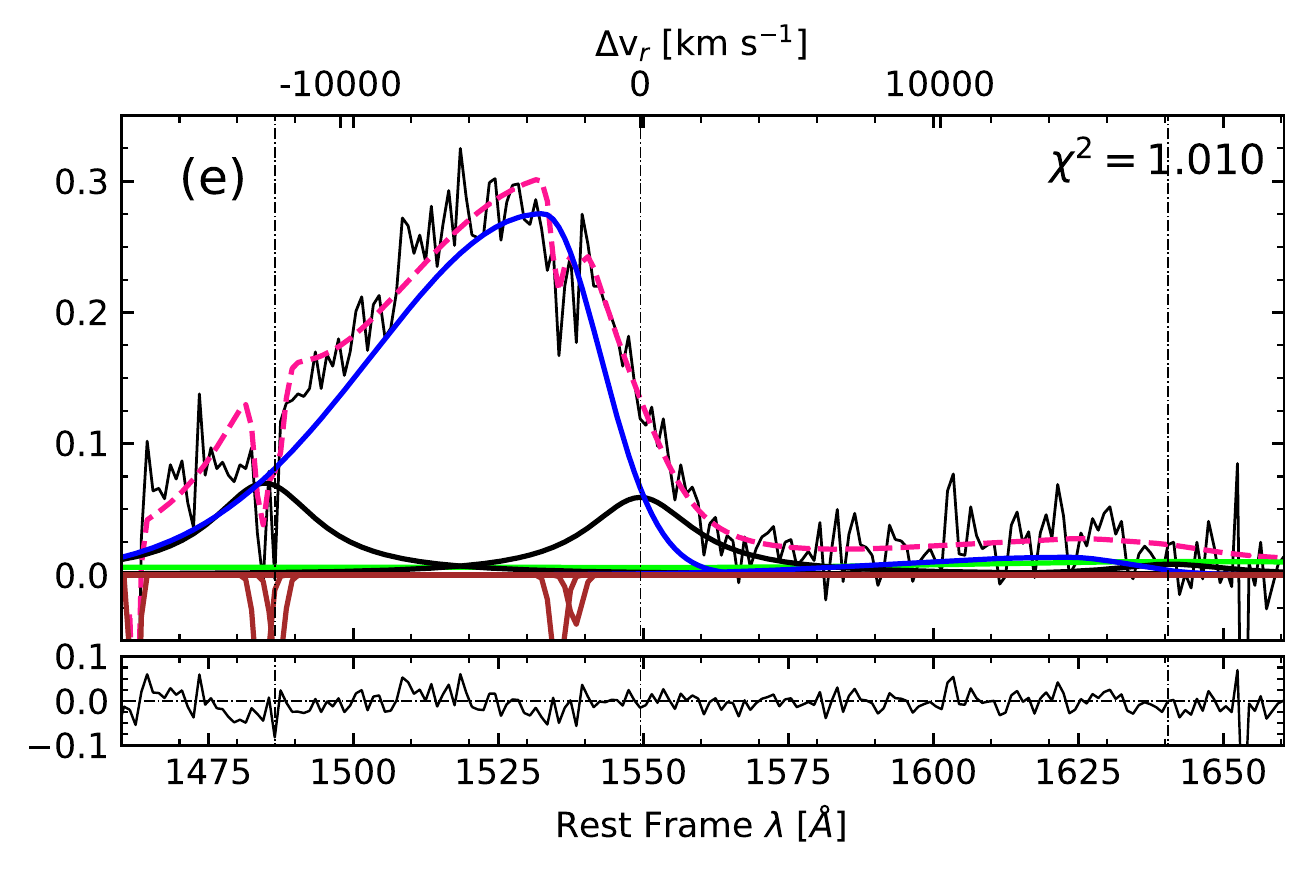}
    \includegraphics[width=0.335\linewidth]{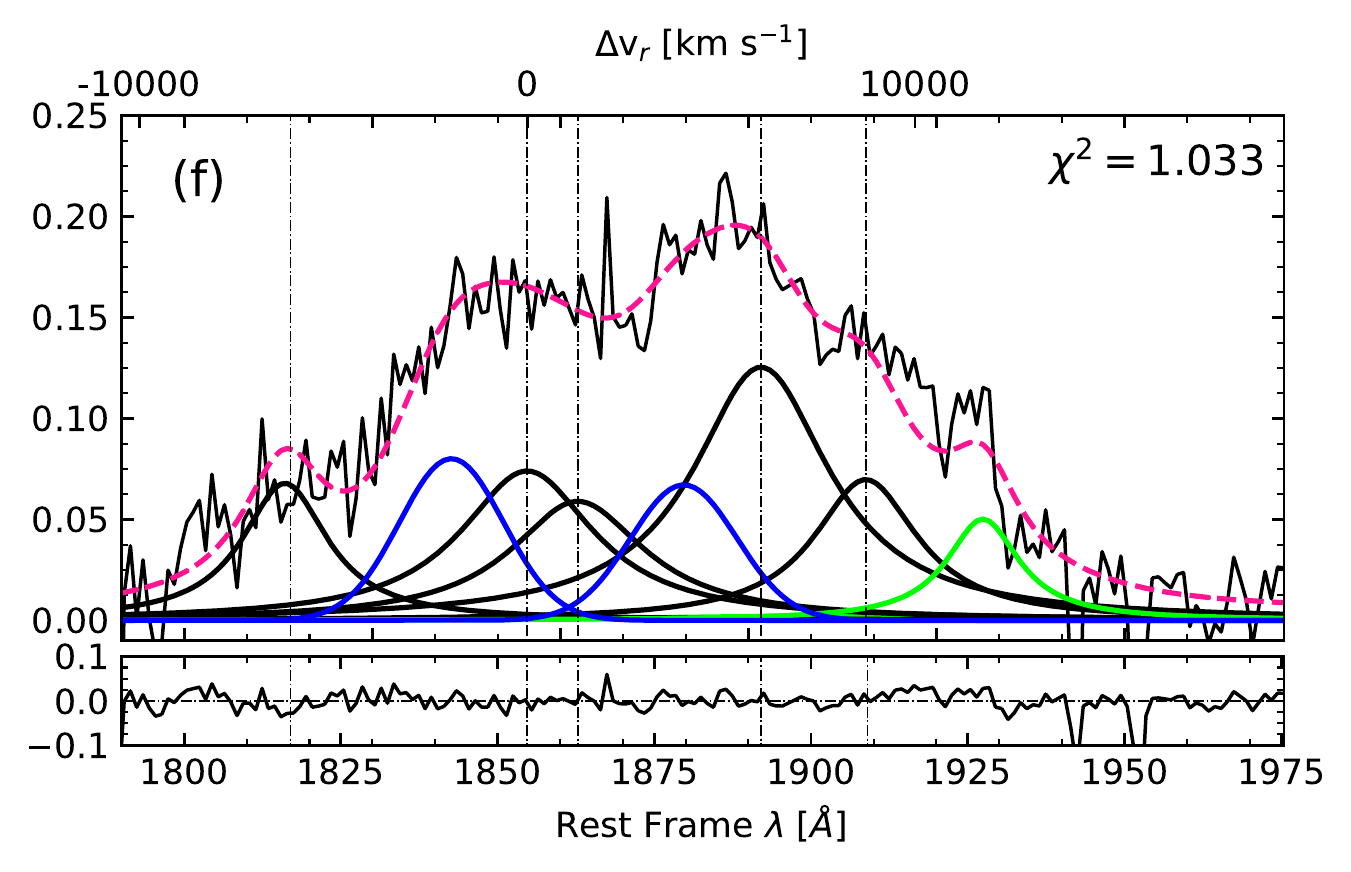}
    \caption{SDSSJ210831.56-063022.5. Same as Figure \ref{fig:1143_UV}.}
    \label{fig:2108_UV}
\end{figure}

\par This is another case in which the \ion{Fe}{II} contribution is estimated through the red side of the spectrum. It is difficult to set the \oiii{} NC at rest-frame since the region close to these lines is dominated by the \ion{Fe}{II} multiplets. Differently from the other Pop. A3 of the sample (SDSSJ005700.18+143737.7 and SDSSJ210524.49+000407.3), for this quasar it was not possible to reach a good $\chi^2$ without including blueshifted components for \aliii{} and \ion{Si}{III}]$\lambda$1892. SDSSJ210831.56-063022.5 is a source that presents one of the widest \hb\ BC of the sample. It could be that in this case the \hb\ BLUE component has been underestimated, since we do find strong blueward asymmetries in the three  fitted UV profiles.

\clearpage

%--------------------------------------------
\subsection{SDSSJ212329.46-005052.9}
\label{SDSSJ2123}

\begin{figure}[h!]
    \centering
    \includegraphics[width=0.685\linewidth]{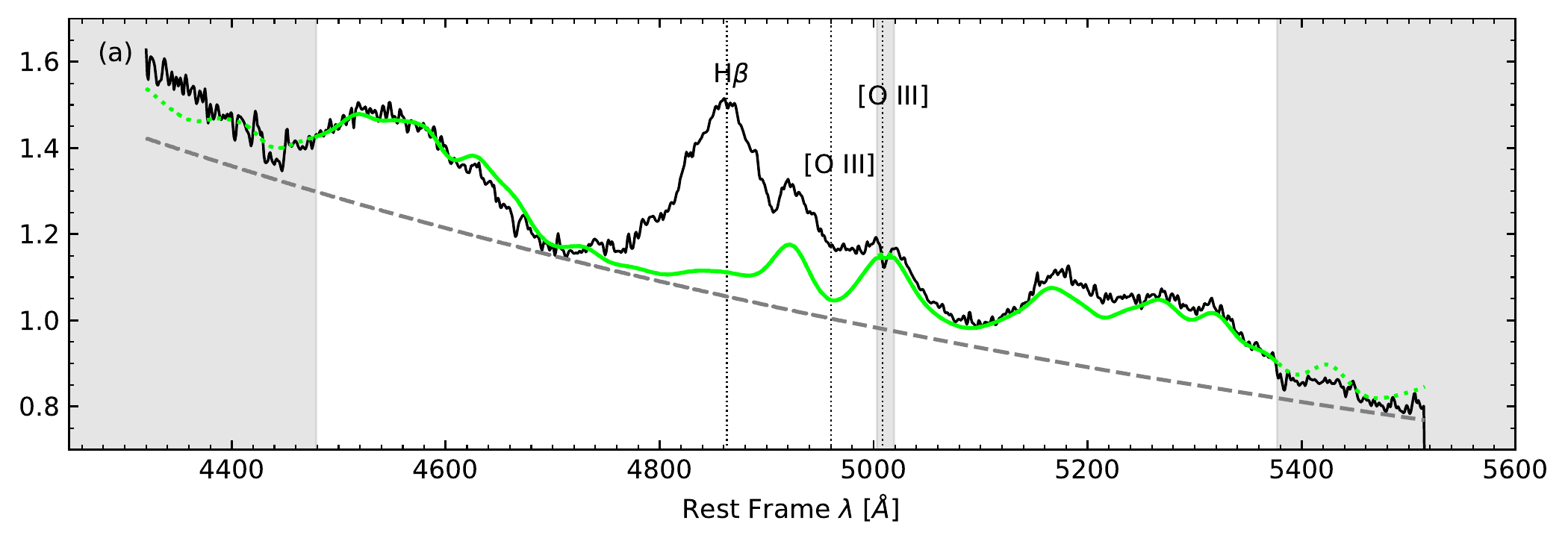}
    \includegraphics[width=0.30\linewidth]{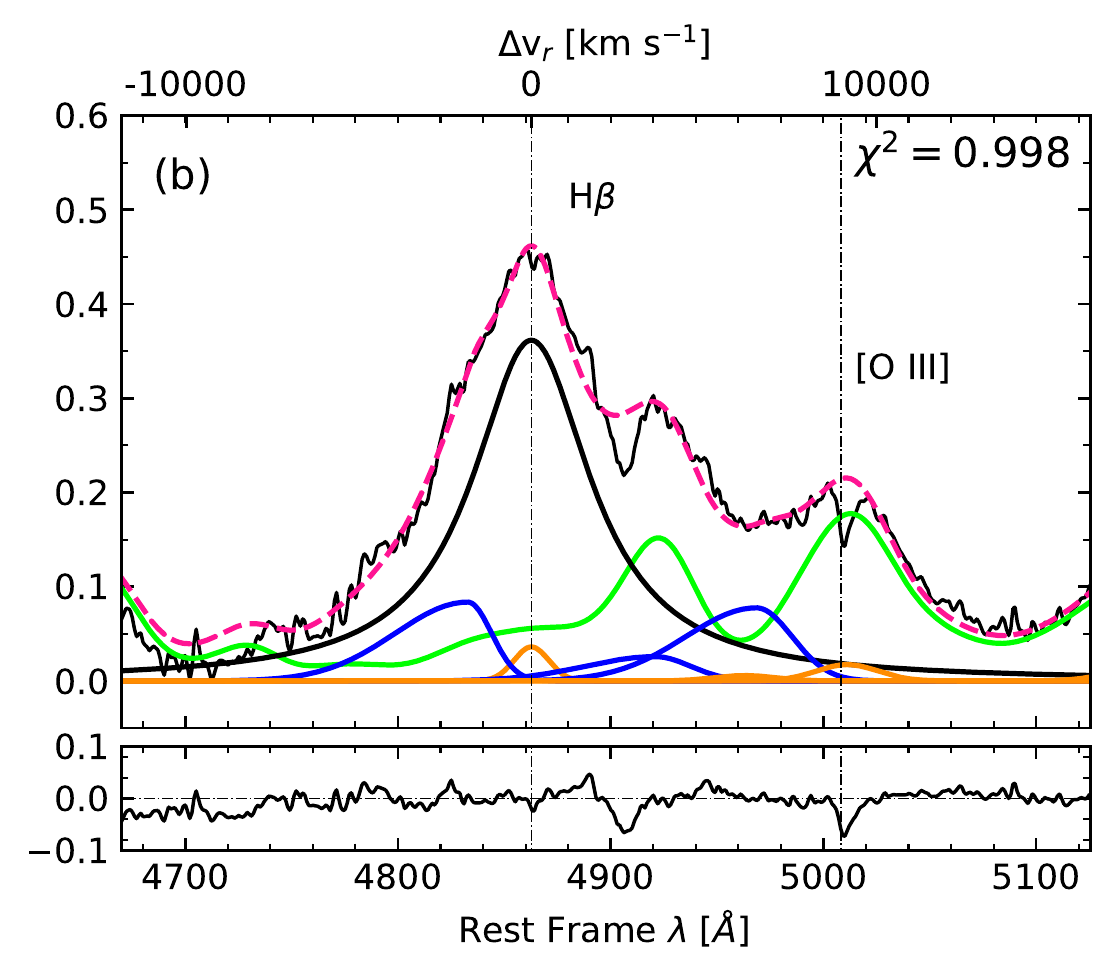}
    \\
    \centering
    \includegraphics[width=\linewidth]{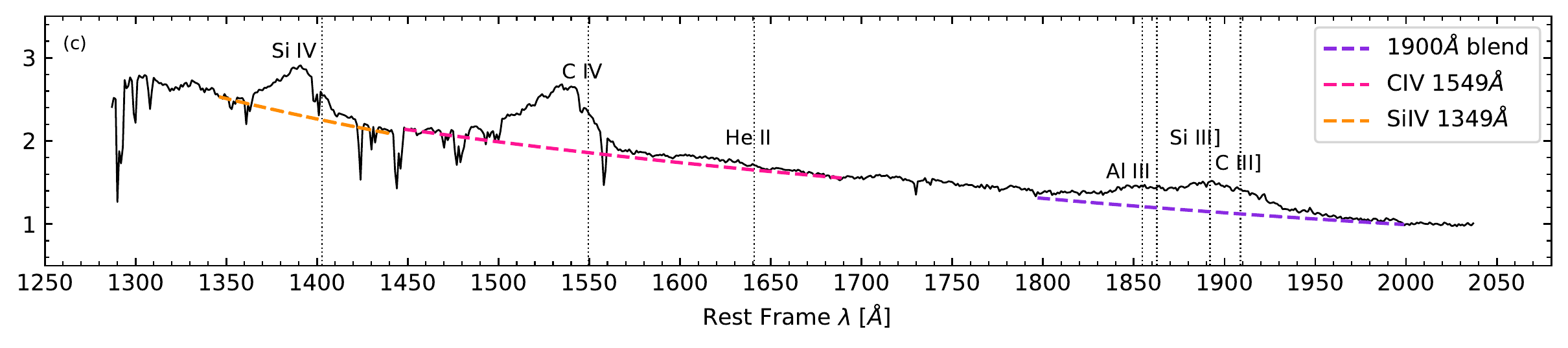}
    \includegraphics[width=0.323\linewidth]{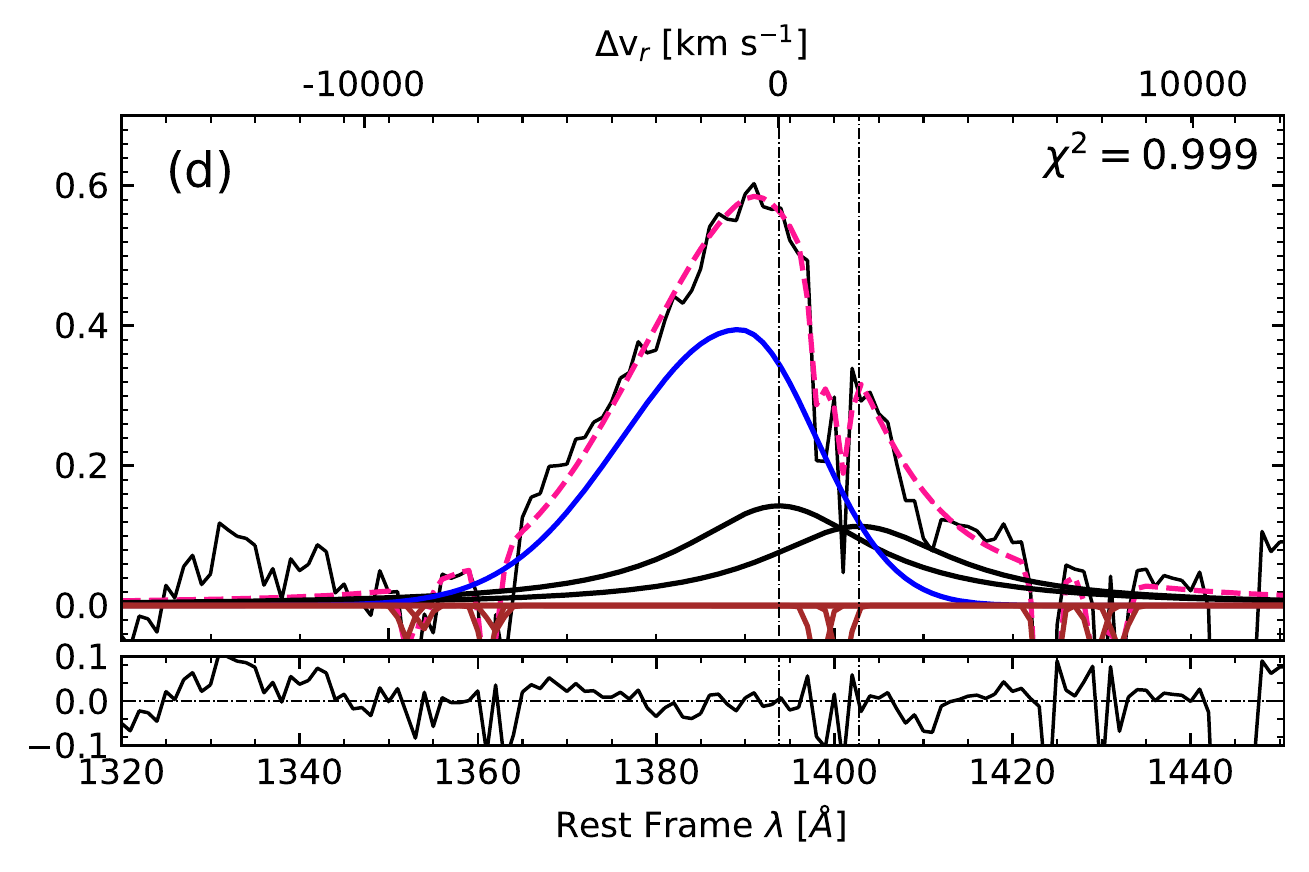}
    \includegraphics[width=0.33\linewidth]{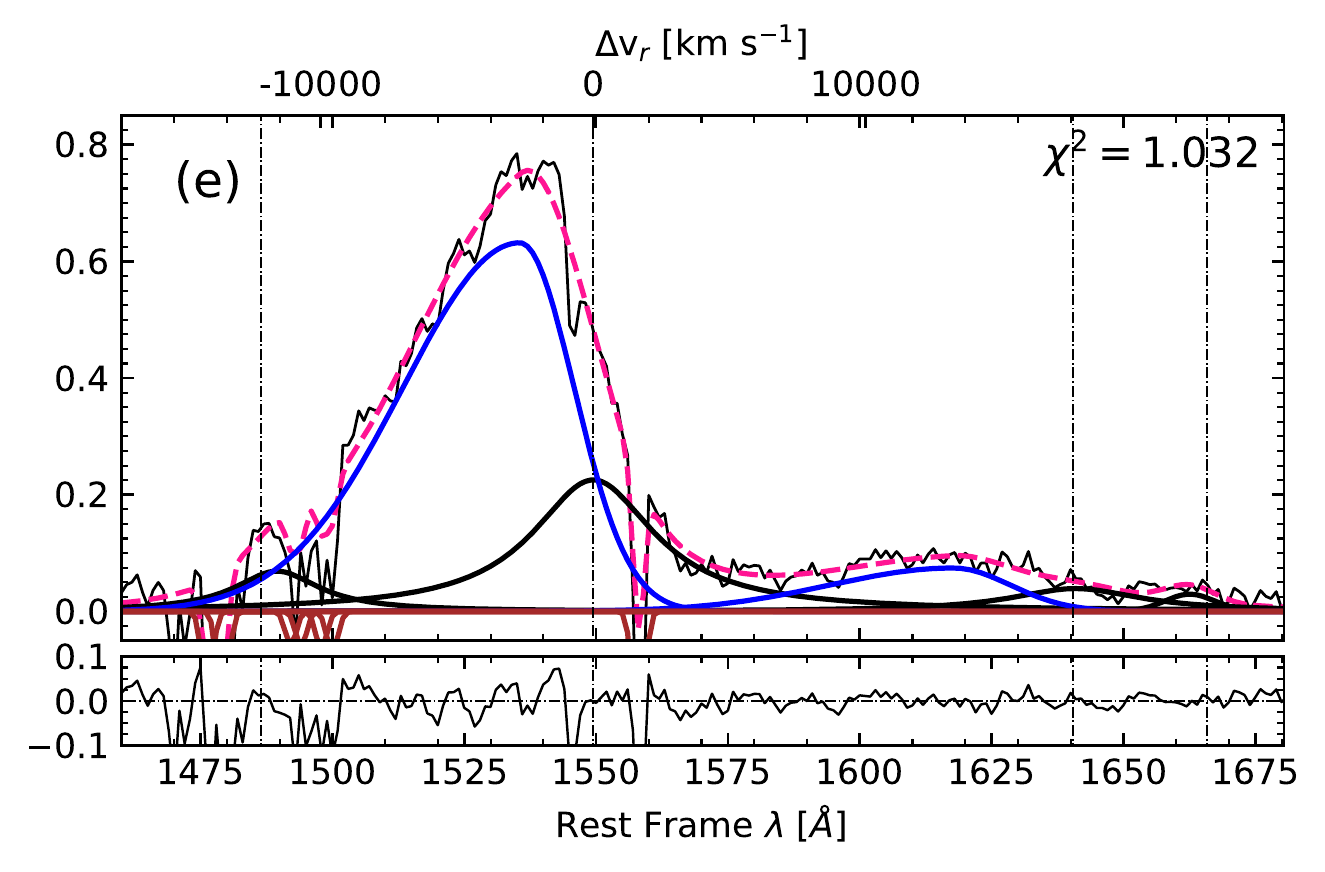}
    \includegraphics[width=0.327\linewidth]{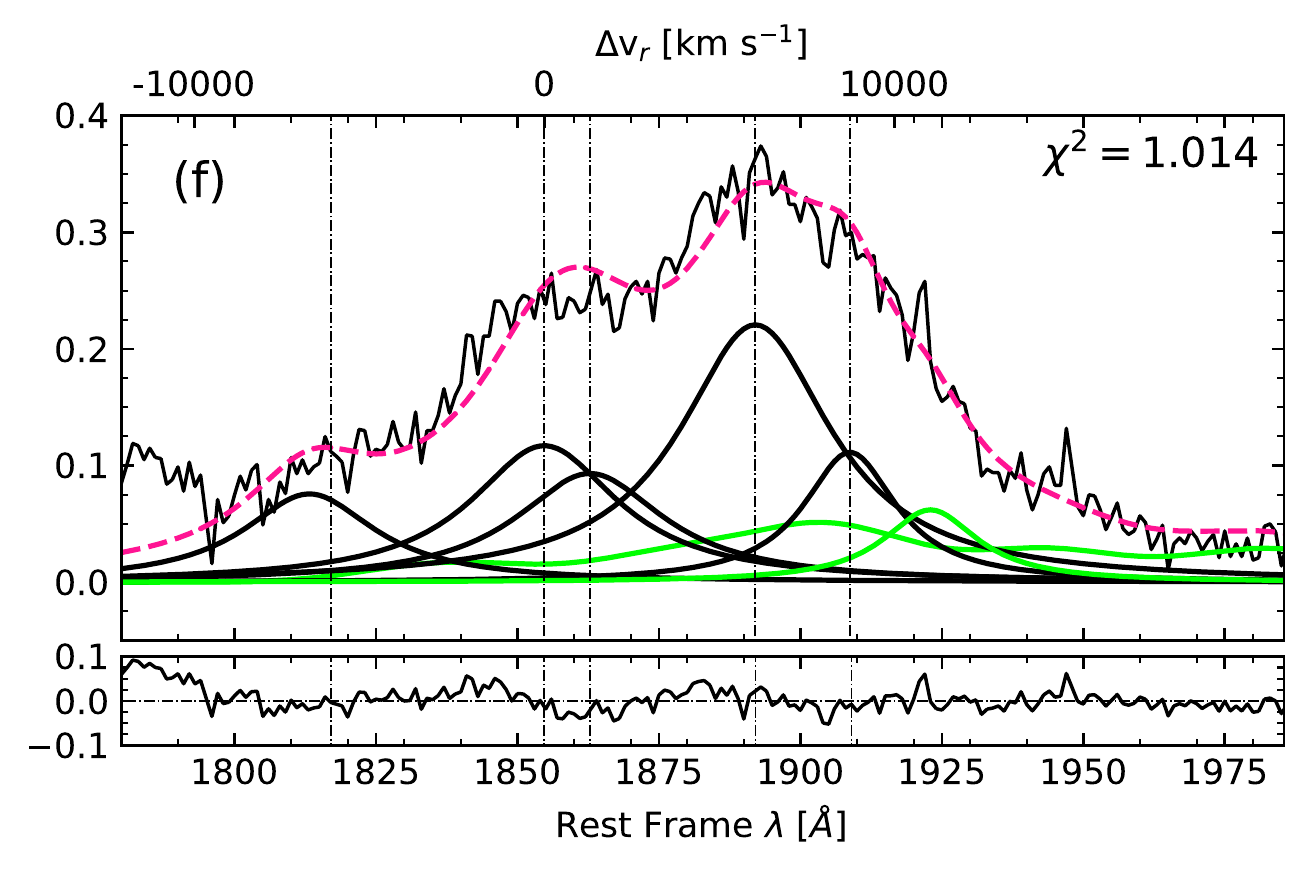}
    \caption{SDSSJ212329.46-005052.9. Same as Figure \ref{fig:1143_UV}.}
    \label{fig:2123_UV}
\end{figure}
\clearpage

%--------------------------------------------
\subsection{PKS 2126-15}
\label{PKS2126}

\begin{figure}[h!]
    \centering
    \includegraphics[width=0.69\linewidth]{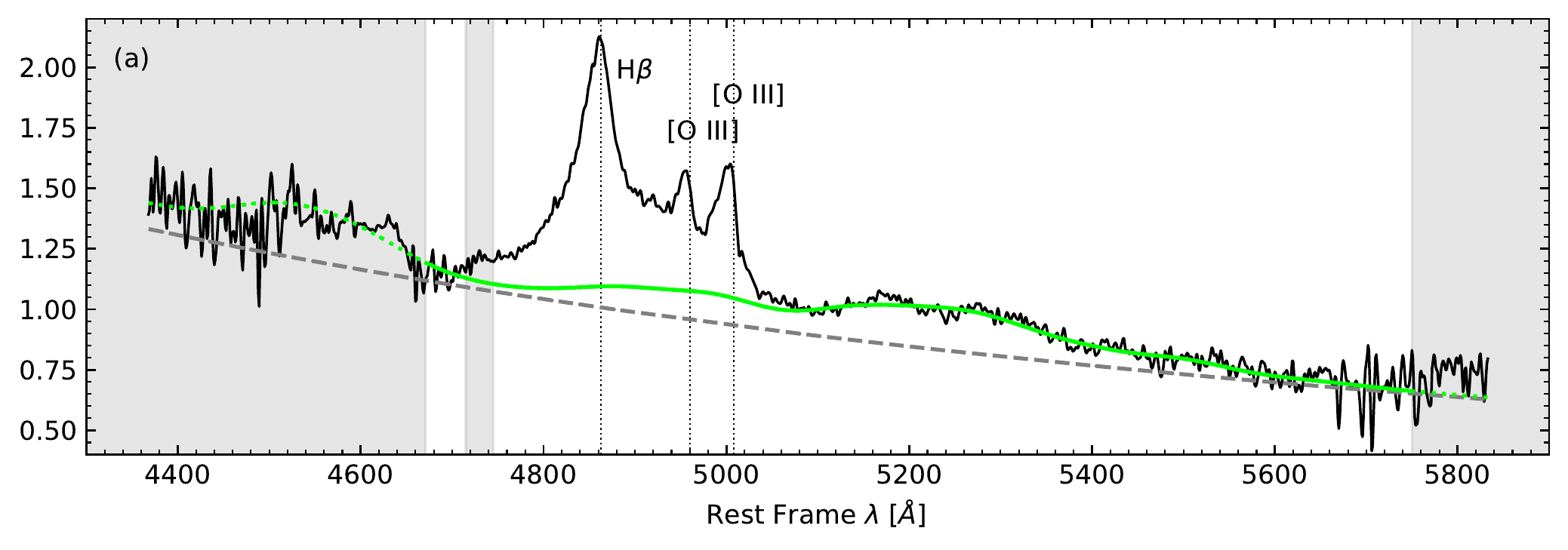}
    \includegraphics[width=0.305\linewidth]{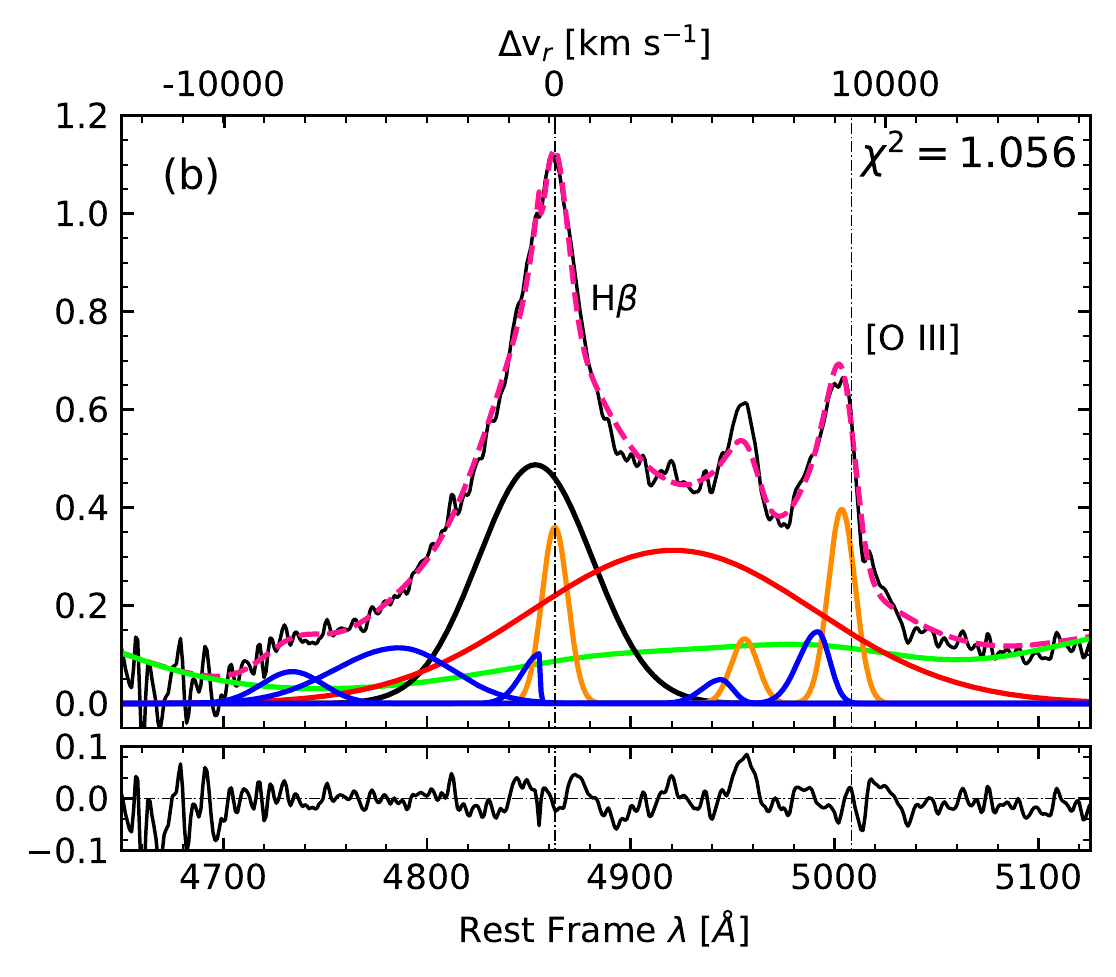}
    \caption{PKS 2126-15. Colours and lines as in Figure \ref{fig:0001_HB}.}
    \label{fig:2126_HB}
\end{figure}

\par This object has been observed in the two runs. Since both observations can be used and present good S/N, we create a new spectrum by combining the two of them through the IRAF task \texttt{scombine}. The \ion{Fe}{II} multiplets were fitted based on the shape seen in the red side of \hb\ due to the fact that the blue side of \hb\ is affected by noise and is almost at the border of the spectrum. In order to account for the \oiii{} contributions, it was necessary to include two blueshifted components, as can be seen in Fig. \ref{fig:2126_HB}.  Some UV spectra were found but they do not include the regions we are interested in. %For the \oiii{} analysis, it was necessary to consider two blueshifted components to represent the full profile. 

PKS 2126-15 is also one of the four radio-loud quasars from the sample.

\newpage
%--------------------------------------------
\subsection{SDSSJ235808.54+012507.2}
\label{SDSSJ2358}

\begin{figure}[h!]
    \centering
    \includegraphics[width=0.68\linewidth]{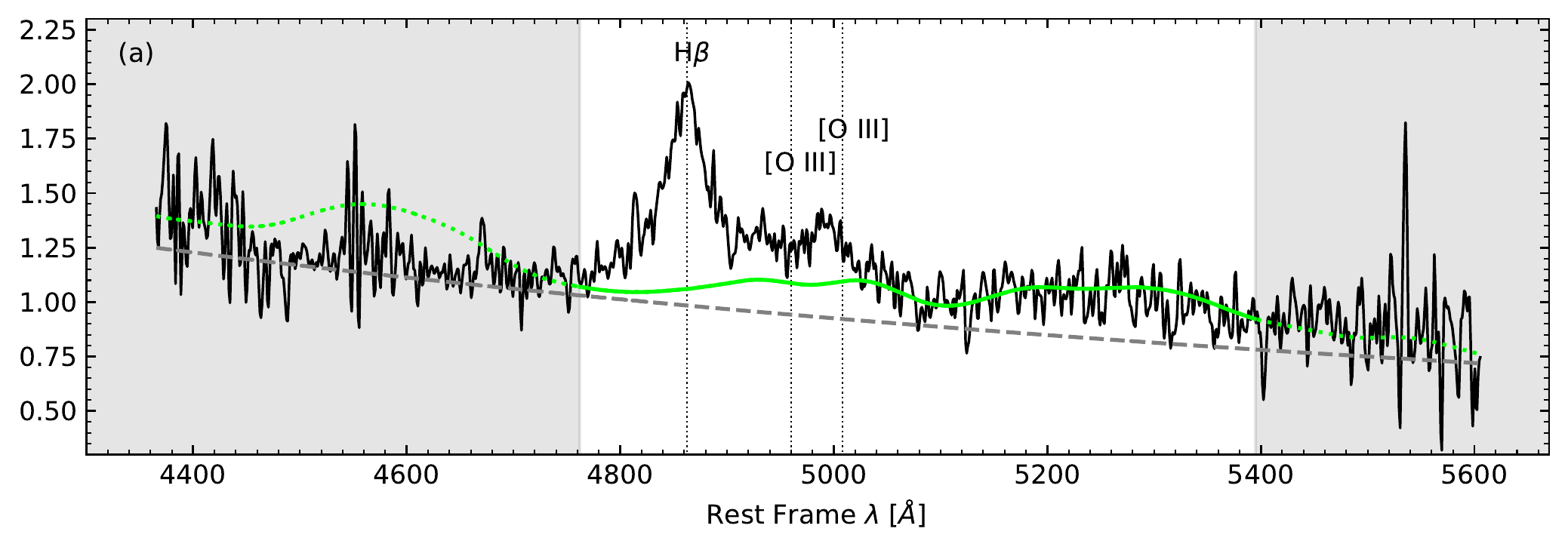}
    \includegraphics[width=0.30\linewidth]{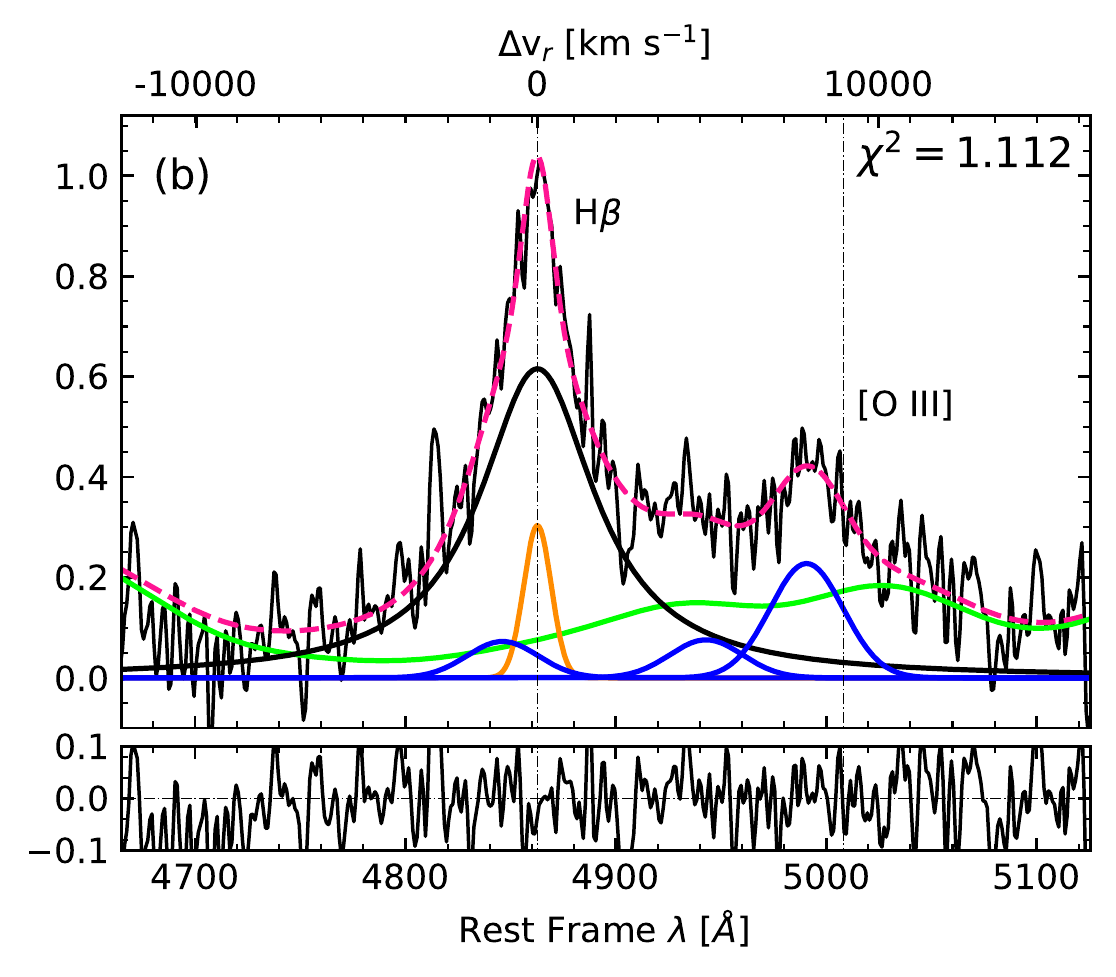}
    \\
    \centering
    \includegraphics[width=\linewidth]{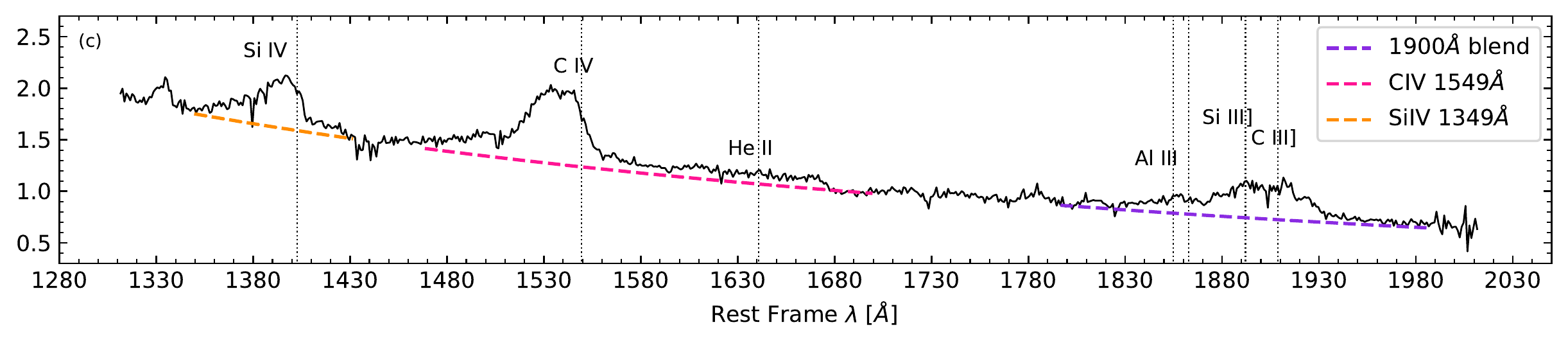}
    \includegraphics[width=0.325\linewidth]{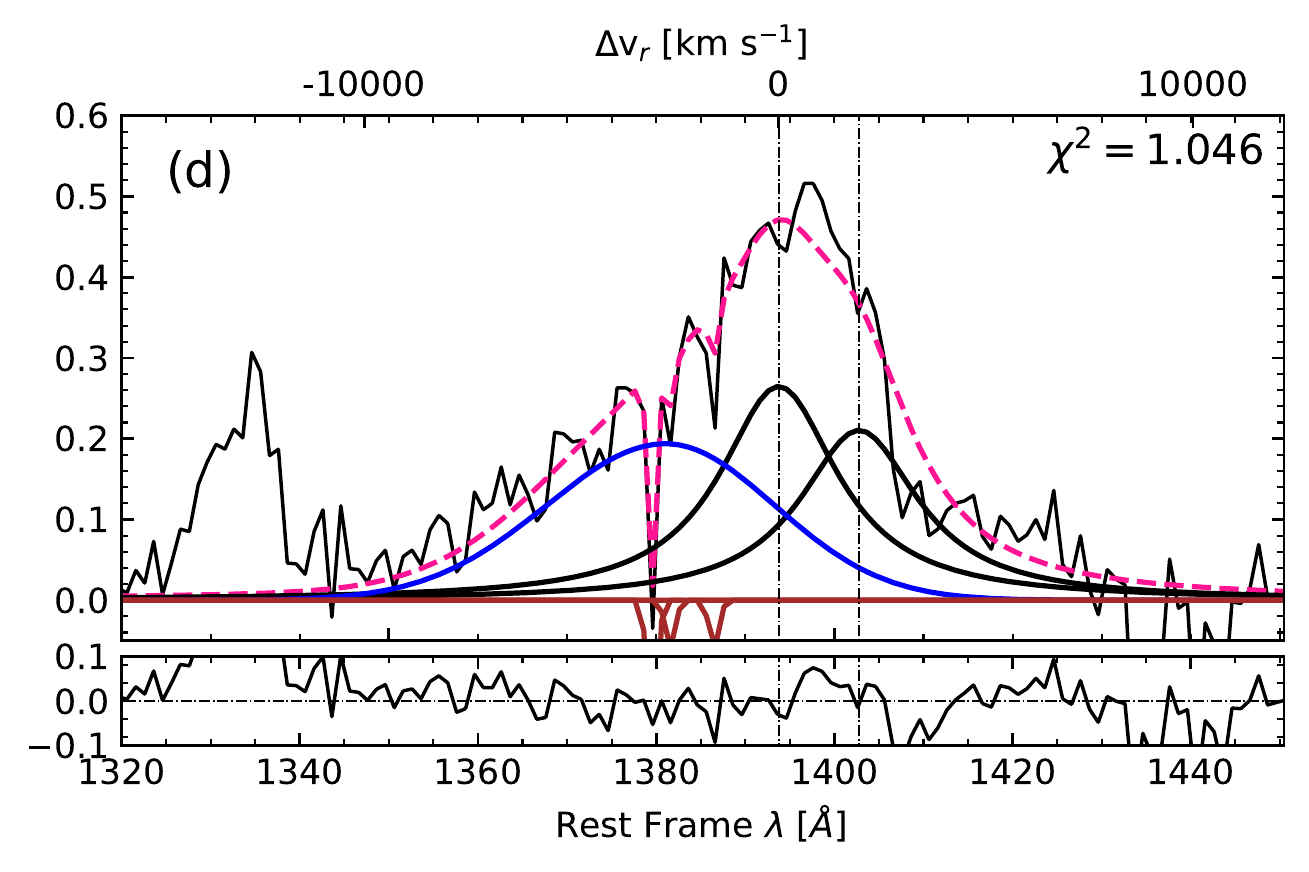}
    \includegraphics[width=0.325\linewidth]{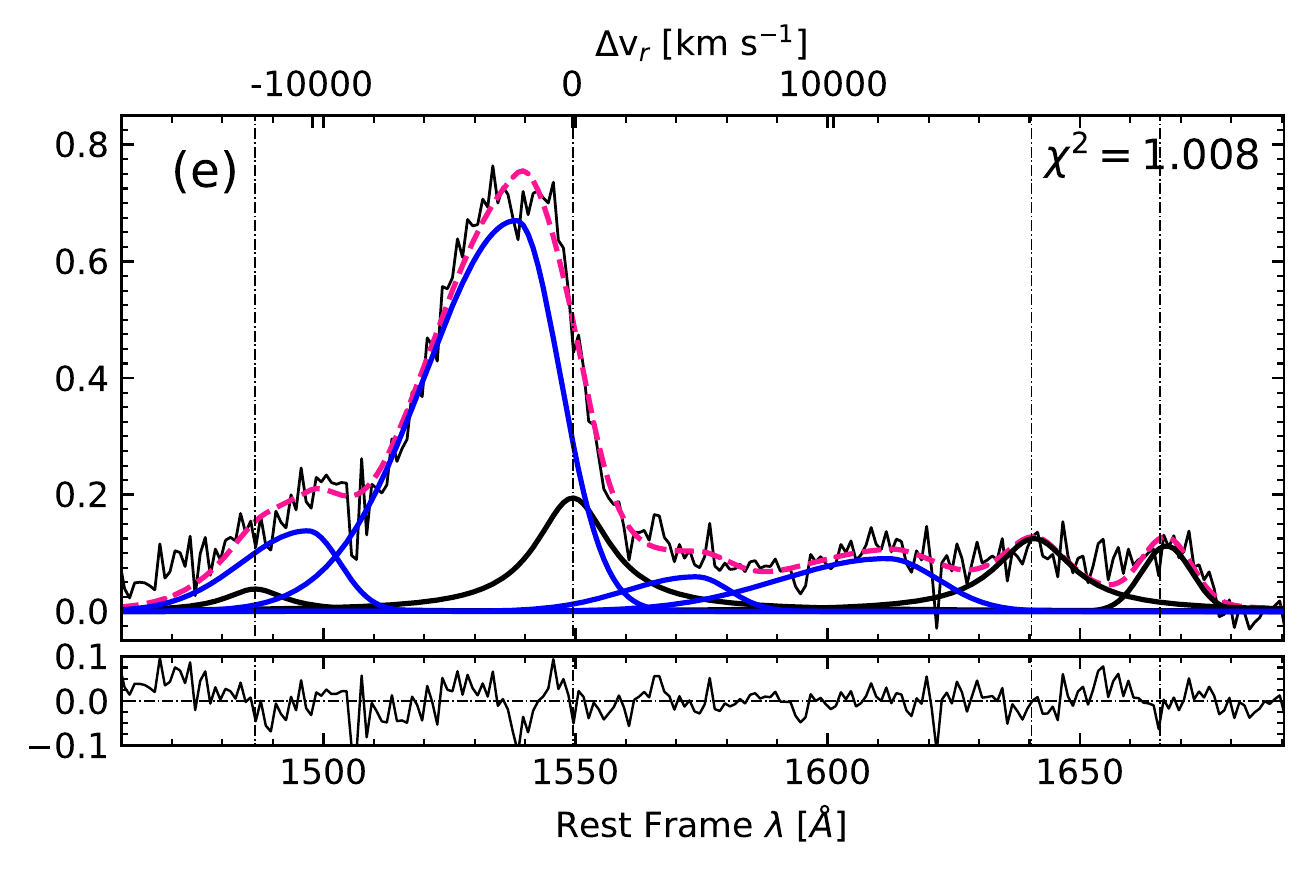}
    \includegraphics[width=0.335\linewidth]{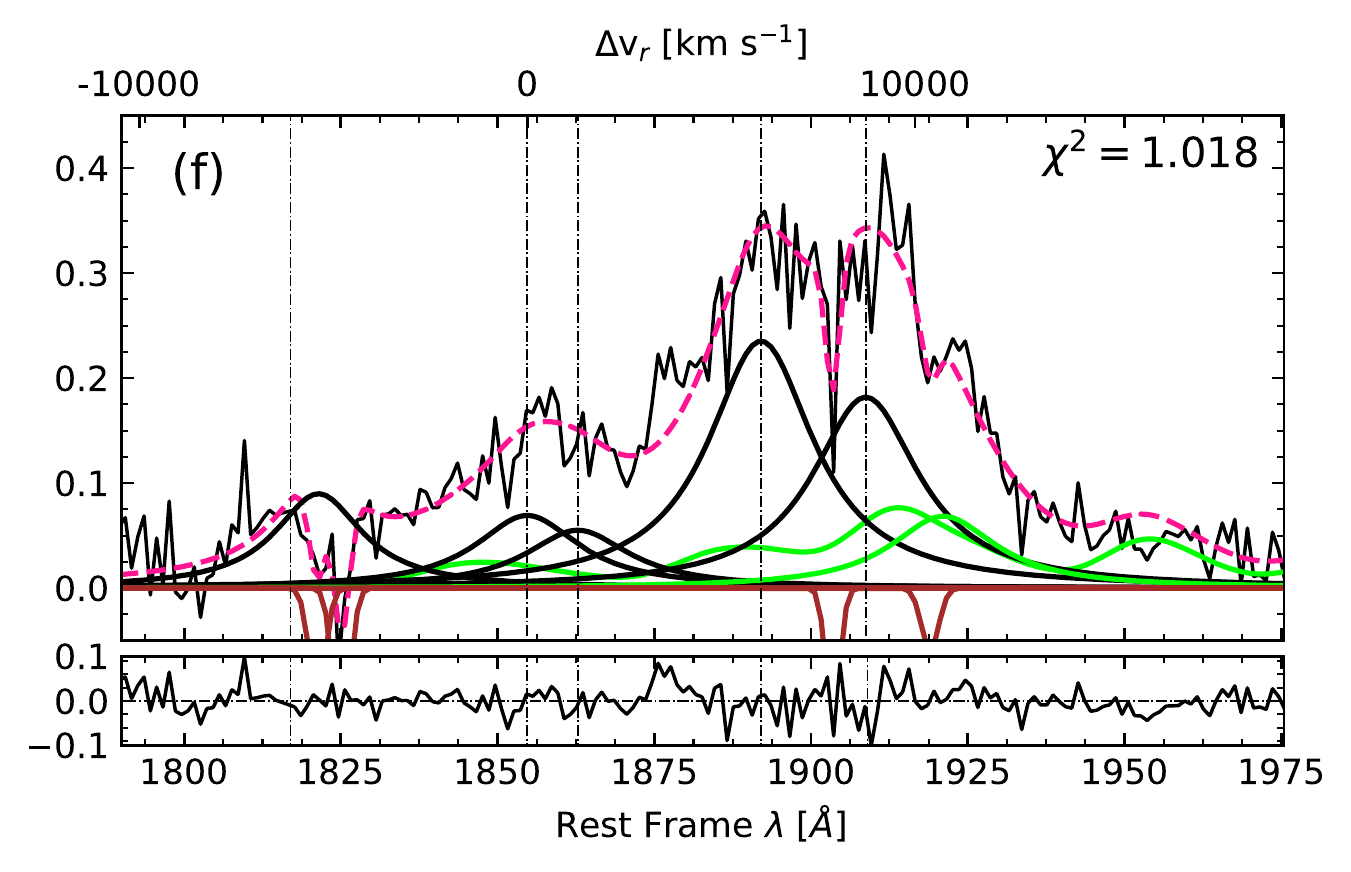}    
    \caption{SDSSJ235808.54+012507.2. Same as Figure \ref{fig:1143_UV}.}
    \label{fig:2355_UV}
\end{figure}

\par The spectrum of this source is of low S/N when compared with the other sources of the sample, which makes it difficult to find the correct intensity of the continuum. As a consequence, caution should be taken when considering the location of the continuum and the \feii\ contribution especially in the blue part of the spectrum.

%It is difficult to set the \ion{Fe}{II} contribution in the blue part of the spectrum due to the fact that the continuum was set at the same intensity of the spectrum. \paolaQ{Unclear}  The measurement of the \ion{Fe}{II} multiplets is performed at $\lambda \ge $ 5000\r{A}.

\end{appendix}

\end{document}